\documentclass[a4paper,11pt]{article}
\pdfoutput=1 

\usepackage{jheppub}
\usepackage[T1]{fontenc} 

\usepackage[dvipsnames]{xcolor}

\usepackage{enumitem}
\usepackage{mathrsfs}

\newcommand{\bei}{\begin{itemize}}
\newcommand{\eei}{\end{itemize}}
\newcommand{\bee}{\begin{enumerate}}
\newcommand{\eee}{\end{enumerate}}
\newcommand{\beeL}{\begin{enumerate}[label=(\Alph*)]}
\newcommand{\beel}{\begin{enumerate}[label=(\alph*)]}
\newcommand{\beeR}{\begin{enumerate}[label=(\Roman*)]}
\newcommand{\beer}{\begin{enumerate}[label=(\roman*)]}
\newcommand{\beeLd}{\begin{enumerate}[label=\Alph*.]}
\newcommand{\beeld}{\begin{enumerate}[label=\alph*.]}
\newcommand{\beeRd}{\begin{enumerate}[label=\Roman*.]}
\newcommand{\beerd}{\begin{enumerate}[label=\roman*.]}

\newcommand{\arcsinh}{\text{arcsinh}}

\newcommand{\bal}{\begin{equation}\begin{aligned}}
\newcommand{\eal}{\end{aligned}\end{equation}}

\newcommand{\ov}{\over}
\newcommand{\g}{\gamma}

\newcommand{\lint}{\int\limits}
\newcommand{\xbr}{\xi}

\def\sgn{{\text{sgn}}}

\definecolor{grey}{rgb}{0.4,0.4,0.5}
\definecolor{darkgreen}{rgb}{0,0.5,0}
\definecolor{darkred}{rgb}{0.6,0.0,0}
\definecolor{lightbrown}{rgb}{1,0.9,0.8}
\definecolor{brown}{rgb}{0.6,0.3,0.3}
\definecolor{darkblue}{rgb}{0,0,0.5}
\definecolor{darkmagenta}{rgb}{0.5,0,0.5}

\def\tx{{\tilde x}}

\def\tg{{\tilde\gamma}}

\def\tPsi{{\tilde\Psi}}
\def\tPhi{{\tilde\Phi}}

\def\barnes{{\text{\tiny odd}}}
\def\besratio{{\text{\tiny even}}}

\def\bQ{{\overline Q}}
\def\bY{{\overline Y}}
\def\bR{{\mathbb R}}

\newcommand{\la}{\label}
\def\a {\alpha}
\def\b {\beta}

\def\bes{{\text{\tiny BES}}}
\def\afs{{\text{\tiny AFS}}}
\def\hl{{\text{\tiny HL}}}
\def\tp{{\widetilde p}}
\def\tE{\widetilde\E}
\def\br {\bar\rho}

\newcommand{\E}{\mathcal E}

\newcommand{\B}{{\scriptscriptstyle\text{B}}}

\renewcommand{\L}{{\scriptscriptstyle\text{L}}}
\newcommand{\R}{{\scriptscriptstyle\text{R}}}

\def\ka {{\kappa}}

\def\pa {\partial}

\def\cE{{\cal E}}
\def\cF{{\cal F}}

\def\cO{{\cal O}}

\def\cR{{\cal R}}

\def\cY{{\cal Y}}

\def\eps{{\epsilon}}

\newcommand{\A}{{\scriptscriptstyle\text{A}}}
\renewcommand{\B}{{\scriptscriptstyle\text{B}}}

\def\adso{{$ AdS_{3} \times S^{3} \times T^{4} $ }}

\def\ta{{\tilde\alpha}}

\title{\boldmath 
Dressing Factors and Mirror Thermodynamic Bethe Ansatz for mixed-flux AdS$_3$/CFT$_2$}

\author[1,4]{Sergey Frolov,}
\author[2]{Davide Polvara,}
\author[3,4]{Alessandro Sfondrini}

\affiliation[1]{Hamilton Mathematics Institute and School of Mathematics Trinity College, Dublin 2, Ireland.}
\affiliation[2]{II. Institut f\"ur Theoretische Physik,  Universit\"at Hamburg, Luruper Chaussee 149, 22761 Hamburg, Germany.}
\affiliation[3]{Dipartimento di Fisica e Astronomia, Universit\`a degli Studi di Padova, via Marzolo 8,
35131 Padova, Italy.}
\affiliation[4]{Istituto Nazionale di Fisica Nucleare, Sezione di Padova, via Marzolo 8, 35131 Padova,
Italy.}

\emailAdd{frolovs@maths.tcd.ie}
\emailAdd{davide.polvara@gmail.com}
\emailAdd{alessandro.sfondrini@unipd.it}

\abstract{We complete the derivation of the dressing factors for the $AdS_3\times S^3\times T^4$ S~matrix with mixed Ramond--Ramond and Neveu-Schwarz--Neveu-Schwarz flux, in the ``string'' and ``mirror'' kinematics. Using these, we propose the mirror Thermodynamic Bethe Ansatz equations which describe the spectrum of the model at any string tension.
}

\begin{document} \begin{flushright}\small{ZMP-HH/25-12}\end{flushright}
\maketitle

\section{Introduction}
\label{sec:introduction}
The $AdS_3/CFT_2$ holographic correspondence~\cite{Maldacena:1997re} is an important instance of holography which has been studied quite extensively since the heydays of holography.%
\footnote{See~\cite{David:2002wn} for an early review of $AdS_3/CFT_2$ and~\cite{Seibold:2024qkh} and references therein for a more recent review centred around the small-tension  limit of the duality.
}
With respect to the correspondence between type-IIB $AdS_5\times S^5$ strings and $\mathcal{N}=4$ supersymmetric Yang--Mills theory, the $AdS_3/CFT_2$ correspondence is less supersymmetric and therefore it presents a much richer dynamics. The simplest instance of this duality is the one involving the string background  $AdS_3\times S^3\times T^4$. Despite some superficial similarities with $AdS_5\times S^5$, one striking feature of this background is that it can be supported by a mixture of Ramond--Ramond (RR) and Neveu--Schwarz--Neveu--Schwarz (NSNS) fluxes. In particular, the supergravity background involves a $B$ field and an NSNS three-form flux $H_3=\text{d}B$ whose coefficient is quantised in terms of an integer~$k$. At large string tension~$T$, one finds that the tension is sourced by~$k$ and by a continuous parameter~$h$
\begin{equation}
    T=\sqrt{h^2+\frac{k^2}{4\pi^2}}\,,\qquad h\geq0\,,\quad k=0,1,2\dots\,.
\end{equation}
The parameter~$h\geq0$ is related to the RR flux of the background and it is continuous in perturbative string theory; it is similar to the 't~Hooft coupling $\sqrt{\lambda}$ of $AdS_5/CFT_4$. 
In the special case where $h=0$, the string model can be studied in the Ramond--Neveu-Schwarz formalism in terms of a supersymmetric WZW model based on $SL(2,\mathbb{R})$ and $SU(2)$~\cite{Maldacena:2000hw}, where $k$ plays the role of the level.
Away from these special points, it is much harder to do so~\cite{Cho:2018nfn}.
Remarkably, the classical Green--Schwarz (GS) action is integrable for any $k,h$~\cite{Cagnazzo:2012se}. This raised the hope of studying the model by quantising the GS action in lightcone gauge and exploiting integrability at the quantum level, like it was done for $AdS_5\times S^5$, see~\cite{Arutyunov:2009ga} for a review, and it spurred a very intense activity aimed at constructing the worldsheet S~matrix of the mixed-flux model~\cite{Hoare:2013pma,Hoare:2013lja,Lloyd:2014bsa,Frolov:2023lwd,OhlssonSax:2023qrk,Frolov:2024pkz}.%
\footnote{In parallel, much effort was devoted to studying integrability for pure--RR backgrounds, that is for $k=0$ and $h\geq0$. This case turns out to be relatively similar to $AdS_5\times S^5$, see~\cite{Demulder:2023bux} for a pedagogical review.}

The integrability program on the string worldsheet starts by constructing the S~matrix in infinite volume. Much of that construction boils down to understanding the symmetries of the model, which determine the two-particle S~matrix up to handful of functions --- the dressing factors. For mixed-flux $AdS_3\times S^3\times T^4$ strings, the S~matrix was determined up to the dressing factors in~\cite{Lloyd:2014bsa}.%
\footnote{The multi-particle S~matrix then follows by factorisation and by the Yang--Baxter equation, much like in relativistic QFT~\cite{Zamolodchikov:1978xm}.}
Finding the dressing factors requires solving the crossing equations, which are similar to the ones in~\cite{Janik:2006dc}, in such a way as to preserve the other symmetries of the theory (unitarity, parity, etc.). Due to the rather involved analytic structure of the mixed-flux model~\cite{Frolov:2023lwd}, a proposal for these dressing factors was put forward in an earlier paper of ours only quite recently~\cite{Frolov:2024pkz}. This proposal was discussed in full detail  for what concerns the scattering of massive worldsheet excitations in~\cite{Frolov:2025uwz}. The aim of this paper is to complete this construction in the case of massless excitations (i.e., excitations whose dispersion relation has no mass gap). This sector of the model is especially subtle; in particular, it was recently argued~\cite{Frolov:2025ozz} that such massless particles might obey non-trivial exchange relations, which modify the crossing equation and therefore its solution.

As this paper completes the construction of the dressing factors of mixed-flux $AdS_3\times S^3\times T^4$ strings, it is a natural starting point for the next step in the integrability program --- the construction of the equations which determine the finite-volume spectrum of the model. This can  be done by means of the `mirror' thermodynamic Bethe ansatz (TBA) approach~\cite{Arutyunov:2007tc} in analogy with what is done for relativistic models~\cite{Zamolodchikov:1989cf}. In this case, we find that the structure of the mirror TBA follows closely that of the pure-RR model ($k=0$)~\cite{Frolov:2021bwp} but that several subtleties arise related to the more involved analytic structure and non-unitarity of the mirror model.

This paper is structured as follows. In Section~\ref{sec:kinematics} we briefly recap the kinematics of the model, which was already discussed in~\cite{Frolov:2023lwd}. In Section~\ref{sec:properties} we list the various  properties that the S~matrix should obey (such as crossing, parity and so on) both in the kinematics of the string-worldsheet model and in the mirror; in formulating the crossing equations, we allow for non-trivial exchange relations~\cite{Frolov:2025ozz}, which turns out to be crucial for the massless--massless scattering. In Section~\ref{sec:proposal} we put forward our proposal for the dressing factors; like in the massive case~\cite{Frolov:2025uwz} it turns out that it is easy to start this construction in the mirror kinematics, and analytically continue it to the string kinematics. In Section~\ref{sec:checks} we check our proposal against all symmetries of the model (in the string and mirror kinematics) and against all existing perturbative computations. In Section~\ref{sec:proposalTBA} we put forward our proposal for the mirror TBA equations from which the spectrum of the model may be computed. Finally, in Section~\ref{sec:conclusions} we present our conclusion and discuss the next steps in this integrability program. 

\section{Kinematics}
\label{sec:kinematics}
We begin by briefly recalling the kinematics of the model, which is discussed in detail in~\cite{Frolov:2025uwz}, see also~\cite{Frolov:2023lwd}. Different kinematical properties were also analysed in~\cite{OhlssonSax:2023qrk}; however, in that paper the authors adopt a slightly different cut structure compared to the one used here, which instead refers to the previous construction of~\cite{Frolov:2023lwd}.
The dispersion relation is~\cite{Hoare:2013lja,Lloyd:2014bsa}
\begin{equation}
\label{eq:dispersion}
    E(p)=\sqrt{\left(\frac{k}{2\pi}p+M\right)^2+4h^2\sin^2\left(\frac{p}{2}\right)}\,,
\end{equation}
which is not parity-invariant.  For $k\ne 0$, in the mirror model, the mirror energy~$\tE(\tp)$ is complex for real~$\tp\,$~\cite{Baglioni:2023zsf}.
We introduce Zhukovsky variables obeying~\cite{Hoare:2013lja,Frolov:2023lwd,Frolov:2025uwz} 
\begin{equation}
\label{eq:zhukovskymap}
    u_{a}(x) = x+\frac{1}{x}-\frac{\kappa_a}{\pi}\ln x\,,\qquad a=\text{L},\text{R}\,,
\end{equation}
where $\kappa_\L=\kappa$ and $\kappa_\R=-\kappa$, with $\kappa=k/h$.
Inverting the Zhukovsky map~\eqref{eq:zhukovskymap} we can define the string and mirror variables
\bal
x_a^{\pm m}=x_a\big(u\pm i \frac{m}{h}\big)\,,\qquad \tx_a^{\pm m}=\tx_a\big(u\pm i \frac{m}{h}\big)\,,
\eal
as discussed in~\cite{Frolov:2023lwd,Frolov:2025uwz}, where $m=|M|$. We then express momentum and energy as
\begin{equation}
\label{eq:def_en_mom_string}
\begin{aligned}
p_a^{(m)}&=i\ln x_a^{-m} -i\ln x_a^{+m}\,,\\
E_a^{(m)}&={ih\ov2}\left( x_a^{-m} -{1\ov x_a^{-m}} - x_a^{+m}+{1\ov x_a^{+m}} \right)\,.
\end{aligned}
\end{equation}
In the mirror kinematics these become
\begin{equation}
\begin{aligned}
\tE_a^{(m)} &= \ln \tx_a^{-m} -\ln \tx_a^{+m} = \ln x_a^{-m} +\ln x_{\bar a}^{+m}\,,\\
\tp_a^{(m)} &= {h\ov2}\left( \tx_a^{-m} -{1\ov \tx_a^{-m}} - \tx_a^{+m} +{1\ov \tx_a^{+m}} \right)
={h\ov2}\left( x_a^{-m} -{1\ov x_a^{-m}} -{1\ov x_{\bar a}^{+m}}+ x_{\bar a}^{+m} \right)\,.
\end{aligned}
\end{equation}
In the rest of this section we define massless Zhukovsky variables as the limit of massive ones with the mass going to zero. We begin by analysing massless particles in the mirror model because the discussion is clearer in that context.

\subsection{Massless particles of the  mirror theory}

\paragraph{Massless mirror energy and momentum.} 
The energy and momentum of mirror massless  particles are obtained by taking $m$ to $0^+$.
We define massless Zhukovsky variables as the limit of massive ones
\bal
\tx_a^{\pm0}(u) = \lim_{m\to0^+} \tx_a\big(u \pm{i\ov h}m\big)\,,\qquad a=\text{L,\,R}\,,
\eal
where $u$ is on a mirror theory cut. If that were not the case, both momentum and energy would vanish.
We recall from~\cite{Frolov:2023lwd} that $\tx_{a}(u)$ has long cuts 
\bal
(-\infty+i \ka_a, -\nu +i \ka_a)  \qquad \text{and} \qquad (\nu, + \infty) \,,
\eal
where
\bal
\label{eq:nu_eta_def}
\nu=2 \cosh \eta -2 \eta \sinh \eta\,, \qquad \eta \equiv \arcsinh \frac{\ka}{2 \pi} \,.
\eal
We refer to the former cut by the name `$\ka_a$-cut' and to the latter by `main mirror cut'.  
As it will be clear in a moment, if $u$ is on the main mirror cut
then the mirror momentum is positive, while if it is on the $\ka_a$-cut
then the mirror momentum is negative. 
In this section we find it convenient to use a real rapidity variable for both positive and negative momentum branches, and we will denote it by $r$, so that $u=r$ or $u=r+i\ka_a$.

The discontinuity equations across the cuts are~\cite{Frolov:2023lwd}
\bal
\tx_{a}(r+i0)=\frac{1}{\tx_{\bar{a}}(r-i0)}\,, \qquad \tx_{a}(-r+i \ka_a+i0)=\frac{1}{\tx_{\bar{a}}(-r-i \ka_a-i0)}\,, \qquad r> \nu \,.
\eal
Therefore, on the main mirror cut the  energy and  momentum of a mirror massless particle are
\bal
 \tE_\L^{(0)}(r) &= \ln \tx_\L(r-i0) -\ln \tx_\L(r+i0)\\
 &= - \ln \tx_{\R}(r+i0) +\ln \tx_\R(r-i0)=\tE_\R^{(0)}(r)>0
 \,,
\\
\tp_\L^{(0)}(r) &= {h\ov2}\left( \tx_\L(r-i0)  -{1\ov \tx_\L(r-i0) } - \tx_\L(r+i0) +{1\ov \tx_\L(r+i0)} \right)
\\
&= {h\ov2}\left( {1\ov \tx_{\R}(r+i0)}  -{ \tx_{\R}(r+i0) } -{1\ov \tx_\R(r-i0)} + \tx_\R(r-i0) \right)
=\tp_\R^{(0)}(r)>0\,,
\eal
where $r\ge \nu$, and 
\bal
 \tE_\L^{(0)}(r) &= \ln \tx_\L(r+i\ka_\L -i0) -\ln \tx_\L(r+i\ka_\L+i0) 
 \\
 &= - \ln \tx_{\R}(r-i\ka_\R+i0) +\ln \tx_\R(r-i\ka_\R-i0)
 =\tE_\R^{(0)}(r)>0
 \,,
\\
\tp_\L^{(0)}(r) &= {h\ov2}\left( \tx_\L(r+i\ka_\L-i0)  -{1\ov \tx_\L(r+i\ka_\L-i0) } - \tx_\L(r+i\ka_\L+i0) +{1\ov \tx_\L(r+i\ka_\L+i0)} \right)
\\
&= {h\ov2}\left( {1\ov \tx_{\R}(r-i\ka_\R+i0)}  -{ \tx_{\R}(r-i\ka_\R+i0) } -{1\ov \tx_{\R}(r-i\ka_\R-i0)}+{ \tx_{\R}(r-i\ka_\R-i0)} \right)
\\
&
=\tp_\R^{(0)}(r)<0\,,
\eal
where $r\le -\nu$. 

Thus, for both momentum branches
\bal
\tp_\L^{(0)}(r) =\tp_\R^{(0)}(r)\,,\quad \tE_\L^{(0)}(r) =\tE_\R^{(0)}(r) \quad \text{for} \quad r\ge \nu \quad \text{or } \quad r\le -\nu
\,,
\eal
and in what follows we drop the indices L,R and use instead indices $\pm$ to distinguish between the positive and negative momentum branches. The two branches are related by the parity transformation which for any $u$ is $u\to -u+i\ka_a$~\cite{Frolov:2025uwz}. Since 
\bal\label{eq:u-reflection}
\tx_a(-u+i\ka_a)=-{1\ov \tx_a(u)}\quad \Longleftrightarrow\quad \tx_a(-u)=-{1\ov \tx_a(u+i\ka_a)}\,,
\eal
we get for real $r\le - \nu$
\bal
  \tp_-^{(0)}(r)  = -\tp_+^{(0)}(-r)  \,, \quad\tE_-^{(0)}(r)  = \tE_+^{(0)}(-r) \,.
\eal
The momentum and energy can be easily analytically continued to any complex value of the rapidity variable, and we get
\bal\label{tEp0}
 \tE_+^{(0)}(u) &= -\ln \tx_\L(u) -\ln \tx_\R(u) 
 \,,
 \\
 \tp_+^{(0)}(u) &= {h\ov2}\left( {1\ov \tx_\L(u)}- \tx_\L(u)-\tx_\R(u) +{1\ov \tx_\R(u)}  \right) 
 \,,
\eal
where  real energy and positive momentum correspond to $u$ on the upper edge of the main mirror cut, $u=r+i0$, $r\ge \nu$.
Similarly,
\bal\label{tEm0}
 \tE_-^{(0)}(u) 
 &= +2\pi i+ \ln \tx_{\L}(u+i\ka) +\ln \tx_\R(u-i\ka) = \tE_+^{(0)}(-u) \,,
 \\
\tp_-^{(0)}(u) 
&= {h\ov2}\left(  \tx_\L(u +i\ka) -{1\ov \tx_\L(u +i\ka)}-{1\ov \tx_{\R}(u- i\ka)}  +{ \tx_{\R}(u- i\ka) }  \right)
=-\tp_+^{(0)}(-u)\,,
\eal
where real energy  and negative momentum  correspond to $u+i\ka_a$ on the lower edge of the $\ka_a$-cut, $u=r-i0$, $r\le -\nu$. Note that the two momentum branches \textit{are not related} to each other through any analytic continuation. This is expected as the string-model dispersion relation~\eqref{eq:dispersion} is not analytic for~$M=0$, and the same is therefore also true for the mirror model.

\paragraph{Massless mirror gamma's.}  Let us recall that the mirror $\tg$-rapidities are defined by~%
\cite{Frolov:2025uwz}\footnote{In the RR case it differs from $\g$-rapidities introduced in \cite{Fontanella:2019baq} (see also~\cite{Beisert:2006ib}) by constant shifts.}
\bal\label{mirrorgamma1b}
\tx_a(\tg)&={\xi_a-i\,e^{\tg} \ov 1 + i\,\xi_a e^{\tg} }\,,
\qquad
\tg_a(\tx) = \ln\left({\xi_a -\tx \ov \tx\,\xi_a +1} \right)-{i\pi\ov2}\,,
 \eal
 where
 \bal
\xi_\L=\frac{1}{\xi_\R}=e^\eta \,,
 \eal
 with $\eta$ given in~\eqref{eq:nu_eta_def}.
This obeys
\begin{equation}
\tg_a(1/\tx) = \tg_{\bar a}(\tx) +i\pi\,\sgn\left(\Im(\tx)\right)\,.
\end{equation} 
The cuts  for $\tg_a(\tx)$ are  $(-\infty,-1/\xi_a)$  and $(\xi_a,+\infty)$, and in the mirror region $\Im(\tx)<0$.
The massless mirror $\tg$'s are defined for positive momentum as
\bal
\tg_a^{\pm 0} = \tg_a(r\pm i0)\,,\quad \tg_a^{- 0}=\tg_{\bar a}^{+0} +i\pi \,, \quad r>\nu\,,
\eal
and for negative momentum as
\bal
\tg_a^{\pm 0} = \tg_a(r+i\ka_a\pm i0)\,,\quad \tg_a^{- 0}=\tg_{\bar a}^{+0} +i\pi \,, \quad r<-\nu\,,
\eal
where $\tg_a$ is considered as a function of $u$
\bal\label{mirrorgamma3b}
\tg_a(u) &=  \ln\left({\xi_a -\tx_a(u) \ov \tx_a(u)\,\xi_a +1} \right)-{i\pi\ov2}\,.
\eal
The massless mirror $\tg^{\pm0}_a$ for both momentum branches have different real parts but opposite constant imaginary parts
\bal
\tg_a^{\pm 0} = \Re(\tg_a^{\pm0}) \mp {i\pi\ov2} \pm i0\quad \Rightarrow\quad \left(\tg_a^{\pm 0} \right)^* = \tg_a^{\pm 0}  \pm i\pi\,,
\eal
where $\pm i0$ indicates that $\tx_a^{\pm 0}$ have infinitesimally small negative imaginary parts.

The analytic continuation of $\tg_a^{+0}$  to any complex value on the mirror $u$-plane (without crossing any cut) just gives $\tg_a(u)$. On the other hand $\tg_a^{- 0}$  becomes  $\tg_{\bar a}(u)+i\pi$ because 
$u-i0$ moves to the anti-mirror $u$-plane.

\paragraph{Crossing transformation for mirror massless particles.} 

We follow the pure-RR case~\cite{Frolov:2021fmj} and define the crossing transformation for massless mirror particles with positive momentum as the analytic continuation through the  main mirror cut from below. Then, the 
massless Zhukovsky variables transform  in the usual way
\bal\la{crxtpm}
\tx_a^{\pm 0}\ \xrightarrow{\text{massless crossing}\,,\, \tp>0} \ {1\ov \tx_{\bar a}^{\pm 0}} = \tx_a^{\mp 0} +i0\,,
\eal
where at the last step we used that the $u$ rapidity variable is on the   main mirror cut, and added $+i0$ to stress that the crossing takes $\tx_a^{\pm 0}$ to the upper half-plane. 

Then, the $\tg$-rapidities  transform as
\bal\la{crgtpm}
\tg_a^{+0}\equiv \tg_{a}(\tx_a^{+0})\ \xrightarrow{\text{massless crossing}\,,\, \tp>0}\ \tg_a(1/\tx_{\bar a}^{+0}) +2i\pi= \tg_{\bar a}(\tx_{\bar a}^{+0}) +i\pi&= \tg_{\bar a}^{+0} +i\pi 
\\&= \tg_{a}^{-0}
\,,
\\
\tg_a^{-0}\equiv \tg_{a}(\tx_a^{-0})\ \xrightarrow{\text{massless crossing}\,,\, \tp>0}\ \tg_a(1/\tx_{\bar a}^{-0})+2i\pi = \tg_{\bar a}(\tx_{\bar a}^{-0}) +i\pi &= \tg_{\bar a}^{-0}+i\pi
\\&= \tg_{a}^{+0}+2i\pi\,,
\eal
where we used that under the crossing $\tx_a^{\pm 0}$ cross the half-line $x>\xi_a$ which is  the cut of~$\tg_a(x)$. 

Similarly, for massless mirror particles with negative momentum we  define the crossing transformation as the analytic continuation through the  mirror $\ka_a$-cut from below. It is easy to check that the massless Zhukovsky variables and $\tg$-rapidities  transform in the same way as in \eqref{crxtpm} and  \eqref{crgtpm}. 
Thus, for massless mirror particles the crossing transformation can be written in the form
\bal\la{crgtpm2}
\tg_a^{+0}\to \tg_{\bar a}^{+0} + i\pi = \tg_{a}^{-0}\,,\qquad \tg_a^{-0}\to \tg_{\bar a}^{-0} + i\pi = \tg_{a}^{+0}+2i\pi\,.
\eal

\subsection{Massless particles of  string theory}
\label{sec:massless_in_string}

 \paragraph{Massless string energy and momentum.} We are going to get the energy and momentum of massless string particles by an analytic continuation of the mirror ones. We have to distinguish the positive momentum branch with momentum in the range $0<p<2\pi$, and the negative one with momentum in the range $-2\pi<p<0$.
 
 \medskip

\begin{figure}[t]
    \centering
\includegraphics[width=0.49\linewidth]{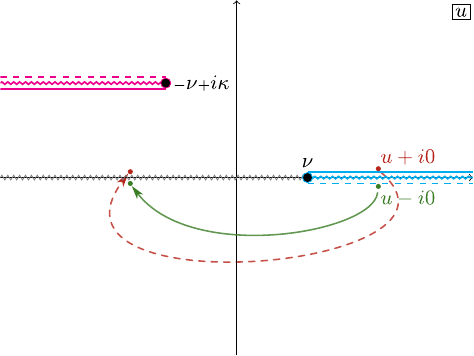}~\includegraphics[width=0.49\linewidth]{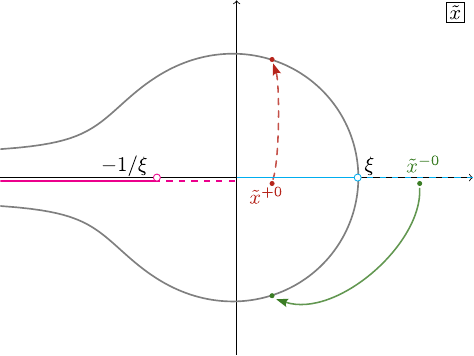}
    \caption{The analytic continuation from the mirror region to the positive-string-momentum region. In the left panel we depict the $u$-plane; the point $u+i0$ crosses the main cut from above, and then $u\pm i0$ ends up on the cut of the (string) $x_{\L}(u)$ function. Correspondingly, in the $\tx$-plane (right panel), $\tx^{+0}$ crosses the interval $(0,\xi)$, and then $\tx^{\pm0}\to x_{\L}^{\pm0}$ which are complex and conjugate. The gray contour on the $\tx$-plane is the image of the edges of the gray cut of $x_{\L}(u)$ on the $u$-plane.}
    \label{fig:positive-p-path}
\end{figure}

\noindent {\bf I.} For the positive momentum branch we  analytically continue the mirror  energy and positive momentum  \eqref{tEp0} through the main mirror cut from above. On the $x$-plane it corresponds to moving $\tx^{+0}_a$ through the interval $(0,\xi_a)$ to a point $x^{+0}_a$ on the upper boundary of the left string region, and $\tx^{-0}_a$ to a point $x_a^{-0}$ on the lower boundary of the left string region.
The sketch of the continuation is reported in figure~\ref{fig:positive-p-path}.

Thus, the  energy and momentum  of  massless string particles  are 
 \bal\la{Epstrp0a}
E_+^{(0)}(u)&={ih\ov2}\left(\tx_\R(u)- {1\ov \tx_\R(u)}-{1\ov \tx_\L(u)} +\tx_\L(u)  \right)  \,,
\\
p_+^{(0)}(u)&= +i\ln \tx_\R(u) +i\ln \tx_\L(u) 
\,,
\eal
where $u$ can take any complex value. The positive energy and momentum in the range $(0,2\pi)$ are obtained by taking $u$ to be on  the string main cut, i.e. $u<\nu$. Assuming, following the RR case, that $u$ is on its upper edge,\footnote{Since $\tx_a(u)$ does not have a cut for $u<\nu$ one can choose either the upper or lower edge of the string main cut.} and therefore  $\Im(u)>0$, we can rewrite \eqref{Epstrp0a} in terms of the string Zhukovsky variables $x_a(u)$
 \bal\la{Epstrp0b}
E_+^{(0)}(u)&={ih\ov2}\left( {1\ov x_\L(u)} -{ x_\L(u)}- x_\R(u) +{1\ov x_\R(u) }\right) \,,
\\
p_+^{(0)}(u)&= -i\ln x_\L(u) -i\ln x_\R(u) 
\,,
\eal
defined everywhere on the string $u$-plane. Taking now $u$ to be the upper edge of the string main cut, we can cast \eqref{Epstrp0b} into the form 
 \bal
E_+^{(0)}(u+i0)
&={ih\ov2}\left({1\ov x_\L^{+0}}- x_\L^{+0} -{1\ov x_\L^{-0}}+ x_\L^{-0}  \right) ={ih\ov2}\left( { x_\R^{-0}}- {1\ov x_\R^{-0}} -{x_\R^{+0}}+{1\ov x_\R^{+0}}  \right)\,,
\\
p_+^{(0)}(u+i0)
&=i\ln x_\L^{-0} -i\ln x_\L^{+0}=-i\ln x_\R^{+0} +i\ln x_\R^{-0}\,,\qquad u<\nu
\,.
\eal
Thus, we could have defined the energy and momentum  of  massless string particles with momentum in the range $(0,2\pi)$ as the limit of massive ones with the mass going to zero.

The momentum and energy  of a massless string particle with  momentum in the range $(0,2\pi)$ are related to the ones of a right $k$-particle bound state through the reflection  transformation $u\mapsto - u$ on the $u$-plane. 
To check this fact we recall 
the following relations from~\cite{Frolov:2025uwz}
\bal
\label{eq:app_minu_string_Zh}
x^{\pm m}_{\L}(u)=-x^{\mp (k+m)}_{\R}(-u) \,,\qquad x^{\pm m}_{\R}(u)=-\frac{1}{x^{\pm (k-m)}_{\R}(-u)}\,,
\eal
which are valid for $0<m<k$ and follows from~\eqref{eq:u-reflection}.
Then, using~\eqref{eq:def_en_mom_string} we see that
\bal
\label{eq:app_relations_u_minus_u}
&E^{(m)}_{\L}(u)=E^{(k+m)}_{\R}(-u)\,, \qquad p^{(m)}_{\L}(u)=2 \pi - p^{(k+m)}_{\R}(-u)\,,\\
&E^{(m)}_{\R}(u)=E^{(k-m)}_{\R}(-u)\,, \qquad p^{(m)}_{\R}(u)=2 \pi - p^{(k-m)}_{\R}(-u)\,,
\eal
where again $0<m<k$. The massless case corresponds to the limit $m\to 0^+$ with $u<\nu$. In this manner the massless energies and momenta take values on the main string theory cut and the $k$-particle bound states energies and momenta are away from the $\ka$-cuts. As expected, in this limit, the first and second line in the equation above become equivalent.
Performing the computation explicitly, in the massless case we have
\bal\la{pE0pEk}
 p_+^{(0)}(u+i0) &= -i\ln x_\L(u+i0) -i\ln x_\R(u+i0) 
 \\
 &=2\pi -i\ln x_\R(-u-i\ka -i0)+i\ln x_\R(-u+i\ka-i0)=2\pi -p_\R^{(k)}(-u) \,,
\\
 E_+^{(0)}(u+i0) &=
{ih\ov2}\left( {1\ov x_\R(u+i0)} -{ x_\R(u+i0)}- x_\L(u+i0) +{1\ov x_\L(u+i0) }\right) 
=E_\R^{(k)}(-u)
 \,.
\eal
As already remarked, for $u$ on the main string cut the points $-u\pm i\ka$ are not on a cut. Note that we would get the same result for any $u$ with $0<\Im(u)<\ka$.

\medskip

\noindent {\bf II.}  Massless string particles with  momenta in the range $(-2\pi,0)$ can be obtained from mirror ones with negative momenta in two steps. Let us recall that  the mirror $\tx_a^{\pm 0}$ of massless mirror particles with negative momenta are defined as
\bal
\tx_a^{\pm 0}(u') = \tx_a(u' +i\ka_a\pm i0)={1\ov \tx_{\bar a}(u' -i\ka_a\mp i0)}={1\ov \tx_{\bar a}^{\mp 0}}\,,\quad u' <-\nu\,.
\eal
The analytic continuation to negative string momentum is done in two steps. First, 
we move the $u'$-rapidity through the upper edge of the $\ka_a$-cut to a point $u>-\nu$. Thus, the massless particle $\tx_a^{+0}$ moves to a point  in the upper half-plane through the interval $-1/\xi_a <x<0$.  As a result, we get ($u>-\nu$)
\bal
\tx_\L^{-0}(u')&\to \tx_\L(u+i\ka) =  {1\ov x_\R^{+k}(u) } \,,
\quad
\tx_\L^{+0}(u')\to{1\ov \tx_\R(u-i\ka)} =  {1\ov x_\R^{-k}(u) } 
\,,
\\
\tx_\R^{-0}(u')&\to \tx_\R(u-i\ka) =  x_\R^{-k}(u) 
\,,\quad
\tx_\R^{+0}(u')\to{1\ov \tx_\L(u+i\ka)} =  x_\R^{+k}(u) 
\,.
\eal
\begin{figure}[t]
    \centering
\includegraphics[width=0.49\linewidth]{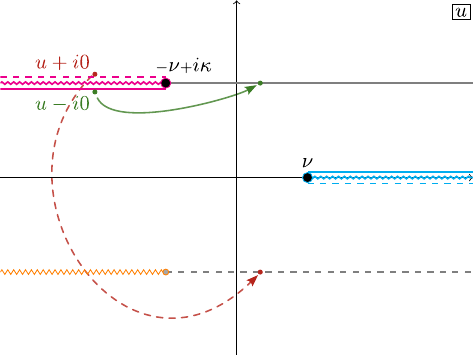}~\includegraphics[width=0.49\linewidth]{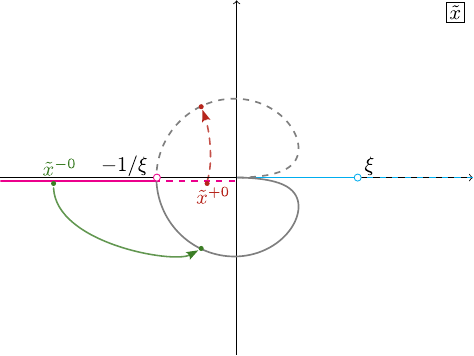}
    \caption{The analytic continuation from the mirror region to the negative-string-momentum region. In the left panel we represent the $u$ plane with the cuts of $\tx$ in blue and magenta; we additionally depict an (orange) cut at $-i\kappa$, which is the cut of $x_{\R}(u)$. In this case, $u+i0$ crosses the cut at $+i\kappa$ and it is mapped to a different $u$-plane, that of $x_{\R}(u)$, and it is continued to a point where $u>-\nu$. In this process, $u-i0$ does not cut any cut. The same transformation is easier to visualise on the $\tx$-plane (right) where $\tx^{+0}$ crosses the cut between $(-1/\xi,0)$; then, $\tx^{\pm0}$ end up on complex and conjugate points on a curve which is the image of the horizontal gray lines by the map $1/x_{\R}(u)$. }
    \label{fig:negative-p-path}
\end{figure}
After the first step the mirror  energy and negative momentum \eqref{tEm0}  become
 \bal
 i\tE_-^{(0)}(u') &\to -2\pi + i\ln \tx_{\L}(u+i\ka) +i\ln \tx_\R(u-i\ka)  
 \\
 &=-2\pi - i\ln x_{\R}^{+k}+i\ln x_\R^{-k}= p_\R^{(k)}(u) -2\pi <0 \,,
\\
i\tp_-^{(0)}(u') &\to 
 {ih\ov2}\left(  \tx_\L(u +i\ka) -{1\ov \tx_\L(u +i\ka)}-{1\ov \tx_{\R}(u- i\ka)}  +{ \tx_{\R}(u- i\ka) }  \right)
\\
&= {ih\ov2}\left( - x_\R^{+k} +{1\ov x_\R^{+k}}-{1\ov x_{\R}^{-k}}  +{ x_{\R}^{-k}}  \right)
= E_\R^{(k)}(u)>0
\,.
\eal
Note, that because of our definition of $\tE_-^{(0)}(u)$ and $ \tp_-^{(0)}(u)$ we do not cross any cut (we have done it already to define them). We see that a massless mirror particle with negative momentum analytically continued through the upper edge of the $\ka_a$-cut to a point $u>-\nu$ can be thought of as a right string $k$-particle bound state with momentum in the interval $(-2\pi,0)$. Now, taking into account \eqref{pE0pEk} which relates a massless string particle with  a right $k$-particle bound state, we perform, as the second step,  the reflection  $u\mapsto - u$ that gives ($u<\nu$)
\bal
 x_\R^{-k}(-u) =- {1\ov \tx_\R(u) } =
\left\{
\begin{array}{ccc}
- {1\ov x_\R(u-i0) }  &=& - {1\ov x_\R^{-0}(u) }      \\
 - x_\L(u+i0) & = & -x_\L^{+0}(u)   
\end{array}
\right.\,,
\eal
\bal
 x_\R^{+k}(-u) 
=-  \tx_\L(u) =
\left\{
\begin{array}{ccc}
- x_\L(u-i0)   &=& -  x_\L^{-0}(u)       \\
 -{1\ov  x_\R(u+i0)}  &= & -{1\ov x_\R^{+0}(u) }     
\end{array}
\right.\,.
\eal
Thus, after the two steps
\bal
\tx_a^{\pm 0}(u')&\to  -{1\ov x_a^{\pm 0}(u) }  \,,
\\
 i\tE_-^{(0)}(u') &\to p_-^{(0)}(u+i0)\equiv p_\R^{(k)}(-u) -2\pi = -p_+^{(0)}(u+i0) \,,
\\
i\tp_-^{(0)}(u') 
&\to  E_-^{(0)}(u+i0)\equiv E_\R^{(k)}(-u)= E_+^{(0)}(u+i0) 
\,,
\eal
see Figure~\ref{fig:negative-p-path}.
It is worth pointing out that as a result of the two steps we obtained a massless string particle  with negative momentum which has the same $u$-rapidity and energy but opposite momentum as a massless string particle with positive momentum.

\paragraph{Massless string gamma's.} The string $\g$-rapidities are defined as follows
\bal
x_a(\g)&={\xi_a+e^{\g} \ov 1 - \xi_a e^{\g} }\,,\qquad  
\g_a(x) = \ln{x - \xi_a\ov x\,\xi_a +1}\,,\qquad   \g_a(1/x) = \g_{\bar a}(x) -i\pi\, \text{sign}\left(\Im\left(x\right)\right) \,,
 \eal
 and the cut  of $\g_a(x)$ is the interval $(-1/\xi_a,\xi_a)$. 
 Then, considering $x$ as a function of $u$, we get the functions $\g_a(u)$ 
\bal
\g_a(u) =  \ln{x_a(u) - \xi_a\ov x_a(u)\, \xi_a +1}\,.
\eal
Clearly, up to a shift the massless $\g_a^{\pm0}$ for the positive momentum branch are 
obtained by analytically continuing $\tg_a^{\pm0}$ through the main mirror cut from above, and are   
given by
\bal
\tg_a^{+0}=\tg_a(u'+i0) &\xrightarrow{\text{to string}}  \tg_{\bar a}(u+i0) -i\pi =\g_a(u+i0) -{3i\pi\ov2} \,,
\\
\tg_a^{-0}=\tg_a(u'-i0) &\xrightarrow{\text{to string}}  \tg_{ a}(u-i0)  =\g_a(u-i0) +{i\pi\ov2}\,.
\eal
Note that it is different from the continuation of massive $\tg$'s 
\bal
\tg_a^{+}&\xrightarrow{\text{to string}}  \g_a^+ +{i\pi\ov2} \,,
\quad
\tg_a^{-}\xrightarrow{\text{to string}}  \g_a^- +{i\pi\ov2}\,.
\eal
Thus,
\bal
\g_a^{+0}(u) = \g_a(u+ i0)= \tg_{\bar a}(u+i0) + {i\pi\ov2} \,,\quad\g_a^{-0}(u) = \g_a(u- i0)= \tg_{ a}(u-i0) - {i\pi\ov2} \,,\eal
\bal
 \g_a^{-0}(u) = \g_{\bar a}^{+0}(u) -i\pi\,,\quad u<\nu\,.
\eal
Therefore, we can  express $ \g_a^{-0}(u)$ in terms of $ \g_{\bar a}^{+0}(u)$, and then drop $+i0$ and use $\g_a(u)$ for any complex $u$ as massless $\g$-rapidity variables. 
The massless string $\g^{\pm0}_a$ are complex conjugate to each other 
but, contrary to the mirror massless $\tg^{\pm0}_a$, their imaginary parts are not constant as $u$ varies along the cut, except in the case $\ka=0$ (i.e., in the pure-RR case).

To get the massless $\g_a^{\pm0}$ for the negative momentum branch we perform the two steps discussed above. First, we analytically continue $\tg_a^{\pm0}$ through the $\ka_a$-mirror cut from above and get
\bal
\tg_\L^{-0}=\tg_\L\big(\tx_\L(u'+i\ka- i0)\big)&\to \tg_\L\big(\tx_\L(u+i\ka)\big) =  \g_\R\big(x_\R(u+i\ka)\big) -{i\pi\ov2} =\g_\R^{+k}(u) -{i\pi\ov2}\,,
\\
\tg_\L^{+0}=\tg_\L\big(\tx_\L(u'+i\ka+ i0)\big)&\to \tg_\L\big({1\ov\tx_\R(u-i\ka)}\big) = \tg_\R\big(\tx_\R(u-i\ka)\big)-i\pi  
\\
&=  \g_\R\big(x_\R(u-i\ka)\big) -{i\pi\ov2} =\g_\R^{-k}(u) -{i\pi\ov2}\,,
\eal
\bal
\tg_\R^{-0}=\tg_\R\big(\tx_\R(u'-i\ka- i0)\big)&\to \tg_\R\big(\tx_\R(u-i\ka)\big) =  \g_\R\big(x_\R(u-i\ka)\big) +{i\pi\ov2} =\g_\R^{-k}(u) +{i\pi\ov2}\,,
\\
\tg_\R^{+0}=\tg_\R\big(\tx_\R(u'-i\ka+ i0)\big)&\to \tg_\R\big({1\ov\tx_\L(u+i\ka)}\big) = \tg_\L\big(\tx_\L(u+i\ka)\big) -i\pi
\\
&=  \g_\R\big(x_\R(u+i\ka)\big) -{3i\pi\ov2} =\g_\R^{+k}(u) -{3i\pi\ov2}\,.
\eal
It is again different from the continuation of massive $\tg$'s to the string theory negative momentum branch
\bal
\tg_\R^\pm\to \g_\R^{\pm(k+1)}+{i\pi\ov2}\,,\quad \tg_\L^\pm\to \g_\L^{\mp(k-1)}\pm {i\pi}\,.
\eal

Next, performing the reflection $u\to -u$, we obtain
\bal
\g_\R^{-k}(-u) &=\g_\R\big(x_\R(-u-i\ka)\big) =\g_\R\big(- {1\ov \tx_\R(u)}\big) 
\\
&=
\left\{
\begin{array}{ccc}
\g_\R\big(- {1\ov x_\R(u-i0)}\big) &=&- \g_\R\big( x_\R(u-i0)\big) -i\pi =- \g_\R^{-0} -i\pi    \\
 \g_\R\big(- x_\L(u+i0)\big) &=&- \g_\L\big( x_\L(u+i0)\big)  =- \g_\L^{+0}    
\end{array}
\right.\,,
\eal
\bal
\g_\R^{+k}(-u) &=\g_\R\big(x_\R(-u+i\ka)\big) =\g_\R\big(- \tx_\L(u)\big) 
\\
&=
\left\{
\begin{array}{ccc}
\g_\R\big(- x_\L(u-i0)\big) &=&- \g_\L\big( x_\L(u-i0)\big) =- \g_\L^{-0}      \\
 \g_\R\big(-{1\ov x_\R(u+i0)}\big)&=&- \g_\R\big( x_\R(u+i0)\big) +{i\pi} =- \g_\R^{+0} +{i\pi}     
\end{array}
\right.\,.
\eal
Thus, after the two steps
\bal
\tg_a^{\pm 0}(u')\to -\g_a^{\pm 0}(u)-{i\pi\ov2}\,.
\eal
For massive $\tg$'s we find 
\bal
\tg_a^{+}\to -\g_a^{+}+{3i\pi\ov2}\,,\quad \tg_a^{-}\to -\g_a^{-}-{i\pi\ov2}\,.
\eal

\paragraph{Crossing transformation for string massless particles.} 
We again follow the pure RR case and define the crossing transformation for massless particles with momenta in the interval $(0,2\pi)$ as the analytic continuation through the string main cut from above. 
The massless Zhukovsky variables transform  in the usual way
\bal\la{crxpm}
x_a^{\pm 0}\ \xrightarrow{\text{massless crossing}\,,\, p>0} \ {1\ov x_{\bar a}^{\pm 0}} = x_a^{\mp 0} \,,
\eal
where at the last step we used that the $u$ rapidity variable is on the   main string cut. 

Then,  $ \g_{a}^{+0}(u)= \g_{a}(x_a^{+0})$ transforms as
\bal
\g_a^{+0}\equiv \g_{a}(x_a^{+0}) \xrightarrow{\text{massless crossing}\,,\, p>0} \g_a(1/x_{\bar a}^{+0}) +2\pi i= \g_{\bar a}(x_{\bar a}^{+0}) +i\pi&= \g_{\bar a}^{+0} +i\pi
\\
&= \g_{ a}^{-0} +2i\pi
\,.
\eal
On the other hand under the crossing $x_a^{-0}$ crosses the semi-line $x>\xi_a$ which is not the cut of $\g_a(x)$, and therefore 
\bal
\g_a^{-0}\equiv \g_{a}(x_a^{-0}) \xrightarrow{\text{massless crossing}\,,\, p>0} \g_a(1/x_{\bar a}^{-0})= \g_{\bar a}(x_{\bar a}^{-0}) +i\pi&= \g_{\bar a}^{-0}+i\pi
\\
&= \g_{ a}^{+0}\,.
\eal
Thus, the crossing transformation of  massless string particles with momenta in the interval $(0,2\pi)$ can be written in the form
\bal\la{crgpm2}
\g_a^{+0}\to \g_{\bar a}^{+0} + i\pi = \g_{a}^{-0}+2i\pi\,,\quad \g_a^{-0}\to \g_{\bar a}^{-0} + i\pi = \g_{a}^{+0}\,.
\eal

As to massless string particles with negative momenta in the interval $(-2\pi,0)$, since their rapidity variables are expressed in terms of those for massless string particles with momenta in the interval $(0,2\pi)$, their crossing transformations are the same.

\section{Properties of the S~matrix}
\label{sec:properties}
The S~matrix is fixed by the light-cone gauge symmetries up to a number of dressing factors. The first check on the dressing factors is that they are such that the S~matrix, as a whole, satisfies all relevant physical properties: crossing symmetry, unitarity, parity- or CP-invariance as appropriate, as well as a suitable notion of analyticity. Below we briefly review these properties for the model at hand, following and expanding on the discussion of~\cite{Frolov:2024pkz}.

\subsection{Highest-weight S-matrix elements}
For any pair of irreducible representations of the light-cone symmetries, there is an undetermined dressing factor. Not all of these dressing factors are independent, as some are simply related by braiding unitarity. Here we list the independent scattering elements for each block, corresponding to the scattering of highest-weight states in a given multiplet.
It turns out that it is simplest to start from the S~matrix in the mirror kinematics, which is what we consider here.

\paragraph{Massive scattering.}
In the case of massive scattering elements, there are%
\footnote{In the language of~\cite{Frolov:2025ozz}, these are elements of the Zamolodchikov-Faddeev (ZF) S~matrix, i.e.~they appear in ZF algebra relations of the type $A^\dagger_{X}(p_1)A^\dagger_{Y}(p_2)=S_{XY}(p_1,p_2)\,A^\dagger_{Y}(p_2)A^\dagger_{X}(p_1)$.
}
\begin{equation}
\label{eq:Sm1m1}
  \mathbf{S}\,\big|Y_{1} Y_{2} \big\rangle=S^{11}_{Y Y}( \tx^{\pm}_{\L1}, \tx^{\pm}_{\L2} )\,
    \big| Y_{1}Y_{2} \big\rangle\,,
    \qquad \mathbf{S}\,\big| Y_{1} \bar{Z}_{2} \big\rangle=
  S_{Y \bar Z}^{11}( \tx^{\pm}_{\L1}, \tx^{\pm}_{\R2} ) \,\big| Y_{1} \bar{Z}_{2} \big\rangle\,.
\end{equation}

\paragraph{Mixed-mass scattering.}
In the case of mixed-mass scattering it is sufficient to consider the following two scattering processes
\bal\la{Sm0m1}
  \mathbf{S}\,\big|\chi^{\dot{A}}_{1} Y_{2} \big\rangle&=S^{01}_{\chi Y}( \tx^{\pm0}_{\L1}, \tx^{\pm}_{\L2} )\,
    \big| \chi_{1}^{\dot{A}}Y_{2} \big\rangle\,,
    \qquad \mathbf{S}\,\big| \chi_{1}^{\dot{A}} \bar{Z}_{2} \big\rangle&=
  S_{\chi \bar Z}^{01}( \tx^{\pm0}_{\L1}, \tx^{\pm}_{\R2} ) \,\big| \chi_{1}^{\dot{A}} \bar{Z}_{2} \big\rangle\,.
 \eal
Here the index $\dot{A}=1,2$ is charged under the so-called $\mathfrak{su}(2)_{\circ}$ symmetry~\cite{Borsato:2014hja,Lloyd:2014bsa}, corresponding to local frame rotations in~$T^4$; clearly, the boundary conditions of~$T^4$ break such a symmetry, though this should not be noticeable when considering the S~matrix. For this reason,  we assume such a symmetry and outright omit the $\dot{A}=1,2$ index on the S-matrix element.

\paragraph{Massless-massless scattering.}
In the case of massless scattering we can simply consider
\bal\la{Sm0m0}
  \mathbf{S}\,\big|\chi_{1}^{\dot{A}} \chi_{2}^{\dot{B}} \big\rangle&=\big(S^{00}_{\chi \chi}( \tx^{\pm0}_{\L1}, \tx^{\pm0}_{\L2} )\big)^{\dot{A}\dot{B}}\,
    \big| \chi_{1}^{\dot{A}}\chi_{2}^{\dot{B}} \big\rangle\,.
 \eal
 In this case $\mathfrak{su}(2)_\circ$ symmetry would dictate that the S-matrix structure is given by two invariant tensors (identity and permutation), see~\cite{Borsato:2014hja} for a more detailed discussion. Interestingly, perturbative results suggest that only the identity tensor on the $\mathfrak{su}(2)_\circ$ representation should appear; this would also seem to fit well with the ``hexagon form factor'' construction~\cite{Eden:2021xhe}. For this reason, in what follows we will generally assume that
 \begin{equation}
     \big(S^{00}_{\chi \chi}( \tx^{\pm0}_{\L1}, \tx^{\pm0}_{\L2} )\big)^{\dot{A}\dot{B}} = \mathbf{1}^{\dot{A}\dot{B}}\,S_{\chi \chi}^{00}( \tx^{\pm0}_{\L1}, \tx^{\pm0}_{\L2} )\,.
 \end{equation}

\subsection{Discrete symmetries}
As we mentioned, some of the discrete symmetries can be thought of as a way to determine some highest-weight--highest-weight scattering process starting from the ones listed above. Yet, these symmetries put further constraints on some of the S-matrix elements. We list those below.

\paragraph{Braiding unitarity.}
Braiding unitarity on mixed-mass and massless elements requires 
\bal
    &S^{01}_{\chi Y}( \tx^{\pm 0}_{\L1}, \tx^{\pm}_{\L2} )\;S^{10}_{Y \chi}( \tx^{\pm}_{\L2}, \tx^{\pm 0}_{\L1} )=1\,,\qquad S^{01}_{\chi \bar{Z}}( \tx^{\pm 0}_{\L1}, \tx^{\pm}_{\R2} )\;S^{10}_{\bar{Z} \chi}( \tx^{\pm}_{\R2}, \tx^{\pm 0}_{\L1} )=1\,,\\
    &S_{\chi \chi}^{00}( \tx^{\pm0}_{\L1}, \tx^{\pm0}_{\L2} )\;S_{\chi \chi}^{00}( \tx^{\pm0}_{\L2}, \tx^{\pm0}_{\L1} )=1\,.
\eal
Similar relations apply to the massive S-matrices.

\paragraph{Physical unitarity in the string kinematics.} In the string kinematics, the S-matrix is unitary, when the momenta are real, and therefore for the diagonal S-matrix elements we have
\begin{equation}
    |S^{11}_{Y Y}( x^{\pm}_{\L1}, x^{\pm}_{\L2} )|=
    |S^{11}_{Y \bar{Z}}( x^{\pm}_{\L1}, x^{\pm}_{\R2} )|=|S^{01}_{\chi Y}( x^{\pm}_{\L1}, x^{\pm}_{\L2} )|=
    |S^{01}_{\chi\bar{Z}}( x^{\pm}_{\L1}, x^{\pm}_{\R2} )|=
    |S^{00}_{\chi\chi}( x^{\pm}_{\L1}, x^{\pm}_{\L2} )|=1
\end{equation}
for real string momenta.

\paragraph{CT invariance of the mirror model.}
The mirror energy is not real, satisfying instead
\begin{equation}
\label{eq:mirrorenergyreality}
    \left(\tE(\tp, M)\right)^* = \tE(\tp, -M)\,,
\end{equation}
for real mirror momentum. Correspondingly, the mirror S~matrix is not unitary, nor it is invariant under time-reversal. However, as~\eqref{eq:mirrorenergyreality} suggests, the mirror S~matrix is invariant under a combination of charge conjugation and time reversal.
This property holds for  the mirror representations as a whole and it is crucial to ensure the reality of wrapping corrections~\cite{Baglioni:2023zsf}. It is possible to check that the matrix part of the S~matrix  is CT invariant, up to the normalisations, which should obey 
\begin{equation}
\label{eq:CTinv}
\begin{aligned}
    \left(S_{YY}^{11}(\tx_{\L1}^\pm,\tx_{\L2}^\pm)\right)^*S_{\bar{Y}\bar{Y}}^{11}(\tx_{\R1}^\pm,\tx_{\R2}^\pm)&=
    \left(S_{Y\bar{Z}}^{11}(\tx_{\L1}^\pm,\tx_{\R2}^\pm)\right)^*S_{\bar{Y}Z}^{11}(\tx_{\R1}^\pm,\tx_{\L2}^\pm)=1\,,\\
    \left(S_{\chi Y}^{01}(\tx_{\L1}^{\pm0},\tx_{\L2}^\pm)\right)^*S_{\tilde\chi\bar{Y}}^{01}(\tx_{\L1}^{\pm0},\tx_{\R2}^\pm)&=
    \left(S_{\chi\bar{Z}}^{01}(\tx_{\L1}^{\pm0},\tx_{\R2}^\pm)\right)^*S_{\tilde\chi Z}^{01}(\tx_{\L1}^{\pm0},\tx_{\L2}^\pm)=1\,,\\
    \left(S_{\chi\chi}^{00}(\tx_{\L1}^{\pm0},\tx_{\L2}^{\pm0})\right)^*S_{\tilde\chi\tilde\chi}^{00}(\tx_{\L1}^{\pm0},\tx_{\L2}^{\pm0})
   &=~1\,.\\
\end{aligned}
\end{equation}

\paragraph{P invariance in the mirror model.}
The mirror model is parity invariant. Combining this condition with braiding unitarity we get%
\footnote{We used that under the transformation $\tp\to-\tp$, $\tx^{\pm0}_{a}\to-1/\tx^{\mp 0}_{a}$, $a=$L,R.}
\begin{equation}
\label{eq:mirror-P}
\begin{aligned}
    &1=S^{11}_{Y Y}( \tx^{\pm}_{\L1}, \tx^{\pm}_{\L2} )S^{11}_{Y Y}\big( -\frac{1}{\tx^{\mp}_{\L1}}, -\frac{1}{\tx^{\mp}_{\L2}} \big),\quad&
    1=S^{11}_{Y\bar{Z}}( \tx^{\pm0}_{\L1}, \tx^{\pm0}_{\R2} )S^{11}_{Y\bar{Z}}\big( -\frac{1}{\tx^{\mp0}_{\L1}}, -\frac{1}{\tx^{\mp0}_{\R2}} \big),\\
    &1=S^{01}_{\chi Y}( \tx^{\pm0}_{\L1}, \tx^{\pm}_{\L2} )S^{01}_{\chi Y}\big( -\frac{1}{\tx^{\mp0}_{\L1}}, -\frac{1}{\tx^{\mp}_{\L2}} \big),\quad
    &1=S^{01}_{\chi\bar Z}( \tx^{\pm0}_{\L1}, \tx^{\pm}_{\R2} )S^{01}_{\chi\bar Z}\big( -\frac{1}{\tx^{\mp0}_{\L1}}, -\frac{1}{\tx^{\mp}_{\R2}} \big),\\
    &1=S^{00}_{\chi\chi}( \tx^{\pm0}_{\L1}, \tx^{\pm0}_{\L2} )S^{00}_{\chi\chi}\big( -\frac{1}{\tx^{\mp0}_{\L1}}, -\frac{1}{\tx^{\mp0}_{\L2}} \big).&&
\end{aligned}
\end{equation}

\paragraph{CP invariance in the string model}

The string theory Lagrangian is invariant under a simultaneous parity transformation on the worldsheet and mapping particles to anti-particles. If this symmetry is preserved at the quantum level then the S-matrix must be invariant under a combination of charge conjugation and parity; we refer to this type of symmetry by CP. In particular it must hold
\bal
\label{eq:CP_equations_main}
S^{0m}_{\chi Y} (p_1, p_2) =S^{m0}_{\bar Y\tilde{\chi}} (-p_2,  -p_1)\,, \qquad S^{0m}_{\chi \bar{Z}} (p_1, p_2) =S^{m0}_{Z \tilde{\chi}} (-p_2,  -p_1)\,.
\eal
The first and second relation in~\eqref{eq:CP_equations_main} are mapped one into the other by a charge conjugation transformation and we expect the validity of the first relation to imply the validity of the second one. For this reason in this paper we will only provide a proof of the first relation above.
Similarly for massless particles we require
\bal
S^{0 0}_{\chi \chi}(p_1, p_2)= S^{0 0}_{\tilde{\chi} \tilde{\chi}}(-p_2, -p_1)\,.
\eal

\subsection{Crossing symmetry}
Crossing symmetry is the most constraining among those discussed so far, because it knows about the physical content of the model and it requires a non-trivial analytic continuation. Crossing symmetry must hold both in the string and in the mirror model.

A subtle point about crossing symmetry is that it depends on the exchange relations of the underlying model. In particular, the crossing equations derived in~\cite{Lloyd:2014bsa} had implicitly assumed bosonic/fermionic exchange relations. However, in two-dimensional QFT, more general exchange relations may and do appear --- for instance, in integrable models like the Chiral Gross--Neveu (CGN) one. In~\cite{Frolov:2025ozz} we discussed in detail how to write ``generalised'' crossing equations, valid for any exchange relations, and the possible modifications for the $AdS_3\times S^3\times T^4$ S~matrix.  In general, these equations take the form
\begin{equation}
\label{eq:generalisedcrossing}
S_{AB}(u_1,u_2)S_{A\bar{B}}(u_1,\bar{u}_2) = S_{AB}^{\text{free}}(u_1,u_2)S_{A\bar{B}}^{\text{free}}(u_1,\bar{u}_2)\, f(\tx_1^\pm,\tx_2^\pm)\,,
\end{equation}
where $S_{AB}^{\text{free}}$ is the value of ZF scattering element in the limit where the interactions disappear.%
\footnote{For a massive relativistic model, this limit is expressed in terms of rapidities as $\theta_1-\theta_2\to+\infty$.}
For a model with trivial exchange relations (featuring only bosons or fermions), the right-hand side reduces  to~$f(\tx_1^\pm,\tx_2^\pm)$, but this is not the case in general. In analogy with what happens with the $SU(2)$ CGN model --- whose S~matrix indeed appears in the $SU(2)_\circ$ part of the massless scattering --- we will assume that massless particles have ``semionic'' statistics, i.e.\ that exchanging any two massless particles produces a factor of $\pm i$. Bearing in mind such exchange relations, we find that the right-hand side of~\eqref{eq:generalisedcrossing} becomes $-f(x_1^\pm,x_2^\pm)$ (notice the minus sign). As a result, the crossing equations are modified by a sign relative to those in~\cite{Lloyd:2014bsa,Frolov:2021fmj}. As for the exchange relations involving massive particles, we see no reason to assume them to be non-trivial; on the contrary, the experience from $AdS_5\times S^5$ suggest that standard bosonic/fermionic exchange relations are correct in this case. We will therefore work with the same crossing equations as in~\cite{Lloyd:2014bsa,Frolov:2021fmj} for both massive and mixed-mass crossing.%
\footnote{Strictly speaking, we could consider non-trivial exchange relations between massive and massless excitations, but we do not see a reason for doing so. This is also based on an ex-post check of the solutions which we obtain with perturbation theory.}

Assuming exchange relations as explained above, the  S~matrix elements \eqref{Sm0m1} satisfy the following crossing equations. First, for the reader's convenience we recap the massive--massive crossing equations,
\begin{equation}
\label{eq:massivemassivecross}
\begin{aligned}
    &S^{11}_{Y Y}(u_1,u_2) S^{11}_{\bar{Z} Y}( \bar{u}_1, u_2) = \frac{ \tx_{\L2}^{-}}{\tx_{\L2}^{+}} \Bigl(\frac{\tx^{+}_{\L1}-\tx^{+}_{\L2}}{\tx^{+}_{\L1}-\tx^{-}_{\L2}} \Bigl)^2\,,\\
    &S^{11}_{\bar{Z} Y}(u_1,u_2) S^{11}_{Y Y}( \bar{u}_1, u_2) = \frac{ \tx_{\L2}^{-}}{\tx_{\L2}^{+}} \Bigl(\frac{1- \tx^{-}_{\R1}\tx^{+}_{\L1}}{1- \tx^{-}_{\R1}\tx^{-}_{\L1}} \Bigl)^2\,.
\end{aligned}
\end{equation}
Turning to mixed-mass scattering, we can cross with respect to massive particles, getting the equations
\bal
\label{mixed_mass_crossing_eq_01a}
&S^{01}_{\chi Y}(u_1,u_2) S^{01}_{\chi \bar Z }( u_1,\bar u_2) = \frac{ \tx_{\L1}^{+0}}{\tx_{\L1}^{-0}} \Bigl(\frac{\tx^{-0}_{\L1}-\tx^{+}_{\L2}}{\tx^{+0}_{\L1}-\tx^{+0}_{\L2}} \Bigl)^2\,,
\\
&S^{01}_{\chi\bar Z }(u_1,u_2) S^{01}_{\chi Y }( u_1,\bar u_2) = {\frac{\tx_{\L1}^{+0}}{ \tx_{\L1}^{-0}}} \Bigl(\frac{1-\tx^{-0}_{\L1}\tx^{-}_{\R2}}{1-\tx^{+0}_{\L1}\tx^{-}_{\R2}} \Bigl)^2\,,
\eal
and with respect to massless particles, obtaining
\bal
\label{mixed_mass_crossing_eq_01b}
&S^{01}_{\chi Y}(u_1,u_2) S^{01}_{\chi Y}(\bar u_1,u_2) = \frac{ \tx_{\L2}^-}{\tx_{\L2}^+} \Bigl(\frac{\tx^{+0}_{\L1}-\tx^+_{\L2}}{\tx^{+0}_{\L1}-\tx^-_{\L2}} \Bigl)^2\,,
\\
&S^{01}_{\chi \bar Z}(u_1,u_2) S^{01}_{\chi \bar Z}(\bar u_1,u_2) = {\frac{\tx_{\R2}^+}{ \tx_{\R2}^-}} \Bigl(\frac{\tx^{+0}_{\L1}\tx^-_{\R2}-1}{\tx^{+0}_{\L1}\tx^+_{\R2}-1} \Bigl)^2\,.
\eal
Finally, the crossing equation for massless particles with semionic statistics is
\begin{equation}
\label{massless_crossing_eq_00}
S^{00}_{\chi \chi}(u_1,u_2) S^{00}_{\chi \chi}(\bar u_1,u_2) = -\frac{ \tx_{\L2}^{-0}}{\tx_{\L2}^{+0}} \Bigl(\frac{\tx^{+0}_{\L1}-\tx^{+0}_{\L2}}{\tx^{+0}_{\L1}-\tx^{-0}_{\L2}} \Bigl)^2\,,
\end{equation}
which differs by a minus sign from~\cite{Lloyd:2014bsa,Frolov:2021fmj}. As we will discuss around eq.~\eqref{eq:aofgamma} below, the solutions of equation~\eqref{massless_crossing_eq_00} are more natural 
than those of the equation with opposite sign. 
This difference in sign (even if restricted to the pure RR case) was also suggested in~\cite{Ekhammar:2024kzp} based on the quantum spectral curve construction of the dressing factors.

\section{Proposal for  dressing factors involving massless particles}
\label{sec:proposal}

Below we detail our proposal for the dressing factors of mixed-mass and massless--massless scattering. As it turns out, taking suitable massless limits of the massive--massive dressing factors of~\cite{Frolov:2025uwz} is sufficient to construct the remaining factors. 

\subsection{Mirror theory S-matrix elements}
\label{sec:proposalmirror}

\paragraph{Mixed-mass S-matrix elements.}

As was discussed in the previous section, in the case of mixed-mass scattering one considers the following two S-matrix elements
\bal
  S^{01}_{\chi Y}( \tx^{\pm0}_{\L1}, \tx^{\pm}_{\L2} )\,,\qquad
  S_{\chi \bar Z}^{01}( \tx^{\pm0}_{\L1}, \tx^{\pm}_{\R2} )\,,
 \eal
 and we assume that the S-matrix elements for massless particles of any chirality have the same functional dependence on rapidity variables.
 
These  S-matrix elements must satisfy the crossing equations \eqref{mixed_mass_crossing_eq_01a} and  \eqref{mixed_mass_crossing_eq_01b}.
Since the crossing equations \eqref{mixed_mass_crossing_eq_01a} are the same as for the scattering elements 
$S_{YY}^{m,1}$, $S_{Y\bar Z}^{m,1}$ of massive particles~\eqref{eq:massivemassivecross} in the limit where the mass $m$ of the first particle is taken to 0, their solutions are given by the same limit of $S_{YY}^{m,1}$, $S_{Y\bar Z}^{m,1}$ up to solutions of homogeneous crossing equations.

Thus, the S-matrix elements are given by
\bal
\label{eq:S01_crossing_splitting}
 S_{\chi Y}^{01}( \tx^{\pm0}_{\L1}, \tx^{\pm}_{\L2} ) &= A_{\chi Y}^{01}( \tx^{\pm0}_{\L1}, \tx^{\pm}_{\L2} )  \left(\Sigma^{01}_{\L\L}( \tx^{\pm0}_{\L1}, \tx^{\pm}_{\L2} ) \right)^{-2} \,,\
 \\
 S_{\chi \bar Z}^{01}( \tx^{\pm0}_{\L1}, \tx^{\pm}_{\R2} ) &= A_{\chi \bar Z}^{01}( \tx^{\pm0}_{\L1}, \tx^{\pm}_{\R2} )  \left(\Sigma^{01}_{\L\R}( \tx^{\pm0}_{\L1}, \tx^{\pm}_{\R2} ) \right)^{-2}\,.
 \eal
 Here $A^{01}$ are simple ``rational'' factors 
 \bal
\tilde A^{01}_{\chi Y }(u_1,u_2)&=H^{01}_{\chi Y }(u_1,u_2)\, \frac{ \tx_{ \L 1}^{+0}}{\tx_{\L 1}^{-0}}\,\frac{\tx_{\L 2}^{-}}{\tx_{\L 2}^{+}} \,\left(\frac{ \tx_{ \L 1}^{-0}-\tx_{\L 2}^{+}}{\tx_{\L 1}^{+0}-\tx_{\L 2}^{-}}\right)^2\,,
\\
A_{\chi \bar Z}^{01}( \tx^{\pm0}_{\L1}, \tx^{\pm}_{\R2} ) &=
  H_{\chi \bar Z}^{01}(u_1,u_2) \, \frac{\tx^+_{\R2}}{\tx^-_{\R2}} \, \frac{1 - \tx^{-0}_{\L1} \tx^{-}_{\R2}}{1 - \tx^{+0}_{\L1} \tx^{+}_{\R2}} \, \frac{1 - \tx^{+0}_{\L1} \tx^{-}_{\R2}}{1 - \tx^{-0}_{\L1} \tx^{+}_{\R2}} \,,
 \eal
 where $H^{01}$ satisfy the homogeneous crossing equations with respect to the second variable
\bal
\label{mixed_mass_crossing_eq_01c}
&H^{01}_{\chi Y}(u_1,u_2) H^{01}_{\chi \bar Z }( u_1,\bar u_2) = 1\,,
\qquad
H^{01}_{\chi\bar Z }(u_1,u_2) H^{01}_{\chi Y }( u_1,\bar u_2) = 1\,.
\eal
Notice that up to these $H$ functions, the dressing factors have been fixed to be the $m\to0^+$ limit of the dressing factors of the massive-massive elements defined in equation (F.3) in~\cite{Frolov:2025uwz}.
The dressing factors $\Sigma^{01}_{ab}$ are obtained from the massive ones by taking the mass of the first particle to 0, and are given by 
\begin{eqnarray}
\label{sigma01ab}
&\displaystyle\left(\Sigma^{01}_{ab}(\tx^{\pm0}_{a1},\tx^\pm_{b2})\right)^{-2} &= \left(\Sigma^{01\barnes}_{ab}(\tg^{\pm0}_{a1},\tg^\pm_{b2})\right)^{-2} \left({\Sigma^{01\bes}_{ab}(\tx^{\pm0}_{a1},\tx^\pm_{b2})\ov \Sigma^{01\hl}_{ab}(\tx^{\pm0}_{a1},\tx^\pm_{b2})}\right)^{-2}\,,\\
\label{sigma01oddll}
&\displaystyle\left(\Sigma^{01\barnes}_{\L\L} (\tx^{\pm0}_{\L1},\tx^\pm_{\L2}) \right)^{-2}&=\frac{R^2(\tg^{-0-}_{\L\L}) R^2(\tg^{+0+}_{\L\L})}{R^2(\tg^{-0+}_{\L\L}) R^2(\tg^{+0-}_{\L\L}) }\,,\\
\label{sigma01oddlr}
&\displaystyle\left( \Sigma^{\barnes}_{\L\R} (\tg^{\pm0}_{\L1}, \tg^\pm_{\R 2}) \right)^{-2}&=\frac{R(\tg^{-0+}_{\L\R}+ i \pi) R(\tg^{-0+}_{\L\R}- i \pi) R(\tg^{+0-}_{\L\R}+ i \pi)R(\tg^{+0-}_{\L\R}- i \pi)}{R(\tg^{-0-}_{\L\R}+ i \pi) R(\tg^{-0-}_{\L\R}- i \pi) R(\tg^{+0+}_{\L\R}+ i \pi) R(\tg^{+0 +}_{\L\R}- i \pi)} \,.
\end{eqnarray}
Since the massless Zhukovsky variables $\tx^{\pm0}_{\L1}$ are real, one has to deform slightly the integration contours in the BES and HL dressing factors. Indeed, in the limit $\lim_{m \to 0^+} \tx^{\pm m}_{\L1}$ the Zhukovsky variables approach the integration contours of BES and HL\footnote{One can also work with underformed contours and introduce a small regulator to assign a mass $\delta \ll 1$ to the massless particle.}.

The S-matrix elements also have to satisfy the crossing equations \eqref{mixed_mass_crossing_eq_01b}. Since the crossing equations for the dressing factors $\Sigma^{01}_{ab}$ can be derived from their explicit form, see appendix \ref{app:crossingm0m1},  the crossing equations \eqref{mixed_mass_crossing_eq_01b} lead to additional crossing equations on $H^{01}$. By using the results in appendix  \ref{app:crossingm0m1}, it is straightforward to show that they are
\bal
\label{mixed_mass_crossing_eq_01d}
&H^{01}_{\chi Y}(u_1,u_2) H^{01}_{\chi Y }( \bar u_1, u_2) = {\ta_{\L2}^-\ov\ta_{\L2}^+}\,,
\qquad
H^{01}_{\chi\bar Z }(u_1,u_2) H^{01}_{\chi \bar Z }( \bar u_1, u_2) = \frac{\ta^+_{\R2}}{\ta^-_{\R2}}\,.
\eal
Here and in what follows we often use the notation ($a=$\,R\,,L\,,\, $j=1,2$)
\bal\label{eq:def_alpha}
\ta_{aj}^{\pm} \equiv \a_a(\tx_{a j}^{\pm})\,,\quad \ta_{aj}^{\pm m} \equiv \a_a(\tx_{a j}^{\pm m})\,,\qquad\a_{aj}^{\pm} \equiv \a_a(x_{a j}^{\pm})\,,\quad \a_{aj}^{\pm m} \equiv \a_a(x_{a j}^{\pm m})\,,
\eal 
where the functions $\a_a(x)$ are defined in \eqref{eq:alphaa}.

The simplest solutions to \eqref{mixed_mass_crossing_eq_01c} and  \eqref{mixed_mass_crossing_eq_01d} that appear to be the correct ones are
\bal
H^{01}_{\chi Y}(u_1,u_2)={\sqrt{ \ta_{\L2}^-}\ov \sqrt{\ta_{\L2}^+}}\,,\qquad  H^{01}_{\chi\bar Z }(u_1,u_2) =\frac{\sqrt{\ta^+_{\R2}}}{\sqrt{\ta^-_{\R2}}}\,.
\eal
 The choice of the square-root branches is such as to ensure good properties under fusion.
 
To show the S-matrix normalisation we are using, and for the reader's convenience let us write the S-matrix elements in a more explicit form 
\bal\la{eq:massless_massive_m0m1}
 S_{\chi Y}^{01}( \tx^{\pm0}_{\L1}, \tx^{\pm}_{\L2} ) &={\sqrt{ \ta_{\L2}^-}\ov \sqrt{\ta_{\L2}^+}}\, \frac{ \tx_{ \L 1}^{+0}}{\tx_{\L 1}^{-0}}\,\frac{\tx_{\L 2}^{-}}{\tx_{\L 2}^{+}} \,\left(\frac{ \tx_{ \L 1}^{-0}-\tx_{\L 2}^{+}}{\tx_{\L 1}^{+0}-\tx_{\L 2}^{-}}\right)^2\,  {1\ov \Sigma^{01}_{\L\L}( \tx^{\pm0}_{\L1}, \tx^{\pm}_{\L2})^{2}} \,,\
 \\
 S_{\chi \bar Z}^{01}( \tx^{\pm0}_{\L1}, \tx^{\pm}_{\R2} ) &=\frac{\sqrt{\ta^+_{\R2}}}{\sqrt{\ta^-_{\R2}}}\, \frac{\tx^+_{\R2}}{\tx^-_{\R2}} \, \frac{1 - \tx^{-0}_{\L1} \tx^{-}_{\R2}}{1 - \tx^{+0}_{\L1} \tx^{+}_{\R2}} \, \frac{1 - \tx^{+0}_{\L1} \tx^{-}_{\R2}}{1 - \tx^{-0}_{\L1} \tx^{+}_{\R2}} \, {1\ov \Sigma^{01}_{\L\R}( \tx^{\pm0}_{\L1}, \tx^{\pm}_{\R2})^{2}}\,.
 \eal
By using fusion, one can also find the S-matrix elements $S_{\chi Y}^{0m}\,,\,  S_{\chi \bar Z}^{0m}$ for the scattering of massless particles with $m$-particle bound states, see appendix \ref{app:S-matrix_elements}. They can also be obtained  from the S-matrix elements $S_{Y Y}^{\delta m}\,,\,  S_{Y \bar Z}^{\delta m}$ by taking the mass $\delta$ of the first particles to 0
\bal\la{eq:Smirroromlimit}
S_{\chi Y}^{0m}( \tx^{\pm0}_{\L1}, \tx^{\pm m}_{\L2} ) &={\sqrt{ \ta_{\L2}^{-m}}\ov \sqrt{\ta_{\L2}^{+m}}}\, \lim_{\delta\to 0}\, S_{Y Y}^{\delta m}( \tx^{\pm \delta}_{\L1}, \tx^{\pm m}_{\L2} )\, ,
\\
S_{\chi \bar Z}^{0m}( \tx^{\pm0}_{\L1}, \tx^{\pm m}_{\R2} ) &=\frac{\sqrt{\ta^{+m}_{\R2}}}{\sqrt{\ta^{-m}_{\R2}}}\,\, \lim_{\delta\to 0}\, S_{Y \bar Z}^{\delta m}( \tx^{\pm \delta}_{\L1}, \tx^{\pm m}_{\R2} )\,.
\eal

\paragraph{Massless-massless S-matrix elements.}

In the case of massless-massless scattering, we assume that the S-matrix elements for massless particles of any chirality have the same functional dependence on rapidity variables. Thus,   we only need to consider the only S-matrix element
\bal
  S^{00}_{\chi \chi}( \tx^{\pm0}_{\L1}, \tx^{\pm 0}_{\L2} )\,.
 \eal
 It satisfies the following crossing equation
 \bal
\label{mixed_mass_crossing_eq_00a}
&S^{00}_{\chi \chi}(u_1,u_2) S^{00}_{\chi \chi}(\bar u_1,u_2) = -\frac{ \tx_{\L2}^{-0}}{\tx_{\L2}^{+0}} \Bigl(\frac{\tx^{+0}_{\L1}-\tx^{+0}_{\L2}}{\tx^{+0}_{\L1}-\tx^{-0}_{\L2}} \Bigl)^2\,,
\eal
where the minus sign appears due to the nontrivial exchange relations for massless particles discussed in the previous section.

Up to the sign the crossing equation \eqref{mixed_mass_crossing_eq_00a} is the same as  \eqref{mixed_mass_crossing_eq_01b} for the scattering element 
$S_{\chi Y}^{01}$ in the limit where the mass of the second particle is taken to 0. 
Therefore, the S-matrix element $ S^{00}_{\chi \chi}$ takes the following form
\bal
\label{eq:massless_massless_m0m0}
  S^{00}_{\chi \chi}(u_1,u_2)&=
    +{\sqrt{\ta_{\L1}^{+0}} \ov \sqrt{\ta_{\L1}^{-0}}} \,  {\sqrt{\ta_{\L2}^{-0}} \ov \sqrt{\ta_{\L2}^{+0}}}  \ \frac{\tx^{+0}_{\L1}}{\tx^{-0}_{\L1}}\, \frac{\tx^{-0}_{\L2}}{\tx^{+0}_{\L2}} \,  \left(\frac{\tx^{-0}_{\L1} - \tx^{+0}_{\L2}}{\tx^{+0}_{\L1} - \tx^{-0}_{\L2}} \right)^2 \left(\Sigma^{00}_{\L\L}(u_1,u_2)\right)^{-2}\,,
\eal
where $\Sigma^{00}_{\L\L}$ is given by \eqref{sigma01ab} and 
\eqref{sigma01oddll} with massive rapidities replaced by massless ones. It can also be obtained from $S^{0m}_{\chi Y}$ by taking $m$ to zero and using \eqref{eq:Smirroromlimit}.

The braiding unitarity of \eqref{eq:massless_massless_m0m0} requires adding  the extra 
$\sqrt{\ta_{\L1}^{+0}}/\sqrt{\ta_{\L1}^{-0}}$ factor.  Remarkably, this factor also produces the minus sign in the crossing equation \eqref{mixed_mass_crossing_eq_00a}. Indeed, the crossing transformation for massless Zhukovsky variables is \eqref{crxtpm}
\bal\la{crxtpm22}
\tx_\L^{\pm 0}\ \xrightarrow{\text{massless crossing}} \  \tx_\L^{\mp 0} +i0\,,
\eal
Since for positive momentum $\tx_\L^{\pm 0}$ do not cross the cut $(0,\xi)$ of $\sqrt{\a_\L}$ while for negative momentum both $\tx_\L^{\pm 0}$  cross the cut $(-\infty,-1/\xi)$, we get  
\bal
{\sqrt{\a_\L(\tx_{\L1}^{+0})} \ov \sqrt{\a_\L(\tx_{\L1}^{-0})}} \ \xrightarrow{\text{massless crossing}} \  {\sqrt{\a_\L(\tx_{\L1}^{-0}+i0)} \ov \sqrt{\a_\L(\tx_{\L1}^{+0}+i0)}} = - {\sqrt{\a_\L(\tx_{\L1}^{-0})} \ov \sqrt{\a_\L(\tx_{\L1}^{+0})}} \,.
\eal
Thus, the S-matrix element \eqref{eq:massless_massless_m0m0} satisfies the crossing equation  \eqref{mixed_mass_crossing_eq_00a}.

It is worth noting that the massless-massless S-matrix can equivalently be obtained by taking the $m\to0^+$ limit of right particles, $\tx^{\pm m}_{\R}\to \tx^{\pm 0}_{\R}$. Then, using that $\tx^{\pm 0}_{\R}=1/\tx^{\mp 0}_{\L}$, the result can be recast in terms of the left-massless kinematics, and it agrees with the one above. This can be seen by using the results of appendix~\ref{app:results_using_LR_massless},
which imply that
\begin{equation}
\lim_{\delta \to 0^+}  \frac{S^{0 \delta}_{\chi \bar{Z}}(\tx^{\pm 0}_{\L1}, \tx^{\pm \delta}_{\R2})}{S^{0 \delta}_{\chi Y}(\tx^{\pm 0}_{\L1}, \tx^{\pm \delta}_{\L2})} = \frac{\ta_{\L1}^{+0}}{\ta_{\L1}^{-0}}\,,
\end{equation}
where after the limit we used $\tx^{\pm 0}_{\R}=1/\tx^{\mp 0}_{\L}$.
It is then natural to identify the massless S-matrix as the following limit of the mixed-mass S~matrices
\begin{equation}
S^{00}_{\chi \chi} (\tx^{\pm 0}_{\L1}, \tx^{\pm 0}_{\L2})
= \frac{\sqrt{\ta_{\L1}^{-0}}}{\sqrt{\ta_{\L1}^{+0}}}\,\lim_{\delta \to 0^+}  S^{0 \delta}_{\chi \bar{Z}}(\tx^{\pm 0}_{\L1}, \tx^{\pm \delta}_{\R2})
= \frac{\sqrt{\ta_{\L1}^{+0}}}{\sqrt{\ta_{\L1}^{-0}}}\,\lim_{\delta \to 0^+}  S^{0 \delta}_{\chi Y}(\tx^{\pm 0}_{\L1}, \tx^{\pm \delta}_{\L2}) \,.
\end{equation}

Finally, let us comment on the plus sign in \eqref{eq:massless_massless_m0m0}. Naively, to guarantee the condition $S^{00}_{\chi \chi}(u,u)=-1$ one would have expected an overall minus sign. Indeed, for equal rapidities the rational factor, and the odd factor are obviously equal to~1 while the BES and HL factors are obtained by exponentiating their phases which are skew-symmetric, and therefore we would expect the phases to vanish if rapidities are equal. 
There is, however, a loophole in this consideration for what concerns the massless HL phase, which is not continuous as $u_2\to u_1$. This is because to define the BES and HL phases for massless particles one has to deform the integration contours in the $\tPhi^\bes$ and $\tPhi^\hl$ functions. For finite $h$, i.e.\ for the BES phase, one can always do the deformation for any values of the rapidities $\tx^{\pm0}_{\L j}=\tx_\L(u_j\pm i0)$, $j=1,2$ the phase depends upon. For infinite $h$, i.e. for the HL phase, the deformation of the integration contours in $\tPhi^\hl(\tx^{+0}_{\L 1},\tx^{-0}_{\L 2})$ and $\tPhi^\hl(\tx^{-0}_{\L 1},\tx^{+0}_{\L 2})$  can be done only for $u_1\neq u_2$. As a result, the $u_2\to u_1$ limit of massless HL phase is ambiguous. The massless HL factor nonetheless is well defined, and it is equal to $-1$ for equal rapidities as it can be checked numerically. In fact, this can be most easily seen by using a different representation of the HL massless factor. Amazingly, in analogy with the HL factor for the RR case (see eq.~\eqref{eq:app_massless_HL_RamRam}), we have found and checked numerically that the massless HL factor can be expressed in terms of Barnes functions as follows 
\bal\la{eq:HLvsBarnes}
\Sigma^{\hl}_{\L\L}(\tx^{\pm0}_{\L 1},\tx^{\pm0}_{\L 2})^{2} &= e^{2i\tilde\theta^\hl(\tx^{\pm0}_{\L 1},\tx^{\pm0}_{\L 2})}
\\
&=- {R(\tg^{+0-0}_{\L\L}+2\pi i)^2 R(\tg^{+0-0}_{\L\L})^2R(\tg^{-0+0}_{\L\L}-2\pi i)^2R(\tg^{-0+0}_{\L\L} )^2\ov R(\tg^{-0-0}_{\L\L})^4 R(\tg^{+0+0}_{\L\L})^4}\,,
\eal
where 
\bal
\tg^{\pm0\pm0}_{\L\L} = \tg^{\pm0}_{\L1} - \tg^{\pm0}_{\L2}\,. 
\eal
The equation \eqref{eq:HLvsBarnes} also agrees with the relativistic limit discussed in section \ref{sec:rel_limit}.

The r.h.s. of the formula \eqref{eq:HLvsBarnes} is well defined for any $\tg$'s and admits a straightforward analytic continuation to any region. Moreover, taking into account that the product of the odd and HL factors gives
\bal\la{eq:HLvsBarnesb}
\Sigma^{00\barnes}_{\L\L}&(\tx^{\pm0}_{\L 1},\tx^{\pm0}_{\L 2})^{-2}\Sigma^{00\hl}_{\L\L}(\tx^{\pm0}_{\L 1},\tx^{\pm0}_{\L 2})^{2} =- {R(\tg^{+0-0}_{\L\L}+2\pi i)^2R(\tg^{-0+0}_{\L\L}-2\pi i)^2\ov R(\tg^{-0-0}_{\L\L})^2 R(\tg^{+0+0}_{\L\L})^2}
\\
&=- {{\ta_{\L1}^{-0}}\ov {\ta_{\L1}^{+0}} } \,  {{\ta_{\L2}^{+0}}\ov {\ta_{\L2}^{-0}}}  \ \frac{\tx^{-0}_{\L1}}{\tx^{+0}_{\L1}}\, \frac{\tx^{+0}_{\L2}}{\tx^{-0}_{\L2}}\left(\frac{\tx^{+0}_{\L1} - \tx^{-0}_{\L2}} {\tx^{-0}_{\L1} - \tx^{+0}_{\L2}}\right)^2 {R(\tg^{+0-0}_{\L\L})^2R(\tg^{-0+0}_{\L\L})^2\ov R(\tg^{-0-0}_{\L\L})^2 R(\tg^{+0+0}_{\L\L})^2}\,,
\eal
we find that the S-matrix element \eqref{eq:massless_massless_m0m0} takes the following simple form
\bal
\label{eq:massless_massless_m0m0b}
  S^{00}_{\chi \chi}(u_1,u_2)&=
    -{\sqrt{\ta_{\L1}^{-0}} \ov \sqrt{\ta_{\L1}^{+0}}} \,  {\sqrt{\ta_{\L2}^{+0}} \ov \sqrt{\ta_{\L2}^{-0}}} {R(\tg^{+0-0}_{\L\L})^2R(\tg^{-0+0}_{\L\L})^2\ov R(\tg^{-0-0}_{\L\L})^2 R(\tg^{+0+0}_{\L\L})^2}\left(\Sigma^{00\bes}_{\L\L}(u_1,u_2)\right)^{-2}\,.
\eal

\paragraph{On the crossing equation with trivial exchange relations.}
Had we assumed trivial exchange relations, we would have found the crossing equation
\begin{equation}
\label{eq:oldcrossing}
S^{00}_{\chi \chi}(u_1,u_2) S^{00}_{\chi \chi}(\bar u_1,u_2) = +\frac{ \tx_{\L2}^{-0}}{\tx_{\L2}^{+0}} \Bigl(\frac{\tx^{+0}_{\L1}-\tx^{+0}_{\L2}}{\tx^{+0}_{\L1}-\tx^{-0}_{\L2}} \Bigl)^2\,,    
\end{equation}
which differs from~\eqref{massless_crossing_eq_00} by an overall sign. When $\kappa=0$, this equation can be solved by introducing the function~\cite{Frolov:2021fmj}
\begin{equation}
\label{eq:aofgamma}
    a(\tilde\gamma) = -i\,\tanh\left(\frac{\tilde\gamma}{2}-\frac{i\pi}{4}\right)\,,\qquad a(\tilde\gamma)a(\tilde\gamma+i\pi)=-1\,.
\end{equation}
This however introduces a zero for $\tilde\gamma_1-\tilde\gamma_2=i\pi/2$, so in this sense this solution is ``less minimal'' than the one of equation~\eqref{massless_crossing_eq_00}. This factor is also incompatible with the analytic structure of the quantum spectral cure (as highlighted in~\cite{Ekhammar:2024kzp}), and it complicates the cut structure of the pure-RR Y system~\cite{Cavaglia:toappear}. 
Things are even worse in the $\kappa>0$ case (where one must distinguish $\tilde\gamma_{\L1}^{+0} -\tilde\gamma_{\L2}^{+0}\neq\tilde\gamma_{\L1}^{-0}-\tilde\gamma_{\L2}^{-0}$, etc.). In this case we could not find a solution of the crossing equation~\eqref{eq:oldcrossing} compatible with the symmetries of the model --- in particular, we could find a solution which is however incompatible with $SU(2)_\circ$ symmetry. We regard this as a further indication that the correct crossing equation is that related to semionic statistics~\eqref{massless_crossing_eq_00}.

\subsection{String theory S-matrix elements} The  S-matrix elements \eqref{eq:massless_massive_m0m1} \eqref{eq:massless_massless_m0m0} can be used to derive
the corresponding string theory S-matrix elements. To this end one analytically continues the mirror variables $\tx_\L^{+0}\,,\, \tg_\L^{+0}$ and $\tx_a^{+}\,,\, \tg_a^{+}$ to their string regions where one can use the string variables $x_\L^{+0}\,,\, \g_\L^{+0}$ and $x_a^{+}\,,\, \g_a^{+}$ . As discussed in section~\ref{sec:kinematics}, the analytic continuation path depends on whether one wants to get an S-matrix element with positive momenta in the interval $(0,2\pi)$ or with negative ones in the interval $(-2\pi,0)$. 
One can also use fusion to find the S-matrix elements $S_{\chi Y}^{0m}\,,\,  S_{\chi \bar Z}^{0m}$ for the scattering of massless particles with $m$-particle string bound states.

We will derive explicit expressions for the string S-matrix elements in appendix \ref{app:stringSmatrix}. They make the physical unitarity  manifest.
Here we present some of the results obtained in appendix \ref{app:stringSmatrix} for positive momenta. Relations involving S-matrix elements with negative momenta will be discussed in subsection \ref{sec:cont_arb_mom}. In this subsection we write the string S-matrix elements as functions of particle's momenta which take values in the interval $(0,2\pi)$. 

First, $S^{0m}_{\chi Y }(p_1,p_2)$  is given by 
\bal\la{eq:SchiY0mstr}
S^{0m}_{\chi Y }(p_1,p_2)&=
{ \sqrt{\a_{\L2}^{+m}}\ov \sqrt{\a_{\L2}^{-m}}}\, \left(\frac{ x_{ \L 1}^{+0}}{x_{\L 1}^{-0}}\right)^{m}\,
 \left(\sigma^{0m}_{\L\L}(p_1,p_2)\right)^{-2}
\,,
\eal
where $\a_{aj}^{\pm m}$ are defined in \eqref{eq:def_alpha}  and $\sigma^{0m}_{\L\L}$ is the string dressing factor, cf (G.12) of \cite{Frolov:2025uwz} 
\bal\la{eq:sigmaLL0mstr}
\sigma^{0m}_{\L\L}(p_1,p_2)^{-2}&=
(-1)^{m-1}
 \left( \frac{x_{\L 1}^{-0} }{x_{\L 1}^{+ 0} }\right)^{m-1} \,  \frac{x_{\L 2}^{-m}}{x_{\L 2}^{+m}}\, \left({ \Gamma\big[{m\ov2}-\tfrac{ih}{2}u_{12}\big]\ov \Gamma\big[{m\ov2}+\tfrac{ih}{2}u_{12}\big]}\right)^2
\frac{u_{12}+ \frac{i}{h}m}{u_{12}- \frac{i}{h}m} \, \Xi^{0m}_{\L\L}(p_1,p_2)^{-2}
\,,
\eal
and
\bal\la{eq:XiLL0mstr}
\Xi^{0m}_{\L\L}(p_1,p_2)^{-2}&=
\Sigma_{\L\L}^\barnes(\g_{\L1}^{\pm0},\g_{\L2}^\pm)^{-2}
  \, e^{-2i\delta^{0m}_{\L\L}}
\,,
\eal
\bal\label{eq:deltaLL0mstr}
2\delta^{0m}_{\L\L}=& +2\tilde\Phi_{\L\L}(x_{\L1}^{\pm 0},x_{\L 2}^{\pm m})-2\tilde\Phi_{\L\L}^\hl(x_{\L1}^{\pm 0},x_{\L 2}^{\pm m})
\\
& -2\tilde\Psi_{\L}(u_1,x_{\L2}^{+ m})+2\tilde\Psi_{\L}(u_1,x_{\L2}^{-m})+\tilde\Psi^+_{\L}(u_2+{i\ov h}m ,x_{\L1}^{+ 0})-\tilde\Psi^+_{\L}(u_2+{i\ov h}m,x_{\L1}^{- 0})
\\
&+\tilde\Psi^{-}_{\L}(u_2-{i\ov h}m,x_{\L1}^{+0})-\tilde\Psi^{-}_{\L}(u_2-{i\ov h}m,x_{\L 1}^{-0})
\,.
\eal
Eqs. \eqref{eq:SchiY0mstr} and \eqref{eq:deltaLL0mstr} do not depend on constituent particles of the bound state, and  \eqref{eq:SchiY0mstr} satisfies the physical unitarity.

Similarly, $S^{0m}_{\chi \bar Z }(p_1,p_2)$  is given by 
\bal\label{eq:SchibarZ0mstr}
S^{0m}_{\chi \bar Z}(p_1,p_2)  &=
   \frac{\sqrt{\a^{-m}_{\R2}}}{\sqrt{\a^{+m}_{\R2}}} \, \left( \frac{x_{\L 1}^{+0} }{x_{\L 1}^{- 0} }\right)^{m+1}\,
    \frac{1 - x^{-0}_{\L1} x^{-m}_{\R2}}{1 - x^{+0}_{\L1} x^{+m}_{\R2}} \,
     \frac{1 - x^{-0}_{\L1} x^{+m}_{\R2}}{1 - x^{+0}_{\L1} x^{-m}_{\R2}} \,
 \left(\sigma^{0m}_{\L\R}(p_1,p_2)\right)^{-2}
\,,
\eal
where $\sigma^{0m}_{\L\R}$ is the string dressing factor, cf (G.22) of \cite{Frolov:2025uwz}
\bal\la{eq:sigmaLR0mstr}
\sigma^{0m}_{\L\R}(p_1,p_2)^{-2}&=
(-1)^{m-1}
 \left( \frac{x_{\L 1}^{-0} }{x_{\L 1}^{+ 0} }\right)^{m-1} \,  \frac{x_{\R 2}^{-m}}{x_{\R 2}^{+m}}\, \left({ \Gamma\big[{m\ov2}-\tfrac{ih}{2}u_{12}\big]\ov \Gamma\big[{m\ov2}+\tfrac{ih}{2}u_{12}\big]}\right)^2
\frac{u_{12}+ \frac{i}{h}m}{u_{12}- \frac{i}{h}m} \, \Xi^{0m}_{\L\R}(p_1,p_2)^{-2}
\,,
\eal
and
\bal\la{eq:XiLR0mstr}
\Xi^{0m}_{\L\R}(p_1,p_2)^{-2}&=
\Sigma_{\L\R}^\barnes(\g_{\L1}^{\pm0},\g_{\R2}^\pm)^{-2}
  \, e^{-2i\delta^{0m}_{\L\L}}
\,,
\eal
\bal\label{eq:deltaLR0mstr}
2\delta^{0m}_{\L\R}=& +2\tilde\Phi_{\L\R}(x_{\L1}^{\pm 0},x_{\R 2}^{\pm m})-2\tilde\Phi_{\L\R}^\hl(x_{\L1}^{\pm 0},x_{\R 2}^{\pm m})
\\
& -2\tilde\Psi_{\R}(u_1,x_{\R 2}^{+ m})+2\tilde\Psi_{\R}(u_1,x_{\R 2}^{-m})+\tilde\Psi^+_{\L}(u_2+{i\ov h}m ,x_{\L1}^{+ 0})-\tilde\Psi^+_{\L}(u_2+{i\ov h}m,x_{\L1}^{- 0})
\\
&+\tilde\Psi^{-}_{\L}(u_2-{i\ov h}m,x_{\L1}^{+0})-\tilde\Psi^{-}_{\L}(u_2-{i\ov h}m,x_{\L 1}^{-0})
\,.
\eal
Equation~\eqref{eq:SchibarZ0mstr} also satisfies the physical unitarity.

When both particles are massless, we have
\bal\label{eq:S00_string_with_deltaLL}
S^{00}_{\chi \chi }(p_1,p_2)&= 
 {\sqrt{\a_{\L1}^{-0}} \ov \sqrt{\a_{\L1}^{+0}}} \, { \sqrt{\a_{\L2}^{+0}}\ov \sqrt{\a_{\L2}^{-0}}}\,
 \left(\sigma^{00}_{\L\L}(p_1,p_2)\right)^{-2}
\,,
\eal
where $\sigma^{00}_{\L\L}$ is the string dressing factor, cf (G.12) of \cite{Frolov:2025uwz} 
\bal\la{eq:sigmaLL00str}
\sigma^{00}_{\L\L}(p_1,p_2)^{-2}&=
-
\frac{x_{\L 1}^{+0} }{x_{\L 1}^{- 0} }\,  \frac{x_{\L 2}^{-0}}{x_{\L 2}^{+0}}\, \left({ \Gamma\big[-\tfrac{ih}{2}u_{12}\big]\ov \Gamma\big[+\tfrac{ih}{2}u_{12}\big]}\right)^2
\, \Xi^{00}_{\L\L}(p_1,p_2)^{-2}
\,,
\eal
and
\bal\la{eq:XiLL00str}
\Xi^{00}_{\L\L}(p_1,p_2)^{-2}&=
\Sigma_{\L\L}^\barnes(\g_{\L1}^{\pm0},\g_{\L2}^{\pm0})^{-2}
  \, e^{-2i\delta^{00}_{\L\L}}
\,,
\eal
\bal\label{eq:deltaLL00str}
2\delta^{00}_{\L\L}=&+2\tilde\varPhi_{\L\L}(x_{\L1}^{\pm 0},x_{\L 2}^{\pm 0})-2\tilde\varPhi_{\L\L}^\hl(x_{\L1}^{\pm 0},x_{\L 2}^{\pm 0})
-2\tilde\Psi_{\L}(u_1,x_{\L2}^{+ 0})+2\tilde\Psi_{\L}(u_1,x_{\L2}^{-0})
\\
& +2\tilde\Psi_{\L}(u_2 ,x_{\L1}^{+ 0})-2\tilde\Psi_{\L}(u_2 ,x_{\L1}^{- 0})
\,.
\eal
We can also use the formula \eqref{eq:massless_massless_m0m0b} to do the analytic continuation to the string region. Then, we get 
\bal\la{eq:S00str4}
S^{00}_{\chi \chi }(p_1,p_2)=& -
 {\sqrt{\a_{\L1}^{+0}} \ov \sqrt{\a_{\L1}^{-0}}} \, { \sqrt{\a_{\L2}^{-0}}\ov \sqrt{\a_{\L2}^{+0}}}\, \frac{x^{+0}_{\L1}}{x^{-0}_{\L1}}\, \frac{x^{-0}_{\L2}}{x^{+0}_{\L2}}\left(\frac{x^{-0}_{\L1} - x^{+0}_{\L2}} {x^{+0}_{\L1} - x^{-0}_{\L2}}\right)^2 
 \\
& \times 
\frac{R^2(\g^{+0-0}_{\L\L}) R^2(\g^{-0+0}_{\L\L})}{R^2(\g^{-0-0}_{\L\L}) R^2(\g^{+0+0}_{\L\L}) }  \,
{1\ov \Sigma^{\bes}_{\L\L}(x_{\L1}^{\pm 0},x_{\L 2}^{\pm 0})^2}
\,,
\eal
where $\Sigma^{\bes}_{\L\L}(x_{\L1}^{\pm 0},x_{\L 2}^{\pm 0})$ is the analytically continued BES factor, see \eqref{eq:BESstring00}.
The expressions in~\eqref{eq:S00_string_with_deltaLL} and \eqref{eq:S00str4} also manifestly satisfy both physical and braiding unitarity, and the condition $S^{00}_{\chi \chi }(p,p)=-1$.

Similar to the mirror mixed-mass and massless S-matrix elements, the string S-matrix elements turn out to be the limit of the string massive elements $S_{Y Y}^{\delta m}\,,\,  S_{Y \bar Z}^{\delta m}$:
\bal\la{eq:Sstringomlimit}
S_{\chi Y}^{0m}( x^{\pm0}_{\L1}, x^{\pm m}_{\L2} ) &={\sqrt{ \a_{\L2}^{+m}}\ov \sqrt{\a_{\L2}^{-m}}}\, \lim_{\delta\to 0^+}\, S_{Y Y}^{\delta m}( x^{\pm \delta}_{\L1}, x^{\pm m}_{\L2} )\, ,
\\
S_{\chi \bar Z}^{0m}( x^{\pm0}_{\L1}, x^{\pm m}_{\R2} ) &=\frac{\sqrt{\a^{-m}_{\R2}}}{\sqrt{\a^{+m}_{\R2}}}\,\, \lim_{\delta\to 0^+}\, S_{Y \bar Z}^{\delta m}( x^{\pm \delta}_{\L1}, x^{\pm m}_{\R2} )\,.
\eal
Note that the factors of $\sqrt{\a_a}$ in \eqref{eq:Sstringomlimit} are inverted in comparison to \eqref{eq:Smirroromlimit}. This is because the analytic continuation paths are different for massless and massive particles.

\medskip

In the next sections we will discuss various checks of the proposed S-matrix elements.

\section{Verifying the proposal}
\label{sec:checks}

Here we check that our proposal passes all available self-consistency checks and that it reproduces the existing perturbative results.

\subsection{Crossing symmetry and braiding unitarity}

Invariance under crossing as well as    braiding unitarity was a guiding principle in the construction of the dressing factor. Braiding unitarity of the S~matrix is manifest, given the form of the dressing factors. As for crossing, the construction in the previous section of the phases from a limit of the massive ones~\cite{Frolov:2025uwz} makes it manifest that they satisfy crossing in the mirror kinematics. Crossing in the string kinematics then follows by analytic continuation of the crossing equations as a whole.

\subsection{Parity in the mirror kinematics}
The parity transformation in the mirror model corresponds to sending $\tp_a\to-\tp_a$ with $\tE_a\to\tE_a$, without any analytic continuation. In fact, such an analytic continuation would not even exist for massless particles, as positive- and negative-momentum branches are not connected, as remarked in section~\ref{sec:kinematics}.
We can analyse the behaviour of the various pieces of the S-matrix elements under parity. The transformation $\tx_a^{\pm}\to-1/\tx_a^{\mp}$ (and similarly for massless variables) can be easily implemented on the rational pre-factors; note in particular that
\begin{equation}
    \frac{\sqrt{\alpha_\L(\tx_\L^+)}}{\sqrt{\alpha_\L(\tx_\L^-)}}\quad\longrightarrow\quad \frac{\sqrt{\alpha_\L(\tx_\L^-)}}{\sqrt{\alpha_\L(\tx_\L^+)}}\,,
\end{equation}
under such a transformation. In this way, the rational pieces cancel out  by themselves in the equations~\eqref{eq:mirror-P}. Similarly, recalling the transformation property of mirror rapidities
\begin{equation}
    \tg(\tx)\quad\longrightarrow\quad -\tg(\tx)\,,\qquad \Im(\tx)<0\,,
\end{equation}
and the fact that the ``odd'' dressing factors are of difference form in the rapidities, we get a straightforward cancellation in~\eqref{eq:mirror-P}. Finally, for BES and HL phases it was verified in~\cite{Frolov:2025uwz} that
\begin{equation}
    \tilde\theta_{ab}(\tx_{a1}^\pm,\tx_{b2}^\pm)=\tilde\theta_{ba}\big(-\frac{1}{\tx_{b2}^\mp},-\frac{1}{\tx_{a1}^\mp}\big)\,.
\end{equation}
This identity makes no assumption on the value of the mass: it remains true if one or both particles' masses are taken to zero. This observation completes the check of~\eqref{eq:mirror-P}.

\subsection{CT invariance in the mirror kinematics}
\label{sec:CT}

To show CT invariance~\eqref{eq:CTinv} it is useful to note that, for the mixed-mass phases,
\bal
&\left(\Sigma^{01\besratio}_{\L a} (\tx^{\pm0}_{\L1},\tx^\pm_{a2})^{-2} \right)^*=\Sigma^{01\besratio}_{\L \bar a} (\tx^{\pm0}_{\L1},\tx^\pm_{\bar a 2})^{+2}\,,\qquad a=\text{\small R,L}\,,
\eal
Using the relations \eqref{eq:normS0mtildechiZ} and 
\eqref{eq:normS0mtildechibarY} between the highest-weight and lowest-weight S-matrix elements, and noting that under conjugation mirror massive excitations transform as
\bal
\left(\tx^{\pm}_a\right)^*= \frac{1}{\tx^{\mp}_{\bar{a}}} \,,\qquad  \left(\tg^{\pm}_a\right)^*= \tg^{\mp}_{\bar{a}} \,,
\eal
and mirror massless ones as
\bal
\left(\tx^{\pm0}_a\right)^*= \tx^{\pm0}_a + i0\,,\qquad  \left(\tg^{\pm0}_a\right)^*= \tg^{\pm0}_a \pm i\pi \,
\eal
it is easy to check the CT conditions~\eqref{eq:CTinv}. It is worth noting that the would-be unitarity condition
\bal
\label{eq:massless_massless_unitarity}
  S^{00}_{\chi \chi}(u_1,u_2)\, S^{00}_{\chi \chi}(u_1,u_2)^*&=
   \frac{\tx^{+0}_{\L1}}{\tx^{-0}_{\L1}}\, \frac{\tx^{-0}_{\L2}}{\tx^{+0}_{\L2}} \, \left(\frac{\tx^{-0}_{\L1} - \tx^{+0}_{\L2}}{\tx^{+0}_{\L1} - \tx^{-0}_{\L2}} \right)^2\stackrel{?}{=}1\,.
\eal
is not satisfied unless $\kappa=0$. Instead, the corresponding CT condition
\bal
\label{eq:massless_massless_unitarity2}
  S^{00}_{\chi \chi}(u_1,u_2)\, S^{00}_{\tilde\chi \tilde\chi}(u_1,u_2)^*&=
   1\,,
\eal
is valid for any~$\kappa$.

\subsection{Continuation to arbitrary momentum in the string kinematics}
\label{sec:cont_arb_mom}
As discussed in Section~\ref{sec:proposal}, it is possible to obtain the string dressing factor by analytically continuing the mirror one. Because the model is massless, there are different branches that are given by non-equivalent choices of analytic continuation. Following the various paths outlined in Figures~\ref{fig:positive-p-path}~and~\ref{fig:negative-p-path} we obtain various expressions which happen to satisfy simple relations with the dressing factors in the $0<p_j<2\pi$ region, which were given in eqs.~\eqref{eq:SchiY0mstr}, \eqref{eq:SchibarZ0mstr} and~\eqref{eq:S00_string_with_deltaLL}. In the rest of this section we assume that any momentum $p\in(0,2\pi)$, explicitly writing $p-2\pi$ to denote negative momentum in the interval $(-2\pi,0)$.

\paragraph{Massless--left scattering.}
We can  express the S-matrix element $S^{0m}_{\chi Y }$ with one or both negative momenta in terms of S-matrix elements with positive momenta as follows
\begin{equation}
\label{eq:arb_mom_massless_left1}
\begin{aligned}
S^{0m}_{\chi Y} (p_1-2 \pi, p_2) &= {\sqrt{ \a_{\L2}^{-m}}\ov \sqrt{\a_{\L2}^{+m}}}\,S^{km}_{\bar{Z}Y} (p_1, p_2) \,,\qquad &&0<m \\
S^{0m}_{\chi Y} (p_1, p_2-2 \pi) &=  {\a_{\L1}^{+0}\ov \a_{\L1}^{-0}} S^{0,k- m}_{\chi\bar{Z}} (p_1, p_2)\,,\qquad &&0<m<k \,,
\\
S^{0,m+k}_{\chi Y}(p_1,p_2-2\pi)&= - S^{0m}_{\chi Y}(p_1,p_2)\,,\quad &&0<m\,,
\end{aligned}
\end{equation}\begin{equation}
\label{eq:arb_mom_massless_left2}
\begin{aligned}
S^{0m}_{\chi Y} (p_1-2 \pi, p_2-2 \pi) &= {\sqrt{ \a_{\R2}^{+(k-m)}}\ov \sqrt{\a_{\R2}^{-(k-m)}}}\,\frac{\a_{\R 1}^{-k}}{\alpha_{\R 1}^{+k}}\  S^{k,k-m}_{\bar{Z} \bar{Z}} (p_1, p_2)\,,\quad &&0<m<k\,,
\\
S^{0,m+k}_{\chi Y} (p_1-2 \pi, p_2-2 \pi) &=- {\sqrt{ \a_{\L2}^{-m}}\ov \sqrt{\a_{\L2}^{+m}}}\,S^{km}_{\bar{Z}Y} (p_1, p_2)\,,\qquad &&0<m\,.
\end{aligned}
\end{equation}

\paragraph{Massless--right scattering.}
For  $S^{0m}_{\chi \bar Z }$ with negative momenta we find
\begin{equation}
\label{eq:arb_mom_massless_right}
\begin{aligned}
S_{\chi \bar Z}^{0m}( p_1-2\pi, p_2) &= 
     \frac{\sqrt{\a^{+m}_{\R2}}}{\sqrt{\a^{-m}_{\R2}}} \,
S_{\bar Z \bar Z}^{km}( p_1, p_2)\,,\qquad&&0<m\,,
\\
S_{\chi\bar Z}^{0m}(p_1,p_2-2\pi) &= - S_{\chi\bar Z}^{0,m+k}(p_1,p_2)\,, \qquad&&0<m\,,
\\
S_{\chi \bar Z}^{0m}( p_1-2\pi, p_2-2\pi) &= 
   - \frac{\sqrt{\a^{+(m+k)}_{\R2}}}{\sqrt{\a^{-(m+k)}_{\R2}}} \,
S_{\bar Z \bar Z}^{k,m+k}( p_1, p_2) \,,\qquad&&0<m\,.
\end{aligned}
\end{equation}
 
\paragraph{Massless--massless scattering.}
Finally, for massless--massless scattering we have
\bal
S^{00}_{\chi \chi}(p_1-2 \pi,p_2)&=  {\sqrt{\a_{\L2}^{-0}} \ov \sqrt{\a_{\L2}^{+0}}} \, S_{\bar{Z} \chi}^{k0}(p_1, p_2)\,,\\
S^{00}_{\chi \chi}(p_1,p_2-2\pi)&=  {\sqrt{\a_{\L1}^{+0}} \ov \sqrt{\a_{\L1}^{-0}}} \, S_{\chi\bar{Z}}^{0k}(p_1, p_2)\,,\\
S^{00}_{\chi \chi}(p_1-2\pi,p_2-2\pi)&= {\sqrt{\a_{\R1}^{-k}} \ov \sqrt{\a_{\R1}^{+k}}} {\sqrt{\a_{\R2}^{+k}} \ov \sqrt{\a_{\R2}^{-k}}} \, S^{kk}_{\bar{Z} \bar{Z}} (p_1, p_2) \,.
\eal
These formulae clearly show that the massless particles with momentum in the interval $(-2\pi,0)$ are related to the bound states of $k$ right particles.  

\paragraph{Scattering of left $k$-particle bound states with $p\in(-2\pi,0)$.}
There is also a similar relation of a left $k$-particle bound state with momentum in the interval $(-2\pi,0)$ to a massless particle with momentum in $(0,2\pi)$. This is a consequence of the following identities
\bal
 S_{YY}^{km}(p_{1}-2\pi,p_{2}) &= { \sqrt{\a_{\L2}^{-m}}\ov \sqrt{\a_{\L2}^{+m}}}\,
 S_{\chi Y}^{0m}(p_{1},p_{2})\,,\quad &&0<m \,,
 \\
 S_{ Y  \bar Z}^{km}(p_{1}-2\pi,p_{2}) &= { \sqrt{\a_{\R2}^{+m}}\ov \sqrt{\a_{\R2}^{-m}}}\,
 S_{\chi \bar Z}^{0m}(p_{1},p_{2}) \,,\quad &&0<m
 \,.
\eal
The analytic continuation of these formulae in $p_2$ gives
\bal 
 S_{YY}^{km}(p_{1}-2\pi,p_{2}-2\pi) &= -{ {\a_{\L1}^{+0}}\ov {\a_{\L1}^{-0}}}\, { \sqrt{\a_{\R2}^{+(k-m)}}\ov \sqrt{\a_{\R2}^{-(k-m)}}}\,
 S_{\chi \bar Z}^{0,k-m}(p_{1},p_{2}) \,,\quad &&0<m<k
 \,,
 \\
  S_{YY}^{k,m+k}(p_{1}-2\pi,p_{2}-2\pi) &= { \sqrt{\a_{\L2}^{-m}}\ov \sqrt{\a_{\L2}^{+m}}}\,
 S_{\chi Y}^{0m}(p_{1},p_{2})\,,\quad &&0<m\,,
 \\
 S_{ Y  \bar Z}^{km}(p_{1}-2\pi,p_{2}-2\pi) &= { \sqrt{\a_{\R2}^{+(m+k)}}\ov \sqrt{\a_{\R2}^{-(m+k)}}}\,
 S_{\chi \bar Z}^{0,m+k}(p_{1},p_{2}) \,,\quad &&0<m
 \,.
\eal
Taking the limit $m\to 0$  of the relations above, we also obtain
\bal\la{eq:rel3}
S_{ \chi  Y}^{0k}(p_{1},p_{2}-2\pi) &= -{ \sqrt{\a_{\L1}^{+0}}\ov \sqrt{\a_{\L1}^{-0}}}\,
 S_{\chi \chi}^{00}(p_{1},p_{2})\,,
 \\
   S_{ \chi  Y}^{0k}(p_{1}-2\pi,p_{2}-2\pi) &= -{ \sqrt{\a_{\R1}^{-k}}\ov \sqrt{\a_{\R1}^{+k}}}\,{ \sqrt{\a_{\L2}^{-0}}\ov \sqrt{\a_{\L2}^{+0}}}\,
 S_{\bar Z \chi}^{k0}(p_{1},p_{2})\,,
 \\
  S_{YY}^{kk}(p_{1}-2\pi,p_{2}-2\pi) &={ \sqrt{\a_{\L1}^{+0}}\ov \sqrt{\a_{\L1}^{-0}}} { \sqrt{\a_{\L2}^{-0}}\ov \sqrt{\a_{\L2}^{+0}}}\,
 S_{\chi \chi}^{00}(p_{1},p_{2})\,.
 \eal
Note that the first two relations in \eqref{eq:rel3} can also be obtained as the limit $m\to0$ of the relations for $S^{0,m+k}_{\chi Y}(p_1,p_2-2\pi)$ and $S^{0,m+k}_{\chi Y}(p_1-2\pi,p_2-2\pi)$.

\subsection{CP in the string kinematics}

As for the massive case~\cite{Frolov:2025uwz} the check of CP for the mixed-mass and massless S-matrices is complicated since it connects different regions of the string theory (associated with $0<p<2\pi$ and $-2 \pi<p<0$) and the S-matrix matrix must be continued to these regions in a non-trivial way.

Let us sketch first the check of CP for the mixed-mass case. 
There are two different kinematical configurations one should analyse: the case involving particles with the same sign of the momenta and the case involving particles with opposite momentum signs. These two configurations are reached by performing two separate analytic continuations of the mixed-mass S-matrix.
To write the two CP constraints explicitly we consider $p_1$ and $p_2$ to be in the interval $(0, 2 \pi)$ and check that
\bal
\label{eq:sec_CP_relations_mixed}
&S^{0m}_{\chi Y} (p_1-2 \pi, p_2-2 \pi) =S^{m0}_{\bar Y\tilde{\chi}} (2 \pi-p_2, 2 \pi-p_1)  \,,\\
&S^{0m}_{\chi Y} (p_1-2 \pi, p_2) =S^{m0}_{\bar Y\tilde{\chi}} (-p_2, 2\pi- p_1)\,.
\eal
Two similar kinematical configurations must be considered also for massless S-matrices where instead we need to check
\bal
\label{eq:sec_CP_relations_massless}
&S^{0 0}_{\tilde{\chi} \tilde{\chi}} (2 \pi - p_2, 2 \pi - p_1)=S^{0 0}_{\chi \chi} (p_1 -2 \pi, p_2 -2 \pi) \,,\\
&S^{0 0}_{\chi \chi} (2 \pi -p_1, p_2-2 \pi) =S^{0 0}_{\tilde{\chi} \tilde{\chi}} (2 \pi - p_2, p_1 - 2\pi)\,.
\eal
Making use of the relations listed in section~\ref{sec:cont_arb_mom} the constraints~\eqref{eq:sec_CP_relations_mixed} and~\eqref{eq:sec_CP_relations_massless} in the $u$ plane are given by\footnote{Actually the second equation in \eqref{eq:sec_CP_relations_mixed} should not lead to a new constraint since it corresponds to the analytic continuation of the first equation under $p_2 \to p_2 +2 \pi$. We will consider the two equations independently anyway.} 
\bal
&S^{k,k-m}_{\bar{Z} \bar{Z}} (u_1, u_2)={\sqrt{ \a_{\R2}^{-(k-m)}}\ov \sqrt{\a_{\R2}^{+(k-m)}}}\,\frac{\alpha_{\R 1}^{+k}}{\alpha_{\R 1}^{-k}} S^{m0}_{\bar Y\tilde{\chi}} (-u_2, -u_1)\,,\\
&S^{k m}_{\bar{Z} Y} (u_1, u_2)=  -\frac{\sqrt{\a^{+m}_{\L2} }}{\sqrt{\a^{-m}_{\L2} }} S^{m+k ,0}_{\bar Y\tilde{\chi}} (-u_2, - u_1) \,,
\eal
and
\bal
&S^{0 0}_{\tilde{\chi} \tilde{\chi}} (-u_2, -u_1)=+{\sqrt{\a_{\R1}^{-k}} \ov \sqrt{\a_{\R1}^{+k}}} {\sqrt{\a_{\R2}^{+k}} \ov \sqrt{\a_{\R2}^{-k}}} \, S^{kk}_{\bar{Z} \bar{Z}} (u_1, u_2)\,,\\
&S^{0 k}_{\chi \bar{Z}} (-u_1, -u_2) = -{\sqrt{\a_{\R1}^{+k}} \ov \sqrt{\a_{\R1}^{-k}}}  \,   {\sqrt{\a_{\L2}^{+0}} \ov \sqrt{\a_{\L2}^{-0}}}   \,  S^{0k}_{\tilde{\chi} \bar{Y}} (u_2, u_1) \,.
\eal
respectively.
We will explicitly show how to connect the CP constraints in momentum and $u$ plane parameterisations in appendix~\ref{app:checking_CP}. In the same appendix we show that all CP constraints are verified by our proposal for the dressing factors.

\subsection{Near-BMN expansion}

It is possible to expand the S-matrix elements at large string tensions, both in the mirror and string kinematics. These results can be compared, at least at tree level, with the perturbative S-matrix, which has been computed in~\cite{Hoare:2013pma} for the string model and in~\cite{Baglioni:2023zsf} for the mirror model.
One-loop results for the mixed-mass and massless S-matrices were obtained in~\cite{Sundin:2015uva,Sundin:2016gqe} for the pure Ramond-Ramond case. In the presence of mixed-flux, only S-matrices between massive excitations have been computed to higher loop orders~\cite{Sundin:2014ema,Bianchi:2014rfa,Roiban:2014cia} while interactions with massless modes are known only at tree-level~\cite{Hoare:2013pma}.
However, already in the pure-RR setup the computations are both infrared and ultraviolet divergent, and it is not clear whether the regularisation schemes adopted e.g.\ in~\cite{Sundin:2015uva,Sundin:2016gqe} are compatible with integrability, or whether some additional counterterms may be needed. This makes the one-loop perturbative S-matrices not completely reliable. A more extended discussion can be found in~\cite{Frolov:2021fmj}.

In either kinematics, we take the string tension $T\gg1$ with~$\ka$ fixed. More precisely, we have
\begin{equation}
 \frac{k}{2\pi}= q\, T\,,\qquad h=\sqrt{1-q^2}\,T  + \cO(1)\,,\qquad    \frac{k}{2\pi h}=\frac{\ka}{2\pi}\xrightarrow{T\to\infty}\frac{q}{\sqrt{1-q^2}}\,,
\end{equation}
where $0\leq q\leq1$ interpolates between the pure-RR backgrounds ($q=0$) and the pure-NSNS ones ($q=1$). Moreover, we will take the string momentum (respectively, the mirror energy) to be small, of order $\mathcal{O}(T^{-1})$.

\paragraph{Physical vs.~ZF S~matrix.}
Before proceeding with the comparison, it is important to remark a distinction between the S~matrix $S_{ij}^{kl}(p_1,p_2)$ which appears in the ZF algebra, and the physical S~matrix~$\mathbb{S}_{ij}^{kl}(p_1,p_2)$ which is defined as the operator allowing to express \textit{in} asymptotic states on a basis of \textit{out} asymptotic states~\cite{Smirnov:1990pr,Frolov:2025ozz}. This distinction is important in the presence of non-trivial exchange relations, like in this case. In fact, in our case the exchange relations are diagonal,
\begin{equation}
\label{eq:exchangerel}
    a^\dagger_{i}(p_1)\;a^\dagger_{j}(p_2)
    =
    e^{-2\pi i\, s_{ij}\,\text{sgn}(v_1-v_2)}\,
    a^\dagger_{j}(p_2)\;a^\dagger_{i}(p_1)\,,
\end{equation}
where $a^{\dagger}$ are \textit{in} (respectively, \textit{out}) creation operators, and $v=\partial \omega/\partial p$ is the group velocity of either excitation. Eq.~\eqref{eq:exchangerel} is written in the string kinematics, but analogous formulae hold in the mirror kinematics.%
\footnote{Strictly speaking, the mirror velocity $\tilde{v}=\partial \tilde\omega/\partial \tilde{p}$ can be complex, but we assume all exchange relations which involve massive particles to be trivial, i.e.~$s_{jk}=0$ or $s_{jk}=1/2$ when at least one of the particles is massive, in which case there is no need to introduce the sign function $\text{sgn}(v_1-v_2)$.}
In this case, the physical S~matrix~$\mathbb{S}_{ij}^{kl}$ is related to the ZF S~matrix~$S_{ij}^{kl}$ as
\begin{equation}
\label{eq:physicalS}
    \mathbb{S}_{ij}^{kl}(p_1,p_2)= e^{+2\pi i\,s_{kl}\,\text{sgn}(v_1-v_2)}\,S_{ij}^{kl}(p_1,p_2)\,,
\end{equation}
and similarly for the mirror kinematics. 
Note in particular that the physical S~matrix can be expanded around the identity, $\mathbb{S}=\mathbb{I}+\dots$, but does not necessarily obey the Yang--Baxter equation; conversely, the ZF S~matrix $S$ obeys the Yang--Baxter equation, but reduces to something nontrivial in the ``free'' limit.
Referring the reader to~\cite{Frolov:2025ozz} and references therein for a more detailed discussion, we emphasise that when comparing with perturbative results we must consider the physical S~matrix~$\mathbb{S}$. In the case at hand we conjecture that 
\begin{equation}
    s_{ij}=\begin{cases}
    \frac{1}{4}&\text{if both }i,j\text{ are massless,}\\
    \frac{1}{2}F_iF_j &\text{else,}
    \end{cases}
\end{equation}
where $F_i=1$ for excitations of type $\psi,\bar{\psi},\chi,\tilde\chi$ and $F_i=0$ for excitations of type $Z,\bar{Z},Y,\bar{Y},T$ --- that is to say, exchange relations are standard apart from massless-massless ones.

\paragraph{Mirror kinematics.}
The mirror energy is given by the positive branch of
\begin{equation}
    \tp^2=\frac{k^2}{4 \pi^2} \tE^2 + 4 h^2 \sinh^2\frac{\tE}{2}\,,
\end{equation}
which in the near-BMN limit where $\tE=\tilde\omega/T$ yields the dispersion
\begin{equation}
    \tilde\omega(\tp) = \big|\tp\big|+\mathcal{O}(T^{-2})\,.
\end{equation}
As expected, this dispersion has two branches, which originate from the main cut of the $u$-plane (for $\tp>0$) and from the $\ka_a$ cut (or more specifically, from the $\ka_{\L}$ cut, since we typically describe massless excitations in the ``left'' kinematics). It is worth noting that the images of $u$-plane branch points go to a finite limit as $T\to\infty$
\begin{equation}
    \xi_\L=\frac{1}{\xi_\R}\equiv\xi \to \sqrt{\frac{1+q}{1-q}}\,.
\end{equation}
The massless Zhukovsky variables take values in the vicinity of the images of the branch points, namely
\begin{equation}
\label{eq:expansion_zhuk_var_massless}
\begin{aligned}
\tx^{\pm 0}_a&=+\xi_a \left(1 \mp \frac{\tilde{p}_a}{2T} +\frac{(3-q)\tilde{p}^2}{24T^2}\right) -i0\, \qquad &&\tp_a>0\,,\\
\tx^{\pm 0}_a&=-\frac{1}{\xi_a}  \left(1 \pm \frac{\tilde{p}_a}{2T} +\frac{(3+q)\tilde{p}^2}{24T^2}\right) -i0 \,, \qquad &&\tp_a<0\,.
\end{aligned}
\end{equation}
It is possible to derive similar formulae for massive particles and for the massive or massless rapidities. They are collected in appendix~\ref{app:BMN_limit} where they are used to expand the dressing factors. Much like in~\cite{Frolov:2025uwz}, the computation of the phases boils down to taking the double derivatives of the kernels. Of course it is necessary to ensure that the two particles which we scatter have group velocity $v_1>v_2$, which in the massless--massless scattering means $v_1>0>v_2$. Since in~\cite{Baglioni:2023zsf} the perturbative S matrix was computed for bosons, it is convenient to check the following (diagonal) S-matrix elements
\begin{equation}
\begin{aligned}
    \mathbb{S}_{T Y}^{01}(\tx_{\L1}^{\pm0},\tx_{\L2}^{\pm0})=1-\frac{i}{T}\tp_1\tp_2+\mathcal{O}(T^{-2})\,,\\
    \mathbb{S}_{T\bar{Z}}^{01}(\tx_{\L1}^{\pm0},\tx_{\R2}^{\pm})=1-\frac{i}{T}\tp_1\tp_2+\mathcal{O}(T^{-2})\,,\\
    \mathbb{S}_{TT}^{00}(\tx_{\L1}^{\pm0},\tx_{\L1}^{\pm0})=1-\frac{i}{T}\tp_1\tp_2+\mathcal{O}(T^{-2})\,.
\end{aligned}
\end{equation}
These results match the perturbative scattering results of~\cite{Baglioni:2023zsf}. For the massless-massless scattering it was crucial to keep track of an additional factor of $i$ coming from the nontrivial exchange relations~\cite{Frolov:2025ozz}, cf.\ eq.~\eqref{eq:physicalS}.

\paragraph{String kinematics.}
Similar considerations apply to the case of the string kinematics. One has to be careful to reproduce the correct branch of the string momentum. For instance, to reproduce the small and \textit{positive} momentum region, we proceed as outlined in Section~\ref{sec:kinematics} and obtain
\begin{equation}
    \tilde{x}_{\L1}^{+0}\to \frac{1}{\tilde{x}_{\R1}^{+0}}={x}_{\L1}^{+0}\,,\qquad
    \tilde{x}_{\L1}^{-0}\to \tilde{x}_{\L1}^{-0}={x}_{\L1}^{-0}\,,
\end{equation}
where the rapidity $u_1$ has gone trough the main mirror cut from above and accordingly $\tx^{+0}_{\L1}$ has crossed the real-$x$ line from below. Instead, to obtain \textit{negative} and small momentum we  start with Zhukovsky variables evaluated on the opposite edges of the $+ \ka$ cuts
\bal
\tx^{\pm 0}_{\L2}=\tx_{\L2}(u_2 + i \ka \pm i0)\,, \qquad u_2< - \nu\,.
\eal
Then we drag $u_2$ slightly below the cut, so that  $\tx^{+0}_{\L2}$ must be continued,
\bal
&\tx^{-0}_{\L2}(u_2 + i \ka )=\tx_{\L2}(u_2 + i \ka - i0)\,, \qquad \ u_2< - \nu\,,\\
&\tx^{+0}_{\L2}(u_2 + i \ka )=\frac{1}{\tx_{\R2}(u_2 - i \ka - i0)}\,, \qquad u_2< - \nu\,.
\eal
Then we move $u_2$ horizontally to the region $u_2 > -\nu$. The Zhukovsky variables after the continuation take the following expressions
\bal
&\tx^{-0}_{\L2}(u_2 + i \ka ) \to \frac{1}{x^{+ k}_{\R2}(u_2)}\,, \qquad  u_2> -\nu\,,\\
&\tx^{+0}_{\L2}(u_2 + i \ka )=\frac{1}{x^{-k}_{\R2}(u_2)}\,, \qquad \ u_2> -\nu\,,
\eal
in terms of the rapidity of a $k$-particle bound state. 
Having crossed the log-cut of the mirror energy, we get
\bal
\tE_{\L 2}^{(0)}= -\log x^{+k}_{\R}(u_2) + \log x^{-k}_{\R}(u_2) + 2 \pi i\,,
\eal
or in the string kinematics
\bal
p^{(0)}_{\L 2}= i \tE^{(0)}_{\L2}= i \left( \log x^{-k}_{\R}(u_2)- \log x^{+k}_{\R}(u_2) \right)-2 \pi= p^{(k)}_{\R 2} - 2 \pi  \in (-2 \pi, 0) \,.
\eal
To make the string momenta small, as it is needed for the near-BMN limit, we must set
\begin{equation}
    p_{\L1}^{(0)}=\frac{p_1}{T}+\mathcal{O}(T^{-1}),\qquad
    p_{\R2}^{(k)}=2\pi+\frac{p_2}{T}+\mathcal{O}(T^{-1})\,,\qquad p_1>0>p_2\,.
\end{equation}
Once again, the Zhukovsky variables take values near the images of the branch points, namely
\begin{equation}
\begin{aligned}
x_{\L1}^{\pm0}&=+\xi \left(1\pm \frac{i p_1}{2T} - (3-q) \frac{p^2_1}{24 T^2}\right)+ \mathcal{O}(T^{-3})\,,\\
\frac{1}{x^{\mp k}_{\R2}}&= -\frac{1}{\xi} \left( 1 \pm \frac{i p_2}{2 T} - (3+q) \frac{p^2_2}{24 T^2} \right)+ \mathcal{O}(T^{-3})\,.
\end{aligned}
\end{equation}
In this way, and taking care of performing the necessary analytic continuation of the dressing factors, we can evaluate the string S-matrix, as detailed in appendices~\ref{app:BMN_string_mixed} and~\ref{app:BMN_string_massless}. Again, both for the mixed-mass and massless scattering, we find agreement with the perturbative results~\cite{Hoare:2013pma}.

\subsection{Relativistic expansion}
\la{sec:rel_limit}

In~\cite{Frolov:2023lwd}, a particular relativistic limit of the model was considered, where $h \ll1$, $k$ was kept fixed, and the momentum was expanded around the minimum of the dispersion relation with fluctuations of order $h$. A similar limit was earlier considered in~\cite{Fontanella:2019ury}, where $h$ was assumed to be much smaller than the momentum fluctuation (as a consequence of this fact all modes became massless). Here we refer to the limit considered in~\cite{Frolov:2023lwd}, which better reproduces the bound state structure of the model.
As already shown in~\cite{Frolov:2025uwz}, the non-perturbative S-matrix for massive excitations matches the one bootstrapped in~\cite{Frolov:2023lwd} in the limit. This provides a nontrivial confirmation of the correctness for the proposal of the massive-massive S-matrices advanced in~\cite{Frolov:2025uwz}. 
In this section we check that this is also the case for S-matrix elements involving the scattering of massless modes, up to removing the factor
\bal
a(\theta)=-i \tanh \left( \frac{\theta}{2} -i\frac{\pi}{4} \right)
\eal
from the solution of~\cite{Frolov:2023lwd}\footnote{As in the pure Ramond-Ramond case, this factor was introduced in the relativistic study of the theory considered in~\cite{Frolov:2023lwd} to solve the crossing equations with an opposite sign compared to the one considered here. This is due to the triviality of the exchange relations assumed in that paper.}.
In particular, we will focus on the scattering of massless particles with positive velocities. The scattering of massless particles with negative velocities can be obtained by a parity transformation, and all the other processes have trivial S-matrices in the limit~\cite{Frolov:2023lwd}. Therefore this case provides a good representative of the limit.

In~\cite{Frolov:2023lwd} it was found that two highest-weight massless particles of the same chirality scatter with the following S-matrix element
\bal
S^{\rm rel}_{\chi\chi} (\theta)= a(\theta) \frac{R^2(\theta - i\pi)R^2(\theta + i\pi)}{R^4(\theta)}\,,
\eal
where $\theta=\theta_{12}=\theta_1-\theta_2$ is the difference of the rapidities of the scattered particles. Removing the factor $a(\theta)$, which with the current sign of the crossing equations is unnecessary, we expect the following relativistic limit of the S-matrix element
\bal
\label{eq:app_dressing_rel_paper}
S^{\rm rel}_{\chi\chi} (\theta)= -\frac{R^2(\theta - i\pi)R^2(\theta + i\pi)}{R^4(\theta)}\,.
\eal 
The overall sign of the S-matrix is not fixed by relativistic crossing, but we have chosen it so that
\bal
S^{\rm rel}_{\chi\chi} (0)=-1\,.
\eal
In this way, we have a well-behaved Bethe wave function\footnote{If we had solved the crossing equations with trivial exchange relations, we would have expected a factor $a(\theta)$, which would have also lead to $S^{\rm rel}_{\chi\chi} (0)=-1$ because $a(0)=-1$.}.
In appendix~\ref{app:rel_limit} we show that in the limit the proposal for the massless-massless S-matrix of this paper agrees with~\eqref{eq:app_dressing_rel_paper}.

\subsection{Pure-RR limit}
As a final check of our proposal, we study the $\ka \to 0$ limit of our solution and compare it with the results in~\cite{Frolov:2021fmj}. Strictly speaking, $k$ is a quantised parameter; there is then no strict reason why the limit should be smooth. It is anyway desirable (and a further indication of the correctness of our proposal) to have agreement with the pure-RR case.

For simplicity, we perform the limit directly in the mirror theory. The computation is performed in detail in appendix~\ref{appendix:pureRRcomparison}.
For the mixed-mass S~matrices, we find an exact agreement with the results in~\cite{Frolov:2021fmj,Frolov:2021bwp} (this is also the case for the massive-massive S~matrices, as found in~\cite{Frolov:2025uwz}). For the massless-massless case, we find an agreement with~\cite{Frolov:2021fmj,Frolov:2021bwp} up to the factor $a(\g)$, which in this paper was replaced by $-1$ (we refer to the discussion in section~\ref{sec:proposalmirror} for the reasoning of removing this term). A further observation is that the limit of the ratios of $\a$ functions appearing in the massless-massless S~matrix is $+1$ or $-1$ depending on whether the scattered particles have equal or opposite chiralities (more comments on this can be found in appendix~\ref{appendix:pureRRcomparison}). This leads to another small difference compared with the pure-RR case, where the sign was chosen to be the same in both chirality sectors. This sign does not affect the mirror TBA equations for the ground state, and it can only be fixed by studying the equations for excited states.

\section{Mirror Thermodynamic Bethe Ansatz}
\label{sec:proposalTBA}

In this section we conjecture the ground state TBA equations for the mixed-flux $AdS_3\times S^3\times T^4$ superstring. Since the mirror theory has a structure very similar to the pure RR one, we expect that the TBA equations have the same form as well. The main difference in comparison to the RR case is that the mixed-flux mirror theory is not unitary, and in particular for a real $u$-rapidity, both the mirror energy and momentum of massive particles are not real. Moreover, as we will see in a moment, for $m<k$ and real $u$ the real part of a mirror momentum is bounded from below. 
This suggests that the integration contours appearing in TBA equations are not real intervals on the $u$-plane.

\subsection{Real mirror momentum contour}

\begin{figure}[tbp]
\begin{center}
\includegraphics*[width=0.45\textwidth]{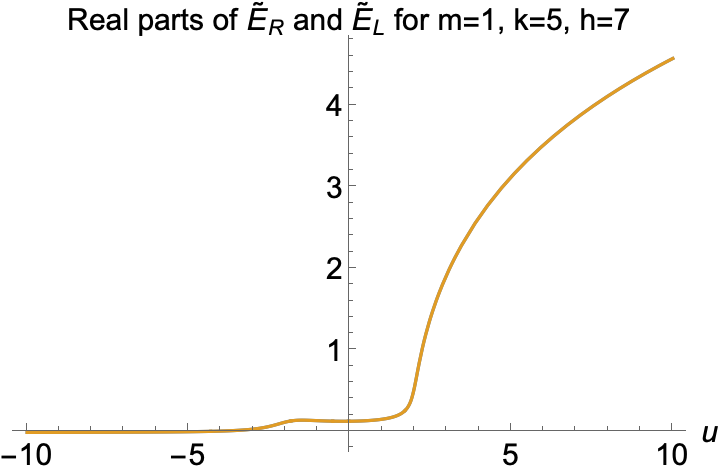} \quad \includegraphics*[width=0.45\textwidth]{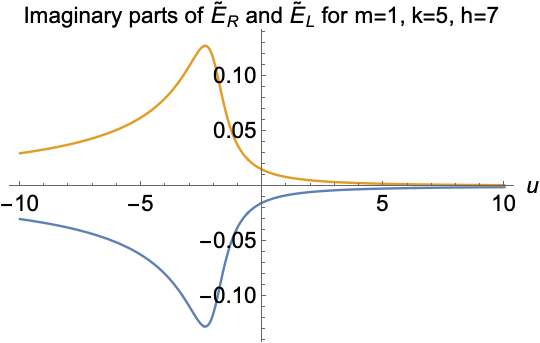} 
\end{center}
\begin{center}
\includegraphics*[width=0.45\textwidth]{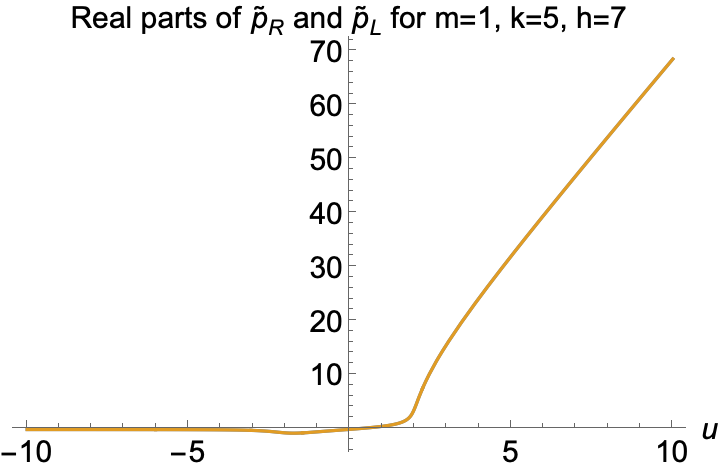} \quad \includegraphics*[width=0.45\textwidth]{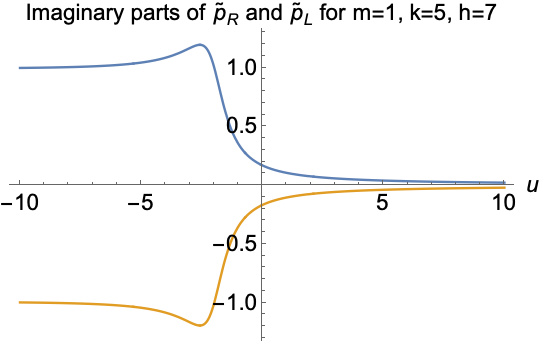} 
\caption{Graphs of real and imaginary parts of mirror energy and momenta as functions of real $u$ rapidity for $m<k$. The real part of mirror momentum is bounded from below. }
\la{fig:appTBA1}
\end{center}
\end{figure}

Recall that the mirror energy and momentum are given by
\begin{equation}
\begin{aligned}
\label{eq:mirrorEandp}
\tE_a^{(m)} &= \ln \tx_a^{-m} -\ln \tx_a^{+m}\,,\qquad
\tp_a^{(m)} = {h\ov2}\left( \tx_a^{-m} -{1\ov \tx_a^{-m}} - \tx_a^{+m} +{1\ov \tx_a^{+m}} \right)
\,.
\end{aligned}
\end{equation}
They satisfy the complex conjugation conditions
\begin{equation}
\begin{aligned}
\left(\tE_a^{(m)}(u) \right)^* &= \tE_{\bar a}^{(m)}(u^*)\,, \qquad &\left(\tp_a^{(m)}(u) \right)^* &= \tp_{\bar a}^{(m)}(u^*)\,,
\end{aligned}
\end{equation}
where, as usual, $\bar{\text{L}}=\text{R}$, $\bar{\text{R}}=\text{L}$. As a result, neither the mirror momentum nor the mirror energy are real for real~$u$. Instead, the images of the real-$u$ line  under $\tE_a^{(m)}(u)$ and $\tp_a^{(m)}(u)$ are curves in the complex plane, whose shape depends on $m$. For $m<k$ the real part of $\tE_\R(u)$ and $\tp_\R(u)$  (equal to the ones of $\tE_\L(u)$ and $\tp_\L(u)$ ) asymptote to 0 as $u\to-\infty$, and to $+\infty$ as  $u\to +\infty$ but they are not monotonic as  functions of $u$. The real part of $\tp_\R(u)$ (and $\tp_\L(u)$)  is bounded from below and asymptotes to 0. The imaginary parts of $\tE_\R(u)$, $\tE_\L(u)$ (and $\tp_\R(u)$, $\tp_\L(u)$) are conjugate to each other and bounded from above and below. They all asymptote to 0 as $u\to +\infty$. In the limit $u\to -\infty$ the imaginary parts of $\tE_\R(u)$, $\tE_\L(u)$ approach 0 while the imaginary parts of $\tp_\R(u)$, $\tp_\L(u)$ asymptote to $\pm m$, as depicted in Figure~\ref{fig:appTBA1}.  
The  dependence of $\tE_a$ and $\tp_a$ on $u$ is very different from the pure RR case ($k=0)$. The reason is that for $m<k$ and real $u$, both points $u+i{m\ov h}$ and $u-i{m\ov h}$ are either below or above a $\ka$-cut while in the RR case they are on opposite sides of the mirror cuts. 

For $m>k$ pictures of real parts of $\tE_a(u)$ and of $\tp_a(u)$ are similar to the ones for the $k=0$ case because $u\pm i{m\ov h}$ are on opposite sides of the two mirror cuts. Their imaginary parts are similar to the ones for $m<k$ case, see  Figure~\ref{fig:appTBA2}.
\begin{figure}[t]
\begin{center}
\includegraphics*[width=0.45\textwidth]{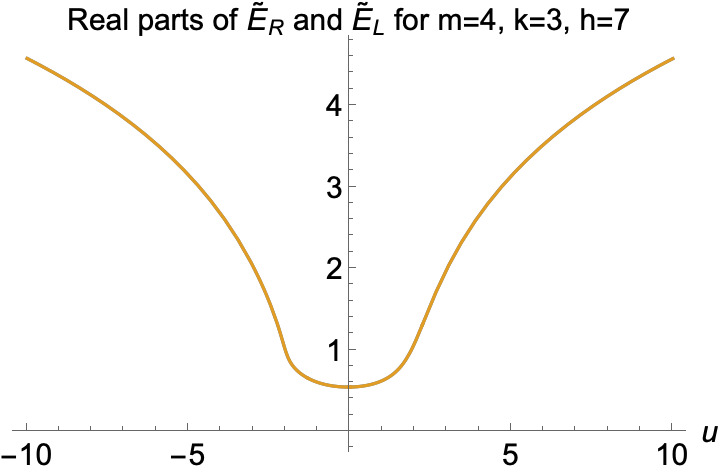} \quad \includegraphics*[width=0.45\textwidth]{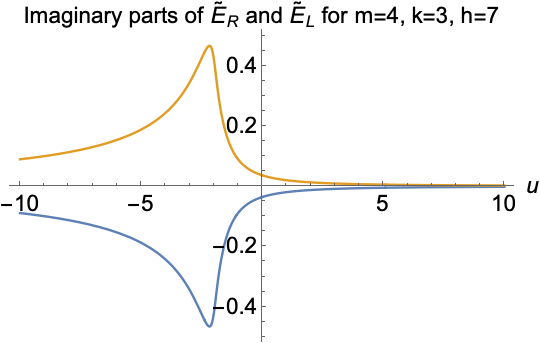} 
\end{center}
\begin{center}
\includegraphics*[width=0.45\textwidth]{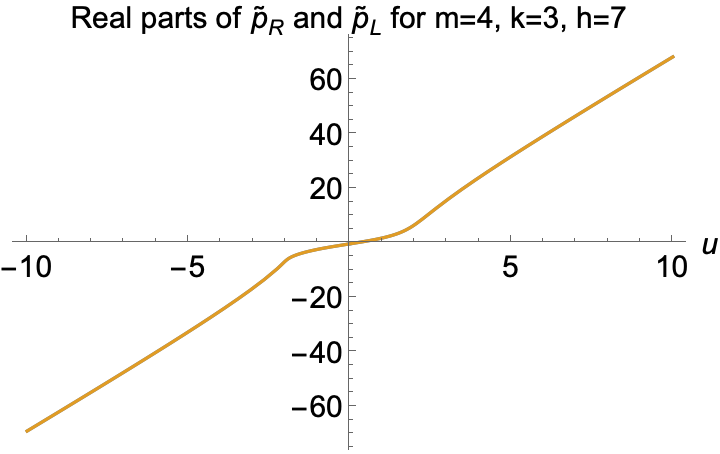} \quad \includegraphics*[width=0.45\textwidth]{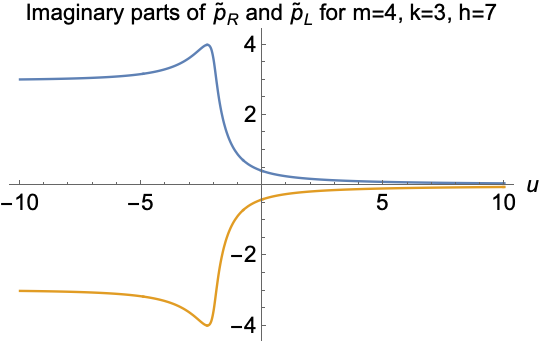} 
\caption{Graphs of real and imaginary parts of mirror energy and momenta as functions of real $u$ rapidity for $m>k$. The real part of mirror momentum runs from $-\infty$ to $+\infty$. }
\la{fig:appTBA2}
\end{center}
\end{figure}

\begin{figure}[ht]
\centering
\includegraphics[width=7cm]{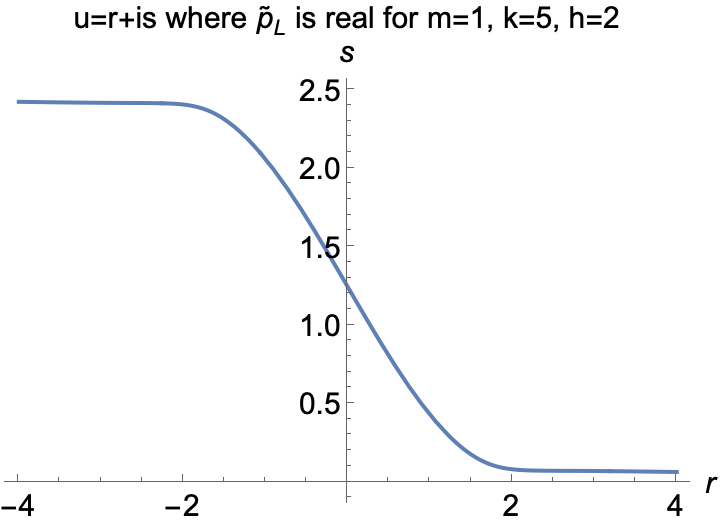}%
\hspace{5mm}%
\includegraphics[width=7cm]{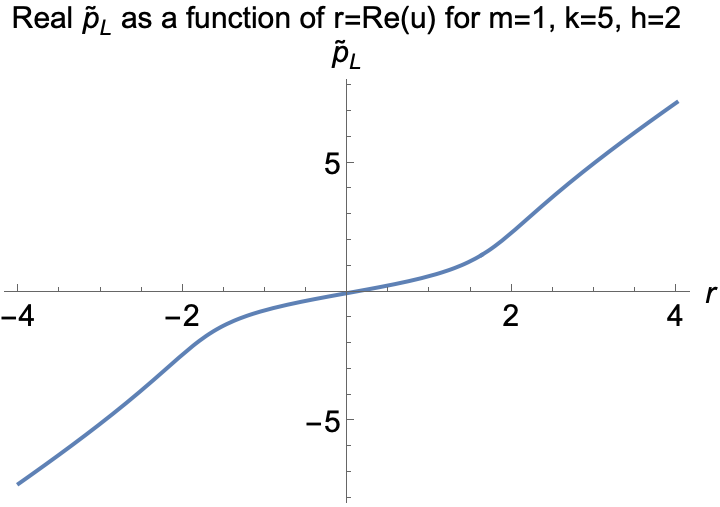}
\caption{\label{fig:realPtcurve}%
Left:  The curve on the $u$-plane where left mirror momentum $\tp_\L$ is real. The curve where  $\tp_\R$ is real is a reflection of this one about the real line. Right: The real $\tp_\L(r)=\tp_\R(r)$ as a function of $r=\Re(u)$. 
}
\end{figure}

\begin{figure}[ht]
\centering
\includegraphics[width=7cm]{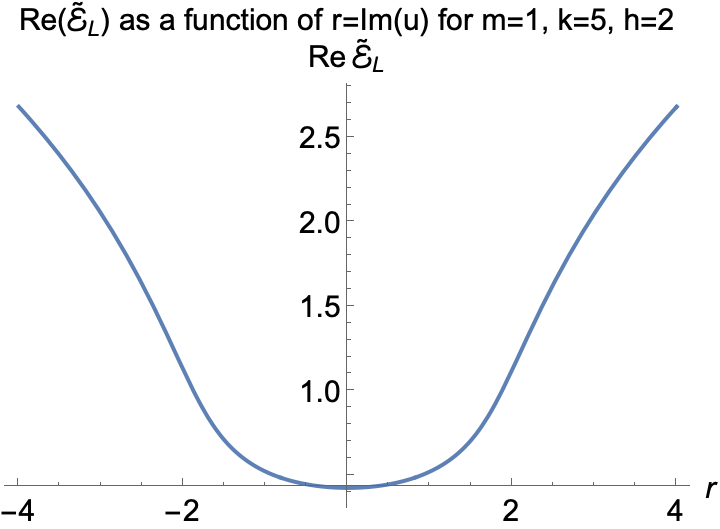}%
\hspace{5mm}%
\includegraphics[width=7cm]{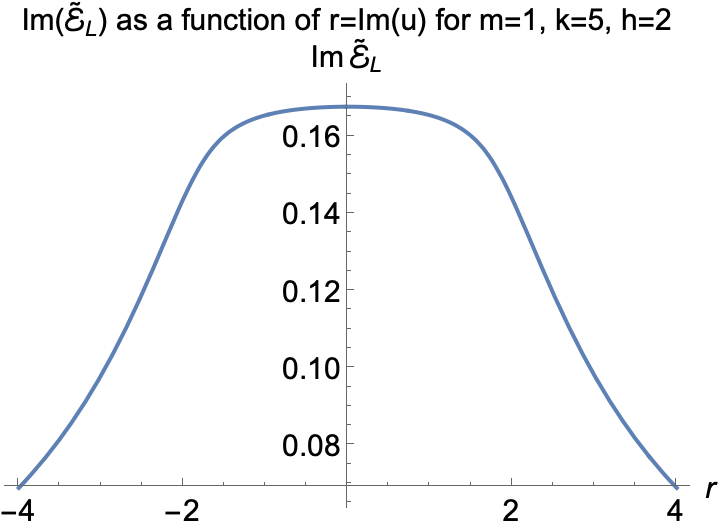}%
\caption{\label{fig:realEtcurve}%
Left: The real part of the mirror energy $\tE_\L(r)=\tE_\R(r)$ as a function of $r=\Re(u)$  for points on the  $u$-plane curves where mirror momenta are real. Right: The imaginary part of the left mirror energy $\tE_\L(r)$. The graph of $\Im\big(\tE_\L(r)\big)$ is a reflection of this one about the real line. 
}
\end{figure}

Such a dependence of $\tE_a$ and $\tp_a$ on $u$ for $m<k$ indicates that one should consider the values of $u$ for which the real part of mirror momentum runs from $-\infty$ to $+\infty$.  Since the parity transformation in the mirror theory corresponds to the map $u\to - u +i\ka_a$, and for any $m$ the real part of $\tp_a$ approaches $+\infty$ as $u\to+\infty$,  it is clear that the corresponding curve should asymptote to the main cut for $\Re(u)\to +\infty$ and to a $\ka_a$-cut for $\Re(u)\to -\infty$. Indeed, the reality of $\tp_a$ follows if $u=r+ i s$, $r,s\in\bR$ satisfies the following  equation
\begin{equation}
\Im\left( \tx_a(r + i s -i{m\ov h}) -{1\ov \tx_a(r + i s-i{m\ov h})} - \tx_a(r + i s +i{m\ov h}) +{1\ov \tx_a(r + i s+i{m\ov h})} \right) =0\,.
\end{equation}
Finding numerically $s$ as a function of real $r$, one can plot the real mirror momentum curve on the $u$-plane, and the graphs of mirror energy and momentum as functions of $r$,
see Figures~\ref{fig:realPtcurve}~and~\ref{fig:realEtcurve}.

Since the mixed-flux mirror model is not unitary the real mirror momentum curve is not the one that would appear in the TBA equations. Nevertheless, we expect that a TBA integration contour would approach the cuts as $r=\Re(u)$ goes to $\pm\infty$, and it would have similar shape.
As discussed in section~\ref{sec:kinematics}, for massless mirror particles the real mirror momentum and energy correspond to $u$ taking values on the mirror cuts. 

\subsection{Mirror Bethe-Yang equations}

In the following we label by $N_1$ the number of fundamental particles of type left, by $N_{\bar1}$ the number of fundamental particles of type right and by $N_0$ the total number of massless particles. Finally, we label by $N^{(\a)}_y$ the number of auxiliary roots of type $\a$ (this is an $\mathfrak{su}(2)_\bullet$ index for the different types of excitations). 

The Bethe-Yang equations are a straightforward generalisation of the equations written in~\cite{Frolov:2021bwp} for the pure RR-case.
For the fundamental particles we have
\bal
+1&=e^{i \tp_{k} R} \prod^{N_{1}}_{j \ne k} S^{1 1}_{Y Y } (u_{k}, u_{j}) \prod^{N_{\bar1}}_{j \ne k} S^{1 1}_{Y \bar{Z}}(u_k, u_j)  \prod^{N_0}_{j =1} S^{1 0}_{Y \chi} (u_k , u_j) \prod^{2}_{\a =1} \prod^{N^{(\a)}_y}_{j =1}  S^{1y}(u_k,y_j^{(\a)})\,,\\
+1&=e^{i \tp_{k} R} \prod^{N_{\bar{1}}}_{j \ne k} S^{1 1}_{\bar{Z} \bar{Z}}(u_k, u_j) \prod^{N_1}_{j \ne k} S^{1 1}_{\bar{Z} Y}(u_k, u_j) \prod^{N_0}_{j =1} S^{1 0}_{\bar{Z} \chi} (u_k , u_j) \prod^{2}_{\a =1} \prod^{N^{(\a)}_y}_{j =1} \overline{S}^{1y}(u_k,y_j^{(\a)})\,,\\
-1&=e^{i \tp_{k} R} \prod^{N_{1}}_{j \ne k} S^{0 1}_{\chi Y } (u_k, u_j) \prod^{N_{\bar1}}_{j \ne k} S^{0 1}_{\chi \bar{Z}}(u_k, u_j) \prod^{N_0}_{j =1} S^{0 0}_{\chi \chi} (u_k , u_j) \prod^{2}_{\a =1} \prod^{N^{(\a)}_y}_{j =1}  S^{0y}(u_k,y_j^{(\a)})\,.
\eal
The equations for the auxiliary roots are instead given by
\bal
-1= \prod^{N_{1}}_{j=1} S^{y1}(y_k^{(\a)},u_j) \, \prod^{N_{\bar1}}_{j=1} \overline{S}^{y1}(y_k^{(\a)},u_j) \,\prod^{N_0}_{j=1} S^{y0}(y_k^{(\a)},u_j)\,,
\eal
where the auxiliary S-matrices are defined by
\bal\label{eq:auxiliary_Smatrices}
S^{1y}(u,y)&= 
\frac{
\sqrt{\tx^-_{\L}}
}
{
\sqrt{\tx^+_{\L}}
} \,  \frac{\tx^+_{\L} - y}{\tx^-_{\L} - y}\,,\qquad &&S^{y1}(y,u)={1\ov S^{1y}(u,y)}\,,
\\
\overline{S}^{1y}(u,y)&= \frac{\sqrt{\tx^+_{\R}}}{\sqrt{\tx^-_{\R}}} \,  \frac{\tx^-_{\R} - 1/y}{\tx^+_{\R} - 1/y}\,,\qquad &&\overline{S}^{y1}(y,u)={1\ov \overline{S}^{1y}(u,y)}\,,
\\
S^{0y}(u,y)&= \frac{\sqrt{\tx^{-0}_{\L}}}{\sqrt{\tx^{+0}_{\L}}} \,  \frac{\tx^{+0}_{\L} - y}{\tx^{-0}_{\L} - y}\,,\qquad &&S^{y0}(y,u)={1\ov S^{0y}(u,y)}\,,
\eal
The bound states equations can be obtained from the equations above following the same route of~\cite{Frolov:2021bwp}.
For any model the fused Bethe-Yang equations can be written in the following concise form
\bal\label{eq:fused_BE}
 (-1)^{\varphi_\A}=e^{i \delta_\A \tp_{\A k}R}\,\prod_\B\prod_{n=1}^{N_\B}\, S_{\A\B}(u_{\A k},u_{\B n})\,.
\eal
Here the indices A, B label Bethe strings (bound states) of different kind, then  $\delta_\A=1$ for Bethe strings carrying momentum and $\delta_\A=0$ otherwise, and $\varphi_\A$ are real constants which in our case can be equal to either 0 or 1 depending on the type of a Bethe string.
$S_{\A\B}$ is the scattering matrix of an A-string with a B-string, and we label by $N_B$ the number of strings of kind $B$. 

\subsection{Mirror TBA equations}

In this subsection we conjecture the mixed-flux \adso TBA equations, and discuss some subtle points which would have to be addressed in a thorough derivation of these equations. Since the bound-state structure and symmetry algebra are similar to those of the pure RR case, we expect that the mixed-flux  TBA equations
are of the same form as the ones proposed in~\cite{Frolov:2021bwp}.

The derivation of the TBA equations from the fused Bethe-Yang equations \eqref{eq:fused_BE} follows the standard route. In the thermodynamic limit  Bethe roots become dense, and introducing the densities of particles and holes, one finds that they satisfy the following integral  equations
\bal\label{eq:densities}
\rho_\A + \br_\A = {R\ov 2\pi}\delta_\A{d\tp_\A\ov du} +K_{\A\B}\star\rho_\B\,.
\eal
Here the convolution kernels are determined by the associated S~matrices\footnote{Note that for non-momentum-carrying auxiliary Bethe strings of type~A the S matrices should be chosen so that the r.h.s. of  \eqref{eq:densities} is positive. If the auxiliary S~matrix $S_{\A\A}$ is constant, e.g. 1 or $-1$, so that $K_{\A\A}=0$, then the real parts of the kernels $K_{\A\B}$ must be positive.
}
\begin{equation}\label{eq:kernels_def}
	K_{\A\B}(u,v) = \dfrac{1}{2 \pi i} \frac{d}{d u} \log S_{\A\B}(u,v)\,,
\end{equation}
and the $\star$-operation is defined as follows
\bal
K_{\A\B}\star\rho_\B(v)   &\equiv \sum_{\B} \int_{C_\B} d u \,  K_{\A\B}(v,u)\rho_\B(u)\,.
\eal
The shape of the integration contour $C_\B$ for the B-string depends on the string and the model under consideration, and we will discuss the contours for the mixed-flux \adso superstring in a due course.

Introducing then a $\cY$-function for each Bethe string
\bal
\cY_\A =  {\rho_\A\ov\br_\A}\,,
\eal
and finding an extremum of the free energy, one gets the canonical TBA equations
\bal
\log\cY_\A =-L\,
\tE_{\A} +\log\left(1+\cY_\B \right)\star K_{\B\A}\,,
\eal
where for any function $f(\cY_\B)$ we define
\bal
f\big(\cY_{\B}\big) \star K_{\B\A} (v) &\equiv \sum_{\B} \int_{C_\B} d u \, f\big(\cY_{\B}(u)\big) K_{\B\A}(u,v)\,.
\eal
The free energy of the model is given by
\bal 
\cF(L) =-{R\ov L}\sum_{\A}{}'\int \, {du\ov 2\pi}\,{d\tp_\A\ov du}\log\left(1+\cY_\A \right)\,,  
\eal
where $\sum_{\A}{}' \equiv \sum_{\A}\delta_\A$. 

\medskip

Let us now specify the formulae above to the 
 \adso TBA system. It involves various convolution kernels $K_{\A\B}$ to be defined later, and the following $Y$-functions
\bee
\item $Y_Q$-functions of left $Q$-particle bound states:\ \ $Y_Q =  {\rho_Q\ov\br_Q}$\,,\ $Q=1,2,\ldots$
\item  $\bY_Q$-functions of right $Q$-particle bound states:\ \ $ \bY_Q =  {\rho_\bQ\ov\br_\bQ}$\,,\ $Q=1,2,\ldots$
\item $Y_0$-function of massless particles:\ \ $Y_0 =  {\rho_0\ov\br_0}$ 
\item $Y_\pm^{(\a)}$-functions of auxiliary $y_\pm$-particles:\ \ $Y_\pm^{(\a)}
= -  e^{i \mu_\a}\, {\bar\rho_{y^\pm}^{(\a)}\ov\rho_{y^\pm}^{(\a)}}$ 
\eee
Here 
$\mu_\a=(-1)^\a\mu$ where $\mu$ is a twist parameter which if $\mu\neq0$ breaks supersymmetry of the light-cone string theory. 

The left, right and massless $Y$-functions for momentum-carrying Bethe roots satisfy the following  equations\footnote{Note that contrary to  \cite{Frolov:2021bwp,Frolov:2023wji} we have chosen the left particles to be in the $su(2)$ sector while the right ones in the $sl(2)$.} 
\begin{equation}\label{TBA_L}
	\begin{aligned}
		\log {Y}_Q & =-L \widetilde{\mathcal{E}}_Q+\log \left(1+{Y}_{Q^{\prime}}\right) \star K_{ Y Y}^{Q^{\prime} Q}  +\log \left(1+ \overline{Y}_{Q^{\prime}}\right) \star {K}_{\bar Z Y}^{Q^{\prime} Q} + \log \left(1+Y_0\right) \check{\star} {K}^{0 Q}_{\chi  Y}\\
		& +\sum_{\alpha=1,2} \log \left(1-\frac{e^{i \mu_\alpha}}{Y_{-}^{(\alpha)}}\right) \hat{\star} {K}_{\L\L}^{y Q}+\sum_{\alpha=1,2} \log \left(1-\frac{e^{i \mu_\alpha}}{Y_{+}^{(\alpha)}}\right) \hat{\star} {K}_{\R\L}^{y Q} \,,
	\end{aligned}
\end{equation}
\begin{equation}\label{TBA_R}
	\begin{aligned}
		\log  \overline{Y}_Q & =-L \widetilde{\mathcal{E}}_Q + \log \left(1+ \overline{Y}_{Q^{\prime}}\right) \star K_{\bar Z\bar Z}^{Q^{\prime} Q}  +\log \left(1+{Y}_{Q^{\prime}}\right) \star {K}_{ Y  \bar Z}^{Q^{\prime} Q}  +\log \left(1+Y_0\right) \check{\star} K^{0 Q}_{\chi  \bar{Z}} \\
		& +\sum_{\alpha=1,2} \log \left(1-\frac{e^{i \mu_\alpha}}{Y_{-}^{(\alpha)}}\right) \hat{\star} {K}_{\L\R}^{y Q}+\sum_{\alpha=1,2} \log \left(1-\frac{e^{i \mu_\alpha}}{Y_{+}^{(\alpha)}}\right) \hat{\star} {K}_{\R\R}^{y Q}\,,
	\end{aligned}
\end{equation}
\begin{equation}\label{TBA_0}
	\begin{aligned}
		\log Y_0^{(\dot{\alpha})} & =-L \widetilde{\mathcal{E}}_0 + \log \left(1+Y_0\right) \check{\star} K^{00}_{\chi\chi}  +\log \left(1+Y_Q\right) \star K^{Q 0}_{Y\chi}+\log \left(1+\overline{Y}_Q\right) \star {K}^{Q 0}_{\bar Z\chi} \\
		& +\sum_{\alpha=1,2} \log \left(1-\frac{e^{i \mu_\alpha}}{Y_{-}^{(\alpha)}}\right) \hat{\star} K_{\L\L}^{y 0}+\sum_{\alpha=1,2} \log \left(1-\frac{e^{i \mu_\alpha}}{Y_{+}^{(\alpha)}}\right) \hat{\star} K_{\R\L}^{y 0}\,,
	\end{aligned}
\end{equation}
and for auxiliary particles $ {y}_{-} $ and $ {y}_{+} $ the following coupled pair appears
\begin{equation}\label{TBA_ym}
\begin{aligned}
	-\log Y_{-}^{(\alpha)}&=\log \left(1+Y_Q\right) \star K_{\L\L}^{Q y}+\log \left(1+\overline{Y}_Q\right) \star {K}_{\R\L}^{Q y}+ \log \left( 1+Y_0\right) \check{\star} K_{\L\L}^{0 y} \,,
\end{aligned}
\end{equation}
\begin{equation}\label{TBA_yp}
\begin{aligned}
	-\log Y_{+}^{(\alpha)} &= + \log \left(1+Y_Q\right) \star K_{\L\R}^{Q y} + \log \left(1+\overline{Y}_Q\right) \star {K}_{\R\R}^{Q y}+ \log \left( 1+Y_0\right) \check{\star} K_{\L\R}^{0 y} \,.
\end{aligned}
\end{equation}
The mirror energy $ \tilde{\cE}_{Q} $ and momentum $\widetilde{p}_Q$ are given by \eqref{eq:mirrorEandp} with $m$ replaced by $Q$.
The energy of the ground state of the light-cone \adso string theory is given by
\bal 
E(L)= \lim_{R\to\infty}{L\ov
R}\cF(L)= &- \sum_{Q=1}^\infty\int\,  {du\ov 2\pi}{d\tp_Q\ov
du}\log\left(1+Y_Q \right)- \sum_{Q=1}^\infty\int\,  {du\ov 2\pi}{d\tp_Q\ov du}\log\left(1+\overline{Y}_Q \right)
\\
&-\int\,  {du\ov 2\pi}{d\tp_0\ov
du}\log\left(1+Y_0 \right)\,. 
\eal
The kernels are related to S~matrices as in \eqref{eq:kernels_def}.
The S~matrices for momentum-carrying excitations have been discussed in \cite{Frolov:2024pkz,Frolov:2025uwz} and this paper.
The remaining S-matrices are given by
\bal\label{eq:new_aux_kernels}
S_{\L\L}^{Q y}(u, v)&= S^{Q y}(u, \tx_\L(v))\,,\qquad &&S_{\L\L}^{yQ}(u, v)= {1\ov S^{yQ}(\tx_\L(u),v)} = { S_{\L\L}^{Q y}(v, u)}\,,
\\
S_{\L\R}^{Q y}(u, v)&= S^{Q y}(u, {1\ov \tx_\R(v)})\,,\qquad &&S_{\R\L}^{yQ}(u, v)= S^{yQ}({1\ov \tx_\R(u)},v) = {1\ov S_{\L\R}^{Q y}(v, u)}\,,
\\
S_{\R\L}^{Q y}(u, v)&= \overline{S}^{Q y}(u, \tx_\L(v))\,,\qquad &&S_{\L\R}^{yQ}(u, v)= {1\ov \overline{S}^{yQ}(\tx_\L(u),v)} = {{S}_{\R\L}^{Q y}(v, u)}\,,
\\
S_{\R\R}^{Q y}(u, v)&= \overline{S}^{Q y}(u, {1\ov \tx_\R(v)})\,,\qquad &&{S}_{\R\R}^{yQ}(u, v)= \overline{S}^{yQ}({1\ov \tx_\R(u)},v) = {1\ov {S}_{\R\R}^{Q y}(v, u)}\,,
\eal
where $Q=0,1,\ldots$ and the auxiliary S matrices are given by \eqref{eq:auxiliary_Smatrices} with $\tx_a^\pm$ replaced by $\tx_a^{\pm Q}$. Note that the difference in the definition of $S_{\L\L}^{yQ}$ and $S_{\L\R}^{yQ}$ which is related to the requirement of the kernel positivity, see footnote 15. 

The auxiliary S matrices and kernels we are using here differ from the ones used in the RR case in 
\cite{Arutyunov:2009ur,Frolov:2021bwp}. For the reader's convenience, we discuss the relation between the new and old kernels in appendix~\ref{app:aux_kernels}.

\medskip

The $\star$-,  $\check{\star}$- and  $\hat{\star}$-convolutions distinguish integration contours for different $Y_\A$ functions, and, as was mentioned above, one should use the following definition
\bal
f\big(Y_{\A}\big) \star K_{\A\B} (v) &\equiv \sum_{\A} \int_{C_\A} d u \, f\big(Y_{\A}(u)\big) K_{\A\B}(u,v)\,,
\eal
where $f\big(Y_{\A}\big)$ is any function of $Y_\A$ and $C_\A$ is the integration contour for the $Y_\A$-function.
In the pure-RR case the integration contours are  intervals of the real line, and they are divided into three groups which explains three different convolutions.  In the mixed-flux case the integration contours are found from the requirement that the densities of particles and holes are real. Since the mirror model is non-unitary it is a highly nontrivial requirement. Indeed, the densities of particles and holes satisfy in the thermodynamic limit the equations \eqref{eq:densities}.  Since $K_{\A\B}$ are not real, the reality of $\rho_\A$ and  $\br_\A$ can be achieved only by choosing the  integration contours so that the imaginary part of the r.h.s. of the equation vanishes. 

Thus, finding the integration contours is a complicated problem by itself. We believe, however, that this may not be necessary to write and solve the mirror TBA equations. Indeed, in the mirror TBA, the contours can be deformed without changing a solution, as long as the deformation does not encounter a singularity (conversely, such singularities are usually related to describing different excited states~\cite{Dorey:1996re}).
We believe that it should be possible to deform the contours to a simple and more convenient shape. In particular, it seems reasonable to expect that for $Q$-particle $Y$-functions one could choose the real mirror momentum contour, or even the contours which run from $-\infty + i\ka_a$ to the vertical line $\Re(u)=0$, then along the vertical line  to the horizontal line $\Im(u)=0$, and finally along the horizontal line to $+\infty$. 
For $Y_\pm$ functions we expect the contours are similar, only they run from a $\ka_a$-branch point to the main one. The contour for massless $Y_0$-function is basically the same as for the RR case. It runs over the lower edge of the $\ka$-cut and over the upper edge of the main mirror cut. All the contours are parity-invariant. Whether this choice of contours is correct requires substantially more careful analysis which will be done in a future publication.

\section{Conclusions}
\label{sec:conclusions}

In this work we have completed the S-matrix bootstrap program for mixed-flux $AdS_3\times S^3\times T^4$ by constructing massless dressing factors which solve crossing and satisfy all expected properties of the model. Interestingly, much of the construction follows as $m\to0^+$ limit of that of massive dressing factors. Having fixed the S~matrix, we have conjectured the mirror TBA equations which should describe the spectrum of the model for any value of the RR/NSNS flux (and hence, of the string tension). This completes the picture for the spectral problem of the model: the case of pure-NSNS models was understood long ago in the framework of the RNS string~\cite{Maldacena:2000hw} (it can be studied by mirror TBA too~\cite{Baggio:2018gct,Dei:2018mfl} starting from the GS string); the opposite case of pure-RR background was recently understood in terms of the mirror TBA~\cite{Frolov:2021bwp,Frolov:2023wji,Brollo:2023pkl,Brollo:2023rgp} and of the QSC~\cite{Ekhammar:2021pys,Cavaglia:2021eqr,Cavaglia:2022xld,Ekhammar:2024kzp}. This work fills the gap between the two cases.
It would be interesting to see if it is possible to derive the Y system, TQ and QQ relations from this TBA, which would yield a derivation of a QSC for the mixed-flux model, along the lines of the ongoing effort for the pure-RR case.

Our results open the way to several exciting explorations. They allow us to start the numerical exploration of the spectrum from the mirror TBA. In the mixed-flux case, it seems natural to consider the limit where~$k$ is fixed and $h\ll1$, where one should recover the pure-NSNS result~\cite{Maldacena:2000hw} (and the pure-NSNS mirror TBA~\cite{Dei:2018mfl}) and its perturbation~\cite{Cho:2018nfn}. The perturbation around $k=1$ is particularly interesting because it can be compared with the perturbed symmetric-product orbifold CFT of~$T^4$~\cite{Giribet:2018ada,Gaberdiel:2018rqv,Eberhardt:2018ouy}, see~\cite{Seibold:2024qkh} for a review. In the case $k=1$, the massless-massless sector is especially important owing to the periodicity~\eqref{eq:2pishift-nomonodromy2} (see also~\cite{Frolov:2023pjw}) which allows to relate all masses to the $m=0$ case in the string kinematics. It is encouraging that the HL phase admits a particularly simple representation in this case. We expect the most leading correction to appear already at order~$\mathcal{O}(h)$~\cite{Fabri:2025rok}, even before what expected from the Bethe--Yang equations.

Another interesting direction is to generalise the integrability approach to other $AdS_3$ backgrounds. Among those there are many interesting integrable deformations of the type reviewed in~\cite{Seibold:2024qkh}, but perhaps even more interestingly there is the family of $AdS_3\times S^3\times S^3\times S^1$ backgrounds. They enjoy the same amount of supersymmetry as $AdS_3\times S^3\times T^4$, but in this case the Killing spinors generate two copies of the $\mathfrak{d}(2,1;\alpha)$ algebra, where $0<\alpha<1$ determines the ratio between the radii of the two three-spheres. As it turns out, the structure of the S~matrix of this model is very close to that of $AdS_3\times S^3\times T^4$~\cite{Borsato:2012ud,Borsato:2015mma}, and the background can be supported by a mixture of RR and NSNS flux.
However, the presence of the additional parameter~$\alpha$ makes the model substantially more involved.%
\footnote{The case where $\alpha =1/2$, where the two spheres have the same radius, is simpler. Then the symmetries are given by two copies of $\mathfrak{osp}(4|2)$ and there is exciting ongoing work to study the spectrum by the QSC, in the pure-RR case.}
Nonetheless, it may be possible to build on the insights presented here to solve the crossing equations of that model too. We hope to return to these problems in the near future.

\section*{Acknowledgments}

We thank the participants of the Workshop \textit{Workshop on Higher-d Integrability} in Favignana (Italy) in June 2025 for stimulating discussions, and in particular M.~de Leeuw, P.~Dorey, G.~Mussardo, N.~Primi, and R.~Tateo. We also thank A.~Cavagli\`a, V.~Kazakov, and G.~Korchemsky for useful related discussion.
A.S.~acknowledges support from the PRIN Project n.~2022ABPBEY, from the CARIPLO Foundation under grant n.~2022-1886, and from the CARIPARO Foundation Grant under grant n.~68079.
This work has received funding from the Deutsche Forschungsgemeinschaft (DFG, German Research Foundation) -- SFB-Gesch\"aftszeichen 1624 -- Projektnummer 506632645. 
S.F.~acknowledges support from the INFN under a Foreign Visiting Fellowship.

\appendix

\section{Definitions and useful formulae}

\subsection{Properties of \texorpdfstring{$\sqrt{\a_a(x)}$}{sqrt(alpha)}}

The S matrix elements we have constructed involve 
the function $\a_a(x)$ defined by
\bal\label{eq:alphaa}
   \alpha_a(x)=\left(1-\frac{\xi_a}{x}\right)\left(x+\frac{1}{\xi_a}\right)\,,
\eal
and its square root. We recall the definitions
\bal
\xi_\L=\frac{1}{\xi_\R}=e^\eta\,, \qquad \eta \equiv \arcsinh \frac{\ka}{2 \pi}\,.
\eal
Here we summarise some of their properties.
The function $\a_a(x)$ transforms as follows under the three discrete symmetries of the mirror region
\bal
 \alpha_a(1/x) =  -\alpha_{\bar a}(x)\,,\quad  \alpha_a(-x) =  -\alpha_{\bar a}(x)\,,\quad  \alpha_a(-1/x) =  \alpha_{a}(x)\,.\quad 
\eal
Then, taking into account that 
\bal
\sgn\left( \Im\left(\alpha_{a}(x)\right)\right) = \sgn\left( \Im(x)\right)\,,
\eal
we get that 
\bal
\sqrt{\a_a(-x)}= -\sgn\left( \Im(x)\right)\,i\,\sqrt{\a_{\bar a}(x)}\,,\quad \sqrt{\a_a(1/x)}= -\sgn\left( \Im(x)\right)\,i\,\sqrt{\a_{\bar a}(x)}\,,
\eal
Using that the cuts of $\sqrt{\a_a(x)}$ are the intervals $(-\infty, -1/\xi_a)$ and $(0, \xi_a)$,
we find that 
moving $x$ to $1/x$  through the cuts of $\sqrt{\a_a(x)}$ gives
\bal
\sqrt{\a_a(x)}\ \ \xrightarrow{x\to 1/x\ \text{through a cut}}\ \  +\sgn\left( \Im(x)\right)\,i\,\sqrt{\a_{\bar a}(x)}\,.
\eal

\subsection{Function \texorpdfstring{$R(\gamma)$}{R(gamma)}}

The odd dressing factors are expressed in terms of the following combination of Barnes Gamma functions $G$:
\bal
\label{eq:app_Rfunct_definition}
R (\g)\equiv {G(1- \frac{\g}{2\pi i})\ov G(1+ \frac{\g}{2\pi i}) } =   \left({e\ov 2\pi}\right)^{+\frac{\gamma}{2\pi i}}\prod_{\ell=1}^\infty \frac{\Gamma(\ell+\frac{\gamma}{2\pi i})}{\Gamma(\ell-\frac{\gamma}{2\pi i})}\,e^{-\frac{\gamma}{\pi i}\,\psi(\ell) }\,,
\eal
where $\psi(\ell) = \frac{d}{d \ell} \log \Gamma(\ell)$.
It satisfies the following monodromy relations
\bal
\label{eq:monodromy_R}
R (\g \pm2\pi i) = i\left({ \sinh{\g\ov2}\ov \pi }\right)^{\pm1}R(\g)\,,\qquad R (\g+\pi i) =  { \cosh{\g\ov2}\ov \pi }R(\g-\pi i)\,,
\eal
and the ``unitarity'' conditions
\bal
\label{eq:brunit_unit_R}
R(\g)R(-\g)=1\,, \qquad R(\g)^* R(\g^*)=1\,.
\eal

\subsection{Functions \texorpdfstring{$\tPhi$}{tildePhi} and \texorpdfstring{$\tPsi$}{tildePsi}}
The mirror BES phases are expressed in terms of 
$\tPhi$-functions defined for all values of $x_1$ and $x_2$ by the double integral introduced in~\cite{Frolov:2024pkz,Frolov:2025uwz}
\begin{equation}
\label{eq:app_tPhidef}
    \tPhi_{ab}^{\alpha \beta}(x_{1},x_{2}) 
= -  \lint_{{ \pa\cR_\alpha}} \frac{{\rm d} w_1}{2\pi i} \lint_{{ \pa\cR_\beta}} \frac{{\rm d} w_2}{2\pi i} {1\ov w_1-x_{1}}{1\ov w_2-x_{2}} K^\bes(u_a(w_1)-u_b(w_2))\,, \quad \a, \b= \pm \,,
\end{equation}
where we use the BES kernel
\begin{equation}
    K^\bes(v)= i \log \frac{\Gamma \left(1+\frac{ih}{2}v \right)}{\Gamma \left(1-\frac{ih}{2}v \right)}\,.
\end{equation}
The integration paths $\pa\cR_+$ and $\pa\cR_-$ correspond to the upper and lower edges of the real line (it is important to distinguish them since $u_a(w)$ has a log cut for $w<0$).
The BES $\tPhi$-functions are defined in terms of $\tPhi_{ab}^{\alpha \beta}$-functions as follows
\bal
\label{eq:appdef_tPhiab}
 \tPhi_{aa}(x_{1},x_{2}) =  {\tPhi_{aa}^{--}(x_{1},x_{2}) + \tPhi_{aa}^{+ +}(x_{1},x_{2}) \ov 2}\,,\quad \tPhi_{a\bar a}(x_{1},x_{2}) =  {\tPhi_{a\bar a}^{-+}(x_{1},x_{2}) + \tPhi_{a\bar a}^{+-}(x_{1},x_{2}) \ov 2}\,.
\eal
Expanding the BES kernel at large-$h$ gives the AFS kernel, the HL kernel, and subleading terms
\bal
K^\bes(v)=h K^\afs(v)+K^\hl(v) + \mathcal{O}(1/h)\,,
\eal
with
\bal\la{eq:HLkernel}
K^\afs(v) &= -\frac{1}{2} v\big( \log ( -i v)+\log ( +i v)\big)  -  v \log \frac{h}{2 e},
\\
K^{\hl}(v)&=-{\pi\ov2}\sgn\big(\Re(v)\big)=\lim_{\eps\to0} \frac{\ln(\epsilon-iv)-\ln(\epsilon+iv)}{2i}\,.
\eal
The double integrals for the HL phase can be reduced to single integrals, see appendix C.2 of \cite{Frolov:2025uwz}.

In the mirror theory $x_1=\tx_{a1}^{\pm m_1}$, $x_2=\tx_{b2}^{\pm m_2}$, and 
for any of these $\tPhi$-functions we define the following combination\footnote{We use the same definition of $\tilde\varPhi$-functions with $\tx_{aj}^{\pm m_j}$ being anywhere on the $x$-plane.}
\bal
\label{eq:appdef_vartPhiab}
\tilde{\varPhi}_{ab}^{\text{any}}(\tx^{\pm m_1}_{a1},\tx^{\pm m_2}_{b2})&\equiv 
\tPhi^{\text{any}}_{ab}(\tx_{a1}^{+ m_1},\tx_{b2}^{+ m_2})-\tPhi^{\text{any}}_{ab}(\tx_{a1}^{+ m_1},\tx_{b2}^{- m_2})
\\
&-\tPhi^{\text{any}}_{ab}(\tx_{a1}^{- m_1},\tx_{b2}^{+ m_2})+\tPhi^{\text{any}}_{ab}(\tx_{a1}^{- m_1},\tx_{b2}^{- m_2})\,,
\eal
Then, the mirror BES and HL phase and dressing factors are 
\bal
\label{eq:app_def_Sigma_theta_any}
\tilde{\theta}_{ab}^{\text{any}}(\tx^{\pm m_1}_{a1},\tx^{\pm m_2}_{b2})&= 
\tilde{\varPhi}_{ab}^{\text{any}}(\tx^{\pm m_1}_{a1},\tx^{\pm m_2}_{b2})\,,
\ \ 
\Sigma^{\text{any}}_{ab}(\tx^{\pm m_1}_{a1},\tx^{\pm m_2}_{b2})=\exp\left[{i\, \tilde{\theta}_{ab}^{\text{any}}(\tx^{\pm m_1}_{a1},\tx^{\pm m_2}_{b2})}\right].
\eal
For $m_j\neq 0$ the mirror Zhukovsky variables are below the integration contours, and the double integrals are well-defined. The massless variables are obtained by taking the limit $m\to0$, and they approach the integration contours from below. Thus, in the massless case the contours have to be slightly deformed that is always possible for finite $h$. For infinite $h$, i.e. for the HL phase, and in the case where both particles are massless the contour deformation leads to an ambiguity for  
$u_1=u_2$ which makes the HL phase discontinuous at $u_1=u_2$. 

If we move $x_1$ from the lower to the upper half plane, the analytic continuation of $\tPhi(x_1,x_2)$ function produces $\tPsi$-functions
\begin{equation}
\label{PsiBESkamirror}
\begin{aligned}
&\tPsi_{b}^{\beta}(x_{1},x_{2}) \equiv -\lint_{\pa\cR_\beta}\frac{{\rm
d}w_2}{2\pi i}\,\frac{1}{w_2-x_{2}} K^\bes(u_a(x_{1})-u_b(w_2))\,.
\end{aligned}
\end{equation}
If we write $x_{1}$ as $x_{1}=\tx_a(u_1)$ or $x_{1}=x_a(u_1)$ then $u_a(x_{1})=u_1$, and therefore $\tPsi_{b}^{\beta}$ depends only on $u_1$. Because of that with a slight abuse of notation we use interchangeably $\tPsi_{b}^{\beta}(x_{1},\tx_{b2})$ and $\tPsi_{b}^{\beta}(u_{1},\tx_{b2})$ for one and the same $\tPsi_{b}^{\beta}$ function.

We also recall from~\cite{Frolov:2025uwz} the following combinations of $\tPsi$ functions
\bal
\label{eq:app_Delta_def}
\Delta^\b_b(u_1 \pm \frac{i}{h}m, x^{\pm m_2}_{b2})=&+\tPsi_{b}^{\beta}(u_1+\frac{i}{h}m_1, x^{+ m_2}_{b2})+\tPsi_{b}^{\beta}(u_1-\frac{i}{h}m_1, x^{- m_2}_{b2})\\
&-\tPsi_{b}^{\beta}(u_1+\frac{i}{h}m_1, x^{- m_2}_{b2})-\tPsi_{b}^{\beta}(u_1-\frac{i}{h}m_1, x^{+ m_2}_{b2}) \,.
\eal
This quantity can be explicitly computed both for $x^{\pm m_2}_{b2}$ in the mirror or string region. We refer to the results in appendix E.2 of~\cite{Frolov:2025uwz} for the explicit value of the expression above in the different kinematical configurations.

\subsection{Normalisation of lowest-weight elements}
We find the following relations of the lowest-weight mixed-mass S-matrix elements to the highest-weight ones
 \bal\la{eq:normS0mtildechiZ}
S^{0m}_{ \tilde \chi Z} (p_1, p_2) &=  {x_{\L1}^{-0}\ov x_{\L1}^{+0}}\,{x_{\L2}^{+m}\ov x_{\L2}^{-m}}\, \left({x_{\L1}^{+0}-x_{\L2}^{-m}\ov x_{\L1}^{-0}-x_{\L2}^{+m}} \right)^2 S^{0m}_{ \chi Y} (p_1, p_2)\,,
\\
S^{m0}_{Z \tilde  \chi} (p_1, p_2) &= {x_{\L1}^{-m}\ov x_{\L1}^{+m}}\,  {x_{\L2}^{+0}\ov x_{\L2}^{-0}}\,\left({x_{\L1}^{+m}-x_{\L2}^{-0}\ov x_{\L1}^{-m}-x_{\L2}^{+0}} \right)^2 S^{m0}_{ Y \chi} (p_1, p_2)\,,
\eal
and
 \bal\la{eq:normS0mtildechibarY}
S^{0m}_{ \tilde \chi \bar Y} (p_1, p_2) &=  {x_{\L1}^{-0}\ov x_{\L1}^{+0}}\,{x_{\R2}^{-m}\ov x_{\R2}^{+m}}\, \left({1-x_{\L1}^{+0}x_{\R2}^{+m}\ov 1- x_{\L1}^{-0}x_{\R2}^{-m}} \right)^2 S^{0m}_{ \chi \bar Z} (p_1, p_2)\,,
\\
S^{m0}_{  \bar Y \tilde \chi} (p_1, p_2) &=  {x_{\R1}^{+m}\ov x_{\R1}^{-m}}\, {x_{\L2}^{+0}\ov x_{\L2}^{-0}}\, \left({1-x_{\R1}^{-m}x_{\L2}^{-0}\ov 1- x_{\R1}^{+m}x_{\L2}^{+0}} \right)^2 S^{m0}_{ \bar Z \chi} (p_1, p_2)\,.
\eal
The massless-massless elements are related as follows
\bal\la{eq:normS00tildechitildechi}
S^{00}_{ \tilde \chi \tilde \chi} (p_1, p_2) &=  {x_{\L1}^{-0}\ov x_{\L1}^{+0}}\,{x_{\L2}^{+0}\ov x_{\L2}^{-0}}\, \left({x_{\L1}^{+0}-x_{\L2}^{-0}\ov x_{\L1}^{-0}-x_{\L2}^{+0}} \right)^2 S^{00}_{ \chi \chi} (p_1, p_2)\,.
\eal

\section{Crossing equations for mixed-mass dressing factors}
\la{app:crossingm0m1}

In this appendix we sketch the derivation of the crossing equations for the mixed-mass dressing factors. Just as in our previous papers, we perform the crossing transformation with respect to the first particle, and therefore we consider as $\Sigma^{10}_{a\L}$ as $\Sigma^{01}_{\L a}$, $a=$R,L. 

\subsection{Crossing equations wrt a massive particle}  
\label{app:wrtmassivecrossing}
Let us derive the crossing equations for the BES, HL  and odd factors with respect to a massive particle. So, we consider
\bal 
\tPhi_{ab}^{\alpha \beta}(\tx_{a1}^{\pm},\tx_{b2}^{\pm0}) =& -  \lint_{{ \pa\cR_\alpha}} \frac{{\rm d} w_1}{2\pi i} \lint_{{ \pa\cR_\beta^\cap}} \frac{{\rm d} w_2}{2\pi i} {1\ov w_1-\tx_{a1}^{\pm}}{1\ov w_2-\tx_{b2}^{\pm0}} K^\bes(u_a(w_1)-u_b(w_2))\,,
\eal
where $a,b=$R,L,  $\alpha, \beta=\pm$, and $\pa\cR_\beta^\cap$ denotes a slightly deformed integration contour to avoid a pole due to the reality of $\tx_{b2}^{\pm0}$.

The derivation of the crossing equation for two massive particles discussed in \cite{Frolov:2025uwz} goes through without any change, and we get (see definition~\eqref{eq:app_Delta_def})
 \bal\label{creqthmir}
 \tilde{\theta}_{ab}^{\a\b}({1\ov \tx_{\bar a 1}^\pm}, \tx_{b2}^{\pm0}) + \tilde{\theta}_{\bar a b}^{-\a \b}( \tx_{\bar a 1}^\pm, \tx_{b2}^{\pm0})
=&-\Delta^{\b}_{b} (u_1 \pm \frac{i}{h}, \tx_{b2}^{\pm0})+{ 1\ov i}\log
\frac{\tx_{b 1}^+-\tx_{b2}^{-0}}{\tx_{b 1}^+-\tx_{b2}^{+0}}\,
 \frac{ \tx_{\bar b 1}^+\tx_{b2}^{+0}-1}{ \tx_{\bar b 1}^+\tx_{b2}^{-0}-1}\,.
 \eal
By using the identities for $\tPsi$'s listed in appendix E.2 of~\cite{Frolov:2025uwz}, we get
\bal
\exp\Big(-i\Delta^{-}_{b} (u_1 \pm \frac{i}{h}, \tx_{b2}^{\pm0})\Big)={\tx_{b2}^{-0}\ov \tx_{b2}^{+0}}\, \frac{ \tx_{ b 1}^--\tx_{b2}^{+0}}{\tx_{b 1}^--\tx_{b2}^{-0}}\, \frac{ \tx_{ b 1}^+-\tx_{b2}^{+0}}{\tx_{b 1}^+-\tx_{b2}^{-0}}\,,
\eal
 \bal\label{creqthmir2}
e^{i \tilde{\theta}_{ab}^{\a-}({1\ov \tx_{\bar a 1}^\pm}, \tx_{b2}^{\pm0}) +i \tilde{\theta}_{\bar a b}^{-\a  -}( \tx_{\bar a 1}^\pm, \tx_{b2}^{\pm0})}
= {\tx_{b2}^{-0}\ov \tx_{b2}^{+0}}\, \frac{ \tx_{ b 1}^--\tx_{b2}^{+0}}{\tx_{b 1}^--\tx_{b2}^{-0}}\, 
 \frac{ \tx_{\bar b 1}^+\tx_{b2}^{+0}-1}{ \tx_{\bar b 1}^+\tx_{b2}^{-0}-1}\,,
 \eal
and
\bal
\exp\Big(-i\Delta^{+}_{b} (u_1 \pm \frac{i}{h}, \tx_{b2}^{\pm0})\Big)={\tx_{b2}^{+0}\ov \tx_{b2}^{-0}}\, \frac{ \tx_{ b 1}^-\tx_{b2}^{-0}-1}{\tx_{b 1}^-\tx_{b2}^{+0}-1}\, \frac{ \tx_{\bar b 1}^+\tx_{b2}^{-0}-1}{\tx_{\bar b 1}^+\tx_{b2}^{+0}-1}\,,
\eal
 \bal\label{creqthmir3}
e^{i \tilde{\theta}_{ab}^{\a+}({1\ov \tx_{\bar a 1}^\pm}, \tx_{b2}^{\pm0}) +i \tilde{\theta}_{\bar a b}^{-\a  +}( \tx_{\bar a 1}^\pm, \tx_{b2}^{\pm0})}
= {\tx_{b2}^{+0}\ov \tx_{b2}^{-0}}\,\frac{\tx_{b 1}^+-\tx_{b2}^{-0}}{\tx_{b 1}^+-\tx_{b2}^{+0}}\, \frac{ \tx_{\bar b 1}^-\tx_{b2}^{-0}-1}{\tx_{\bar b 1}^-\tx_{b2}^{+0}-1} \,.
 \eal
Thus, the crossing equations for the BES factors are
\bal\label{creqthmirBES}
e^{2i \tilde{\theta}_{aa}^{\bes}({1\ov \tx_{\bar a 1}^\pm}, \tx_{a2}^{\pm0}) +2i \tilde{\theta}_{\bar a a}^{\bes}( \tx_{\bar a 1}^\pm, \tx_{a2}^{\pm0})}&=e^{2i \tilde{\theta}_{\bar aa}^{\bes}({1\ov \tx_{ a 1}^\pm}, \tx_{a2}^{\pm0}) +2i \tilde{\theta}_{ a a}^{\bes}( \tx_{ a 1}^\pm, \tx_{a2}^{\pm0})}
\\
&=\frac{ \tx_{ a 1}^--\tx_{a2}^{+0}}{\tx_{a 1}^--\tx_{a2}^{-0}}\, 
 \frac{ \tx_{\bar a 1}^+\tx_{a2}^{+0}-1}{ \tx_{\bar a 1}^+\tx_{a2}^{-0}-1}\,\frac{\tx_{a 1}^+-\tx_{a2}^{-0}}{\tx_{a 1}^+-\tx_{a2}^{+0}}\, \frac{ \tx_{\bar a 1}^-\tx_{a2}^{-0}-1}{\tx_{\bar a 1}^-\tx_{a2}^{+0}-1} \,.
 \eal
 The equations can be found from the bound-state crossing equations of~\cite{Frolov:2025uwz} by taking the mass of the second particle to zero.
The crossing equations for the HL factors have the same form as for the massive particles with the replacement $\tx_{a2}^{\pm}\to \tx_{a2}^{\pm0}$, and therefore the ratio of the BES and HL factors satisfies the homogeneous crossing equation
\bal
\left({ \Sigma_{aa}^{\bes}({1\ov \tx_{\bar a 1}^\pm}, \tx_{a2}^{\pm0})\ov  \Sigma_{aa}^{\hl}({1\ov \tx_{\bar a 1}^\pm}, \tx_{a2}^{\pm0})}\right)^2\left({ \Sigma_{\bar aa}^{\bes}({ \tx_{\bar a 1}^\pm}, \tx_{a2}^{\pm0})\ov  \Sigma_{\bar aa}^{\hl}({ \tx_{\bar a 1}^\pm}, \tx_{a2}^{\pm0})}\right)^2=\left({ \Sigma_{\bar aa}^{\bes}({1\ov \tx_{ a 1}^\pm}, \tx_{a2}^{\pm0})\ov  \Sigma_{\bar aa}^{\hl}({1\ov \tx_{ a 1}^\pm}, \tx_{a2}^{\pm0})}\right)^2\left({ \Sigma_{ aa}^{\bes}({ \tx_{ a 1}^\pm}, \tx_{a2}^{\pm0})\ov  \Sigma_{ aa}^{\hl}({ \tx_{ a 1}^\pm}, \tx_{a2}^{\pm0})}\right)^2=1\,.
\eal
The odd factors also satisfy the massive particles crossing equations  with the replacement $\tx_{a2}^{\pm}\to \tx_{a2}^{\pm0}$
\bal
\left( \Sigma^{10\barnes}_{\R \L}(\bar{u}_1, u_2) \right)^{-2} \left( \Sigma^{10\barnes}_{\L\L}(u_1, u_2) \right)^{-2}&=  \frac{\left(\tx^-_{\L 1}-\tx^{-0}_{\L 2} \right) \left(\tx^+_{\L 1} -\tx^{+0}_{\L 2} \right) }{\left(\tx^-_{\L 1} -\tx^{+0}_{\L 2} \right) \left(\tx^+_{\L 1} -\tx^{-0}_{\L 2} \right) }\,,\\
\left( \Sigma^{10\barnes}_{\L\L}(\bar{u}_1, u_2) \right)^{-2} \left( \Sigma^{10\barnes}_{\R \L} (u_1, u_2)  \right)^{-2}&= \frac{\left(\tx^+_{\R 1} \tx^{-0}_{\L 2} -1\right) \left(\tx^-_{\R1} \tx^{+0}_{\L 2} -1\right) }{\left(\tx^+_{\R 1} \tx^{+0}_{\L 2} -1\right) \left(\tx^-_{\R 1} \tx^{-0}_{\L 2} -1\right) }\,,
\eal
due to the monodromy property \eqref{eq:monodromy_R} of $R$ functions.

By using these formulae, it is straightforward to check that to solve crossing the factors $H^{01}$ must satisfy the homogeneous equations
\eqref{mixed_mass_crossing_eq_01c}, and that the S-matrix elements 
\eqref{eq:massless_massive_m0m1} satisfy the crossing equations \eqref{mixed_mass_crossing_eq_01a}.

\subsection{Crossing equations wrt a massless particle}  
\label{app:wrtmasslesscrossing}

\paragraph{Crossing equations for BES factors.}
Let us derive the crossing equations for the BES factors with respect to a massless particle. We consider
\bal 
\tPhi_{ab}^{\alpha \beta}(\tx_{a1}^{\pm0},\tx_{b2}^{\pm}) =& -  \lint_{{ \pa\cR_\alpha^\cap}} \frac{{\rm d} w_1}{2\pi i} \lint_{{ \pa\cR_\beta}} \frac{{\rm d} w_2}{2\pi i} {1\ov w_1-\tx_{a1}^{\pm0}}{1\ov w_2-\tx_{b2}^{\pm}} K^\bes(u_a(w_1)-u_b(w_2))\,,
\eal
and use the crossing transformation \eqref{crxtpm22} for massless particles
\bal\la{crxtpm23}
\tx_\L^{\pm 0}\ \xrightarrow{\text{massless crossing}} \  \tx_\L^{\mp 0} +i0\,.
\eal
The crossing path is chosen so that one does not cross any cut of resulting $\tPsi$-functions. It is easy to see that the total contribution from $\tPsi$-functions vanish, and therefore  the BES factors satisfy the homogeneous crossing equations with respect to a massless particle
\bal
 \Sigma^{10\bes}_{aa}(\bar{u}_1, u_2) \Sigma^{10\bes}_{\bar a a}({u}_1, u_2) = \Sigma^{10\bes}_{\bar a a}(\bar{u}_1, u_2) \Sigma^{10\bes}_{ a a}({u}_1, u_2) =1\,.
\eal

\paragraph{Crossing equations for  HL  factors}

To define the HL phases we need to take $h\to\infty$. This requires extra deformation of integration contours such that on the $v$-planes the cuts of $K^\hl(v_{12})$ would not intersect the deformed contours. Then, repeating the calculation done in \cite{Frolov:2025uwz} for the massive HL phases we get the same expressions but with deformed integration contours
\bal 
\tilde\theta_{ab}^{\hl }(\tx_{a1}^{\pm0},\tx_{b2}^\pm) = I_{ab}^{\hl }(\tx_{a1}^{+0},\tx_{b2}^+)+I_{ab}^{\hl }(\tx_{a1}^{-0},\tx_{b2}^-)-I_{ab}^{\hl }(\tx_{a1}^{+0},\tx_{b2}^-)-I_{ab}^{\hl }(\tx_{a1}^{-0},\tx_{b2}^+)\,,
\eal
where
\bal
\label{eq:app_I_HL}
I^\hl_{aa}(x_{1},x_{2}) &\equiv - \frac{1}{4 \pi}   \lint_{\widetilde{ \rm cuts}^\cap} {\rm d} v \frac{\tx'_a(v)}{\tx_a(v) - x_{1}} \left( \log\left(\tx_a(v - i \eps)-x_{2} \right) - \log\left(\tx_a(v + i \eps)-x_{2} \right) \right)\,,\\
I^\hl_{\bar a a}(x_{1},x_{2})&\equiv +\frac{1}{4 \pi}   \lint_{\widetilde{ \rm cuts}^\cap} {\rm d} v \frac{\tx'_{\bar{a}}(v)}{\tx_{\bar{a}}(v) - x_{1}} \left( \log\left(\frac{1}{\tx_{\bar{a}}(v-i \eps)}-x_{2} \right) - \log\left(\frac{1}{\tx_{\bar{a}}(v+ i \eps)}-x_{2} \right) \right) \, ,
\eal
and $\widetilde{ \rm cuts}^\cap$ denotes the deformed integration contours. By using the formulae, one then finds that the HL  dressing factors satisfy the following crossing equations 
\bal
\label{eq:crossing_HL}
&\left( \Sigma^{01\hl}_{\L \L}(\bar{u}_1, u_2) \right)^{2} \left( \Sigma^{01\hl}_{\L\L}(u_1, u_2) \right)^{2}=\left(\frac{  \tx^{+0}_{\L 1} - \tx^+_{\L 2}  }{ \tx^{+0}_{\L 1} - \tx^-_{\L 2}}\ 
\frac{ \tx^{-0}_{\L 1} - \tx^-_{\L 2}  }{ \tx^{-0}_{\L 1} - \tx^+_{\L 2}}
\right)^2 \,,
\\
&\left( \Sigma^{01\hl}_{\L \R}(\bar{u}_1, u_2) \right)^{2} \left( \Sigma^{01\hl}_{\L\R}(u_1, u_2) \right)^{2}=\left( \frac{\tx^{+0}_{\L1} \tx^-_{\R 2}-1 }{ \tx^{+0}_{\L 1} \tx^+_{\R 2}-1} \ 
\frac{ \tx^{-0}_{\L1} \tx^+_{\R 2}-1 }{\tx^{-0}_{\L1} \tx^-_{\R 2}-1} 
\right)^2
\,.
\eal

\paragraph{Crossing equations for odd factors.}
To find  the crossing equations for the odd factors we use that 
under crossing 
\bal
\tg_a^{+0}\to \tg_a^{-0}\,,\quad \tg_a^{-0}\to \tg_a^{+0} +2i\pi\,,
\eal
and therefore
\bal
\left( \Sigma^{01\barnes}_{\L \L}(\bar{u}_1, u_2) \right)^{-2} \left( \Sigma^{01\barnes}_{\L\L}(u_1, u_2) \right)^{-2}&=    {\sinh^2{\tg_{\L\L}^{+0-} \ov2}  \ov \sinh^2{\tg_{\L\L}^{+0+} \ov2} } =  {\ta_{\L2}^+\ov \ta_{\L2}^-}\,  {\tx_{\L2}^+\ov \tx_{\L2}^-}\, 
\left({\tx_{\L1}^{+0}-\tx_{\L2}^-\ov \tx_{\L1}^{+0}-\tx_{\L2}^+}\right)^2\,,\\
\left( \Sigma^{01\barnes}_{\L\R}(\bar{u}_1, u_2) \right)^{-2} \left( \Sigma^{01\barnes}_{\R \L} (u_1, u_2)  \right)^{-2}&= { \cosh^2{\tg_{\L\R}^{+0+} \ov2} \ov \cosh^2{\tg_{\L\R }^{+0-} \ov2} } =  {\ta_{\R2}^-\ov \ta_{\R2}^+}\,  {\tx_{\R2}^-\ov \tx_{\R2}^+}\, 
\left({\tx_{\L1}^{+0}\tx_{\R2}^+ -1\ov \tx_{\L1}^{+0}\tx_{\R2}^- -1}\right)^2\,.
\eal
By using the formulae above, it is easy to verify that in order for the S-matrix elements 
\eqref{eq:S01_crossing_splitting} to satisfy the crossing equations \eqref{mixed_mass_crossing_eq_01b}
the factors $H^{01}$ must satisfy the crossing equations
\eqref{mixed_mass_crossing_eq_01d}.
Solutions to the crossing equations are then provided by the expression in~\eqref{eq:massless_massive_m0m1}.

\section{S-matrix normalisation using left and right Zhukovsky variables}
\label{app:S-matrix_elements}

There are two separate representations for massless particles spanned by a $\mathfrak{su}(2)_\circ$ index $\dot{\alpha}=1,2$. The particles of the two representations are connected one to another under charge conjugation `C' (up to some sign $\epsilon^{\dot{1}\dot{2}}=-\epsilon^{\dot{2}\dot{1}}$)
\bal
&|\chi^{\dot{1}} \rangle= | \phi_\L^{B} \rangle \otimes |\phi_\L^{F} \rangle \ \longleftarrow \text{C} \longrightarrow \ |\tilde{\chi}^{\dot{2}} \rangle= | \varphi_\R^{B} \rangle \otimes |\varphi_\R^{F} \rangle\,,\\
&|\chi^{\dot{2}} \rangle= | \phi_\R^{F} \rangle \otimes |\phi_\R^{B} \rangle \ \longleftarrow \text{C} \longrightarrow \ |\tilde{\chi}^{\dot{1}} \rangle= | \varphi_\L^{F} \rangle \otimes |\varphi_\L^{B} \rangle \,.
\eal
The two representations are isomorphic and in the main text (as well as in~\cite{Lloyd:2014bsa}) we parametrised all massless particles with left Zhukovsky variables.
However, we can equally parameterise the massless kinematics using right variables.
In this appendix, we parameterise the representation associated with the index $\dot{1}$ using left Zhukovsky variables and the representation associated with the index $\dot{2}$ using right Zhukovsky variables. 
This is particularly convenient if one want to solve the crossing equations in the same way as is done for massive particles in~\cite{Frolov:2025uwz}, and then take the mass to zero.
It is possible to show that the two representations are equivalent for massless particles, and there is no distinction between the two indices. As a consequence of this fact massless particles can be parameterised either using left or right Zhukovsky variables. We will show this explicitly in appendix~\ref{app:Massless_crossing_using_LR}. This is a nontrivial check of the correctness of our proposal for the dressing factors.

\subsection{S-matrix normalisation in the mirror region}

\paragraph{Massless-massive elements in mirror theory.}

In section~\ref{sec:proposal} we proposed the S-matrix elements $S^{01}_{\chi Y}\equiv S^{01}_{\chi^{\dot{1}} Y}$ and $S^{01}_{\chi \bar{Z}}\equiv S^{01}_{\chi^{\dot{1}} \bar{Z}}$. Here we associate the index $\dot{1}$ to these solutions since they are expressed in terms of massless left Zhukovsky variables. Once these elements are known, one can apply a charge conjugation tranformation to obtain the elements $S^{01}_{\tilde{\chi}^{\dot{2}} \bar{Y}}$ and $S^{01}_{\tilde{\chi}^{\dot{2}} Z}$, and from them get $S^{01}_{\chi^{\dot{2}} \bar{Z}}$ and $S^{01}_{\chi^{\dot{2}} Y}$ by changing normalisation from the scattering of lowest weight state to highest weight states (this is realised by multiplying the S-matrix element by $F(p_1,p_2)^{-2}$ in the notation of appendix~B in~\cite{Frolov:2023lwd}).
The charge conjugation transformation corresponds to mapping left Zhukovsky variables to right Zhukovsky variables and vice versa. 
Starting from the solutions in~\eqref{eq:massless_massive_m0m1} we can therefore generate all the remaining S-matrix elements for the scattering of highest weight states in the mixed-mass sector. For bound states of arbitrary mass we find
\begin{equation}
\label{eq:massless_massive_normmir_bstates}
    \begin{aligned}
    \mathbf{S}\,\big| \chi^{\dot{1}}_{1} Y^m_{2} \rangle&=&
    {\sqrt{\ta_{\L2}^{-m}} \ov \sqrt{\ta_{\L2}^{+m}}}  \ \frac{\tx^{-m}_{\L2}}{\tx^{+m}_{\L2}} \, \frac{\tx^{+0}_{\L1}}{\tx^{-0}_{\L1}} \, \left(\frac{\tx^{-0}_{\L1} - \tx^{+m}_{\L2}}{\tx^{+0}_{\L1} - \tx^{-m}_{\L2}} \right)^2 \left(\Sigma^{0m}_{\L\L}(u_1,u_2)\right)^{-2}\,
    \big| \chi^{\dot{1}}_{1} Y^m_{2} \rangle,\\
    \mathbf{S}\,\big| \chi^{\dot{1}}_{1} \bar{Z}^m_{2} \big\rangle&=&
    \frac{\sqrt{\ta^{+m}_{\R2}}}{\sqrt{\ta^{-m}_{\R2}}} \, \frac{\tx^{+m}_{\R2}}{\tx^{-m}_{\R2}} \, \frac{1 - \tx^{-0}_{\L1} \tx^{-m}_{\R2}}{1 - \tx^{+0}_{\L1} \tx^{+m}_{\R2}} \, \frac{1 - \tx^{+0}_{\L1} \tx^{-m}_{\R2}}{1 - \tx^{-0}_{\L1} \tx^{+m}_{\R2}} \left(\Sigma^{0 m}_{\L\R}(u_1,u_2)\right)^{-2} \,\big| \chi^{\dot{1}}_{1} \bar{Z}^m_{2} \big\rangle,\\
    \mathbf{S}\,\big|\chi^{\dot{2}}_{1} Y^m_{2}\big\rangle&=&
 {\sqrt{\ta_{\L2}^{+m}}\ov \sqrt{\ta_{\L2}^{-m}}}  \, \frac{\tx^{-0}_{\R1}}{\tx^{+0}_{\R1}}   \, \frac{1-\tx^{+0}_{\R1} \tx^{-m}_{\L2}}{1-\tx^{-0}_{\R1} \tx^{+m}_{\L2}} \, 
 \frac{1-\tx^{+0}_{\R1} \tx^{+m}_{\L2}}{1-\tx^{-0}_{\R1} \tx^{-m}_{\L2}}
  \left(\Sigma^{0m}_{\R\L}(u_1,u_2)\right)^{-2}\,
    \big|\chi^{\dot{2}}_{1} Y^m_{2}\big\rangle,\\
    \mathbf{S}\,\big| \chi^{\dot{2}}_{1} \bar{Z}^m_{2} \big\rangle&=&
    {\sqrt{\ta_{\R2}^{-m}} \ov \sqrt{\ta_{\R2}^{+m}}} \left(\Sigma^{0 m}_{\R\R}(u_1,u_2)\right)^{-2} \,\big| \chi^{\dot{2}}_{1} \bar{Z}^m_{2} \big\rangle .
    \end{aligned}
\end{equation}
Bound states in the mirror theory have been obtained by fusing particles of type $Z$ and $\bar{Z}$, with the constraint on the constituents
\bal
\label{eq:bound_state_condition_mir}
\tx^{+m}_{\R}(u)=\tx^{+}_{\R1}\,, \quad \tx^{-}_{\R1}=\tx^{+}_{\R2}\,, \quad \tx^{-}_{\R2}=\tx^{+}_{\R3} \, \dots\,, \quad \tx^{-}_{\R m}=\tx^{-m}_{\R}(u)\,.
\eal
We refer to~\cite{Frolov:2025uwz} for further details on the construction of bound states.
The elements above are manifestly left-right symmetric.

In the main text we worked using only the first two rows in the expression above due to the fact that the representations for massless particles are equivalent, which is
\bal
S^{0 m}_{\chi^{\dot{1}} Y }(u_1, u_2)=S^{0 m}_{\chi^{\dot{2}} Y }(u_1, u_2) \,, \qquad S^{0 m}_{\chi^{\dot{1}} \bar{Z} }(u_1, u_2)=S^{0 m}_{\chi^{\dot{2}} \bar{Z} }(u_1, u_2) \,. 
\eal 
These equalities are not implied by left-right symmetry and need to be checked. Remarkably after mapping left variables to right variables $\tx^{\pm 0}_{\L1}=\frac{1}{\tx^{\mp 0}_{\R1}}$ the equalities above turn out to be correct. This is a nontrivial fact supporting the validity of our proposal.
We will return to this in appendix~\ref{app:Massless_crossing_using_LR}.

The massive-massless S-matrix elements can be immediately obtained by applying braiding unitarity on the elements in~\eqref{eq:massless_massive_normmir_bstates}.

\paragraph{Massless-massless elements in mirror theory.}

The massless-massless element $S_{\chi^{\dot{1}} \chi^{\dot{1}}}$ was proposed in~\eqref{eq:massless_massless_m0m0} (again, we associate an index $\dot{1}$ to the massless particles of this element since we use left Zhukovsky variables).
From $S_{\chi^{\dot{1}} \chi^{\dot{1}}}$ by mapping left Zhukovsky variables to right Zhukovsky variables we obtain all the remaining elements $S_{\chi^{\dot{1}} \chi^{\dot{2}}}$, $S_{\chi^{\dot{2}} \chi^{\dot{1}}}$ and $S_{\chi^{\dot{2}} \chi^{\dot{2}}}$.
We obtain
\begin{equation}
\label{eq:massless_massless_normmir}
    \begin{aligned}
    \mathbf{S}\,\big| \chi^{\dot{1}}_{1} \chi^{\dot{1}}_{2} \rangle&=&
    +{\sqrt{\ta_{\L1}^{+0}} \ov \sqrt{\ta_{\L1}^{-0}}} \,  {\sqrt{\ta_{\L2}^{-0}} \ov \sqrt{\ta_{\L2}^{+0}}}  \ \frac{\tx^{-0}_{\L2}}{\tx^{+0}_{\L2}} \, \frac{\tx^{+0}_{\L1}}{\tx^{-0}_{\L1}} \, \left(\frac{\tx^{-0}_{\L1} - \tx^{+0}_{\L2}}{\tx^{+0}_{\L1} - \tx^{-0}_{\L2}} \right)^2 \left(\Sigma^{00}_{\L\L}(u_1,u_2)\right)^{-2}\,
    \big| \chi^{\dot{1}}_{1} \chi^{\dot{1}}_{2} \rangle,\\
    \mathbf{S}\,\big| \chi^{\dot{1}}_{1} \chi^{\dot{2}}_{2} \big\rangle&=& -{\sqrt{\ta_{\L1}^{-0}} \ov \sqrt{\ta_{\L1}^{+0}}}
    \frac{\sqrt{\ta^{+0}_{\R2}}}{\sqrt{\ta^{-0}_{\R2}}} \, \frac{\tx^{+0}_{\R2}}{\tx^{-0}_{\R2}} \, \frac{1 - \tx^{-0}_{\L1} \tx^{-0}_{\R2}}{1 - \tx^{+0}_{\L1} \tx^{+0}_{\R2}} \, \frac{1 - \tx^{+0}_{\L1} \tx^{-0}_{\R2}}{1 - \tx^{-0}_{\L1} \tx^{+0}_{\R2}} \left(\Sigma^{00}_{\L\R}(u_1,u_2)\right)^{-2} \,\big| \chi^{\dot{1}}_{1} \chi^{\dot{2}}_{2} \big\rangle,\\
    \mathbf{S}\,\big|\chi^{\dot{2}}_{1} \chi^{\dot{1}}_{2}\big\rangle&=& -{\sqrt{\ta_{\R1}^{-0}} \ov \sqrt{\ta_{\R1}^{+0}}}
 {\sqrt{\ta_{\L2}^{+0}}\ov \sqrt{\ta_{\L2}^{-0}}}  \, \frac{\tx^{-0}_{\R1}}{\tx^{+0}_{\R1}}   \, \frac{1-\tx^{+0}_{\R1} \tx^{-0}_{\L2}}{1-\tx^{-0}_{\R1} \tx^{+0}_{\L2}} \, 
 \frac{1-\tx^{+0}_{\R1} \tx^{+0}_{\L2}}{1-\tx^{-0}_{\R1} \tx^{-0}_{\L2}}
  \left(\Sigma^{00}_{\R\L}(u_1,u_2)\right)^{-2}\,
    \big|\chi^{\dot{2}}_{1} \chi^{\dot{1}}_{2}\big\rangle,\\
    \mathbf{S}\,\big| \chi^{\dot{2}}_{1} \chi^{\dot{2}}_{2} \big\rangle&=& +{\sqrt{\ta_{\R1}^{+0}} \ov \sqrt{\ta_{\R1}^{-0}}}
    {\sqrt{\ta_{\R2}^{-0}} \ov \sqrt{\ta_{\R2}^{+0}}} \left(\Sigma^{00}_{\R\R}(u_1,u_2)\right)^{-2} \,\big| \chi^{\dot{2}}_{1} \chi^{\dot{2}}_{2} \big\rangle ,
    \end{aligned}
\end{equation}
It can be shown that the normalisation above solves braiding unitarily, it is left-right symmetric, and that all the elements are equal after mapping left variables to right variables:
\bal
\label{eq:app_equivalent_masslessmassless_S_mat}
S_{\chi^{\dot{1}} \chi^{\dot{1}}}(u_1, u_2)=S_{\chi^{\dot{1}} \chi^{\dot{2}}}(u_1, u_2)=S_{\chi^{\dot{2}} \chi^{\dot{1}}}(u_1, u_2)=S_{\chi^{\dot{2}} \chi^{\dot{2}}}(u_1, u_2)\,.
\eal

\subsection{S-matrix normalisation in the string region}

Starting from the mirror region it is possible to continue the S-matrix elements to the string region. This can be done either to the region where the momenta are positive ($0<p <2\pi$) or to the region of negative momenta ($-2\pi<p< 0$). S-matrices with string particles of negative momenta can also be obtained by using the quasi-periodicity under shifts $\{m \to m\pm k, \, p \to p\mp 2\pi\}$. This periodicity was already discussed in~\cite{Frolov:2023lwd, OhlssonSax:2023qrk} and analysed in full detail on the S-matrix elements for massive particles in~\cite{Frolov:2025uwz}. This extends also to massless particles as we discuss in appendix~\ref{app:stringSmatrix}. It is then enough to write down the normalisation for particles having both momenta positive and in the region $(0, 2\pi)$. All the other regions can be obtained by using the periodicity under shifts of $2 \pi$.

\paragraph{Massless-massive elements in string theory.}
From the continuation of the dressing factors, we obtain the following normalisation for the massless-massive S-matrix elements in the string region of momenta $0<p_1<2 \pi$, $0<p_2<2 \pi$:
\begin{equation}
\label{eq:massless_massive_normstr_bstates}
    \begin{aligned}
    \mathbf{S}\,\big| \chi^{\dot{1}}_{1} Y^m_{2} \rangle=&
    {\sqrt{\a_{\L2}^{+m}}\ov \sqrt{\a_{\L2}^{-m}}}  \, \left(\frac{x^{+0}_{\L1}}{x^{-0}_{\L1}} \right)^m \, \left(\frac{x^{-0}_{\L1} - x^{+m}_{\L2}}{x^{+0}_{\L1} - x^{+m}_{\L2}} \right)^2 \,\\
    &\times \prod^{m-1}_{j=1}\left(\frac{x^{-0}_{\L1} - x^{+}_{\L2 j}}{x^{+0}_{\L1} - x^{+}_{\L2 j}} \right)^2 \left(\Sigma^{0m}_{\L\L}(u_1,u_2)\right)^{-2}\,
    \big| \chi^{\dot{1}}_{1} Y^m_{2} \rangle,\\
    \mathbf{S}\,\big| \chi^{\dot{1}}_{1} \bar{Z}^m_{2} \big\rangle=&
    \frac{\sqrt{\a^{-m}_{\R2}}}{\sqrt{\a^{+m}_{\R2}}}  \, \left( \frac{x^{-0}_{\L1}}{x^{+0}_{\L1}} \right)^{m-1} \frac{1 - x^{-0}_{\L1} x^{-m}_{\R2}}{1- x^{+0}_{\L 1} x^{-m}_{\R 2}} \, \frac{1- x^{+0}_{\L 1} x^{+m}_{\R 2}}{1 - x^{-0}_{\L1} x^{+m}_{\R2}}\\
    &\times \prod_{j=1}^{m-1} \left(\frac{1-x^{+0}_{\L1} x^+_{\R2 j}}{1-x^{-0}_{\L1} x^+_{\R2 j}} \right)^2 \,  \left(\Sigma^{0 m}_{\L\R}(u_1,u_2)\right)^{-2} \,\big| \chi^{\dot{1}}_{1} \bar{Z}^m_{2} \big\rangle,\\
    \mathbf{S}\,\big|\chi^{\dot{2}}_{1} Y^m_{2}\big\rangle=&
 {\sqrt{\a_{\L2}^{-m}}\ov \sqrt{\a_{\L2}^{+m}}}  \, \left( \frac{x^{-0}_{\R1}}{x^{+0}_{\R1}} \right)^m   \, {x_{\L2}^{-m}\ov x_{\L2}^{+m}}  \, \frac{\left( 1- x^{+0}_{\R 1} x^{+m}_{\L 2} \right)^3}{\left( 1- x^{-0}_{\R 1} x^{+m}_{\L 2} \right) \left( 1- x^{-0}_{\R 1} x^{-m}_{\L 2} \right) \left( 1- x^{+0}_{\R 1} x^{-m}_{\L 2} \right)}\\
  &\times \prod_{j=1}^{m-1} \left(\frac{1-x^{+0}_{\R1} x^+_{\L2 j}}{1-x^{-0}_{\R1} x^+_{\L2 j}} \right)^2 \left(\Sigma^{0 m}_{\R\L}(u_1,u_2)\right)^{-2}\,
    \big|\chi^{\dot{2}}_{1} Y^m_{2}\big\rangle,\\
    \mathbf{S}\,\big| \chi^{\dot{2}}_{1} \bar{Z}^m_{2} \big\rangle&=
    {\sqrt{\a_{\R2}^{+m}} \ov \sqrt{\a_{\R2}^{-m}}} \ {x_{\R2}^{+m}\ov x_{\R2}^{-m}}\,  \left( \frac{x^{+0}_{\R1}}{x^{-0}_{\R1}} \right)^{m-1}
\left({x_{\R1}^{+0}-x_{\R2}^{-m} \ov x_{\R1}^{+0}-x_{\R2}^{+m}} \right)^2\\
&\times \prod^{m-1}_{j=1}\left(\frac{x^{-0}_{\R1} - x^{+}_{\R2 j}}{x^{+0}_{\R1} - x^{+}_{\R2 j}} \right)^2 \,  \left(\Sigma^{0m}_{\R\R}(u_1,u_2)\right)^{-2} \,\big| \chi^{\dot{2}}_{1} \bar{Z}^m_{2} \big\rangle .
    \end{aligned}
\end{equation}
In string theory, the bound states are composed of multiple particles of type $Y$ (or multiple particles of type $\bar{Y}$). The fusion condition in this case is
\bal\label{eq:fusion_condition}
x^{-m}_{\L}(u)=x^-_{\L1}\,, \quad x^+_{\L1}=x^-_{\L2}\,, \quad x^+_{\L2}=x^-_{\L3}\,, \dots\,, \quad x^+_{\L m}=x^{+m}_{\L}(u)\,.
\eal
We fuse assuming all the constituents particles to be in the mirror region but $x^+_{\L m}$ (which we choose in the string region) as discussed in~\cite{Frolov:2025uwz}.
We stress that the odd dressing factors in the expressions above are evaluated at string $\g$ functions. We leave the continuation of the even part of the dressing factors implicit
\bal
\left(\Sigma^{0m}_{ab}(u_1,u_2) \right)^{-2}=\left(\Sigma^{0m, \besratio}_{ab}(u_1,u_2) \right)^{-2} \, \left(\Sigma^{0m, \barnes}_{ab}(\g^{\pm 0}_{a1},\g^{\pm m}_{b2}) \right)^{-2} \,.
\eal
The massive-massless S-matrix elements for string bound states are obtained from the expressions above using braiding unitarity.

\paragraph{Massless-massless elements in string theory.}

Continuing the first expression in~\eqref{eq:massless_massless_normmir} to the string region where both massless particles have positive momenta we obtain
\begin{equation}
\label{eq:S_massless_massless_pos_pos_final}
    \begin{aligned}
    \mathbf{S}\,\big| \chi^{\dot{1}}_{1} \chi^{\dot{1}}_{2} \rangle&=+
    {\sqrt{\a_{\L1}^{-0}} \ov \sqrt{\a_{\L1}^{+0}}} \,  {\sqrt{\a_{\L2}^{+0}} \ov \sqrt{\a_{\L2}^{-0}}}   \,   \left(\Sigma^{\barnes}_{\L \L} (\g^{\pm0}_{\L1}, \g^{\pm0}_{ \L 2}) \right)^{-2} \left(\Sigma^{00, \besratio}_{\L\L}(u_1,u_2)\right)^{-2} \, \big| \chi^{\dot{1}}_{1} \chi^{\dot{1}}_{2} \rangle \, .
    \end{aligned}
\end{equation}
We omit the remaining massless-massless S-matrix elements, since they are all equal to each other.

\section{Crossing using left and right Zhukovsky variables}
\label{app:Massless_crossing_using_LR}

In the main text of this paper, we parameterised  massless kinematics using left Zhukovsky variables --- which makes their $\mathfrak{su}(2)_\circ$ invariance manifest~\cite{Lloyd:2014bsa}. However, as a consequence, the crossing equations for massless particles take a quite different form compared to the crossing equations of massive particles (which instead couple left and right particles). Giving up manifest $\mathfrak{su}(2)_\circ$ invariance, we can parametrise half of the massless excitations in terms of right Zhukovsky variables, in such a way as to make manifest that the massless crossing equations can be obtained as a limit of the massive ones. In this appendix, we show this fact explicitly and check the equivalence between S-matrix elements expressed in terms of left and right variables for massless particles.
Since the crossing equations wrt a massive particle take exactly the same form discussed in~\cite{Frolov:2025uwz} and the massless particle is a spectator, we consider here only crossing equations wrt massless particles, where this relation is nontrivial.

\subsection{Crossing from a limit of the massive case }

Let us consider the crossing equations in~\eqref{mixed_mass_crossing_eq_01b}. Since the massless particles do not distinguish between indices $\dot{1}$ and $\dot{2}$ (i.e. $S^{0 m}_{\chi^{\dot{1}} Y}=S^{0 m}_{\chi^{\dot{2}} Y}$ and $S^{0 m}_{\chi^{\dot{1}} \bar{Z}}=S^{0 m}_{\chi^{\dot{2}} \bar{Z}}$) then we can write down the crossing equations in the following form
\bal
&S^{0 m}_{\chi^{\dot{1}} Y} (u_1, u_2) S^{0 m}_{\chi^{\dot{2}} Y} (\bar{u}_1, u_2) = \frac{\tx^{-m}_{\L2}}{\tx^{+m}_{\L2}} \left( \frac{\tx^{+0}_{\L1} - \tx^{+m}_{\L2}}{\tx^{+0}_{\L1} - \tx^{-m}_{\L2}} \right)^{2}\,,\\
&S^{0 m_2}_{\chi^{\dot{1}} Y} (\bar{u}_1, u_2) S^{0 m_2}_{\chi^{\dot{2}} Y} (u_1, u_2) = \frac{\tx^{-m}_{\L2}}{\tx^{+m}_{\L2}} \left( \frac{1-\tx^{-0}_{\R1} \tx^{+m}_{\L2}}{1-\tx^{-0}_{\R1} \tx^{-m}_{\L2}} \right)^{2}\,,
\eal
\bal
&S^{0 m}_{ \chi^{\dot{2}} \bar{Z}} (u_1, u_2) S^{0 m}_{\chi^{\dot{1}} \bar{Z}} (\bar{u}_1, u_2) = \frac{\tx^{+m}_{\R2}}{\tx^{-m}_{\R2}} \left( \frac{\tx^{-0}_{\R1} - \tx^{-m}_{\R2}}{\tx^{-0}_{\R1} - \tx^{+m}_{\R2}} \right)^{2}\,,\\
&S^{0 m}_{ \chi^{\dot{2}} \bar{Z}} (\bar{u}_1, u_2) S^{0 m}_{\chi^{\dot{1}} \bar{Z}} (u_1, u_2) = \frac{\tx^{+m}_{\R2}}{\tx^{-m}_{\R2}} \left( \frac{1-\tx^{+0}_{\L1} \tx^{-m}_{\R2}}{1-\tx^{+0}_{\L1} \tx^{+m}_{\R2}} \right)^{2}\,.
\eal
These equations correspond exactly to the crossing equations of massive-massive elements in the limit where the first mass becomes zero. Similarly, starting from~\eqref{massless_crossing_eq_00} and assuming~\eqref{eq:app_equivalent_masslessmassless_S_mat} we obtain the following crossing equations for the massless-massless S-matrices
\bal
&S^{00}_{\chi^{\dot 1} \chi^{\dot 1} }(u_1,u_2) S^{00}_{\chi^{\dot 2} \chi^{\dot 1} }(\bar u_1,u_2) =- \frac{ \tx_{\L2}^{-0}}{\tx_{\L2}^{+0}} \Bigl(\frac{\tx^{+0}_{\L1}-\tx^{+0}_{\L2}}{\tx^{+0}_{\L1}-\tx^{-0}_{\L2}} \Bigl)^2\,,
\\
&S^{00}_{\chi^{\dot 2} \chi^{\dot 1} }(u_1,u_2) S^{00}_{\chi^{\dot 1} \chi^{\dot 1} }(\bar u_1,u_2) =- {\frac{\tx_{\L2}^{-0}}{ \tx_{\L2}^{+0}}} \Bigl(\frac{1-\tx^{-0}_{\R1}\tx^{+0}_{\L2}}{1-\tx^{-0}_{\R1}\tx^{-0}_{\L2}} \Bigl)^2\,,\\
&S^{0 0}_{ \chi^{\dot 2} \chi^{\dot 2}} (u_1, u_2) S^{0 0}_{\chi^{\dot 1} \chi^{\dot 2}} (\bar{u}_1, u_2) = -\frac{\tx^{+0}_{\R2}}{\tx^{-0}_{\R2}} \left( \frac{\tx^{-0}_{\R1} - \tx^{-0}_{\R2}}{\tx^{-0}_{\R1} - \tx^{+0}_{\R2}} \right)^{2}\,,\\
&S^{0 0}_{ \chi^{\dot 2} \chi^{\dot 2}} (\bar{u}_1, u_2) S^{0 0}_{\chi^{\dot 1} \chi^{\dot 2}} (u_1, u_2) =- \frac{\tx^{+0}_{\R2}}{\tx^{-0}_{\R2}} \left( \frac{1-\tx^{+0}_{\L1} \tx^{-0}_{\R2}}{1-\tx^{+0}_{\L1} \tx^{+0}_{\R2}} \right)^{2} \,.
\eal
Using that $\tx^{+0}_{\R}={1}/{\tx^{-0}_{\L}}$ and $\tx^{-0}_{\R}={1}/{\tx^{+0}_{\L}}$ it is easy to check that all the crossing equations above are equivalent, and one can solve these equations with the further requirement that all the S-matrix elements are the same and do not depend on the index $\dot{\alpha}$. Using the equations above together with the normalisation in appendix~\ref{app:S-matrix_elements} we obtain the following equations on the dressing factors expressed in terms of left and right variables
\bal
\label{eq:massless_massive_crossing_df}
&\left(\Sigma^{0 m}_{a a}(u_1, u_2) \right)^2 \left(\Sigma^{0 m}_{\bar{a} a}(\bar{u}_1, u_2) \right)^2= \frac{\tx^{+0}_{a1} - \tx^{-m}_{a2}}{\tx^{+0}_{a1} - \tx^{+m}_{a2} } \frac{\tx^{-0}_{a1} - \tx^{+m}_{a2}}{\tx^{-0}_{a1} - \tx^{-m}_{a2} } \,,\\
&\left(\Sigma^{0m}_{a a}(\bar{u}_1, u_2) \right)^2 \left(\Sigma^{0m}_{\bar{a} a}(u_1, u_2) \right)^2= \frac{1-\tx^{+0}_{\bar{a}1} \tx^{+m}_{a2}}{1-\tx^{-0}_{\bar{a}1} \tx^{+m}_{a2} }  \frac{1-\tx^{-0}_{\bar{a}1} \tx^{-m}_{a2}}{1-\tx^{+0}_{\bar{a}1} \tx^{-m}_{a2} } \,,
\eal
\bal
\label{eq:massless_massless_crossing_df}
&\left(\Sigma^{00}_{a a}(u_1, u_2) \right)^2 \left(\Sigma^{00}_{\bar{a} a}(\bar{u}_1, u_2) \right)^2= \frac{\tx^{+0}_{a1} - \tx^{-0}_{a2}}{\tx^{+0}_{a1} - \tx^{+0}_{a2} } \frac{\tx^{-0}_{a1} - \tx^{+0}_{a2}}{\tx^{-0}_{a1} - \tx^{-0}_{a2} } \,,\\
&\left(\Sigma^{00}_{a a}(\bar{u}_1, u_2) \right)^2 \left(\Sigma^{00}_{\bar{a} a}(u_1, u_2) \right)^2=  \frac{1-\tx^{+0}_{\bar{a}1} \tx^{+0}_{a2}}{1-\tx^{-0}_{\bar{a}1} \tx^{+0}_{a2} }  \frac{1-\tx^{-0}_{\bar{a}1} \tx^{-0}_{a2}}{1-\tx^{+0}_{\bar{a}1} \tx^{-0}_{a2} } \,.
\eal
where as usual if $a=\text{L}$ then $\bar{a}=\text{R}$ and vice-versa. To construct the solutions, we can then take the limit of the results in~\cite{Frolov:2025uwz} where the mass of the first (and possibly the second) particle becomes $0^+$.
As for the massive-massive case in the mixed mass sector we have
\bal
\left(\Sigma^{0m}_{a b}(u_1,u_2)\right)^{2}=\left(\Sigma^{0m\besratio}_{a b}(u_1,u_2)\right)^{2} \left(\Sigma^{0m\barnes}_{a b}(u_1,u_2)\right)^{2}
\eal
where
\bal
\label{eq:app_even_dressing_massless_massive}
\left(\Sigma^{\besratio}_{a b} (\tx^{\pm0}_{a1}, \tx^{\pm m}_{b2}) \right)^{-2}=e^{-2 i \tilde{\theta}^{\bes}_{a b}(\tx^{\pm 0}_{a1}, \tx^{\pm m}_{b2})+2 i \tilde{\theta}^{\hl}_{a b}(\tx^{\pm 0}_{a1}, \tx^{\pm m}_{b2})}
\eal
and
\bal
\label{eq:odd_Sigma_mirror}
&\left(\Sigma^{\barnes}_{a a} (\tx^{\pm0}_{a1}, \tx^{\pm m}_{a2}) \right)^{-2}=\frac{R^2(\tg^{-0, -m}_{aa}) R^2(\tg^{+0, +m}_{aa})}{R^2(\tg^{-0, +m}_{aa}) R^2(\tg^{+0, -m}_{aa}) }\,,\\
&\left( \Sigma^{\barnes}_{\bar{a} a} (\tx^{\pm0}_{\bar{a}1}, \tx^{\pm m}_{a2}) \right)^{-2}=\frac{R(\tg^{-0, +m}_{\bar a a}+ i \pi) R(\tg^{-0, +m}_{\bar a a}- i \pi) R(\tg^{+0, -m}_{\bar a a}+ i \pi)R(\tg^{+0, -m}_{\bar a a}- i \pi)}{R(\tg^{-0, -m}_{\bar a a}+ i \pi) R(\tg^{-0, -m}_{\bar a a}- i \pi) R(\tg^{+0, +m}_{\bar a a}+ i \pi) R(\tg^{+0, +m}_{\bar a a}- i \pi)} \,.
\eal
The massless-massless dressing factors are obtain by sending the second mass to zero in the expression above.

\subsection{Recovering the results of the main text}
\label{app:results_using_LR_massless}

We now check that the scattering matrix is insensitive to the indices $\dot{1}$ and $\dot{2}$ for massless particles --- though this is no longer manifest --- and that therefore it agrees with the proposal in the main text.

\paragraph{Equivalence between left and right representations for BES and HL.}

We start checking the equivalence between left and right parameterisation for massless particles on the BES and HL phases.
From~\cite{Frolov:2025uwz}, we know that
\bal
\tPhi^{\alpha \beta}_{ab} (x, y)+\tPhi^{-\alpha \beta}_{\bar{a}b}( \frac{1}{x}, y)= \tPhi^{- \alpha, \beta}_{\bar{a} b}(0,y) \quad  \forall \ x, y\,, \qquad \a, \b = \pm \,.
\eal
If we consider $x=\tx^{\pm \delta}_a$ (with $\delta \ll 1$ being a regulator to assign a mass to the massless particle) then we get
\bal
\tPhi^{\alpha \beta}_{ab} (\tx^{\pm \delta}_a, y)+\tPhi^{-\alpha \beta}_{\bar{a}b}( \frac{1}{\tx^{\pm \delta}_a}, y)= \tPhi^{- \alpha, \beta}_{\bar{a} b}(0,y).
\eal
Note that in the limit of $\delta \to 0^+$ it holds that
\bal
\tPhi^{-\alpha \beta}_{\bar{a}b}( \frac{1}{\tx^{\pm \delta}_a}, y)=\tPhi^{-\alpha \beta}_{\bar{a}b}( \tx^{\mp \delta}_{\bar{a}}, y)+ \tPsi_{b}^\beta(u_1, y)
\eal
where $u_1=u_a(\tx^{\pm \delta}_a)$. The residue $\tPsi_{b}^\beta(u_1, y)$ comes from the fact that $\frac{1}{\tx^{\pm \delta}_a}$ and $\tx^{\mp \delta}_a$ are close but on the opposite sides of the contour. Then we get
\bal
\tPhi^{\alpha \beta}_{ab} (\tx^{\pm \delta}_a, y)=-\tPhi^{-\alpha \beta}_{\bar{a}b}( \tx^{\mp \delta}_{\bar{a}}, y)- \tPsi_{b}^\beta(u_1, y)+\tPhi^{- \alpha, \beta}_{\bar{a} b}(0,y)\,.
\eal
From the relation above and the definition in~\eqref{eq:appdef_vartPhiab} it follows that
\bal
&\tPhi^{\alpha \beta}_{ab}( \tx^{+\delta}_a,y) - \tPhi^{\alpha \beta}_{ab}( \tx^{-\delta}_a,y)=\tPhi^{-\alpha \beta}_{\bar{a}b}( \tx^{+\delta}_{\bar{a}},y) - \tPhi^{-\alpha \beta}_{\bar{a}b}( \tx^{-\delta}_{\bar{a}},y)\\
&\implies \tilde{\varPhi}^{\alpha \beta}_{ab} (\tx^{\pm \delta}_{a1}, \tx^{\pm m}_{b2}) = \tilde{\varPhi}^{-\alpha \beta}_{\bar{a} b} (\tx^{\pm \delta}_{\bar{a} 1}, \tx^{\pm m}_{b2})\,,
\eal
allowing to identify left and right massless particles at the level of the BES phase. Since the HL phase corresponds to the subleading order of BES for large $h$ it also satisfies the same relation.
Recalling the definitions in~\eqref{eq:appdef_tPhiab} and~\eqref{eq:appdef_vartPhiab} 
we conclude that
\bal
\tilde{\theta}^{\text{any}}_{ab} (\tx^{\pm 0}_{a1}, \tx^{\pm m}_{b2}) = \tilde{\theta}^{\text{any}}_{\bar{a}b} (\tx^{\pm 0}_{\bar{a}1}, \tx^{\pm m}_{b2})\,,
\eal
where `any' can either be BES or HL, and therefore
\bal
\label{eq:app_equivalence_LR_massless_even}
\left( \Sigma^{\besratio}_{\bar{a} a} (\tx^{\pm0}_{\bar{a}1}, \tx^{\pm 0}_{a2}) \right)^{-2}=\left( \Sigma^{\besratio}_{a a} (\tx^{\pm0}_{a 1}, \tx^{\pm 0}_{a2}) \right)^{-2} \,.
\eal

\paragraph{Equivalence between left and right representations for the odd phases.}

We start by finding a relation between the different odd phases.
By using 
\bal
\tg_{\bar a}^{\pm0} = \tg_{a}^{\mp0} \mp i\pi\,,
\eal
and~\eqref{eq:odd_Sigma_mirror} we get
\bal
\label{eq:app_mapping_left_right_massless_odd}
\left( \Sigma^{\barnes}_{\bar{a} a} (\tx^{\pm0}_{\bar{a}1}, \tx^{\pm m}_{a2}) \right)^{-2}
&=\frac{R(\tg^{+0, +m}_{ a a}+ 2i \pi) R(\tg^{+0, +m}_{a a}) R(\tg^{-0, -m}_{ a a})R(\tg^{-0, -m}_{ a a}-2 i \pi)}{R(\tg^{+0, -m}_{ a a}+2 i \pi) R(\tg^{+0, -m}_{ a a}) R(\tg^{-0, +m}_{ a a}) R(\tg^{-0, +m}_{\bar a a}- 2i \pi)} 
\\
&= {\sinh{\tg^{+0, +m}_{aa}\ov2} \sinh{\tg^{-0, +m}_{aa}\ov2} \ov \sinh{\tg^{-0, -m}_{aa}\ov2} \sinh{\tg^{+0, -m}_{aa}\ov2} } \frac{R^2(\tg^{-0, -m}_{aa}) R^2(\tg^{+0,+m}_{aa})}{R^2(\tg^{-0,+m}_{aa}) R^2(\tg^{+0,-m}_{aa}) }
\\
&=\frac{\tx_{a2}^{-m}}{\tx_{a2}^{+m}} \, {\a_a(\tx_{a2}^{-m}) \ov \a_a(\tx_{a2}^{+m})} \,{\tx_{a1}^{+m}-\tx_{a2}^{+m}\ov \tx_{a1}^{-0}-\tx_{a2}^{-m}}\,{\tx_{a1}^{-0}-\tx_{a2}^{+m}\ov \tx_{a1}^{+0}-\tx_{a2}^{-m}}\left(\Sigma^{\barnes}_{a a} (\tx^{\pm0}_{a1}, \tx^{\pm m}_{a2}) \right)^{-2}\,.
\eal
Combining this result with~\eqref{eq:app_equivalence_LR_massless_even} and the normalisation it is immediate to show that all S-matrix elements are blind to the index $\dot{\a}$. The results trivially apply also to the case in which the second particle is massless (we only need to take the limit $m \to 0^+$).

For example, using $\tx^{\pm 0}_{\bar{a}}=\frac{1}{\tx^{\mp 0}_{a}}$, after setting $\bar{a}=  \text{L}$ and $a=\text{R}$, we obtain
\bal
S_{\chi^{\dot 1}_1 \chi^{\dot 2}_2}(u_1, u_2)=- {\sqrt{\ta_{\L1}^{-0}} \ov \sqrt{\ta_{\L1}^{+0}}} \, {\sqrt{\ta_{\R2}^{-0}} \ov \sqrt{\ta_{\R2}^{+0}}}
     \left( \Sigma^{00}_{\R \R} (u_1, u_2) \right)^{-2} \,.
\eal
From the last line in~\eqref{eq:massless_massless_normmir} then we see that the following equality holds
\bal
S_{\chi^{\dot 1}_1 \chi^{\dot 2}_2}(u_1, u_2)=-  {\sqrt{\ta_{\L1}^{-0}} \ov \sqrt{\ta_{\L1}^{+0}}} \, {\sqrt{\ta_{\R1}^{-0}} \ov \sqrt{\ta_{\R1}^{+0}}}
     S_{\chi^{\dot 2}_1 \chi^{\dot 2}_2}(u_1, u_2) = S_{\chi^{\dot 2}_1 \chi^{\dot 2}_2}(u_1, u_2) \,.
\eal
A similar check can be done on all the massless-massless S-matrix elements in~\eqref{eq:massless_massless_normmir} to show that the chain of equalities in~\eqref{eq:app_equivalent_masslessmassless_S_mat} is satisfied. 
This is in agreement with the fact that the $\mathfrak{su}(2)_\circ$ S-matrix acting on the massless representations is trivial and left and right massless representations are equivalent. A similar check can be performed on the mixed-mass S-matrices.

\section{String S-matrix elements}
\label{app:stringSmatrix}

In this appendix we often write the string S-matrix elements as functions of the particles' momenta. We take any $p_k$ to be in the interval $(0,2\pi)$, and we continue to $p_k-2\pi$ to get negative momentum in the interval $(-2\pi,0)$. To derive the formulae in the appendix we use identities for $\tPsi$-functions from appendix E.2 of \cite{Frolov:2025uwz}.

\subsection{String \texorpdfstring{$S_{\chi Y}^{0m}(p_1,p_2)$}{SchiY0mp1p2} }

We begin with the mirror S-matrix element $S^{01}_{\chi Y}$ given by \eqref{eq:massless_massive_m0m1}
\bal
\label{eq:app_massless_massive_string_S_mat}
S^{01}_{\chi Y}(\tx_{\L1}^{\pm0},\tx_{\L2}^{\pm}) =  A^{01}_{\chi Y}(\tx_{\L1}^{\pm0},\tx_{\L2}^{\pm})\left(\Sigma^{01}_{\L\L}(\tx_{\L1}^{\pm0},\tx_{\L2}^{\pm})\right)^{-2}\,.
\eal
We want to continue it to the string region, and use fusion  to get the string $S_{\chi Y}^{01}$ with positive momenta $p_1$ and $p_2$. The fusion condition \eqref{eq:fusion_condition} has the following solution
\bal
x_{\L 2,j}^- &=  \tx_{\L2}^{2j-m-2} \,,\quad x_{\L 2,j}^+ =  \tx_{\L2}^{2j-m} \,,\quad j=1,\ldots, m-1 \,,
\\
x_{\L 2,m}^- &=  \tx_{\L2}^{m-2}  \,,\quad x_{\L 2}^{-m}=x_{\L 2,1}^-\,,\quad   x_{\L 2}^{+m}=x_{\L 2,m}^{+} =  {1\ov \tx_{\L2}^{+m} }\,.
\eal
The fusion of $A^{01}_{\chi Y }$ gives 
\bal
  A^{0m}_{\chi Y }(x_{\L1}^{\pm0},x_{\L2}^{\pm m})=
{ \sqrt{\a_{\L2}^{-m}}\ov \sqrt{\a_{\L2}^{+m}}}\, \left(\frac{ x_{ \L 1}^{+0}}{x_{\L 1}^{-0}}\right)^{m}\,
\frac{x_{\L 2}^{-m}}{x_{\L 2}^{+m}} \,
\left( \frac{ x_{ \L 1}^{-0}-x_{\L 2}^{+m}}{x_{\L 1}^{+0}-x_{\L 2}^{-m}}\right)^2
 \prod_{j=1}^{m-1}\left(\frac{x_{\L  1}^{-0}-\tx_{\L 2}^{2j-m}}{ x_{\L 1}^{+0}-\tx_{\L 2}^{2j-m} }\right)^2\,,
\eal
where we used that the cuts of $\sqrt{\a_a(x)}$ are the intervals $(-\infty, -1/\xi_a)$ and $(0, \xi_a)$.

To find the fusion of the dressing factors we have to consider  two cases
\bee
\item $ j\le m-1$.  We analytically continue $\tx_{\L 1}^{+0} \to x_{\L 1}^{+0}$, and get 
\bal\la{iBES0js}
\tilde\theta_{\L\L}^\bes(x_{\L 1}^{\pm 0},x_{\L 2j}^\pm)  &=  \tilde\Phi_{\L\L}(x_{\L 1}^{+0},x_{\L 2j}^+)+ \tilde\Phi_{\L\L}(x_{\L 1}^{-0},x_{\L 2j}^-)- \tilde\Phi_{\L\L}(x_{\L 1}^{+0},x_{\L 2j}^-) - \tilde\Phi_{\L\L}(x_{\L 1}^{-0},x_{\L 2j}^+)
\\
&-\tilde\Psi_{\L}(u_1,x_{\L 2j}^+)+\tilde\Psi_{\L}(u_1,x_{\L 2j}^-)\,.
\eal
The analytic continuation of the HL phase gives
\bal\la{iHL0js}
2\tilde\theta_{\L\L}^{\hl}(\tx_{\L1}^{\pm0},x_{\L2j}^\pm) &\quad \xrightarrow{\tx_{\L1}^{\pm0}\to x_{\L1}^{\pm0}}\quad 2\tilde\theta_{\L\L}^{\hl}(x_{\L1}^{\pm0},x_{\L2j}^\pm)
\\
&=2\tilde\varPhi_{\L\L}^{\hl}(x_{\L1}^{\pm0},x_{\L2j}^\pm) 
 -{1\ov i}\ln{x_{\L2j}^+\ov x_{\L2j}^-}
   -{1\ov i}  \ln(\tx_{\L1}^{+0}-x_{\L2j}^+)  +{1\ov i} \ln(x_{\L1}^{+0}-x_{\L2j}^+)
 \\
 &+{1\ov i}  \ln(\tx_{\L1}^{+0}-x_{\L2j}^-)  -{1\ov i} \ln(x_{\L1}^{+0}-x_{\L2j}^-)  
\,,
\eal

\item $ j=m\, $. We analytically continue $\tx_{\L 1}^{+0} \to x_{\L 1}^{+0}$, $\tx_{\L2}^{+m}\to x_{\L 2,m}^+= x_{\L2}^{+m}$, and get
\bal\la{iBESm1m2}
\hspace{-1cm}&\tilde\theta_{\L\L}^\bes(x_{\L1}^{\pm 0},x_{\L2}^{+ m},x_{\L2,m}^{-})   =  \tilde\Phi_{\L\L}(x_{\L1}^{+ 0},x_{\L2}^{+ m}) + \tilde\Phi_{\L\L}(x_{\L1}^{-0},x_{\L2,m}^{- })  - \tilde\Phi_{\L\L}(x_{\L1}^{+ 0},x_{\L2,m}^{- }) \\
& - \tilde\Phi_{\L\L}(x_{\L1}^{- 0},x_{\L2}^{+ m}) 
-\tilde\Psi_{\L}(u_1,x_{\L2}^{+ m})+\tilde\Psi_{\L}(u_1,x_{\L2,m}^{-})+\tilde\Psi_{\L}(u_2+{i\ov h}m ,x_{\L1}^{+ 0})
\\
&-\tilde\Psi_{\L}(u_2+{i\ov h}m,x_{\L1}^{- 0})
-K^\bes(u_{12}-{i\ov h}m)
\,,
\eal
\bal\la{iHLm1m2}
\hspace{-2cm}2 &\tilde\theta_{\L\L}^\hl(x_{\L1}^{\pm 0},x_{\L2}^{+ m},x_{\L2,m}^{-})   =  \tilde\varPhi_{\L\L}^\hl(x_{\L1}^{\pm 0},x_{\L2}^{+ m},x_{\L2,m}^{-}) 
+{1\ov i}\ln{x_{\L1}^{+0}\ov x_{\L1}^{-0}}+{1\ov i}\ln{ x_{\L2,m}^-\ov x_{\L2}^{+m}}
  \\
 & -{1\ov i}  \ln(\tx_{\L2}^{+m}-x_{\L1}^{+0}) -{1\ov i}  \ln(\tx_{\L1}^{+0}-x_{\L2}^{+m})  +{1\ov i} \ln(x_{\L1}^{+0}-x_{\L2}^{+m})^2
 \\
 &+{1\ov i}  \ln(\tx_{\L1}^{+0}-x_{\L2,m}^-)  -{1\ov i} \ln(x_{\L1}^{+0}-x_{\L2,m}^-)  
 +{1\ov i}  \ln(\tx_{\L2}^{+m}-x_{\L1}^{-0}) -{1\ov i} \ln(x_{\L2}^{+m}-x_{\L1}^{-0}) 
\,.
\eal

\eee
The fusion of $\tilde\Phi$'s is trivial. Next, consider
\bal
 \sum_{j=1}^{m-1}&\left(-\tilde\Psi(u_1,x_{\L 2j}^+)+\tilde\Psi(u_1,x_{\L 2j}^-)\right) 
=-\tilde\Psi(u_1,x_{\L2,m}^-)+\tilde\Psi(u_1,x_{\L2}^{-m})\,.
\eal
Adding all the terms we get
\bal\la{iBESm1m2fused}
\tilde\theta_{\L\L}^\bes(x_{\L1}^{\pm 0},x_{\L2}^{\pm m})   &=  \tilde\Phi_{\L\L}(x_{\L1}^{+ 0},x_{\L2}^{+ m}) + \tilde\Phi_{\L\L}(x_{\L1}^{-0},x_{\L2}^{- m})  - \tilde\Phi_{\L\L}(x_{\L1}^{+ 0},x_{\L2}^{- m})  - \tilde\Phi_{\L\L}(x_{\L1}^{- 0},x_{\L2}^{+ m}) 
\\
&-\tilde\Psi_{\L}(u_1,x_{\L2}^{+ m})+\tilde\Psi_{\L}(u_1,x_{\L2}^{-m})+\tilde\Psi_{\L}(u_2+{i\ov h}m ,x_{\L1}^{+ 0})-\tilde\Psi_{\L}(u_2+{i\ov h}m,x_{\L1}^{- 0})
\\
&-K^\bes(u_{12}-{i\ov h}m)\,.
\eal
We rewrite the combination of $\tPsi$ functions in it as follows (see~\eqref{eq:app_Delta_def} for the definition of $\Delta^-_{\L}$)
\bal
&-2\tilde\Psi_{\L}(u_1,x_{\L2}^{+ m})+2\tilde\Psi_{\L}(u_1,x_{\L2}^{-m})+2\tilde\Psi_{\L}(u_2+{i\ov h}m ,x_{\L1}^{+ 0})-2\tilde\Psi_{\L}(u_2+{i\ov h}m,x_{\L1}^{- 0})
\\
=&-2\tilde\Psi_{\L}(u_1,x_{\L2}^{+ m})+2\tilde\Psi_{\L}(u_1,x_{\L2}^{-m})+\tilde\Psi^+_{\L}(u_2+{i\ov h}m ,x_{\L1}^{+ 0})-\tilde\Psi^+_{\L}(u_2+{i\ov h}m,x_{\L1}^{- 0})
\\
&+\tilde\Psi^{-}_{\L}(u_2-{i\ov h}m,x_{\L1}^{+0})-\tilde\Psi^{-}_{\L}(u_2-{i\ov h}m,x_{\L 1}^{-0})
+\Delta_{\L}^{-}(u_2\pm {i\ov h}m,x_{\L 1}^{+ 0},x_{\L 1}^{- 0})
\,.
\eal
By using identities for $\tPsi$ functions, we find
\bal
e^{-i\Delta_{\L}^{-}(u_2\pm {i\ov h}m,x_{\L 1}^{+ 0},x_{\L 1}^{- 0})}
&=\left(\frac{h}{2}\right)^{+2m}\prod_{j=1}^m\left(u_{12}-\frac{i(2j-m)}{h}\right) \left(u_{12}+\frac{i(2j-m)}{h}\right)
\\
&\times 
\left( \frac{x_{\L 1}^{-0} }{x_{\L 1}^{+ 0} }\right)^m \, \frac{ (x_{\L 2}^{-m}-x_{\L 1}^{+0}) (\tx_{\L 2}^{+m}-x_{\L 1}^{+0}) }{
   (x_{\L 2}^{-m}-x_{\L 1}^{- 0}) (\tx_{\L 2}^{+m}-x_{\L  1}^{-0})}\left(\prod_{j=1}^{m-1}\frac{ \tx_{\L 2}^{+m-2j}-x_{\L 1}^{+0} }{\tx_{\L 2}^{+m-2j}-x_{\L  1}^{-0}}\right)^2\,.
\eal
We also have 
\bal
&e^{2iK^\bes(u_{12}-{i\ov h}m)}\left(\frac{h}{2}\right)^{+2m}\prod_{j=1}^m\left(u_{12}-\frac{i(2j-m)}{h}\right) \left(u_{12}+\frac{i(2j-m)}{h}\right)
\\
&=(-1)^m\left({ \Gamma\big[{m\ov2}-\tfrac{ih}{2}u_{12}\big]\ov \Gamma\big[{m\ov2}+\tfrac{ih}{2}u_{12}\big]}\right)^2
\frac{u_{12}+ \frac{i}{h}m}{u_{12}- \frac{i}{h}m}  \prod_{j=1}^{m-1}\left( \frac{u_{12}+ \frac{i}{h}(2j-m)}{u_{12}- \frac{i}{h}(2j-m)} \right)^4
\,. 
\eal
Next, for the HL phase we  get
\bal
 {1\ov i}\sum_{j=1}^{m-1}\log \frac{x_{\L2j}^-}{x_{\L2j}^+}  \frac{x_{\L 1}^{+ 0} -x_{\L2j}^+}{x_{\L1}^{+ 0} -x_{\L2j}^-}   \frac{\tx_{\L1}^{+ 0} -x_{\L2j}^-}{\tx_{\L1}^{+ 0} -x_{\L2j}^+} = {1\ov i}\log \frac{x_{\L2}^{-m}}{x_{\L2,m}^{-}}  \frac{x_{\L 1}^{+ 0} -x_{\L2,m}^{-}}{x_{\L1}^{+ 0} -x_{\L2}^{-m}}   \frac{\tx_{\L1}^{+ 0} -x_{\L2}^{-m}}{\tx_{\L1}^{+ 0} -x_{\L2,m}^-} \,.
\eal
Adding all the terms, we obtain
\bal\la{iHLm1m2fused}
e^{2i\tilde\theta_{\L\L}^{\hl}(x_{\L 1}^{\pm0},x_{\L2}^{\pm m})} =&  e^{2i\tilde\varPhi_{\L\L}^{\hl}(x_{\L1}^{\pm0},x_{\L2}^{\pm m})}
\\
&\times (-1)\,{x_{\L1}^{+0}\ov x_{\L1}^{-0}}\, {x_{\L2}^{-m}\ov x_{\L2}^{+m}}\, 
{(x_{\L1}^{+0}-x_{\L2}^{+m})^2\ov  (x_{\L1}^{-0}-x_{\L2}^{+m})^2}\, {(x_{\L1}^{-0}-x_{\L2}^{-m})(\tx_{\L2}^{+m}-x_{\L1}^{-0}) \ov (x_{\L1}^{+0}-x_{\L2}^{-m})  
(\tx_{\L2}^{+m}-x_{\L1}^{+0})} 
\,.
\eal
where we used
\bal
\tx_{\L1}^{+0} = {1\ov x_{\R1}^{+0}} = x_{\L1}^{-0}\quad \text{if}\quad u_1<\nu,
\eal
Thus, ($u_1,u_2$ are on the string $u$-plane)
\bal
\left({\Sigma^{0m\bes}_{\L\L}(u_1,u_2)\ov \Sigma^{0m\hl}_{\L\L}(u_1,u_2)}\right)^{-2}
 &=
 (-1)^{m-1}
 \left( \frac{x_{\L 1}^{-0} }{x_{\L 1}^{+ 0} }\right)^{m-1} \,  {x_{\L2}^{-m}\ov x_{\L2}^{+m}}\, 
{(x_{\L1}^{+0}-x_{\L2}^{+m})^2\ov  (x_{\L1}^{-0}-x_{\L2}^{+m})^2}\, \left(\prod_{j=1}^{m-1}\frac{ \tx_{\L 2}^{+m-2j}-x_{\L 1}^{+0} }{\tx_{\L 2}^{+m-2j}-x_{\L  1}^{-0}}\right)^2
\\
&\times\left({ \Gamma\big[{m\ov2}-\tfrac{ih}{2}u_{12}\big]\ov \Gamma\big[{m\ov2}+\tfrac{ih}{2}u_{12}\big]}\right)^2
\frac{u_{12}+ \frac{i}{h}m}{u_{12}- \frac{i}{h}m} \, e^{-2i\delta^{0m}_{\L\L}}
\,,
\eal
where
\bal
2\delta^{0m}_{\L\L}=& +2\tilde\varPhi_{\L\L}(x_{\L1}^{\pm 0},x_{\L 2}^{\pm m})-2\tilde\varPhi_{\L\L}^\hl(x_{\L1}^{\pm 0},x_{\L 2}^{\pm m})
\\
& -2\tilde\Psi_{\L}(u_1,x_{\L2}^{+ m})+2\tilde\Psi_{\L}(u_1,x_{\L2}^{-m})+\tilde\Psi^+_{\L}(u_2+{i\ov h}m ,x_{\L1}^{+ 0})-\tilde\Psi^+_{\L}(u_2+{i\ov h}m,x_{\L1}^{- 0})
\\
&+\tilde\Psi^{-}_{\L}(u_2-{i\ov h}m,x_{\L1}^{+0})-\tilde\Psi^{-}_{\L}(u_2-{i\ov h}m,x_{\L 1}^{-0})
\,.
\eal
The fusion  of the odd factor is trivial, and we get
\bal
\left(\Sigma^{01\barnes}_{\L\L} (\tg^{\pm0}_{\L1},\tg^\pm_{\L2}) \right)^{-2}=\left(\Sigma^{01\barnes}_{\L\L} (\g^{\pm0}_{\L1},\g^\pm_{\L2}) \right)^{-2}  {\sinh^2{\g_{\L\L}^{+0-m} \ov2}  \ov \sinh^2{\g_{\L\L}^{+0+m} \ov2} }\,,
\eal
where
\bal
\g^{\pm 0\pm m}_{\L\L}= \g^{\pm 0}_{\L}-\g^{\pm m}_{\L}
\,.
\eal
Taking into account that 
\bal
{\sinh^2{\g_{\L\L}^{+0-m} \ov2}  \ov \sinh^2{\g_{\L\L}^{+0+m} \ov2} } =  {\a_{\L2}^{+m}\ov \a_{\L2}^{-m}}\,  {x_{\L2}^{+m}\ov x_{\L2}^{-m}}\, 
{(x_{\L1}^{+0}-x_{\L2}^{-m})^2\ov (x_{\L1}^{+0}-x_{\L2}^{+m})^2} \,,
\eal
we find that
the fusion of $S^{01}_{\chi Y }$ gives 
\bal
S^{0m}_{\chi Y }(p_1,p_2)&=
{ \sqrt{\a_{\L2}^{+m}}\ov \sqrt{\a_{\L2}^{-m}}}\, \left(\frac{ x_{ \L 1}^{+0}}{x_{\L 1}^{-0}}\right)^{m}\,
\frac{R^2(\g^{-0-m}_{\L\L}) R^2(\g^{+0+m}_{\L\L})}{R^2(\g^{-0+m}_{\L\L}) R^2(\g^{+0-m}_{\L\L}) }  
\\
& \times (-1)^{m-1}
 \left( \frac{x_{\L 1}^{-0} }{x_{\L 1}^{+ 0} }\right)^{m-1} \,  \frac{x_{\L 2}^{-m}}{x_{\L 2}^{+m}}\, \left({ \Gamma\big[{m\ov2}-\tfrac{ih}{2}u_{12}\big]\ov \Gamma\big[{m\ov2}+\tfrac{ih}{2}u_{12}\big]}\right)^2
\frac{u_{12}+ \frac{i}{h}m}{u_{12}- \frac{i}{h}m} \, e^{-2i\delta^{0m}_{\L\L}}
\,,
\eal
The expression above satisfies the physical unitarity.

In the pure RR case we have the formula
\bal
\left({\sigma^\bes(x_1^{\pm m_1},x_2^{\pm m_2})\ov \sigma^\hl(x_1^{\pm m_1},x_2^{\pm m_2})}\right)^{-2}
&=e^{-2i\delta^{m_1 m_2}}
\left(-1\right)^{m_1-m_2+1} 
  \left(\frac{x_{ 1}^{-m_1}}{x_{ 1}^{+m_1}}\right)^{m_2-1}
  \left( \frac{x_{ 2}^{+m_2}}{x_{ 2}^{-m_2}}\right)^{m_1-1}
\\
&
\times  \frac{u_{12}+ \frac{i}{h} (m_1+m_2)}{u_{12}- \frac{i}{h} (m_1+m_2)} 
  \left({ \Gamma\big[{m_1+m_2\ov2}-\tfrac{ih}{2}u_{12}\big]\ov \Gamma\big[{m_1+m_2\ov2}+\tfrac{ih}{2}u_{12}\big]}\right)^2
\,.
\eal
Setting $m_1=0$ and $m_2=m$, we get
\bal
\left({\sigma^\bes(x_1^{\pm 0},x_2^{\pm m})\ov \sigma^\hl(x_1^{\pm 0},x_2^{\pm m})}\right)^{-2}
&=
\left(-1\right)^{m-1} 
  \left(\frac{x_{ 1}^{-0}}{x_{ 1}^{+0}}\right)^{m-1}
  \frac{x_{ 2}^{-m}}{x_{ 2}^{+m}}
\,  \frac{u_{12}+ \frac{i}{h} m}{u_{12}- \frac{i}{h} m} 
  \left({ \Gamma\big[{m\ov2}-\tfrac{ih}{2}u_{12}\big]\ov \Gamma\big[{m\ov2}+\tfrac{ih}{2}u_{12}\big]}\right)^2e^{-2i\delta^{0m}}
\!,
\eal
and, therefore, 
we have an agreement with the RR case.

In what follows we sometimes use the following normalisation of the string S-matrix element
\bal
S^{0m}_{\chi Y}(x_{\L1}^{\pm0},x_{\L2}^{\pm m})=  B^{0m}_{\chi Y}(x_{\L1}^{\pm0},x_{\L2}^{\pm m})\left(\Sigma^{0m}_{\L\L}(x_{\L1}^{\pm0},x_{\L2}^{\pm m})\right)^{-2}\,,
\eal
\bal\la{B0mYYstring}
B^{0m}_{\chi Y }(x_{\L1}^{\pm0},x_{\L2}^{\pm m})&=
{ \sqrt{\a_{\L2}^{+m}}\ov \sqrt{\a_{\L2}^{-m}}}\, \left(\frac{ x_{ \L 1}^{+0}}{x_{\L 1}^{-0}}\right)^{m} \,
\left( \frac{ x_{ \L 1}^{-0}-x_{\L 2}^{+m}}{x_{\L 1}^{+0}-x_{\L 2}^{+m}}\right)^2\,
 \prod_{j=1}^{m-1}\left(\frac{x_{\L  1}^{-0}-\tx_{\L 2}^{+m-2j}}{ x_{\L 1}^{+0}-\tx_{\L 2}^{+m-2j} }\right)^2\,,
\eal
\bal
\left(\Sigma^{0m}_{\L\L}(x_{\L1}^{\pm0},x_{\L2}^{\pm m})\right)^{-2} &= \left(\Sigma^{0m\,\barnes}_{\L\L}(\g_{\L1}^{\pm0},\g_{\L2}^{\pm m})\right)^{-2} \left({\Sigma^{0m\,\bes}_{\L\L}(x_{\L1}^{\pm0},x_{\L2}^{\pm m})\ov \Sigma^{0m\,\hl}_{\L\L}(x_{\L1}^{\pm0},x_{\L2}^{\pm m})}\right)^{-2}\,,
\eal
\bal
&\left(\Sigma^{0m\,\barnes}_{\L\L} (\g_{\L1}^{\pm0},\g_{\L2}^{\pm m}) \right)^{-2}=\frac{R^2(\g^{-0-}_{\L\L}) R^2(\g^{+0+}_{\L\L})}{R^2(\g^{-0+}_{\L\L}) R^2(\g^{+0-}_{\L\L}) }\,,
\eal
and $\Sigma^{0m\bes}_{\L\L}$ and $\Sigma^{0m\hl}_{\L\L}$ are the analytically continued BES and HL factors.

\subsection{String 
\texorpdfstring{$S_{\chi Y}^{0m}(p_1-2\pi,p_2)$}{SchiY0mp1m2pip2} and \texorpdfstring{$S_{\chi Y}^{0m}(p_1-2\pi,p_2-2\pi)$}{SchiY0mp1m2pip2m2pi}}

We begin with the mirror S-matrix element
\bal
S^{01}_{\chi Y}(u_1,u_2) =  A^{01}_{\chi Y}(u_1,u_2)\left(\Sigma^{01}_{\L\L}(u_1,u_2)\right)^{-2}\,.
\eal
We first move $\tx_{\L2}^\pm$ to $x_{\L2}^\pm$, and then $\tx_\L^{\pm 0}(u')$ to $ {1/ x_\R^{\mp k}(u) } $.

\subsubsection*{Step I:  $\tx_{\L 2}^{\pm}\to x_{\L 2}^{\pm}$ }
The first step is to move $\tx_{\L 2}^{+m}$ to $x_{\L 2}^{+m}$ through $x>\xi$ that gives
\bal
 A^{01}_{\chi Y }(u_1,u_2)\to+{\sqrt{ \a_{\L2}^-}\ov \sqrt{\a_{\L2}^+}}\, \frac{ \tx_{ \L 1}^{+0}}{\tx_{\L 1}^{-0}}\,\frac{x_{\L 2}^{-}}{x_{\L 2}^{+}} \,\left(\frac{ \tx_{ \L 1}^{-0}-x_{\L 2}^{+}}{\tx_{\L 1}^{+0}-x_{\L 2}^{-}}\right)^2\,,
\eal
\bal
\tilde{\theta}_{\L b}^{\bes}(\tx^{\pm 0}_{\L 1}, x^\pm_{b2})&=\tilde\varPhi_{\L b}(\tx^{\pm 0}_{\L 1},x^\pm_{b2})+\tPsi_{\L}(x^+_{b2}, \tx^{+ 0}_{\L 1})-\tPsi_{\L}(x^+_{b2},\tx^{- 0}_{\L 1})
\,,
\eal
\bal 
2\tilde\theta_{\L \L}^\hl(\tx^{\pm 0}_{\L 1}, x^\pm_{\L2})  &=  2\tilde\Phi^\hl(\tx^{\pm 0}_{\L 1}, x^\pm_{\L2})
+{1\ov i} \log {\tx^{+ 0}_{\L 1}\ov \tx^{- 0}_{\L 1}}\, \frac{\tx^{+ 0}_{\L 1}-x_{\L2}^{+}}{\tx^{- 0}_{\L 1}-x_{\L2}^{+}}\,\frac{\tx^{-0}_{\L 1}x_{\R2}^{+}-1}{\tx^{+0}_{\L 1}x_{\R2}^{+}-1}  \,.
\eal
\bal
\Sigma_{\L\L}^\barnes(\tx_{\L1}^{\pm0},x^{\pm }_{\L2})=\frac{R^2(\g^{-0-}_{\L\L}-{i\pi\ov2}) R^2(\g^{+0+}_{\L\L}-{i\pi\ov2})}{R^2(\g^{-0+}_{\L\L}-{i\pi\ov2}) R^2(\g^{+0-}_{\L\L}-{i\pi\ov2}) }  \,,
\eal
where
\bal
\g^{\pm 0\pm }_{\L\L}= \tg^{\pm 0}_{\L}-\g^{\pm}_{\L}
\,.
\eal

\subsubsection*{Step II:  $\tx_{\L 1}^{\pm 0}\to  {1/ x_{\R1}^{\mp k} } $ }

Next, we move $\tx_{\L 1}^{\pm 0}$ to  ${1/ x_{\R1}^{\mp k} } $ through 
 the upper edge of the $\ka$-cut to a point $u_1>-\nu$. Thus, the massless particle $\tx_a^{+0}$ moves to a point  in the upper half-plane through the interval $-1/\xi_a <x<0$ without crossing any cut of resulting $\tPsi$ functions. We get 
 
 \noindent {\bf The rational factor }
 \bal
 A^{01}_{\chi Y }(u_1,u_2)\to+{\sqrt{ \a_{\L2}^-}\ov \sqrt{\a_{\L2}^+}}\, \frac{x_{\R1}^{- k}}{x_{\R1}^{+ k}}\,\frac{x_{\L 2}^{-}}{x_{\L 2}^{+}} \,\left(\frac{ x_{\R1}^{+ k}x_{\L 2}^{+}-1}{x_{\R1}^{- k}x_{\L 2}^{-}-1}\right)^2\,,
\eal

 \noindent {\bf The odd factor }
 \bal
\Sigma_{\L\L}^\barnes(\tx_{\L1}^{\pm0},x^{\pm }_{\L2})&\to \frac{R^2(\g^{+k-}_{\R\L}-{i\pi}) R^2(\g^{-k+}_{\R\L}-{i\pi})}{R^2(\g^{+k+}_{\R\L}-{i\pi}) R^2(\g^{-k-}_{\R\L}-{i\pi}) } 
\\
&={\cosh{\g^{+k+}_{\R\L}\ov2}\cosh{\g^{-k-}_{\R\L}\ov2}\ov \cosh{\g^{+k-}_{\R\L}\ov2}\cosh{\g^{-k+}_{\R\L}\ov2} }\,\Sigma_{\R\L}^\barnes(x_{\R1}^{\pm k},x^{\pm}_{\L2})
\\
&={x_{\R1}^{- k}x^{-}_{\L2}-1\ov x_{\R1}^{- k}x^{+}_{\L2}-1}\,{x_{\R1}^{+ k}x^{+}_{\L2}-1\ov x_{\R1}^{+ k}x^{-}_{\L2}-1}\,\Sigma_{\R\L}^\barnes(x_{\R1}^{\pm k},x^{\pm}_{\L2})
 \,,
\eal
where
\bal
\g^{\pm k\pm }_{\R\L}= \g^{\pm k}_{\R1}-\g^{\pm}_{\L2}
\,.
\eal

 \noindent {\bf The  BES phase}
 \bal
\tPhi_{\L\L}^{--}( \tx_{\L 1}^{+0},x^{\pm}_{\L 2})\to &+\tPhi_{\L\L}^{--}({1\ov x_{\R 1}^{- k}},x_{\L 2}^{\pm}) - \tPsi_{\L}^{-}(u_1+{i\ov h}k,x_{\L 2}^{\pm})\,,
\\
\tPhi_{\L\L}^{++}( \tx_{\L 1}^{+0},x^{\pm}_{\L 2})\to &+\tPhi_{\L\L}^{++}({1\ov x_{\R 1}^{- k}},x_{\L 2}^{\pm}) - \tPsi_{\L}^{+}(u_1-{i\ov h}k,x_{\L 2}^{\pm})\,,
\eal
\bal
\tPsi_{\L}^-(x^+_{\L2}, \tx^{+ 0}_{\L 1})\to & \tPsi_{\L}^-(x^+_{\L2}, {1\ov x_{\R 1}^{- k}}) -K(u_{12}+{i\ov h}(k-1))\,,
\\
\tPsi_{\L}^+(x^+_{\L2}, \tx^{+ 0}_{\L 1})\to & \tPsi_{\L}^+(x^+_{\L2}, {1\ov x_{\R 1}^{- k}}) -K(u_{12}-{i\ov h}(k+1))\,,
\eal
\bal
2\tilde\theta_{\L\L}^\bes&(\tx_{\L 1}^{\pm 0},x_{\L 2}^{\pm})\to 
2\tilde\theta_{\R\L}^{k1}({ x_{\R 1}^{\pm k}},{x_{\L 2}^{\pm }})
+\Delta_{\L}^{+}(u_1\pm{i\ov h}k,x_{\L 2}^{\pm})
\\
&+K^\bes(u_{12}+{i\ov h}(k-1))
-K^\bes(u_{12}-{i\ov h}(k+1))
\,.
\eal
\bal
\exp&\big(-i\,\Delta_{\L}^{+}(u_1\pm {i\ov h}k,x_{\L 2}^{\pm})\big) 
\\
&={\left(\frac{h}{2}\right)^{+2 k}}{\prod
   _{j=1}^{k} \left(u_{12}+\frac{i \left(2
   j-1-k\right)}{h}\right) \left(u_{12}+\frac{i \left(-2
   j-1+k\right)}{h}\right)}
   \\
   &\times  
\left(\frac{x_{\L 2}^{+} }{x_{\L 2}^{- } }\right)^{k}
 \frac{ (\tx_{\R 1}^{-k}x_{\L 2}^{- }-1) (x_{\L 1}^{+k}-x_{\L 2}^{- }) } {
  (\tx_{\R 1}^{-k}x_{\L 2}^{+ }-1)(x_{\L 1}^{+k}-x_{\L 2}^{+ }) }
\prod_{j=1}^{k-1}
 \left( 
 \frac{ \tx_{\R1}^{k-2j}x_{\L 2}^{- }-1 } {
 \tx_{\R1}^{k-2j}x_{\L 2}^{+ }-1}\right)^2
 \,,
\eal
\bal
&\exp[i(K^\bes(u_{12}-{i\ov h} -{i\ov h}k)-K^\bes(u_{12} -{i\ov h}+{i\ov h}k))]
\\
&=\left(\frac{h}{2}\right)^{-2 k}{1\ov \prod
   _{j=1}^{k} \left(u_{12}+\frac{i \left(2
   j-1-k\right)}{h}\right) \left(u_{12}+\frac{i \left(-2
   j-1+k\right)}{h}\right)}
\,,
\eal

\bal
e^{-2i\tilde\theta_{\L\L}^\bes(\tx_{\L 1}^{\pm 0},x_{\L 2}^{\pm})}&\to 
e^{-2i\tilde\theta_{\R\L}^{k1}({ x_{\R 1}^{\pm k}},{x_{\L 2}^{\pm }})}
\\
   &\times 
\left(\frac{x_{\L 2}^{+} }{x_{\L 2}^{- } }\right)^{k}
 \frac{ (x_{\R 1}^{-k}x_{\L 2}^{- }-1) (x_{\L 1}^{+k}-x_{\L 2}^{- }) } {
  (x_{\R 1}^{-k}x_{\L 2}^{+ }-1)(x_{\L 1}^{+k}-x_{\L 2}^{+ }) }
\prod_{j=1}^{k-1}
 \left( 
 \frac{ \tx_{\R1}^{k-2j}x_{\L 2}^{- }-1 } {
 \tx_{\R1}^{k-2j}x_{\L 2}^{+ }-1}\right)^2
\eal

 \noindent {\bf The  HL phase}
 \bal 
e^{2i\tilde\theta_{\L \L}^\hl(\tx^{\pm 0}_{\L 1}, x^\pm_{\L2}) } &\to e^{2i\tilde\theta_{\R\L}^{\hl}({x_{\R 1}^{\pm k}},x_{\L 2}^{\pm})}
\,\frac{x^{- k}_{\R 1}x_{\L2}^{+}-1}{x^{- k}_{\R 1}x_{\L2}^{-}-1}\,
\frac{(x^{+k}_{\R 1}x_{\L2}^{-}-1)^2}{(x^{+ k}_{\R 1}x_{\L2}^{+}-1)^2}\,
\frac{x^{+ k}_{\L 1}-x_{\L2}^{+}}{x^{+ k}_{\L 1}-x_{\L2}^{-}} \,.
\eal
For the ratio of BES and HL we find
\bal
&e^{-2i\tilde\theta_{\L\L}^\bes(\tx_{\L 1}^{\pm 0},x_{\L 2}^{\pm})+2i\tilde\theta_{\L \L}^\hl(\tx^{\pm 0}_{\L 1}, x^\pm_{\L2}) }\to 
 e^{-2i\tilde\theta_{\R\L}^{k1}({ x_{\R 1}^{\pm k}},{x_{\L 2}^{\pm }})+2i\tilde\theta_{\R\L}^{\hl}({x_{\R 1}^{\pm k}},x_{\L 2}^{\pm})}
\\
   &\times 
\left(\frac{x_{\L 2}^{+} }{x_{\L 2}^{- } }\right)^{k}
\frac{(x^{+k}_{\R 1}x_{\L2}^{-}-1)^2}{(x^{+ k}_{\R 1}x_{\L2}^{+}-1)^2}\,
\prod_{j=1}^{k-1}
 \left( 
 \frac{ \tx_{\R}(u_1+{i\ov h}(k-2j))x_{\L 2}^{- }-1 } {
 \tx_{\R}(u_1+{i\ov h}(k-2j))x_{\L 2}^{+ }-1}\right)^2\,.
\eal

\paragraph{$S^{01}_{\chi Y}$ after two steps.} Combining the pieces, we get
\bal
S^{01}_{\chi Y }&(u_1,u_2)\to {\sqrt{ \a_{\L2}^-}\ov \sqrt{\a_{\L2}^+}}\, \frac{x_{\R1}^{- k}}{x_{\R1}^{+ k}}\, \left(\frac{x_{\L 2}^{+} }{x_{\L 2}^{- } }\right)^{k-1}
\frac{x^{+k}_{\R 1}x_{\L2}^{-}-1}{x^{- k}_{\R 1}x_{\L2}^{+}-1}\,{x_{\R1}^{+ k}x^{+}_{\L2}-1\ov x_{\R1}^{- k}x^{-}_{\L2}-1}
\\
&\times
\prod_{j=1}^{k-1}
 \left( 
 \frac{ \tx_{\R}(u_1+{i\ov h}(k-2j))x_{\L 2}^{- }-1 } {
 \tx_{\R}(u_1+{i\ov h}(k-2j))x_{\L 2}^{+ }-1}\right)^2\,\Sigma^{k1}_{\R\L}(u_1,u_2)^{-2}
 ={\sqrt{ \a_{\L2}^-}\ov \sqrt{\a_{\L2}^+}}\, S^{k1}_{\bar Z Y }(u_1,u_2)\,
 \,.
\eal
This can be written in the form
\bal
&S^{01}_{\chi Y} (p_1-2 \pi, p_2) = {\sqrt{ \a_{\L2}^-}\ov \sqrt{\a_{\L2}^+}}\,S^{k1}_{\bar{Z}Y} (p_1, p_2) \,,
\eal
and it is similar to the one for massive particles. By using fusion, we also get
\bal
&S^{0m}_{\chi Y} (p_1-2 \pi, p_2) = {\sqrt{ \a_{\L2}^{-m}}\ov \sqrt{\a_{\L2}^{+m}}}\,S^{km}_{\bar{Z}Y} (p_1, p_2)  
\,.
\eal

\medskip

Now, by using ($m_1=1,2,\ldots,k-1$)
\bal
&S^{m_1 \, m_2}_{Y \bar{Z}} (p_1-2 \pi, p_2) = 
\frac{\alpha_{\R 2}^{+m_2}}{\alpha_{\R 2}^{-m_2}}\  S^{k-m_1, m_2}_{\bar{Z} \bar{Z}} (p_1, p_2) \,,
\eal
we find
\bal
&S^{0m}_{\chi Y} (p_1-2 \pi, p_2-2 \pi) = {\sqrt{ \a_{\R2}^{+(k-m)}}\ov \sqrt{\a_{\R2}^{-(k-m)}}}\,\frac{\alpha_{\R 1}^{-k}}{\alpha_{\R 1}^{+k}}\  S^{k,k-m}_{\bar{Z} \bar{Z}} (p_1, p_2)\,,
\eal
For $m>k$ we simply get
\bal
&S^{0m}_{\chi Y} (p_1-2 \pi, p_2-2 \pi) = -{\sqrt{ \a_{\L2}^{-(m-k)}}\ov \sqrt{\a_{\L2}^{+(m-k)}}}\,S^{k,m-k}_{\bar{Z}Y} (p_1, p_2)\,,
\eal
where the minus sign comes because of the analytic continuation of $x_{\L2}^{+m}$ through the cut of $\sqrt{\a_\L(x)}$.

\subsection{String 
\texorpdfstring{$S_{\chi \bar Z}^{0m}(p_1,p_2)$}{SchibZ0mp1p2}}

Let us consider the analytic continuation to the string region of the following mirror S-matrix element
\bal\la{eq:SchibarZ01}
S_{\chi \bar Z}^{01}( \tx^{\pm0}_{\L1}, \tx^{\pm}_{\R2} ) = A_{\chi \bar Z}^{01}( \tx^{\pm0}_{\L1}, \tx^{\pm}_{\R2} )  \left(\Sigma^{01}_{\L\R}( \tx^{\pm0}_{\L1}, \tx^{\pm}_{\R2} ) \right)^{-2}\,,
 \eal
\bal
A_{\chi \bar Z}^{01}( \tx^{\pm0}_{\L1}, \tx^{\pm}_{\R2} ) &=
    \frac{\sqrt{\ta^+_{\R2}}}{\sqrt{\ta^-_{\R2}}} \, \frac{\tx^+_{\R2}}{\tx^-_{\R2}} \, \frac{1 - \tx^{-0}_{\L1} \tx^{-}_{\R2}}{1 - \tx^{+0}_{\L1} \tx^{+}_{\R2}} \, \frac{1 - \tx^{+0}_{\L1} \tx^{-}_{\R2}}{1 - \tx^{-0}_{\L1} \tx^{+}_{\R2}} \,,
 \eal
\bal
&\left(\Sigma^{01}_{\L\R}(\tx^{\pm0}_{\L1}, \tx^\pm_{\R 2})\right)^{2}=\left(\Sigma^{\besratio}_{\L\R}(\tx^{\pm0}_{\L1}, \tx^\pm_{\R 2})\right)^{2} \left(\Sigma^{\barnes}_{\L\R}(\tg^{\pm0}_{\L1}, \tg^\pm_{\R 2})\right)^{2} \,,
\eal
\bal
&\left( \Sigma^{\barnes}_{\L\R} (\tg^{\pm0}_{\L1}, \tg^\pm_{\R 2}) \right)^{-2}=\frac{R(\tg^{-0+}_{\L\R}+ i \pi) R(\tg^{-0+}_{\L\R}- i \pi) R(\tg^{+0-}_{\L\R}+ i \pi)R(\tg^{+0-}_{\L\R}- i \pi)}{R(\tg^{-0-}_{\L\R}+ i \pi) R(\tg^{-0-}_{\L\R}- i \pi) R(\tg^{+0+}_{\L\R}+ i \pi) R(\tg^{+0 +}_{\L\R}- i \pi)} \,.
\eal
To do the analytic continuation to positive $p_1$, $p_2$ we use $ \tx^{\pm0}_{\L1}\to  x^{\pm0}_{\L1}$, $ \tx^{\pm}_{\R2}\to x^{\pm}_{\R2}$, and get
\bal
A_{\chi \bar Z}^{01}( x^{\pm0}_{\L1}, x^{\pm}_{\R2} ) &=
    \frac{\sqrt{\a^+_{\R2}}}{\sqrt{\a^-_{\R2}}} \, \frac{x^+_{\R2}}{x^-_{\R2}} \, \frac{1 - x^{-0}_{\L1} x^{-}_{\R2}}{1 - x^{+0}_{\L1} x^{+}_{\R2}} \, \frac{1 - x^{+0}_{\L1} x^{-}_{\R2}}{1 - x^{-0}_{\L1} x^{+}_{\R2}}\,.
    \eal
Then,
\bal
&\left( \Sigma^{\barnes}_{\L\R} (\tg^{\pm0}_{\L1}, \tg^\pm_{\R 2}) \right)^{-2}\to  \frac{{\a^-_{\R2}}}{{\a^+_{\R2}}} \, \frac{x^-_{\R2}}{x^+_{\R2}} \, \frac{(1 - x^{+0}_{\L1} x^{+}_{\R2})^2}{(1 - x^{+0}_{\L1} x^{-}_{\R2})^2} \left( \Sigma^{\barnes}_{\L\R} (\g^{\pm0}_{\L1}, \g^\pm_{\R 2}) \right)^{-2}
\,.
\eal
Thus, we can write the string S-matrix element in the following form
\bal
S_{\chi \bar Z}^{01}( x^{\pm0}_{\L1}, x^{\pm}_{\R2} ) = B_{\chi \bar Z}^{01}( x^{\pm0}_{\L1}, x^{\pm}_{\R2} )  \left(\Sigma^{01}_{\L\R}( x^{\pm0}_{\L1}, x^{\pm}_{\R2} ) \right)^{-2}\,,
 \eal
\bal
B_{\chi \bar Z}^{01}( x^{\pm0}_{\L1}, x^{\pm}_{\R2} ) &=
    \frac{\sqrt{\a^-_{\R2}}}{\sqrt{\a^+_{\R2}}} \,
    \frac{1 - x^{-0}_{\L1} x^{-}_{\R2}}{1 - x^{-0}_{\L1} x^{+}_{\R2}} \,
     \frac{1 - x^{+0}_{\L1} x^{+}_{\R2}}{1 - x^{+0}_{\L1} x^{-}_{\R2}} \,,
 \eal
\bal
&\left(\Sigma^{01}_{\L\R}(x^{\pm0}_{\L1}, x^\pm_{\R 2})\right)^{2}=\left(\Sigma^{\besratio}_{\L\R}(x^{\pm0}_{\L1}, x^\pm_{\R 2})\right)^{2} \left(\Sigma^{\barnes}_{\L\R}(\g^{\pm0}_{\L1}, \g^\pm_{\R 2})\right)^{2} \,.
\eal
The fusion then gives the S-matrix element for a massless particle and a string $m$-particle bound state
\bal\la{eq:SchibarZom}
S^{0m}_{\chi \bar Z}( x^{\pm0}_{\L1}, x^{\pm m}_{\R2} )  = B^{0m}_{\chi \bar Z}( x^{\pm0}_{\L1}, x^{\pm m}_{\R2} ) \left(\Sigma^{0m}_{\L\R}( x^{\pm0}_{\L1}, x^{\pm m}_{\R2} ) \right)^{-2}\,,
\eal
\bal
 B^{0m}_{\chi \bar Z }( x^{\pm0}_{\L1}, x^{\pm m}_{\R2} ) =
\frac{\sqrt{\a^{-m}_{\R2}}}{\sqrt{\a^{+m}_{\R2}}}  \left(\frac{x_{ \L 1}^{-0}}{x_{\L 1}^{+0}}\right)^{m-1} \frac{1 - x^{-0}_{\L1} x^{-m}_{\R2}}{1- x^{+0}_{\L 1} x^{-m}_{\R 2}} \frac{1- x^{+0}_{\L 1} x^{+m}_{\R 2}}{1 - x^{-0}_{\L1} x^{+m}_{\R2}}  \prod_{j=1}^{m-1} \left(\frac{1-x^{+0}_{\L1} \tx_{\R 2}^{m-2j} }{1-x^{-0}_{\L1}\tx_{\R 2}^{m-2j} } \right)^2 \,.
\eal
The fusion of $\Sigma^{\barnes}_{\L\R}$ is trivial. Let us analyse the fusion of the even dressing factor.

The BES dressing phase analytically continued and fused is given by the usual expression
\bal 
\tilde\theta_{\L\R}^\bes(x_{\L1}^{\pm 0},x_{\R2}^{\pm m})   &=  \tilde\varPhi_{\L\R}(x_{\L1}^{\pm 0},x_{\R2}^{\pm m}) 
-\tilde\Psi_{\R}(u_1,x_{\R2}^{+ m})+\tilde\Psi_{\R}(u_1,x_{\R2}^{-m})
\\
&+\tilde\Psi_{\L}(u_2+{i\ov h}m ,x_{\L1}^{+ 0})-\tilde\Psi_{\L}(u_2+{i\ov h}m,x_{\L1}^{- 0})
-K^\bes(u_{12}-{i\ov h}m)\,.
\eal
We rewrite the combination of $\tPsi$ functions in it as follows
\bal
&-2\tilde\Psi_{\R}(u_1,x_{\R 2}^{+m})+2\tilde\Psi_{\R}(u_1,x_{\R 2}^{-m})+2\tilde\Psi_{\L}(u_2+{i\ov h}m,x_{\L1}^{+0})-2\tilde\Psi_{\L}(u_2+{i\ov h}m,x_{\L 1}^{-0})
\\
=&-2\tilde\Psi_{\R}(u_1,x_{\R 2}^{+m})+2\tilde\Psi_{\R}(u_1,x_{\R 2}^{-m})+\tilde\Psi^{-}_{\L}(u_2+{i\ov h}m,x_{\L1}^{+0})-\tilde\Psi^{-}_{\L}(u_2+{i\ov h}m,x_{\L 1}^{-0})
\\
&+\tilde\Psi^{+}_{\L}(u_2-{i\ov h}m,x_{\L1}^{+0})-\tilde\Psi^{+}_{\L}(u_2-{i\ov h}m,x_{\L 1}^{-0})
+\Delta_{\L}^{+}(u_2\pm {i\ov h}m,x_{\L 1}^{+ 0},x_{\L 1}^{- 0})
\,.
\eal
Using a proper identity for $\tPsi$ functions, we find
\bal
e&^{-i\Delta_{\L}^{+}(u_2\pm {i\ov h}m,x_{\L 1}^{+ 0},x_{\L 1}^{- 0})}
\\
&=\left(\frac{h}{2}\right)^{+2 m}
\prod_{j=1}^{m} \left(u_{12}-\frac{i \left(2 j-m\right)}{h}\right) \left(u_{12}-\frac{i \left(-2j+m\right)}{h}\right) 
   \\
   &\times \left(\frac{x_{\L 1}^{+0} }{x_{\L 1}^{-0} }\right)^{m}
   \frac{ (x_{\L 1}^{-0}x_{\R 2}^{- m}-1) }{
 (x_{\L 1}^{+ 0}x_{\R 2}^{-m}-1)} 
    \frac{ (x_{\L 1}^{- 0}\tx_{\R 2}^{+m}-1) }{ (x_{\L  1}^{+ 0}\tx_{\R 2}^{+m}-1)}
   \prod_{j=1}^{m-1}
 \left( \frac{x_{\L  1}^{- 0}\tx_{\R 2}^{+(m-2j)}-1 }{ x_{\L 1}^{+ 0} \tx_{\R 2}^{+(m-2j)}-1}\right)^2\,.
\eal
We also have 
\bal
&e^{2iK^\bes(u_{12}-{i\ov h}m)}\left(\frac{h}{2}\right)^{+2m}\prod_{j=1}^m\left(u_{12}-\frac{i(2j-m)}{h}\right) \left(u_{12}+\frac{i(2j-m)}{h}\right)
\\
&=(-1)^m\left({ \Gamma\big[{m\ov2}-\tfrac{ih}{2}u_{12}\big]\ov \Gamma\big[{m\ov2}+\tfrac{ih}{2}u_{12}\big]}\right)^2
\frac{u_{12}+ \frac{i}{h}m}{u_{12}- \frac{i}{h}m} \,. 
\eal
The HL phase gives
\bal
2\tilde\theta_{\L\R}^{\hl}(\tx_{\L1}^{\pm0},\tx_{\R 2}^{\pm m})&\ \to\  2\tilde\varPhi_{\L\R}^{\hl}({x_{ \L1}^{\pm0}},x_{\R 2}^{\pm m}) +\pi 
\\
&+ {1\ov i}\log  \frac{ x_{\L1}^{+0}}{  x_{\L1}^{-0}}\, \frac{  x_{\R 2}^{-m}}{ x_{\R 2}^{+m}}\,\frac{ ( x_{\L1}^{-0}x_{\R 2}^{+m}-1)^2}{( x_{\L1}^{+0} x_{\R 2}^{+m}-1)^2}
\, \frac{  x_{\L1}^{+0}  x_{\R 2}^{-m}-1}{ x_{\L1}^{-0}x_{\R 2}^{-m}-1}
\, \frac{  x_{\L1}^{+0} \tx_{\R 2}^{+m}-1}{  x_{\L1}^{-0}  \tx_{\R 2}^{+m}-1}
\,.
\eal
Thus,
\bal
\left({\Sigma^{0m\bes}_{\L\R}(\tx_{\L1}^{\pm0},\tx_{\R 2}^{\pm m})\ov \Sigma^{0m\hl}_{\L\R}(\tx_{\L1}^{\pm0},\tx_{\R 2}^{\pm m})}\right)^{-2}
 &=
 \left( \frac{x_{\L 1}^{+0} }{x_{\L 1}^{- 0} }\right)^{m+1} \,  {x_{\R2}^{-m}\ov x_{\R2}^{+m}}\, 
\frac{ ( x_{\L1}^{-0}x_{\R 2}^{+m}-1)^2}{( x_{\L1}^{+0} x_{\R 2}^{+m}-1)^2}\,  \prod_{j=1}^{m-1}
 \left( \frac{x_{\L  1}^{- 0}\tx_{\R 2}^{+(m-2j)}-1 }{ x_{\L 1}^{+ 0} \tx_{\R 2}^{+(m-2j)}-1}\right)^2
\\
&\times
 (-1)^{m-1}\left({ \Gamma\big[{m\ov2}-\tfrac{ih}{2}u_{12}\big]\ov \Gamma\big[{m\ov2}+\tfrac{ih}{2}u_{12}\big]}\right)^2
\frac{u_{12}+ \frac{i}{h}m}{u_{12}- \frac{i}{h}m} \, e^{-2i\delta^{0m}_{\L\R}}
\,,
\eal
where
\bal
2\delta^{0m}_{\L\R}=& +2\tilde\varPhi_{\L\R}(x_{\L1}^{\pm 0},x_{\R 2}^{\pm m})-2\tilde\varPhi_{\L\R}^\hl(x_{\L1}^{\pm 0},x_{\R 2}^{\pm m})
\\
& -2\tilde\Psi_{\R}(u_1,x_{\R 2}^{+ m})+2\tilde\Psi_{\R}(u_1,x_{\R 2}^{-m})+\tilde\Psi^+_{\L}(u_2+{i\ov h}m ,x_{\L1}^{+ 0})-\tilde\Psi^+_{\L}(u_2+{i\ov h}m,x_{\L1}^{- 0})
\\
&+\tilde\Psi^{-}_{\L}(u_2-{i\ov h}m,x_{\L1}^{+0})-\tilde\Psi^{-}_{\L}(u_2-{i\ov h}m,x_{\L 1}^{-0})
\,.
\eal
Collecting all the terms, we get
\bal
S^{0m}_{\chi \bar Z}( p_1, p_2)  &=
   \frac{\sqrt{\a^-_{\R2}}}{\sqrt{\a^+_{\R2}}} \, \left( \frac{x_{\L 1}^{+0} }{x_{\L 1}^{- 0} }\right)^{m+1}\,
    \frac{1 - x^{-0}_{\L1} x^{-m}_{\R2}}{1 - x^{+0}_{\L1} x^{+m}_{\R2}} \,
     \frac{1 - x^{-0}_{\L1} x^{+m}_{\R2}}{1 - x^{+0}_{\L1} x^{-m}_{\R2}} \,
\Sigma_{\L\R}^\barnes(\g_{\L1}^{\pm0},\g_{\R2}^\pm)^{-2}\, 
\\
& \times (-1)^{m-1}
 \left( \frac{x_{\L 1}^{-0} }{x_{\L 1}^{+ 0} }\right)^{m-1} \,  \frac{x_{\R 2}^{-m}}{x_{\R 2}^{+m}}\, \left({ \Gamma\big[{m\ov2}-\tfrac{ih}{2}u_{12}\big]\ov \Gamma\big[{m\ov2}+\tfrac{ih}{2}u_{12}\big]}\right)^2
\frac{u_{12}+ \frac{i}{h}m}{u_{12}- \frac{i}{h}m} \, e^{-2i\delta^{0m}_{\L\R}}
\,,
\eal
which agrees with the pure RR case. Note that up to the factor $ \frac{\sqrt{\a^-_{\R2}}}{\sqrt{\a^+_{\R2}}} $ it coincides with the limit $\delta\to 0$ of $S^{\delta m}_{Y \bar Z}( x^{\pm\delta}_{\L1}, x^{\pm m}_{\R2} ) $.

\subsection{String 
\texorpdfstring{$S_{\chi \bar Z}^{0m}(p_1-2\pi,p_2)$}{SchibZ0mp1m2pip2} and \texorpdfstring{$S_{\chi \bar Z}^{0m}(p_1-2\pi,p_2-2\pi)$}{SchibZ0mp1m2pip2m2pi}}

We use
again the  mirror S-matrix element \eqref{eq:massless_massive_m0m1}, and continue as
\bal
\tx_{\L 1}^{\pm 0}\ \to\   {1\ov x_{\R1}^{\pm k} } \,,
\quad
\tg_{\L1}^{\mp 0}\ \to \ \g_{\R1}^{\pm k} -{i\pi\ov2}\,,
\eal
\bal
\tx_{\R 2}^{\pm}\ \to\   x_{\R2}^{\pm}  \,,
\quad
\tg_{\R2}^{\pm } 
\ \to\   \g_{\R2}^{\pm } +{i\pi\ov2}\,.
\eal
We get
\bal
A_{\chi \bar Z}^{01}( \tx^{\pm0}_{\L1}, \tx^{\pm}_{\R2} ) &\to A_{\chi \bar Z}^{01}( p_1-2\pi, p_2 )=
    \frac{\sqrt{\a^+_{\R2}}}{\sqrt{\a^-_{\R2}}} \, \frac{x^+_{\R2}}{x^-_{\R2}} \, \frac{ x^{+k}_{\R1} -x^{-}_{\R2}}{x^{+k}_{\R1} -x^{+}_{\R2}} \, \frac{x^{-k}_{\R1}- x^{-}_{\R2}}{x^{-k}_{\R1} - x^{+}_{\R2}} \,,
 \eal
 \bal
\left( \Sigma^{\barnes}_{\L\R} (\tg^{\pm0}_{\L1}, \tg^\pm_{\R 2}) \right)^{-2}&\to \frac{ x^{+k}_{\R1} -x^{-}_{\R2}}{x^{+k}_{\R1} -x^{+}_{\R2}} \, \frac{x^{-k}_{\R1}- x^{+}_{\R2}}{x^{-k}_{\R1} - x^{-}_{\R2}}\left( \Sigma^{\barnes}_{\R\R} (\g^{\pm k}_{\R1}, \g^\pm_{\R 2}) \right)^{-2}\,,
\eal
Thus, we can write the string S-matrix element in the following form
\bal
S_{\chi \bar Z}^{01}( {1\ov x^{\mp k}_{\R1}}, x^{\pm}_{\R2} ) = B_{\chi \bar Z}^{01}( {1\ov x^{\mp k}_{\R1}}, x^{\pm}_{\R2} )  \left(\Sigma^{01}_{\L\R}({1\ov x^{\mp k}_{\R1}}, x^{\pm}_{\R2} ) \right)^{-2}\,,
 \eal
\bal
B_{\chi \bar Z}^{01}( {1\ov x^{\mp k}_{\R1}}, x^{\pm}_{\R2} ) &=
    \frac{\sqrt{\a^+_{\R2}}}{\sqrt{\a^-_{\R2}}} \,
  \, \frac{x^+_{\R2}}{x^-_{\R2}} \, \left(\frac{ x^{+ k}_{\R1} -x^{-}_{\R2}}{x^{+ k}_{\R1} -x^{+}_{\R2}} \right)^2\,,
 \eal
\bal
&\left(\Sigma^{01}_{\L\R}({1\ov x^{\mp k}_{\R1}}, x^\pm_{\R 2})\right)^{2}=\left(\Sigma^{\besratio}_{\L\R}({1\ov x^{\mp k}_{\R1}}, x^\pm_{\R 2})\right)^{2} \left(\Sigma^{\barnes}_{\R\R}(\g^{\pm k}_{\R1}, \g^\pm_{\R 2})\right)^{2} \,.
\eal

 Let us analyse the continuation of the even dressing factor. The BES factor is continued in the same way as for massive particles, and we get
\bal
2\tilde\theta_{\L\R}^\bes&(\tx_{\L 1}^{\pm 0},\tx_{\R 2}^{\pm})\to 
2\tilde\theta_{\R\R}^{k1}({ x_{\R 1}^{\pm k}},{x_{\R 2}^{\pm }})
\\
&+\Delta_{\R}^{-}(u_1\pm{i\ov h}k,x_{\R 2}^{\pm })+K^\bes(u_{12}-{i\ov h}+{i\ov h}k)
-K^\bes(u_{12}-{i\ov h}-{i\ov h}k)\,,
\eal
By using identities for $\tPsi$-functions, and the BES kernel, we obtain
\bal
e^{-2i\tilde\theta_{\L\R}^\bes(\tx_{\L 1}^{\pm 0},\tx_{\R 2}^{\pm})}&\to 
e^{-2i\tilde\theta_{\R\R}^{k1}({ x_{\R 1}^{\pm k}},{x_{\R 2}^{\pm }})} {\left(u_{12}+\frac{i \left(k+1\right)}{h}\right) \ov \left(u_{12}-\frac{i \left(k+1\right)}{h}\right)}{\left(u_{12}+\frac{i \left(k-1\right)}{h}\right)\ov\left(u_{12}-
\frac{i \left(k-1\right)}{h}\right)}
\\
   &\times \left(\frac{x_{\R 2}^{-} }{x_{\R 2}^{+} }\right)^{k}
   \frac{ (x_{\R 1}^{-k}-x_{\R 2}^{+}) (\tx_{\R 1}^{+k}-x_{\R 2}^{+}) }{
   (x_{\R 1}^{-k}-x_{\R 2}^{-}) (\tx_{\R 1}^{+k}-x_{\R 2}^{-})} \prod_{j=1}^{k-1}
 \left( \frac{ \tx_{\R 1}^{+(k-2j)}-x_{\R 2}^{+ } }{\tx_{\R 1}^{+(k-2j)}-x_{\R 2}^{-} }\right)^2\,.
\eal

The HL phase is continued as
\bal
2\tilde\theta_{\L\R}^\hl(\tx_{\L 1}^{\pm 0},\tx_{\R 2}^{\pm})&\ \to\  2\tPhi^{\hl}_{\R\R}({ x_{\R 1}^{\pm k}},{x_{\R 2}^{\pm }}) +\pi 
\\
&+ {1\ov i}\log  \frac{ x_{\R 1}^{+ k}}{  x_{\R 1}^{- k}}\, \frac{  x_{\R 2}^{-}}{ x_{\R 2}^{+}}\,\frac{ ( x_{\R 1}^{+ k}-x_{\R 2}^{+})^2}{( x_{\R 1}^{- k}-x_{\R2}^{+})^2}
\, \frac{ x_{\R 1}^{- k} - x_{\R 2}^-}{ x_{\R 1}^{+ k}-x_{\R 2}^-}
\, \frac{ x_{\R 1}^{- k} -\tx_{\R2}^{+}}{ x_{\R 1}^{+ k}-  \tx_{\R 2}^+}
\,.
\eal
Using that
\bal
2\tilde\theta_{\R\R}^{\hl}({ x_{\R 1}^{\pm k}},{x_{\R 2}^{\pm }})&=2\tPhi_{\R\R}^{\hl}({ x_{\R 1}^{\pm k}},{x_{\R 2}^{\pm }}) +\pi+ {1\ov i}\log {x_{\R1}^{+k}\ov x_{\R1}^{-k}}{x_{\R2}^-\ov x_{\R2}^+}\, 
\frac{x_{ \R1}^{+k} -  x_{\R2}^-}{x_{\R1}^{-k}-x_{\R2}^+}\,  
\frac{\tx_{\R1}^{+k} - x_{\R2}^+ }{\tx_{\R1}^{+k} -  x_{\R2}^-}\,
 \frac{x_{\R1}^{-k}-\tx_{ \R2}^+ }{x_{ \R1}^{+k} - \tx_{\R2}^+  } 
\,,
\eal
we get
\bal
2\tilde\theta_{\L\R}^\hl(\tx_{\L 1}^{\pm 0},\tx_{\R 2}^{\pm})&\ \to\  2\tilde\theta^{\hl}_{\R\R}({ x_{\R 1}^{\pm k}},{x_{\R 2}^{\pm }}) + {1\ov i}\log \frac{ ( x_{\R 1}^{+ k}-x_{\R 2}^{+})^2}{( x_{\R 1}^{+ k}-x_{\R2}^{-})^2}
\, \frac{ x_{\R 1}^{- k} - x_{\R 2}^-}{ x_{\R 1}^{- k}-x_{\R 2}^+}\,\frac{\tx_{\R1}^{+k} - x_{\R2}^- }{\tx_{\R1}^{+k} -  x_{\R2}^+}
\,.
\eal
Thus,  the ratio of BES and HL is given by
\bal
&e^{-2i\tilde\theta_{\L\R}^\bes(\tx_{\L 1}^{\pm 0},\tx_{\R 2}^{\pm})+2i\tilde\theta_{\L\R}^\hl(\tx_{\L 1}^{\pm 0},\tx_{\R 2}^{\pm})}\to 
e^{-2i\tilde\theta_{\R\R}^{k1}({ x_{\R 1}^{\pm k}},{x_{\R 2}^{\pm }})+2i\tilde\theta^{\hl}_{\R\R}({ x_{\R 1}^{\pm k}},{x_{\R 2}^{\pm }}) }
\\
   &\times {u_{12}+\frac{i \left(k+1\right)}{h} \ov u_{12}-\frac{i \left(k+1\right)}{h}}
   \, {u_{12}+\frac{i \left(k-1\right)}{h}\ov u_{12}-\frac{i \left(k-1\right)}{h}}
 \left(\frac{x_{\R 2}^{-} }{x_{\R 2}^{+} }\right)^{k}
  \frac{ ( x_{\R 1}^{+ k}-x_{\R 2}^{+})^2}{( x_{\R 1}^{+ k}-x_{\R2}^{-})^2}\prod_{j=1}^{k-1}
 \left( \frac{ \tx_{\R 1}^{+(k-2j)}-x_{\R 2}^{+ } }{\tx_{\R 1}^{+(k-2j)}-x_{\R 2}^{-} }\right)^2\,.
\eal
We therefore find
\bal
S_{\chi \bar Z}^{01}( p_1-2\pi, p_2) = 
    \frac{\sqrt{\a^+_{\R2}}}{\sqrt{\a^-_{\R2}}} \,
S_{\bar Z \bar Z}^{k1}( p_1, p_2) 
\,.
 \eal
The fusion then gives
\bal\la{eq:SchibarZ0mp1np2p}
S_{\chi \bar Z}^{0m}( p_1-2\pi, p_2) = 
    \frac{\sqrt{\a^{+m}_{\R2}}}{\sqrt{\a^{-m}_{\R2}}} \,
S_{\bar Z \bar Z}^{km}( p_1, p_2)
\,.
 \eal

 \medskip

Now, by using
\bal 
 S_{ \bar Z \bar Z}^{m_1,m_2}(p_{1}-2\pi,p_{2}) = S_{ \bar Z \bar Z}^{m_1+k,m_2}(p_{1},p_{2}) \,.
\eal
we get
\bal
S_{\chi \bar Z}^{0m}( p_1-2\pi, p_2-2\pi) &= 
   - \frac{\sqrt{\a^{+(m+k)}_{\R2}}}{\sqrt{\a^{-(m+k)}_{\R2}}} \,
S_{\bar Z \bar Z}^{k,m+k}( p_1, p_2)\,,
 \eal
 where the minus sign comes from crossing the cut of $\a_\R(x)$.

\subsection{String \texorpdfstring{$S_{\chi Y}^{0m}(p_1,p_2-2\pi)$}{SchiY0mp1p2m2pi} and \texorpdfstring{$S_{\chi \bar Z}^{0m}(p_1,p_2-2\pi)$}{SchibZ0mp1p2m2pi}}

Let us also find the string S-matrix element $S_{\chi Y}^{0m}(p_1,p_2-2\pi)$ with  $p_1,p_2\in (0,2\pi)$. 
This can be done by using the relations \eqref{eq:Sstringomlimit}
\bal
S_{\chi Y}^{0m}( p_1, p_2) &={\sqrt{ \a_{\L2}^{+m}}\ov \sqrt{\a_{\L2}^{-m}}}\, \lim_{\delta\to 0}\, S_{Y Y}^{\delta m}( p_1, p_2 )\,,
\\
S_{\chi \bar Z}^{0m}( p_1, p_2 ) &=\frac{\sqrt{\a^{-m}_{\R2}}}{\sqrt{\a^{+m}_{\R2}}}\,\, \lim_{\delta\to 0}\, S_{Y \bar Z}^{\delta m}( p_1, p_2 )\,,
\eal
their analytic continuation in $p_2$
\bal
S_{\chi Y}^{0m}( p_1, p_2-2\pi) &=+{\sqrt{ \a_{\R2}^{-(k-m)}}\ov \sqrt{\a_{\R2}^{-(k-m)}}}\ \lim_{\delta\to 0}\, S_{Y Y}^{\delta m}( p_1, p_2-2\pi )\,,\qquad &&0<m<k\,,
\\
S_{\chi Y}^{0m}( p_1, p_2-2\pi) &=-{\sqrt{ \a_{\L2}^{+(m-k)}}\ov \sqrt{\a_{\L2}^{-(m-k)}}}\ \lim_{\delta\to 0}\, S_{Y Y}^{\delta m}( p_1, p_2-2\pi )\,,\qquad &&k<m\,,
\\
S_{\chi \bar Z}^{0m}( p_1, p_2-2\pi ) &=-\frac{\sqrt{\a^{-(m+k)}_{\R2}}}{\sqrt{\a^{+(m+k)}_{\R2}}}\,\, \lim_{\delta\to 0}\, S_{Y \bar Z}^{\delta, m}( p_1, p_2 -2\pi)\,,\qquad &&0<m\,,
\eal
and the monodromy relations for $S_{Y Y}^{\delta m}$ from \cite{Frolov:2025uwz}
\bal
&S^{\delta m}_{{Y} {Y}} (p_1, p_2-2 \pi) = \frac{\alpha_{\L 1}^{+\delta}}{\alpha_{\L 1}^{-\delta}}\ S^{\delta , k-m}_{{Y} {\bar Z}} (p_1, p_2)\,,&&\qquad 0<m<k\,,
\\
&S^{\delta m}_{{Y} {Y}} (p_1, p_2-2 \pi) = S^{\delta , m-k}_{{Y} {Y}} (p_1, p_2)\,,&&\qquad k<m
\\
&S^{\delta m}_{{Y} {\bar Z}} (p_1, p_2-2 \pi) = S^{\delta , m+k}_{{Y} {\bar Z}} (p_1, p_2)\,,&&\qquad 0<m
 \,.
\eal
Using the formulae, we find immediately
\bal
S^{0m}_{\chi Y} (p_1, p_2-2 \pi) &=  {\a_{\L1}^{+0}\ov \a_{\L1}^{-0}} S^{0,k- m}_{\chi\bar{Z}} (p_1, p_2)\,,&&\qquad 0<m<k \,,
\eal
which is the same relation as for massive particles. 

Next,
\bal
S^{0m}_{\chi Y}(p_1,p_2-2\pi)& = - S^{0,m-k}_{\chi Y}(p_1,p_2)\,,&&\qquad k<m\,,
\eal
which differs by the minus sign from the corresponding relation for massive particles. 

Finally,
\bal
S_{\chi \bar Z}^{0m}( p_1, p_2-2\pi ) &=- S_{\chi\bar Z}^{0,m+k}(p_1,p_2)\,,\qquad &&0<m\,,
\eal
which again differs by the  sign from the  relation for massive particles.

\subsection{String 
\texorpdfstring{$S_{\chi \chi}^{00}(p_1,p_2)$}{Schichi00p1p2}}

We begin with the mirror S-matrix element $S_{\chi \chi}^{00}$ given by \eqref{eq:massless_massless_m0m0}
\bal
  S^{00}_{\chi \chi}(\tx_{\L1}^{\pm0},\tx_{\L2}^{\pm0})&=A^{00}_{\chi \chi}(\tx_{\L1}^{\pm0},\tx_{\L2}^{\pm0})\left(\Sigma^{00}_{\L\L}(\tx_{\L1}^{\pm0},\tx_{\L2}^{\pm0})\right)^{-2}\,,
  \\
   A^{00}_{\chi \chi}(\tx_{\L1}^{\pm0},\tx_{\L2}^{\pm0}) &=
    {\sqrt{\ta_{\L1}^{+0}} \ov \sqrt{\ta_{\L1}^{-0}}} \,  {\sqrt{\ta_{\L2}^{-0}} \ov \sqrt{\ta_{\L2}^{+0}}}  \ \frac{\tx^{+0}_{\L1}}{\tx^{-0}_{\L1}}\, \frac{\tx^{-0}_{\L2}}{\tx^{+0}_{\L2}} \, \left(\frac{\tx^{-0}_{\L1} - \tx^{+0}_{\L2}}{\tx^{+0}_{\L1} - \tx^{-0}_{\L2}} \right)^2\,.
\eal
The analytic continuation of $ A^{00}_{\chi \chi}$ to the string region on the positive momentum branch is straightforward
\bal
A^{00}_{\chi \chi}(\tx_{\L1}^{\pm0},\tx_{\L2}^{\pm0}) &\to A^{00}_{\chi \chi}(x_{\L1}^{\pm0},x_{\L2}^{\pm0}) =
   {\sqrt{\a_{\L1}^{+0}} \ov \sqrt{\a_{\L1}^{-0}}} \,  {\sqrt{\a_{\L2}^{-0}} \ov \sqrt{\a_{\L2}^{+0}}}   \ \frac{x^{+0}_{\L1}}{x^{-0}_{\L1}} \, \frac{x^{-0}_{\L2}}{x^{+0}_{\L2}} \, \left(\frac{x^{-0}_{\L1} - x^{+0}_{\L2}}{x^{+0}_{\L1} - x^{-0}_{\L2}} \right)^2\,.
\eal
The BES phase is continued as follows 
\bal\la{eq:BESstring00}
\tilde{\theta}_{\L \L}^{\bes}(\tx^{\pm 0}_{\L 1}, \tx^{\pm 0}_{\L 2})\to \tilde{\theta}_{\L \L}^{\bes}(x^{\pm 0}_{\L 1}, x^{\pm 0}_{\L 2})&=\tilde{\varPhi}_{\L \L}(x^{\pm 0}_{\L 1},x^{\pm 0}_{\L 2})-\tPsi_{\L}(u_1, x^{+0}_{\L2})+\tPsi_{\L}(u_1, x^{-0}_{\L2})\\
&+\tPsi_\L(u_2, x^{+0}_{\L1}) -\tPsi_\L(u_2, x^{-0}_{\L1})- K^\bes \left(u_1-u_2\right) \,.
\eal
The continuation of HL can be done using the results in appendix D.3 of~\cite{Frolov:2025uwz}
\bal
2\tilde{\theta}^\hl_{\L\L}(\tx^{\pm 0}_{\L1}, \tx^{\pm 0}_{\L2})\to 2\tilde{\theta}^\hl_{\L\L}(x^{\pm 0}_{\L1}, x^{\pm 0}_{\L2})&= 2\tilde{\varPhi}^\hl_{\L\L}(x^{\pm 0}_{\L1}, x^{\pm 0}_{\L2})+ \frac{1}{ i} \log \frac{x^{+ 0}_{\L1}}{x^{- 0}_{\L1}} \frac{x^{- 0}_{\L2}}{x^{+ 0}_{\L2}} + \pi\,.
\eal
The analytically continued HL dressing factor satisfies the following equality
\bal
\Sigma^{\hl}_{\L\L} (\g^{\pm0}_{\L1}, \g^{\pm0}_{\L1})^2  = -1\,.
\eal
Equivalently, we could have obtained this continuation from the Barnes function representation~\eqref{eq:massless_massless_m0m0b}.
The continuation of the odd factor is  simple, and we get
\bal
\Sigma^{\barnes}_{\L\L} (\tg^{\pm0}_{a1}, \tg^{\pm 0}_{a2})^{-2} \rightarrow \,  \frac{x_{\L1}^{-0}}{x_{\L1}^{+0}} \, \frac{x_{\L2}^{+0}}{x_{\L2}^{-0}} \, \frac{\alpha^{-0}_{\L1}}{\alpha^{+0}_{\L1}} \, \frac{\alpha^{+0}_{\L2}}{\alpha^{-0}_{\L2}} \, \left(\frac{x^{+0}_{\L1} - x^{-0}_{\L2}}{x^{-0}_{\L1} - x^{+0}_{\L2}} \right)^2 \, \Sigma^{\barnes}_{\L\L} (\g^{\pm0}_{\L1}, \g^{\pm0}_{\L2})^{-2}.
\eal
Therefore,  the S-matrix element $S_{\chi \chi}^{00}$ in the string region is given by 
\begin{equation}
\label{eq:S_massless_massless_pos_pos_final2}
    \begin{aligned}
S_{\chi \chi}^{00}(p_1,p_2)=
    {\sqrt{\a_{\L1}^{-0}} \ov \sqrt{\a_{\L1}^{+0}}} \,  {\sqrt{\a_{\L2}^{+0}} \ov \sqrt{\a_{\L2}^{-0}}} \,  \left(\Sigma^{\barnes}_{\L \L} (\g^{\pm0}_{\L1}, \g^{\pm0}_{ \L 2}) \right)^{-2} \left(\Sigma^{00, \besratio}_{\L\L}(u_1,u_2)\right)^{-2} \, .
    \end{aligned}
\end{equation}
It satisfies
\bal
S_{\chi \chi}^{00}(p,p)=-1\,,
\eal
which is the usual condition one imposes if one does not want to have any number of particles with the same momentum.

Note that \eqref{eq:S_massless_massless_pos_pos_final2} can be also written in the  form  
\bal\la{eq:S00str1}
S^{00}_{\chi \chi }(p_1,p_2)&= -
 {\sqrt{\a_{\L1}^{-0}} \ov \sqrt{\a_{\L1}^{+0}}} \, { \sqrt{\a_{\L2}^{+0}}\ov \sqrt{\a_{\L2}^{-0}}}\,\frac{ x_{ \L 1}^{+0}}{x_{\L 1}^{-0}} \,  \frac{x_{\L 2}^{-0}}{x_{\L 2}^{+0}}\,
\frac{R^2(\g^{-0-0}_{\L\L}) R^2(\g^{+0+0}_{\L\L})}{R^2(\g^{-0+0}_{\L\L}) R^2(\g^{+0-0}_{\L\L}) }  \ 
  \left({ \Gamma\big[-\tfrac{ih}{2}u_{12}\big]\ov \Gamma\big[+\tfrac{ih}{2}u_{12}\big]}\right)^2\, e^{-2i\delta^{00}_{\L\L}}
\,,
\eal
where 
\bal
2\delta^{00}_{\L\L}=& +2\tilde\varPhi_{\L\L}(x_{\L1}^{\pm 0},x_{\L 2}^{\pm 0})-2\tilde\varPhi_{\L\L}^\hl(x_{\L1}^{\pm 0},x_{\L 2}^{\pm 0})
-2\tilde\Psi_{\L}(u_1,x_{\L2}^{+ 0})+2\tilde\Psi_{\L}(u_1,x_{\L2}^{-0})
\\
& +2\tilde\Psi_{\L}(u_2 ,x_{\L1}^{+ 0})-2\tilde\Psi_{\L}(u_2 ,x_{\L1}^{- 0})
\,.
\eal
Finally, we can also replace the contribution of the HL factor with a $\g$-dependent term by using that
\bal\la{eq:HLvsBarnesstr}
\Sigma^{\hl}_{\L\L}(x^{\pm0}_{\L 1},x^{\pm0}_{\L 2})^{2} &= -\frac{ x_{ \L 1}^{+0}}{x_{\L 1}^{-0}}\,\frac{x_{\L 2}^{-0}}{x_{\L 2}^{+0}}\, e^{2i\tilde\varPhi^\hl(x^{\pm0}_{\L 1},x^{\pm0}_{\L 2})}
\\
&=- {R(\g^{+0-0}_{\L\L}-2\pi i)^2 R(\g^{+0-0}_{\L\L})^2R(\g^{-0+0}_{\L\L}+2\pi i)^2R(\g^{-0+0}_{\L\L} )^2\ov R(\g^{-0-0}_{\L\L})^4 R(\g^{+0+0}_{\L\L})^4}\,,
\eal
where 
\bal
\g^{\pm0\pm0}_{\L\L} = \g^{\pm0}_{\L1} - \g^{\pm0}_{\L2}\,. 
\eal
Then, eq.\eqref{eq:S00str1} takes the form 
\bal\la{eq:S00str2}
S^{00}_{\chi \chi }(p_1,p_2)=& -
 {\sqrt{\a_{\L1}^{-0}} \ov \sqrt{\a_{\L1}^{+0}}} \, { \sqrt{\a_{\L2}^{+0}}\ov \sqrt{\a_{\L2}^{-0}}}\,
\frac{R^2(\g^{+0-0}_{\L\L}-2\pi) i R^2(\g^{-0+0}_{\L\L}+2\pi i)}{R^2(\g^{-0-0}_{\L\L}) R^2(\g^{+0+0}_{\L\L}) }  
\\
& \times 
  \left({ \Gamma\big[-\tfrac{ih}{2}u_{12}\big]\ov \Gamma\big[+\tfrac{ih}{2}u_{12}\big]}\right)^2\, e^{-2i\check\delta^{00}_{\L\L}}
\,,
\eal
where 
\bal
2\check\delta^{00}_{\L\L}=& +2\tilde\varPhi_{\L\L}(x_{\L1}^{\pm 0},x_{\L 2}^{\pm 0})
-2\tilde\Psi_{\L}(u_1,x_{\L2}^{+ 0})+2\tilde\Psi_{\L}(u_1,x_{\L2}^{-0})
 +2\tilde\Psi_{\L}(u_2 ,x_{\L1}^{+ 0})-2\tilde\Psi_{\L}(u_2 ,x_{\L1}^{- 0})
\,,
\eal
or equivalently
\bal\la{eq:S00str3}
S^{00}_{\chi \chi }(p_1,p_2)=& -
 {\sqrt{\a_{\L1}^{+0}} \ov \sqrt{\a_{\L1}^{-0}}} \, { \sqrt{\a_{\L2}^{-0}}\ov \sqrt{\a_{\L2}^{+0}}}\, \frac{x^{+0}_{\L1}}{x^{-0}_{\L1}}\, \frac{x^{-0}_{\L2}}{x^{+0}_{\L2}}\left(\frac{x^{-0}_{\L1} - x^{+0}_{\L2}} {x^{+0}_{\L1} - x^{-0}_{\L2}}\right)^2 
 \\
& \times 
\frac{R^2(\g^{+0-0}_{\L\L}) i R^2(\g^{-0+0}_{\L\L})}{R^2(\g^{-0-0}_{\L\L}) R^2(\g^{+0+0}_{\L\L}) }  \,
  e^{-2i\theta^{\bes}_{\L\L}(x_{\L1}^{\pm 0},x_{\L 2}^{\pm 0})}
\,,
\eal
where $\theta^{\bes}_{\L\L}(x_{\L1}^{\pm 0},x_{\L 2}^{\pm 0})$ is the analytically continued BES phase \eqref{eq:BESstring00}.

\subsection{String 
\texorpdfstring{$S_{\chi \chi}^{00}(p_1,p_2-2\pi)$}{Schichi00p1p2m2pi} and \texorpdfstring{$S_{\chi \chi}^{00}(p_1-2\pi,p_2-2\pi)$}{Schichi00p1m2pip2m2pi}}

We use again \eqref{eq:massless_massless_m0m0}, and first do
the analytic continuation of $A^{00}_{\chi \chi}$ 
\bal
A^{00}_{\chi \chi}(\tx_{\L1}^{\pm0},\tx_{\L2}^{\pm0}) &\to A^{00}_{\chi \chi}(p_1,p_2-2\pi) =
    {\sqrt{\a_{\L1}^{+0}} \ov \sqrt{\a_{\L1}^{-0}}} \,  {\sqrt{\a_{\R2}^{+k}} \ov \sqrt{\a_{\R2}^{-k}}}  \ \frac{x^{+0}_{\L1}}{x^{-0}_{\L1}} \, \frac{x^{+k}_{\L2}}{x^{-k}_{\R2}} \, \left(\frac{x^{-0}_{\L1} x^{-k}_{\R2}-1}{x^{+0}_{\L1} x^{+k}_{\R2}-1} \right)^2\,,
\eal
and the odd factor
\bal
\label{eq:sec_pos_neg_massless_branch_odd_dressing_final}
(\Sigma^{0 0 \barnes}_{\L\L}(u_1, u_2))^{-2}\to \frac{\a^{-k}_{\R2}}{\a^{+k}_{\R2}}\, \frac{x^{-k}_{\R2}}{x^{+k}_{\R2}} \, \frac{(1-x_{\L1}^{-0} x_{\R2}^{+k})(1-x_{\L1}^{+0} x_{\R2}^{+k})}{(1-x_{\L1}^{+0} x_{\R2}^{-k}) (1-x_{\L1}^{-0} x_{\R2}^{-k})}  \, (\Sigma^{0 k \barnes}_{\L\R} (u_1, u_2))^{-2}\,.
\eal

The BES phase is continued as follows. 
The  continuation of the first variable to  the positive momentum branch of the string region gives
\bal
\tilde{\theta}_{\L \L}^{\bes}(x^{\pm 0}_{\L 1}, \tx^{\pm 0}_{\L2})&=\tilde{\varPhi}_{\L \L}(x^{\pm 0}_{\L 1},\tx^{\pm0}_{\L2})-\tPsi_{\L}(u_1, \tx^{+0}_{\L2})+\tPsi_{\L}(u_1, \tx^{-0}_{\L2})\,.
\eal
Continuing then the second variable to the negative momentum branch, we obtain
\bal\la{eq:temp1}
\tilde{\theta}_{\L \L}^{\bes}(x^{\pm 0}_{\L 1}, \frac{1}{x^{\mp k}_{\R2}})&=\tilde{\varPhi}_{\L \L}(x^{\pm 0}_{\L 1},\frac{1}{x^{\mp k}_{\R2}})\\
&-\tPsi_{\L}(u_1, \frac{1}{x^{-k}_{\R2}})+\tPsi_{\L}(u_1, \frac{1}{x^{+k}_{\R2}})+\frac{1}{2}\tPsi^-_\L(u_2+\frac{i}{h} k, x^{+0}_{\L1}) -\frac{1}{2}\tPsi^-_\L(u_2+\frac{i}{h} k, x^{-0}_{\L1})\\
&+\frac{1}{2}\tPsi^+_\L(u_2-\frac{i}{h} k, x^{+0}_{\L1}) -\frac{1}{2}\tPsi^+_\L(u_2-\frac{i}{h} k, x^{-0}_{\L1})\\
&-\frac{1}{2} K^\bes \left(u_1-u_2-\frac{i}{h} k \right) -\frac{1}{2} K^\bes \left(u_1-u_2+\frac{i}{h} k \right) \,.
\eal
Now we use (up to terms canceled in) the full phase
\bal
\tilde{\varPhi}_{\L \L}(x^{\pm 0}_{\L 1},\frac{1}{x^{\mp k}_{\R2}})=-\tilde{\varPhi}_{\L \R}(x^{\pm 0}_{\L 1},x^{\mp k}_{\R2})=\tilde{\varPhi}_{\L \R}(x^{\pm 0}_{\L 1},x^{\pm k}_{\R2})
\,,\quad 
\tPsi_{\L}(u_1, \frac{1}{x^{\pm k}_{\R2}})=-\tPsi_{\R}(u_1, x^{\pm k}_{\R2})\,,
\eal
and the general formula for the BES phase in the positive-positive branch of the string region
\bal
\tilde{\theta}_{a b}^{\bes}(x^{\pm 0}_{a 1}, x^{\pm m}_{b2})&=\tilde{\varPhi}_{a b}(x^{\pm 0}_{a 1},x^{\pm m}_{b2})-\tPsi_{b}(u_1, x^{+m}_{b2})+\tPsi_{b}(u_1, x^{-m}_{b2})\\
&+\tPsi_{a}(u_2+\frac{i}{h}m, x^{+ 0}_{a 1})-\tPsi_{a}(u_2+\frac{i}{h}m,x^{- 0}_{a 1})- K^\bes(u_1 - u_2-\frac{i}{h}m)\,,
\eal
to simplify \eqref{eq:temp1} as 
\bal
2\tilde{\theta}_{\L \L}^{\bes}(x^{\pm 0}_{\L 1}, \frac{1}{x^{\mp k}_{\R2}}) &=
2\tilde{\theta}_{\L \R}^{\bes}(x^{\pm 0}_{\L 1}, x^{\pm k}_{\R2})- \Delta^+_\L(u_2 \pm \frac{i}{h} k, x^{\pm m}_{\L1})
\\
&+ K^\bes \left(u_{12}-\frac{i}{h} k \right) -  K^\bes \left(u_{12}+\frac{i}{h} k \right)\,.
\eal
By using identities for $\tPsi$-functions, and the BES kernel, we finally obtain
\bal
2\tilde{\theta}_{\L \L}^{\bes}(\tx^{\pm 0}_{\L 1}, \tx^{\pm 0}_{\L2})&\to 2\tilde{\theta}_{\L \R}^{\bes}(x^{\pm 0}_{\L 1}, x^{\pm k}_{\R2})\\
&+\frac{1}{i} \log \left(\frac{x^{+0}_{\L1}}{x^{-0}_{\L1}} \right)^k \, \frac{(x^{-0}_{\L1} x^{-k}_{\R2} -1)}{(x^{+0}_{\L1} x^{-k}_{\R2} -1)} \frac{(x^{-0}_{\L1} -x^{+k}_{\L2} )}{(x^{+0}_{\L1}- x^{+k}_{\L2})} \prod^{k-1}_{j=1} \left( \frac{x^{-0}_{\L1} \tx^{k-2j}_{\R2} -1}{x^{+0}_{\L1} \tx^{k-2j}_{\R2} -1} \right)^2 \,.
\eal

Next, the continuation of HL to the string region is given by
\bal
2\tilde{\theta}_{\L \L}^{\hl}(\tx^{\pm 0}_{\L 1}, \tx^{\pm 0}_{\L2})&\to  2\tilde{\varPhi}^\hl_{\L\R}(x^{\pm 0}_{\L1}, x^{\pm k}_{\R2})+ \frac{1}{ i} \log \frac{x^{+ 0}_{\L1}}{x^{- 0}_{\L1}} \frac{x^{- k}_{\R2}}{x^{+ k}_{\R2}}+ \pi\,.
\eal
Taking into account that  the string HL phase for general particles of positive momenta is 
\bal
&2 \tilde{\theta}^\hl_{ab}(x^{\pm 0}_{a1}, x^{\pm m}_{b2}) = 2 \tilde{\varPhi}^\hl_{ab}(x^{\pm 0}_{a1}, x^{\pm m}_{b2}) \\
&+ \frac{1}{i} \log \left( -\frac{x^{+0}_{a1}}{x^{-0}_{a1}} \frac{x^{-m}_{b2}}{x^{+m}_{b2}} \frac{(x^{+0}_{a1} - x^{+m}_{a2}) (x^{+0}_{b1} - x^{+m}_{b2}) (x^{-0}_{b1} - x^{-m}_{b2}) (x^{-0}_{a1} - \tx^{+m}_{a2})}{(x^{+0}_{b1} - x^{-m}_{b2}) (x^{-0}_{a1} - x^{+m}_{a2}) (x^{+0}_{a1} - \tx^{+m}_{a2}) (x^{-0}_{b1} - x^{+m}_{b2})} \right)\,,
\eal
 we obtain
\bal
2\tilde{\theta}_{\L \L}^{\hl}&(\tx^{\pm 0}_{\L 1}, \tx^{\pm 0}_{\L2})\to2\tilde{\theta}^\hl_{\L\R}(x^{\pm 0}_{\L1}, x^{\pm k}_{\R2})+ \frac{1}{i} \log  \frac{x^{-0}_{\L1} - x^{+k}_{\L2}}{x^{+0}_{\L1} - x^{+k}_{\L2}} \, \frac{1-x^{-0}_{\L1} x^{-k}_{\R2}}{1-x^{+0}_{\L1} x^{-k}_{\R2}} \, \frac{(1-x^{+0}_{\L1} x^{+k}_{\R2})^2}{(1-x^{-0}_{\L1} x^{+k}_{\R2})^2}
\eal

Combining all the terms, we find that the continuation of the full S-matrix element is given by
\bal
S^{00}_{\chi \chi}(\tx_{\L1}^{\pm0},\tx_{\L2}^{\pm0})  \to &\, {\sqrt{\a_{\L1}^{+0}} \ov \sqrt{\a_{\L1}^{-0}}} {\sqrt{\a_{\R2}^{-k}} \ov \sqrt{\a_{\R2}^{+k}}}\,  \left(\frac{x^{-0}_{\L1}}{x^{+0}_{\L1}} \right)^{k-1} \, \frac{(1-x^{+0}_{\L1} x^{+k}_{\R2}) (1-x^{-0}_{\L1} x^{-k}_{\R2})}{(1-x_{\L1}^{+0} x_{\R2}^{-k}) (1-x^{-0}_{\L1} x^{+k}_{\R2})}   \\
&  \, \prod^{k-1}_{j=1} \left( \frac{x^{+0}_{\L1} \tx^{k-2j}_{\R2} -1}{x^{-0}_{\L1} \tx^{k-2j}_{\R2} -1} \right)^2 \, (\Sigma^{0 k}_{\L\R} (u_1, u_2))^{-2}\,.
\eal
We see that up to a factor it is the S-matrix element $S_{\chi\bar{Z}}^{0k}$, and the formula above can be written as
\bal
\label{eq:cont_pos_neg_massless_massless_string_full_Smat}
 S^{00}_{\chi \chi}(p_1,p_2-2\pi)=  {\sqrt{\a_{\L1}^{+0}} \ov \sqrt{\a_{\L1}^{-0}}} \, S_{\chi\bar{Z}}^{0k}(p_1, p_2) \,.
\eal
From the formulas in appendix B of~\cite{Frolov:2023lwd} we also have the following continuation
\bal
(F_{\L\L}(\tx^{\pm 0}_{\L1},\tx^{\pm 0}_{\L2} ) )^2 \to (F_{\L\R}(x^{\pm 0}_{\L1},x^{\pm k}_{\R2} ) )^2
\eal
and the expression above implies that
\bal
\label{eq:cont_pos_neg_massless_massless_string_full_Smat2}
S^{00}_{\tilde{\chi} \tilde{\chi}} (p_1,p_2-2\pi) = {\sqrt{\a_{\L1}^{+0}} \ov \sqrt{\a_{\L1}^{-0}}} \,   S^{0k}_{\tilde{\chi} \bar{Y}} (p_1, p_2) \,.
\eal

\medskip

By using \eqref{eq:SchibarZ0mp1np2p}, we also get 
\bal
\label{eq:cont_neg_neg_massless_massless}
S^{00}_{\chi \chi}(p_1-2\pi,p_2-2\pi)= {\sqrt{\a_{\R1}^{-k}} \ov \sqrt{\a_{\R1}^{+k}}} {\sqrt{\a_{\R2}^{+k}} \ov \sqrt{\a_{\R2}^{-k}}} \, S^{kk}_{\bar{Z} \bar{Z}} (p_1, p_2) \,.
\eal
We see that for equal momenta 
$S^{00}_{\chi \chi}(p-2\pi,p-2\pi)= -1\,.$

\subsection{String 
\texorpdfstring{$S_{Y Y}^{km}(p_1-2\pi,p_2)$}{SYYkmp1m2pp2} and \texorpdfstring{$S_{Y \bar Z}^{km}(p_1-2\pi,p_2)$}{SYbZkmp1m2pp2} for any \texorpdfstring{$m$}{mass}}

We get a left string $k$-particle bound state with negative momentum by doing the usual steps.
First, the continuation of $x_\L^{+k}=1/\tx_{\R}^{+k}$ through the half-line $x>\xi$  in the $x$-plane or through the lower edge of the main mirror cut in the $u$-plane  replaces  $x_\L^{\pm k}$ with  $\tx_\L^{\pm k}$.
Second, the analytic continuation of $\tx_\L^{+k}$ through the semi-line $x<-1/\xi$ in the $x$-plane or through the lower edge of the $+\ka$-cut  in the mirror $u$-plane shifts the string momentum by $-2\pi$, and does not cross the cut of $\g_\L$.

So, we move $\tx_\L^{+k}$ through the $+\ka$-cut, and get
\bal 
\tx_\L^{+k}&
\ \xrightarrow{+\ka\text{-cut}} 
\ {1\ov \tx_\R(u -{i\ov h}k)}\,,
\quad
\tx_\L^{-k}
\ \xrightarrow{+\ka\text{-cut}} 
\ { \tx_\L(u -{i\ov h}k)}\,.
\eal
Then, we shift $u$ by $+ik/h$, and get
\bal 
\tx_\L^{+k}&
 \xrightarrow{+\ka\text{-cut and shift of}\ u} {1\ov \tx_\R(u)}\,,
\qquad
\tx_\L^{-k}
 \xrightarrow{+\ka\text{-cut and shift of}\ u} { \tx_\L(u)} \,.
\eal
Now we take $u$ to be on the upper edge of the string main cut, so that
\bal 
\tx_\L^{+k}&
\ \xrightarrow{+\ka\text{-cut and shift of}\ u} 
\ x_\L^{+0}\,,
\qquad
\tx_\L^{-k}
\ \xrightarrow{+\ka\text{-cut and shift of}\ u} 
\ x_\L^{-0}\,,
\eal
and 
\bal
p_\L( \tx_\L^{\pm k})  \xrightarrow{+\ka\text{-cut and shift of}\ u}  p_\L( x_\L^{\pm 0}) -2\pi \,, \quad
F( \tx_\L^{\pm k}) \xrightarrow{+\ka\text{-cut and shift of}\ u}  F( x_\L^{\pm0}) 
\,,
\eal
where $F= \{ E_\L\,,\, \g_\L \}$.

We see that these transformations are the limit $Q\to k$ with $Q>k$ of the ones for left string $Q$-particle bound state. This implies that the continuation of S-matrix elements with left $k$-particle bound states can be obtained by taking the limit in the continuation of general S-matrix elements. Thus, by using that (see, (5.55) of \cite{Frolov:2025uwz})
\bal
\label{eq:2pishift-nomonodromy2}
&S^{k+\delta \, m}_{{Y} {Y}} (p_1-2 \pi, p_2) = S^{\delta \, m}_{{Y} {Y}} (p_1, p_2)\,,\\
&S^{k+\delta \, m}_{{Y} \bar Z} (p_1-2 \pi, p_2) = S^{\delta \, m}_{{Y} \bar Z} (p_1, p_2) \,,
\eal
and taking the limit $\delta\to0$, we get
\bal\la{eq:SYYkm}
 S_{ Y  Y}^{km}(p_{1}-2\pi,p_{2}) &= { \sqrt{\a_{\L2}^{-m}}\ov \sqrt{\a_{\L2}^{+m}}}\,
 S_{\chi Y}^{0m}(p_{1},p_{2}) \,,
 \\
 S_{ Y  \bar Z}^{km}(p_{1}-2\pi,p_{2}) &= { \sqrt{\a_{\R2}^{+m}}\ov \sqrt{\a_{\R2}^{-m}}}\,
 S_{\chi \bar Z}^{0m}(p_{1},p_{2}) 
 \,,
\eal
where we used the relations \eqref{eq:Sstringomlimit}.
Next, taking the limit $m\to 0$ in \eqref{eq:SYYkm}, we obtain
\bal\la{eq:SYchik0}
 S_{ Y  \chi}^{k0}(p_{1}-2\pi,p_{2}) &= -{ \sqrt{\a_{\L2}^{-0}}\ov \sqrt{\a_{\L2}^{+0}}}\,
 S_{\chi \chi}^{00}(p_{1},p_{2})\,,
 \\
  S_{ \chi  Y}^{0k}(p_{1},p_{2}-2\pi) &= -{ \sqrt{\a_{\L1}^{+0}}\ov \sqrt{\a_{\L1}^{-0}}}\,
 S_{\chi \chi}^{00}(p_{1},p_{2}) \,.
\eal
The analytic continuation in $p_1$ then gives
\bal\la{eq:SchiY0k}
   S_{ \chi  Y}^{0k}(p_{1}-2\pi,p_{2}-2\pi) &= -{ \sqrt{\a_{\R1}^{-k}}\ov \sqrt{\a_{\R1}^{+k}}}\,{ \sqrt{\a_{\L2}^{-0}}\ov \sqrt{\a_{\L2}^{+0}}}\,
 S_{\bar Z \chi}^{k0}(p_{1},p_{2}) \,.
\eal

\section{Checking CP in the string region}
\label{app:checking_CP}

\subsection{Massless S-matrices as limits of massive ones}

To prove that all mixed-mass and massless S-matrix elements are CP symmetric, it is important to notice that in the positive momentum region of the string kinematics, these elements can be obtained as the limit of the massive ones. In particular, let us recall the result in~\eqref{eq:Sstringomlimit}, which is
\bal\la{eq:Sstringomlimithighest}
S_{\chi Y}^{0m}( x^{\pm0}_{\L1}, x^{\pm m}_{\L2} ) &={\sqrt{ \a_{\L2}^{+m}}\ov \sqrt{\a_{\L2}^{-m}}}\, \lim_{\delta\to 0^+}\, S_{Y Y}^{\delta m}( x^{\pm \delta}_{\L1}, x^{\pm m}_{\L2} )\, ,
\\
S_{\chi \bar Z}^{0m}( x^{\pm0}_{\L1}, x^{\pm m}_{\R2} ) &=\frac{\sqrt{\a^{-m}_{\R2}}}{\sqrt{\a^{+m}_{\R2}}}\,\, \lim_{\delta\to 0^+}\, S_{Y \bar Z}^{\delta m}( x^{\pm \delta}_{\L1}, x^{\pm m}_{\R2} )\,.
\eal
Similar relations also apply to the scattering of lowest-weight states, where by introducing the same normalisation factor on the LHS and RHS of the expressions above one finds
\bal\la{eq:Sstringomlimitlowest}
S_{\tilde{\chi} Z}^{0m}( x^{\pm0}_{\L1}, x^{\pm m}_{\L2})&={\sqrt{ \a_{\L2}^{+m}}\ov \sqrt{\a_{\L2}^{-m}}}\, \lim_{\delta\to 0^+}\, S_{Z Z}^{\delta m}( x^{\pm \delta}_{\L1}, x^{\pm m}_{\L2} )\, ,
\\
S_{\tilde{\chi} \bar Y}^{0m}( x^{\pm0}_{\L1}, x^{\pm m}_{\R2} ) &=\frac{\sqrt{\a^{-m}_{\R2}}}{\sqrt{\a^{+m}_{\R2}}}\,\, \lim_{\delta\to 0^+}\, S_{Z \bar Y}^{\delta m}( x^{\pm \delta}_{\L1}, x^{\pm m}_{\R2} )\,.
\eal
Finally, comparing~\eqref{eq:sigmaLL00str} with equation (G.12) in~\cite{Frolov:2025uwz}, we notice that in the positive momentum region of string kinematics we have
\bal\la{eq:Sstringomlimitmassless}
S_{\chi \chi}^{00}(x^{\pm 0}_{\L1}, x^{\pm 0}_{\L2})&= \lim_{\delta \to 0^+} {\sqrt{\a_{\L1}^{-\delta}} \ov \sqrt{\a_{\L1}^{+\delta}}} \, { \sqrt{\a_{\L2}^{+\delta}}\ov \sqrt{\a_{\L2}^{-\delta}}} S_{Y Y}^{\delta \delta}(x^{\pm \delta}_{\L1}, x^{\pm \delta}_{\L2})\,,
\\
S_{\tilde{\chi} \tilde{\chi}}^{00}(x^{\pm 0}_{\L1}, x^{\pm 0}_{\L2})&= \lim_{\delta \to 0^+} {\sqrt{\a_{\L1}^{-\delta}} \ov \sqrt{\a_{\L1}^{+\delta}}} \, { \sqrt{\a_{\L2}^{+\delta}}\ov \sqrt{\a_{\L2}^{-\delta}}} S_{Z Z}^{\delta \delta}(x^{\pm \delta}_{\L1}, x^{\pm \delta}_{\L2})\,.
\eal
In the following, we will use the relations in~\eqref{eq:Sstringomlimithighest}, \eqref{eq:Sstringomlimitlowest} and~\eqref{eq:Sstringomlimitmassless} to prove CP as a limit of the results in~\cite{Frolov:2025uwz}.

To check CP it will be necessary to map certain relations originally expressed in terms of momenta to relations expressed in terms of $u$ variables. To do so we will use the identities in~\eqref{eq:app_minu_string_Zh} and~\eqref{eq:app_relations_u_minus_u}.

\subsection{CP for mixed-mass elements}

\paragraph{CP for $\sgn (p_1) = \sgn(p_2)$.}

Let $p_1$ and $p_2$ be in the interval $(0, 2 \pi)$. Then the CP symmetry implies the following relation
\bal
\label{eq:app_CP_relation_mixed_eqmom}
S^{0m}_{\chi Y} (p_1-2 \pi, p_2-2 \pi) =S^{m0}_{\bar Y\tilde{\chi}} (2 \pi-p_2, 2 \pi-p_1)  \,,
\eal
connecting S-matrix elements with both positive momenta to S-matrix element with both negative momenta.
From~\eqref{eq:arb_mom_massless_left2} we see that
\bal
&S^{0m}_{\chi Y} (p_1-2 \pi, p_2-2 \pi)= {\sqrt{ \a_{\R2}^{+(k-m)}}\ov \sqrt{\a_{\R2}^{-(k-m)}}}\,\frac{\alpha_{\R 1}^{-k}}{\alpha_{\R 1}^{+k}}\  S^{k,k-m}_{\bar{Z} \bar{Z}} (p_1, p_2)\,,
\eal
and the relation~\eqref{eq:app_CP_relation_mixed_eqmom} can be written as follows
\bal
S^{k,k-m}_{\bar{Z} \bar{Z}} (p_1, p_2)={\sqrt{ \a_{\R2}^{-(k-m)}}\ov \sqrt{\a_{\R2}^{+(k-m)}}}\,\frac{\alpha_{\R 1}^{+k}}{\alpha_{\R 1}^{-k}} S^{m0}_{\bar Y\tilde{\chi}} (2 \pi-p_2, 2 \pi-p_1) \,.
\eal
If we now parameterise
\bal
p_1=p_{\R}^{(k)}(u_1)=2 \pi - p_{\L}^{(0)}(-u_1)\,, \qquad p_2=p_{\R}^{(k-m)}(u_2)=2\pi - p_{\R}^{(m)}(-u_2)
\eal
(where we used~\eqref{eq:app_relations_u_minus_u})
then the relation above on the $u$ plane is given by
\bal
\label{eq:app_mixedmass_CP_u_start}
S^{k,k-m}_{\bar{Z} \bar{Z}} (u_1, u_2)={\sqrt{ \a_{\R2}^{-(k-m)}}\ov \sqrt{\a_{\R2}^{+(k-m)}}}\,\frac{\alpha_{\R 1}^{+k}}{\alpha_{\R 1}^{-k}} S^{m0}_{\bar Y\tilde{\chi}} (-u_2, -u_1) \,.
\eal

Let us show in some details how the equality in~\eqref{eq:app_mixedmass_CP_u_start} can be checked. Firs we replace $u_1 \to -u_1$ and $u_2 \to -u_2$. Using that
\bal
\label{eq:app_CP_zhuk_rel_mixedmass}
x^{\pm (k-m)}_{\R}(-u_2)=-\frac{1}{x^{\pm m}_{\R}(u_2)}\,, \qquad x^{\pm k}_{\R}(-u_1)=-x^{\mp 0}_{\L}(u_1)\,.
\eal
we obtain
\bal
\label{eq:app_mixedmass_CP_u_start2}
S^{k,k-m}_{\bar{Z} \bar{Z}} (-u_1, -u_2)={\sqrt{ \a_{\R2}^{-m}}\ov \sqrt{\a_{\R2}^{+m}}}\,\frac{\alpha_{\L 1}^{-0}}{\alpha_{\L 1}^{+0}}\, S^{m0}_{\bar Y\tilde{\chi}} (u_2, u_1) \,.
\eal
Applying braiding unitarity to the second relation in~\eqref{eq:Sstringomlimitlowest} and plugging it into the RHS of the expression above, we obtain
\bal
\lim_{\delta \to 0^+} S^{k+ \delta,k-m}_{\bar{Z} \bar{Z}} (-u_1, -u_2)=\lim_{\delta \to 0^+} \frac{\alpha_{\L 1}^{-\delta}}{\alpha_{\L 1}^{+\delta}} S^{m \delta}_{\bar Y Z} (u_2, u_1) \,.
\eal
This relation follows now from the results in~\cite{Frolov:2025uwz} (see third line in eq. (I.35) of that paper).
This concludes the check of CP in the mixed-mass sector for particles having the same momentum sign.

\paragraph{CP for $\sgn (p_1) = -\sgn(p_2)$.}

Now we want to check the constraint
\bal
\label{eq:app_CP_relation_mixed_oppmom}
&S^{0m}_{\chi Y} (p_1-2 \pi, p_2) =S^{m0}_{\bar Y\tilde{\chi}} (-p_2, 2\pi- p_1)\,.
\eal
Using again the analytic continuation at negative momenta (see results in equations~\eqref{eq:arb_mom_massless_left1}, \eqref{eq:arb_mom_massless_left2} and~\eqref{eq:arb_mom_massless_right}) we have
\bal
&S^{0m}_{\chi Y} (p_1-2 \pi, p_2) = \frac{\sqrt{\a^{-m}_{\L2} }}{\sqrt{\a^{+m}_{\L2}}} S^{k m}_{\bar{Z} Y} (p_1, p_2)\,,\\
&S^{m0}_{\bar Y\tilde{\chi}} (-p_2, 2\pi- p_1)=-S^{m+k ,0}_{\bar Y\tilde{\chi}} (2 \pi-p_2, 2\pi- p_1)
\eal
Plugging these relations into~\eqref{eq:app_CP_relation_mixed_oppmom} we obtain
\bal
S^{k m}_{\bar{Z} Y} (p_1, p_2)= -\frac{\sqrt{\a^{+m}_{\L2}}}{\sqrt{\a^{-m}_{\L2}}} S^{m+k ,0}_{\bar Y\tilde{\chi}} (2 \pi-p_2, 2\pi- p_1) \,.
\eal
Using~\eqref{eq:app_relations_u_minus_u}, in the $u$ plane this equality is given by 
\bal
S^{k m}_{\bar{Z} Y} (u_1, u_2)=- \frac{\sqrt{\a^{+m}_{\L2} }}{\sqrt{\a^{-m}_{\L2}}} S^{m+k ,0}_{\bar Y\tilde{\chi}} (-u_2, - u_1) \,.
\eal
First send $u_1 \to -u_1$ and $u_2 \to -u_2$ so that the relation above becomes
\bal
S^{k m}_{\bar{Z} Y} (-u_1, -u_2)=\frac{\sqrt{\a^{-(k+m)}_{\R2}}}{\sqrt{\a^{+(k+m)}_{\L2}}} S^{m+k ,0}_{\bar Y\tilde{\chi}} (u_2, u_1) \,.
\eal
Then we use the result in the second line of~\eqref{eq:Sstringomlimitlowest}and write this relation as the limit
\bal
\lim_{\delta \to 0^+}
S^{k+ \delta,  m}_{\bar{Z} Y} (-u_1, -u_2)=\lim_{\delta \to 0^+} S^{m+k ,\delta}_{\bar Y Z} (u_2, u_1) \,.
\eal
This equality follows from the second relation in (I.35) of~\cite{Frolov:2025uwz}.

\subsection{CP for massless elements}

In momentum representation the CP condition on massless-massless S-matrices takes the following form
\bal
S^{0 0}_{\chi \chi}(p_1, p_2)= S^{0 0}_{\tilde{\chi} \tilde{\chi}}(-p_2, -p_1)\,.
\eal
This symmetry relates the chiral-chiral branch to the antichiral-antichiral branch, and relates the chiral-antichiral branch to itself.

\paragraph{CP on the same-chirality branch.}
Let us consider the chiral-chiral and antichiral-antichiral branches first. In this case we require
\bal
\label{eq:app_starting_CP_same_chir_massless}
S^{0 0}_{\tilde{\chi} \tilde{\chi}} (2 \pi - p_2, 2 \pi - p_1)=S^{0 0}_{\chi \chi} (p_1 -2 \pi, p_2 -2 \pi) \,, \qquad 0<p_1< 2 \pi, \, 0<p_2< 2 \pi \,.
\eal
Parameterising 
\bal
p_1=p_{\R}^{(k)}(u_1)\,, \qquad p_2=p_{\R}^{(k)}(u_2)
\eal
and recalling from~\eqref{eq:app_relations_u_minus_u} that
\bal
\label{eq:app_p0_mu}
p_{\L}^{(0)}(-u)=2 \pi - p_{\R}^{(k)}(u)
\eal
then to have CP on the same-chirality branch it must hold that 
\bal
\label{eq:appSbcbcmu2mu1_ZbZbZu1u2}
S^{0 0}_{\tilde{\chi} \tilde{\chi}} (-u_2, -u_1)={\sqrt{\a_{\R1}^{-k}} \ov \sqrt{\a_{\R1}^{+k}}} {\sqrt{\a_{\R2}^{+k}} \ov \sqrt{\a_{\R2}^{-k}}} \, S^{kk}_{\bar{Z} \bar{Z}} (u_1, u_2)\,,
\eal
where we replaced the RHS of~\eqref{eq:app_starting_CP_same_chir_massless} with~\eqref{eq:cont_neg_neg_massless_massless}. 
To check the relation above we first send $u_1 \to -u_1$ and $u_2 \to -u_2$, and get
\bal
S^{0 0}_{\tilde{\chi} \tilde{\chi}} (u_2, u_1)={\sqrt{\a_{\L1}^{+0}} \ov \sqrt{\a_{\L1}^{-0}}} {\sqrt{\a_{\L2}^{-0}} \ov \sqrt{\a_{\L2}^{+0}}} \, S^{kk}_{\bar{Z} \bar{Z}} (-u_1, -u_2)\,.
\eal
Using then the second relation in~\eqref{eq:Sstringomlimitmassless} then the expression above implies that
\bal
\lim_{\delta \to 0^+}  S_{Z Z}^{\delta , \delta}(u_2, u_1)= \lim_{\delta \to 0^+}  S^{k+ \delta, k+ \delta}_{\bar{Z} \bar{Z}} (-u_1, -u_2)\,.
\eal
This equality was proven in~\cite{Frolov:2025uwz} (see first relation in eq. (I.35) of that paper) and concludes the check of CP for massless-massless S-matrices involving particles of the same chirality.

\paragraph{CP on the opposite-chirality branch.}

In the chiral-antichiral branch we write the CP condition as follows
\bal
\label{eq:CP_on_chiral_antichiral_starting}
S^{0 0}_{\chi \chi} (2 \pi -p_1, p_2-2 \pi) =S^{0 0}_{\tilde{\chi} \tilde{\chi}} (2 \pi - p_2, p_1 - 2\pi)\,.
\eal
Substituting the LHS and RHS of the expression above with~\eqref{eq:cont_pos_neg_massless_massless_string_full_Smat} and~\eqref{eq:cont_pos_neg_massless_massless_string_full_Smat2}, and 
parameterising
\bal
p_1=p_\R^{(k)}(u_1) \,, \qquad p_2=p_\R^{(k)}(u_2)
\eal
then we obtain
\bal
{\sqrt{\a_{\R1}^{-k}} \ov \sqrt{\a_{\R1}^{+k}}}  \,  S^{0 k}_{\chi \bar{Z}} (-u_1, u_2) = {\sqrt{\a_{\R2}^{-k}} \ov \sqrt{\a_{\R2}^{+k}}} \,  S^{0k}_{\tilde{\chi} \bar{Y}} ( - u_2, u_1)
\eal
As before we wrote the relation in the $u$ plane by making use of~\eqref{eq:app_p0_mu}. Moreover we consider $u_1 > -\nu$ so that the kinematics of the massless particle on the LHS of the expression above is physical (since $-u_1<\nu$ is on the main string theory cut).
We also notice that for this configuration, $u_1 \pm \frac{i}{h}k $ are away from the $\ka$ cuts and the relation on the RHS is well defined. Plugging the second relation in~\eqref{eq:Sstringomlimithighest} into the LHS of the expression above, and the second relation in~\eqref{eq:Sstringomlimitlowest} into the RHS, we obtain
\bal
\lim_{\delta \to 0^+}S^{\delta, k+\delta}_{Y \bar{Z}} (-u_1, u_2) =  \lim_{\delta \to 0^+} S^{\delta,  k+\delta}_{Z \bar{Y}} ( - u_2, u_1)\,.
\eal
The validity of this relation follows from the second line in eq. (I.35) of~\cite{Frolov:2025uwz}.
This concludes the check of CP for massless particles in the chiral-antichiral branch of the kinematics. 
The check on the antichiral-chiral branch is connected to the one above by braiding unitarity and does not require a separate check.

\subsection{CP for k-particle bound states}
\label{app:CP_k_particle_bs}

One can check that, as for massless particles, also S-matrices involving $k$-particle bound states are CP symmetric. Even in this case, this can be proven from the limit of the relations in~\cite{Frolov:2025uwz}.
In the following, we will check the relation
\bal
\label{eq:app_example_CPk0}
S^{0k}_{\chi Y}(p_1-2\pi, p_2-2 \pi)=S^{k0}_{\bar{Y} \tilde{\chi}}(2 \pi-p_2, 2 \pi-p_1)\,, \quad \text{with} \ 0<p_1<2 \pi, \, 0<p_2<2 \pi\,.
\eal
The other cases can be verified similarly.  
Using the second relation in~\eqref{eq:Sstringomlimitlowest} and parameterising
\bal
&p_1=p_{\R}^{(k)}(u_1)\,, \qquad p_2=p_{\L}^{(0)}(u_2)\\
&\implies 2 \pi - p_1=p^{(0)}_{\L}(-u_1), \quad 2 \pi - p_2=p^{(k)}_{\R}(-u_2)
\eal
we can write the RHS of the expression above as
\bal
&S^{k0}_{\bar{Y} \tilde{\chi}}(-u_2, -u_1)= \frac{\sqrt{\a_{\R}(-1/x^{+0}_{\R}(u_2))}}{\sqrt{\a_{\R}(-1/x^{-0}_{\R}(u_2))}}\,\, \lim_{\delta\to 0^+}\, S_{\bar Y Z}^{k \delta}( -u_2, -u_1 )\\
&= \frac{\sqrt{\a_{\R}(-x^{-0}_{\L}(u_2))}}{\sqrt{\a_{\R}(-x^{+0}_{\L}(u_2))}}\,\, \lim_{\delta\to 0^+}\, S_{\bar Y Z}^{k \delta}( -u_2, -u_1 )=- \frac{\sqrt{\a^{-0}_{\L2}}}{\sqrt{\a^{+0}_{\L2}}}\,\, \lim_{\delta\to 0^+}\, S_{\bar Y Z}^{k \delta}( -u_2, -u_1 )\,.
\eal
The LHS is instead obtained by using first the expression in~\eqref{eq:SchiY0k} and then the second relation in~\eqref{eq:Sstringomlimit}:
\bal
S^{0k}_{\chi Y}(p_1-2\pi, p_2-2 \pi)&=-{ \sqrt{\a_{\R1}^{-k}}\ov \sqrt{\a_{\R1}^{+k}}}\,{ \sqrt{\a_{\L2}^{-0}}\ov \sqrt{\a_{\L2}^{+0}}}\,
 S_{\bar Z \chi}^{k0}(u_{1},u_{2})=-{ \sqrt{\a_{\L2}^{-0}}\ov \sqrt{\a_{\L2}^{+0}}}\,
 \lim_{\delta \to 0^+}S_{\bar Z Y}^{k \delta}(u_{1},u_{2})\,.
\eal
Comparing LHS and RHS we end up with the following relation
\bal
\lim_{\delta\to 0^+}\, S_{\bar Y Z}^{k \delta}( -u_2, -u_1 )=\lim_{\delta \to 0^+}S_{\bar Z Y}^{k \delta}(u_{1},u_{2})\,,
\eal
which is equal to the second line in eq. (I.35) of~\cite{Frolov:2025uwz}. This concludes the check of CP for the relation~\eqref{eq:app_example_CPk0} involving $k$ particle bound states.

\section{Comparison with Ramond-Ramond results in mirror region}
\label{appendix:pureRRcomparison}

In this appendix, we check that in the $\ka \to 0$ limit the S-matrix elements proposed in this paper match the ones in~\cite{Frolov:2021bwp} for particles in the mirror region. 

\subsection{Improved HL and Barnes functions}

It is possible to show that in the pure Ramond-Ramond case the improved HL phase, defined as a double integral around the mirror cuts, is connected to the $R$ functions in~\eqref{eq:app_Rfunct_definition} through the following relation
\bal
 \Sigma^\hl(x^{\pm Q}_1, x^{\pm P}_2)^2=\frac{R(\tg_{12}^{++}+ i\pi) R(\tg_{12}^{++}-i \pi) R(\tg_{12}^{--}+ i\pi) R(\tg_{12}^{--}-i \pi)}{R(\tg_{12}^{+-}+ i\pi) R(\tg_{12}^{+-}-i \pi) R(\tg_{12}^{-+}+ i\pi) R(\tg_{12}^{-+}-i \pi)} \frac{R^2(\tg_{12}^{+-}) R^2(\tg_{12}^{-+})}{R^2(\tg_{12}^{++})R^2(\tg_{12}^{--})} \,,
\eal
where the bound state indices $Q$ and $P$ on the RHS of the expression above have been assumed.
In the case in which the second particle becomes massless, that is $P \to 0^+$, we have that
\bal
\tg_2^+=\tg^\circ_2\,, \qquad \tg_2^-=\tg^\circ_2 + i \pi \,.
\eal
Then, using the monodromy properties of the $R$ functions, we get
\bal
\label{eq:app_mixed_HL_RamRam}
 \Sigma^\hl(x^{\pm Q}_1, x^{\pm 0}_2)^2= \frac{\tanh \frac{\tg^{+ \circ}_{12}}{2}}{\tanh \frac{\tg^{- \circ}_{12}}{2}} \,  \frac{R^4(\tg_{12}^{- \circ})}{R^4(\tg_{12}^{+\circ})} \, \frac{R^2(\tg_{12}^{+\circ}- i \pi) }{R^2(\tg_{12}^{- \circ} - i \pi)}\, \frac{R^2(\tg_{12}^{+ \circ}+ i\pi) }{R^2(\tg_{12}^{-\circ}+ i\pi) } \,.
\eal
Finally, in the case in which the first particle is also massless, we end up with
\bal
\label{eq:app_massless_HL_RamRam}
 \Sigma^\hl(x^{\pm 0}_1, x^{\pm 0}_2)^2=  - \frac{R^4(\tg_{12}^{\circ \circ} - i \pi) R^4(\tg_{12}^{\circ \circ}+ i \pi)  }{R^8(\tg_{12}^{\circ \circ})}  \,.
\eal

\subsection{Mirror S-matrices in pure Ramond-Ramond}
\label{app:mirror_crossing_RamondRamond}

\subsubsection*{Mixed-mass S-matrices.}

Let us consider the scattering of a massless and a massive particle. Normalising as in appendix B of~\cite{Frolov:2021bwp}
\begin{equation}
    \begin{aligned}
    \mathbf{S}\,\big|Y_{1}\chi^{\dot{\alpha}}_{2}\big\rangle&=&{\bar S}^{Q 0}(  u_1, u_2) \,\big|Y_{1}\chi^{\dot{\alpha}}_{2}\big\rangle=
     i\,\sqrt{\frac{\tx_1^{+}}{\tx_1^{-}}} \,  \frac{\tx^-_1-\tx_2}{\tx^+_1 \tx_2-1} \big(\hat{\Sigma}^{10} \big)^{-2}\,
    \big|Y_{1}\chi^{\dot{\alpha}}_{2}\big\rangle,\\
    \mathbf{S}\,\big|\bar{Z}_{1} \chi^{\dot{\alpha}}_{2}\big\rangle&=&{S}^{Q 0}(  u_1, u_2)  \,\big|\bar{Z}_{1}\chi^{\dot{\alpha}}_{2}\big\rangle=
 i\, \sqrt{\frac{\tx_1^{-}}{\tx_1^{+}}}   \frac{\tx^+_1 \tx_2-1}{\tx^-_1-\tx_2} \big(\hat{\Sigma}^{1 0} \big)^{-2} \,
    \big|\bar{Z}_{1}\chi^{\dot{\alpha}}_{2}\big\rangle\,,
    \end{aligned}
\end{equation}
then the crossing equations take the following form
\bal
\label{eq:crossing_mixed_mass}
\big(\hat{\Sigma}^{10}(u_1, u_2 \big)^{-2} \, \big(\hat{\Sigma}^{10}(\bar{u}_1, u_2 \big)^{-2}= - \frac{\tx^+_1 - \tx_2}{\tx^-_1 - \tx_2} \, \frac{1-\tx^-_1 \tx_2}{1-\tx^+_1 \tx_2} \,.
\eal
Notice that we put an hat on the dressing factor since this is not the same quantity defined in the rest of this paper. In this appendix we use the notation of~\cite{Frolov:2021bwp}, where a different normalisation was used and the dressing factor was not split into an even and odd part.
After considering the $h \to 0$ limit of the $\tg$ variables
\begin{equation}
\tg(u)=\frac{1}{2} \log \left( \frac{2-u}{2+u} \right)\,, \qquad \tg^\pm(u)=\tg(u\pm \frac{i}{h})\,, \qquad \tg^\circ(u)=\tg(u+ i0)\,,
\end{equation}
then using the monodromy relations of $R$ functions it is simple to check that the crossing equation~\eqref{eq:crossing_mixed_mass} is solved by
\bal
\left(\hat{\Sigma}^{1 0}(u_1, u_2) \right)^{-2}=-i \tanh \left(\frac{\tg^{+ \circ}_{12}}{2} \right) \frac{R^2(\tg_{12}^{-\circ})}{R^2(\tg_{12}^{+\circ})}  \  \frac{R(\tg_{12}^{+\circ}- i \pi) R(\tg_{12}^{+\circ}+ i\pi)}{R(\tg_{12}^{-\circ}- i \pi) R(\tg_{12}^{-\circ}+ i \pi)} \big(\Sigma^\bes(\tx^{\pm}_1, \tx^{\pm 0}_2) \big)^{-2} \,.
\eal
The solution above agrees with the result in~\cite{Frolov:2021bwp}. To make the comparison with the results in~\cite{Frolov:2021bwp}
it is necessary to 
redefine the $\tg$ functions
\begin{equation}
\tg^+ \to \tg^+ - \frac{i \pi}{2} \, , \qquad \tg^- \to \tg^- + \frac{i \pi}{2}\,,
\end{equation}
in such a way that they match the definition (C.6) in~\cite{Frolov:2021bwp}.

We do not write the crossing equation and solution for $\left(\Sigma^{0 1}(u_1, u_2) \right)^{-2}$ since this dressing factor is connected to the one above either by parity or braiding unitarity.

\subsubsection*{Massless S-matrix.}

Normalising the scattering of two massless modes by
\bal
\mathbf{S}\,\big|\chi^{\dot{\alpha}}_{1}\chi^{\dot{\beta}}_{2}\big\rangle =  \big(\Sigma^{00} \big)^{-2}\,
    \big|\chi^{\dot{\alpha}}_{2}\chi^{\dot{\beta}}_{2}\big\rangle
\eal
the massless-massless crossing equations take the form
\bal
\big(\Sigma^{00}(u_1, u_2 \big)^{-2} \, \big(\Sigma^{00}(\bar{u}_1 , u_2) \big)^{-2} =-\Bigl(\frac{\tx_{1}-\tx_{2}}{\tx_{1} \tx_{2} -1} \Bigl)^2 = -\tanh^2 \frac{\tg^{\circ \circ}_{12}}{2}\,.
\eal
The equation above contains a minus sign compared to the crossing equations for massless particles written in the literature (see, e.g., \cite{Frolov:2021fmj}). This sign was discussed in~\cite{Ekhammar:2024kzp} and is justified by the nontrivial exchange relations for massless particles~\cite{Frolov:2025ozz}. 
This sign is necessary to solve crossing with a minimal dressing factor, without introducing any further zero or pole in the physical strip of massless particles.
For massless particles under crossing it holds that
\bal
\g^\circ(\bar{u}) = \g^\circ(u)+ i \pi \,.
\eal 
The improved BES phase satisfies homogeneous crossing equations
and a solution to the crossing equations is given by
\bal
\big(\Sigma^{00}(u_1, u_2 \big)^{-2} = -\frac{R^2( \g^{\circ \circ}_{12} - i \pi) R^2( \g^{\circ \circ}_{12} + i \pi)}{R^4(\g^{\circ \circ}_{12})} \, \big(\Sigma^\bes(x^{\pm 0}_1, x^{\pm 0}_2 \big)^{-2} \,.
\eal
This solution differs from the one in~\cite{Frolov:2021bwp} by the absence of the factor
\bal
\label{eq:a_g}
a(\tg^{\circ \circ}_{12})=-i \tanh \left( \frac{\tg^{\circ \circ}_{12}}{2} - \frac{i \pi}{4} \right)\,,
\eal
which is replaced here by $-1$. The overall sign is fixed by requiring that in the limit $p_1=p_2$ the S-matrix is $-1$, according with fermionic statistics. However, this consideration is valid only for particles of equal chirality. If the particles have opposite chirality (and therefore $u_1$ and $u_2$ lye on different mirror cuts) then they cannot have the same momentum and the dressing factor can have an opposite sign. This is actually what happens in the $h \to 0$ limit of our proposal, as we will show in appendix~\ref{app:RRlimit_massless_mirror}.

\subsection{The Ramond-Ramond limit of mixed-mass S-matrices}

Consider the $k\to 0$ limit of the following S-matrix element
\bal    
S^{Q 0}_{Y \chi^{\dot{1}}}=
    {\sqrt{\ta_{\L1}^{+Q}} \ov \sqrt{\ta_{\L1}^{-Q}}}   \, \frac{\tx^{+Q}_{\L1}}{\tx^{-Q}_{\L1}} \,  \frac{\tx^{-0}_{\L2}}{\tx^{+0}_{\L2}} \, \left(\frac{\tx^{-Q}_{\L1} - \tx^{+0}_{\L2}}{\tx^{+Q}_{\L1} - \tx^{-0}_{\L2}} \right)^2 \left(\Sigma^{Q0}_{\L\L}(u_1,u_2)\right)^{-2} \,,
\eal
which we obtain applying braiding unitarily on the first relation in~\eqref{eq:massless_massive_normmir_bstates}. The dressing factor is given by
\bal
\left(\Sigma^{Q0}_{\L\L}(u_1,u_2)\right)^{-2}=\left(\Sigma^{Q0, \besratio}_{\L\L}(u_1,u_2)\right)^{-2} \, \left(\Sigma^{Q0, \barnes}_{\L\L}(u_1,u_2)\right)^{-2}
\eal
where
\bal
\left(\Sigma^{Q0, \besratio}_{\L \L}(u_1, u_2) \right)^{-2} = \left(\frac{\Sigma^{Q0, \bes}_{\L \L}(u_1, u_2)}{\Sigma^{Q0, \hl}_{\L \L}(u_1, u_2)} \right)^{-2}
\eal
and
\bal
&\left(\Sigma^{Q0, \barnes}_{\L \L} (\tg^{\pm Q}_{\L 1}, \tg^{\pm 0}_{\L 2}) \right)^{-2}=\frac{R^2(\tg^{-Q -0}_{\L \L}) R^2(\tg^{+Q +0}_{\L \L})}{R^2(\tg^{-Q +0}_{\L \L}) R^2(\tg^{+Q -0}_{ \L \L}) } \,.
\eal
In the limit the normalisation becomes
\bal
\label{eq:app_lim_RR_mixedmass_norm}
{\sqrt{\ta_{\L1}^{+Q}} \ov \sqrt{\ta_{\L1}^{-Q}}}   \, \frac{\tx^{+Q}_{\L1}}{\tx^{-Q}_{\L1}} \,  \frac{\tx^{-0}_{\L2}}{\tx^{+0}_{\L2}} \, \left(\frac{\tx^{-Q}_{\L1} - \tx^{+0}_{\L2}}{\tx^{+Q}_{\L1} - \tx^{-0}_{\L2}} \right)^2 &\to {\sqrt{\ta_{1}^{+Q}} \ov \sqrt{\ta_{1}^{-Q}}}   \, \frac{\tx^{+Q}_{1}}{\tx^{-Q}_{1}}  \, \left(\frac{\tx^{-Q}_{1} - \tx_{2}}{\tx^{+Q}_{1} \tx_2 - 1} \right)^2\\
&= \frac{\sinh \frac{\tg^{-Q \circ}_{12}}{2}}{\cosh \frac{\tg^{+Q \circ}_{12}}{2}} \, \sqrt{\frac{\tx^{+Q}_{1}}{\tx^{-Q}_{1}}}  \, \, \frac{\tx^{-Q}_{1} - \tx_{2}}{\tx^{+Q}_{1} \tx_2 - 1} \,.
\eal
Then we have the following limit of $\tg$ functions
\bal
\label{eq:app_RR_limit_g_functions}
\tg^{+0}_{\L} \to \tg^{\circ}\,, \qquad \tg^{-0}_{\L}\to \tg^{\circ}+i \pi \,,
\eal
leading to
\bal
&\Sigma^{Q0, \barnes}_{\L \L} (\tg^{\pm Q}_{\L 1}, \tg^{\pm 0}_{\L 2})^{-2} \to \frac{R^2(\tg^{-Q \circ}_{\L \L} -i \pi) R^2(\tg^{+Q \circ}_{\L \L})}{R^2(\tg^{-Q \circ}_{\L \L}) R^2(\tg^{+Q \circ}_{ \L \L} -i \pi) }\,.
\eal
Combining the even and odd part of the dressing factor, and using relation~\eqref{eq:app_mixed_HL_RamRam}, then the dressing factor becomes
\bal
\Sigma^{Q0}_{\L\L}(u_1,u_2)^{-2} &\to \, \frac{\tanh \frac{\tg^{+Q \circ}_{12}}{2}}{\tanh \frac{\tg^{-Q \circ}_{12}}{2}} \,  \frac{R^2(\tg_{12}^{-Q \circ})}{R^2(\tg_{12}^{+Q \circ})}\, \frac{R^2(\tg_{12}^{+Q \circ}+ i\pi) }{R^2(\tg_{12}^{-Q\circ}+ i\pi) } \, \Sigma^{Q0}_{\bes}(u_1, u_2)^{-2} \\
&= \frac{\sinh \frac{\tg^{+Q \circ}_{12}}{2}}{\sinh \frac{\tg^{-Q \circ}_{12}}{2}} \,  \frac{R^2(\tg_{12}^{-Q \circ})}{R^2(\tg_{12}^{+Q \circ})}\, \frac{R^2(\tg_{12}^{+Q \circ}- i\pi)R(\tg_{12}^{+Q \circ}+ i\pi) }{R(\tg_{12}^{-Q \circ}- i\pi)R(\tg_{12}^{-Q \circ}+ i\pi) } \, \Sigma^{Q0}_{\bes}(u_1, u_2) ^{-2} \,.
\eal
From the limit of the dressing factor and the limit of the normalisation~\eqref{eq:app_lim_RR_mixedmass_norm} we get
\bal
S^{Q 0}_{Y \chi^{\dot{1}}}(u_1, u_2) &\to\sqrt{\frac{\tx^{+Q}_{1}}{\tx^{-Q}_{1}}}  \, \, \frac{\tx^{-Q}_{1} - \tx_{2}}{\tx^{+Q}_{1} \tx_2 - 1} \, \tanh \frac{\tg^{+Q \circ}_{12}}{2} \\
&\qquad\times  \frac{R^2(\tg_{12}^{-Q \circ})}{R^2(\tg_{12}^{+Q \circ})}\, \frac{R^2(\tg_{12}^{+Q \circ}- i\pi)R(\tg_{12}^{+Q \circ}+ i\pi) }{R(\tg_{12}^{-Q \circ}- i\pi)R(\tg_{12}^{-Q \circ}+ i\pi) } \, \Sigma^{\bes}(x^{\pm Q}_1, x^{\pm 0}_2) ^{-2} \,.
\eal
This result agrees with the pure Ramond-Ramond mixed-mass S-matrix at the beginning of section~\ref{app:mirror_crossing_RamondRamond} (which is the result written in~\cite{Frolov:2021bwp} after redefining the $\tg$ functions through shifts of $i \frac{\pi}{2}$).

\subsection{The Ramond-Ramond limit of massless S-matrices}
\label{app:RRlimit_massless_mirror}

Let us now consider the $\ka\to 0$ limit of the massless-massless S-matrix written in the first row of~\eqref{eq:massless_massless_normmir}:
\bal
S_{\chi^{\dot{1}} \chi^{\dot{1}}}(u_1, u_2) =+
    {\sqrt{\ta_{\L1}^{+0}} \ov \sqrt{\ta_{\L1}^{-0}}} \,  {\sqrt{\ta_{\L2}^{-0}} \ov \sqrt{\ta_{\L2}^{+0}}}  \ \frac{\tx^{-0}_{\L2}}{\tx^{+0}_{\L2}} \, \frac{\tx^{+0}_{\L1}}{\tx^{-0}_{\L1}} \, \left(\frac{\tx^{-0}_{\L1} - \tx^{+0}_{\L2}}{\tx^{+0}_{\L1} - \tx^{-0}_{\L2}} \right)^2 \left(\Sigma^{00}_{\L\L}(u_1,u_2)\right)^{-2} \,.
\eal
In this case we have
\bal
\left(\Sigma^{00}_{\L \L}(u_1, u_2) \right)^{-2}=  \left(\Sigma^{00, \besratio}_{\L \L}(u_1, u_2) \right)^{-2} \left(\Sigma^{00, \barnes}_{\L \L}(u_1, u_2) \right)^{-2} \,.
\eal
where
\bal
\left(\Sigma^{00, \besratio}_{\L \L}(u_1, u_2) \right)^{-2} = \left(\frac{\Sigma^{00, \bes}_{\L \L}(u_1, u_2)}{\Sigma^{00, \hl}_{\L \L}(u_1, u_2)} \right)^{-2}\,,
\eal
\bal
&\left(\Sigma^{00, \barnes}_{\L \L} (\tg^{\pm0}_{\L 1}, \tg^{\pm 0}_{\L 2}) \right)^{-2}=\frac{R^2(\tg^{-0 -0}_{\L \L}) R^2(\tg^{+0 +0}_{\L \L})}{R^2(\tg^{-0 +0}_{\L \L}) R^2(\tg^{+0 -0}_{ \L \L}) } \,.
\eal
In the $\ka \to 0$ limit the normalisation becomes 
\bal
{\sqrt{\ta_{\L1}^{+0}} \ov \sqrt{\ta_{\L1}^{-0}}} \,  {\sqrt{\ta_{\L2}^{-0}} \ov \sqrt{\ta_{\L2}^{+0}}}  \ \frac{\tx^{-0}_{\L2}}{\tx^{+0}_{\L2}} \, \frac{\tx^{+0}_{\L1}}{\tx^{-0}_{\L1}} \, \left(\frac{\tx^{-0}_{\L1} - \tx^{+0}_{\L2}}{\tx^{+0}_{\L1} - \tx^{-0}_{\L2}} \right)^2 \to +1
\eal
if the two particles have the same chirality and
\bal
{\sqrt{\ta_{\L1}^{+0}} \ov \sqrt{\ta_{\L1}^{-0}}} \,  {\sqrt{\ta_{\L2}^{-0}} \ov \sqrt{\ta_{\L2}^{+0}}}  \ \frac{\tx^{-0}_{\L2}}{\tx^{+0}_{\L2}} \, \frac{\tx^{+0}_{\L1}}{\tx^{-0}_{\L1}} \, \left(\frac{\tx^{-0}_{\L1} - \tx^{+0}_{\L2}}{\tx^{+0}_{\L1} - \tx^{-0}_{\L2}} \right)^2 \to -1
\eal
if the two particles have opposite chiralities.
Then using~\eqref{eq:app_RR_limit_g_functions} we obtain
\bal
&\left(\Sigma^{\barnes}_{\L \L} (\tg^{\pm0}_{\L 1}, \tg^{\pm 0}_{\L 2}) \right)^{-2} \to \frac{R^4(\tg^{\circ \circ}_{12})}{R^2(\tg_{12}^{\circ \circ}+ i \pi) R^2(\tg_{12}^{\circ \circ} - i \pi )} \,.
\eal
In the limit, we also have that (see~\eqref{eq:app_massless_HL_RamRam})
\bal
\left( \Sigma^{00, \hl}_{\L \L}(u_1, u_2) \right)^2 \to - \frac{R^4(\tg_{12}^{\circ \circ}+ i \pi) R^4(\tg_{12}^{\circ \circ} - i \pi )}{R^8(\tg^{\circ \circ}_{12})}\,.
\eal
Then, combining the pieces above, we end up with
\bal
S_{\chi \chi}(u_1, u_2) \to  \, \varepsilon_{12} \, \frac{R^2(\tg_{12}^{\circ \circ}+ i \pi) R^2(\tg_{12}^{\circ \circ} - i \pi )}{R^4(\tg^{\circ \circ}_{12})} \left(\Sigma^{\bes}(\tx^{\pm 0}_1, \tx^{\pm 0}_2) \right)^{-2} \,,
\eal
where
\bal
\varepsilon_{12}=
\begin{cases}
&-1 \qquad \text{if same chirality}\,,\\
&+1 \qquad \text{if opposite chirality}\,.
\end{cases}
\eal
In principle one could define the S~matrix to be the same in both chirality sectors by introducing different signs in the normalisation~\eqref{eq:massless_massless_m0m0} depending on the chirality of the scattered particles; namely, one could change the sign of the opposite-chirality scattering in the mirror kinematics (and hence in the string kinematics as well). This would require changing the exchange relations by a minus sign in order to reproduce the physical S-matrix~$\mathbb{S}$.
Let us also remark that the sign of the same-chirality  S~matrix is fixed so that $S(u,u)=-1$ for the ZF S~matrix.

\section{Near-BMN limit}
\label{app:BMN_limit}

In this appendix, we show how to compute the near BMN expansion of our conjectured S-matrices, both for mirror and string kinematics.
In both kinematical regions, we find agreement with the results known in the literature at the tree level~\cite{Hoare:2013pma, Baglioni:2023zsf}.

\subsection{Near BMN limit in the mirror region}

\paragraph{Expansion of mirror kinematical variables.}

The expansion of the massive Zhukovsky variables was already considered in~\cite{Frolov:2025uwz} and we recall the formulae from that paper:
\bal
\label{eq:app_expansion_massive_Zhuk_BMN_mirror}
&\tx^\pm_a= \tx_a \left(1 \mp  \frac{\tilde{\omega}_{a}}{2T}\right)\,, \qquad \tx_a \equiv \frac{\sqrt{\tp_a^2-q^2+1} + q_a \tp_a}{(\tp_a+i) \sqrt{1-q^2}}\,,
\eal
where
 \bal
\tilde{\omega}_a=i q_a +\sqrt{\tp_\L^2-q^2+1}\,.
\eal
and $q_\L=-q_\R=1$. The $\tg$ variables are instead given by
 \bal
\tg^{\pm}_a = \tg_a - \frac{i \pi}{2} \mp \frac{i}{2 T} \tilde{\omega}^2_a\,, \qquad \tg_a \equiv \ln \frac{\xi_a - \tx_a}{\xi_a \tx_a +1}\,.
 \eal

For mirror massless particles, we have two different kinematical branches depending on whether the momentum is positive or negative. We obtain 
\bal
\label{eq:app_BMN_massless_Zh_mir}
\tx^{\pm 0}_a&=\xi_a \left(1 \mp \frac{\tilde{p}_a}{2T} +\frac{(3-q)\tilde{p}^2}{24T^2}\right) -i0\, \ \qquad \tp_a>0\,,\\
\tx^{\pm 0}_a&=\frac{1}{\xi_a}  \left(-1 \mp \frac{\tilde{p}_a}{2T} -\frac{(3+q)\tilde{p}^2}{24T^2}\right) -i0 \,, \quad \tp_a<0\,.
\eal
and
\bal
&\tg^{\pm 0}_a= \log \frac{ \xi_a \tp_a}{2 (1+\xi^2_a) T} \mp \frac{i \pi}{2} \pm \frac{\tp_a}{3 T} \frac{\xi^2_a-1}{\xi^2_a+1}+ \mathcal{O}(1/T^2)\,, \qquad &\tp_a>0\,,\\
&\tg^{\pm 0}_a= \log \left(-\frac{2 T (1+\xi^2_a)}{\xi_a \tp_a} \right) \mp \frac{i \pi}{2} \mp \frac{\tp_a}{3T} \frac{\xi^2_a-1}{\xi^2_a+1}+ \mathcal{O}(1/T^2) \,, \qquad &\tp_a<0\,.
 \eal

\paragraph{The limit of BES.}

To compute the AFS order of the BES phases we give a small mass $\delta$ to the massless particles so that $\Im(\tx^{\pm \delta}_a)<0$ and both points are just below the integration contour in the $x$ plane.  
For $\tx_{1}$ and $\tx_{2}$ in the mirror region in~\cite{Frolov:2025uwz} the following results for the mixed derivative of the phases were found\footnote{We follow the same notation of~\cite{Frolov:2025uwz} and define $\stackrel{\prime\prime}{\tPhi}(x,y) \equiv \frac{\partial^2}{\partial x \partial y} \tPhi$.}
\bal 
\label{eq:app_afs_mixed_der_general_masses}
{\stackrel{\prime\prime}{\tPhi}}{}^{\afs}_{a a}(\tx_{a1},\tx_{a2}) =&-\frac{1}{2}\frac{u_a'(\tx_{a1}) u_a'(\tx_{a2})}{u_1-u_2}+\frac{1}{4} \frac{u_a'(\tx_{a2})+u_a'(\tx_{a1})}{\tx_{a1} - \tx_{a2}}+\frac{\tx_{a1} - \tx_{a2}}{4 \tx^2_{a1} \tx^2_{a2}}\,,\\
{\stackrel{\prime\prime}{\tPhi}}{}^{\afs}_{\bar a a}(\tx_{\bar a1},\tx_{a2})=&\frac{1}{2 \tx_{\bar a1} \tx_{ a2} \left(\tx_{\bar a1} \tx_{a2} -1 \right)} \left[ \tx_{\bar a1} - \tx_{a2} - \frac{\ka_{\bar{a}}}{2 \pi} \left(\tx_{\bar a1} \tx_{a2} +1 \right) \right]\,,
\eal
and are valid for arbitrary points in the mirror region (see equation (6.14) in~\cite{Frolov:2025uwz}). Below we evaluate the second derivative of the phase on the different branches.

Suppose the first particle is massless on the branch $\tp_{1}>0$ and parameterise such a particle using left Zhukovsky variables. Then at the leading order we set $\tx_{1}=\xi$ (where $\tx_{1}$ can either be $\tx_{a1}$ or $\tx_{\bar{a} 1}$ depending if we use the first or second relation in~\eqref{eq:app_afs_mixed_der_general_masses}). Depending on the type of the second particle, we set $\tx_{a 2}=\tx_{\L 2}$ or $\tx_{a 2}=\tx_{\R 2}$ (see the definitions in~\eqref{eq:app_expansion_massive_Zhuk_BMN_mirror}).
By doing so we obtain
\bal 
\label{eq:app_AFS_BMN_mixed_mir}
{\stackrel{\prime\prime}{\tPhi}}{}^{\afs}_{\L \L}(\xi,\tx_{\L2}) &=-\frac{1}{2 \tx_{\L 2} \xi \sqrt{1-q^2}} \,, \qquad {\stackrel{\prime\prime}{\tPhi}}{}^{\afs}_{\L \R}(\xi,\tx_{\R2})&=-\frac{1}{2 \tx_{\R 2} \xi \sqrt{1-q^2}} \,.
\eal

On the other hand if both particles are massless and satisfy $\tp_{1}>0>\tp_{2}$ then from~\eqref{eq:app_BMN_massless_Zh_mir} we see that at the leading order we have $\tx_{a1}=-\frac{1}{\tx_{a2}}= \xi$ and
\bal 
\label{eq:afs_mixed_der_massless_massless_CA_same}
{\stackrel{\prime\prime}{\tPhi}}{}^{\afs}_{\L \L}(\xi,-1/\xi) &=\frac{1}{4} \left( \xi + \frac{1}{\xi} \right) = \frac{1}{2} \frac{1}{\sqrt{1-q^2}} \,.
\eal

Let us now derive the BMN limit of the BES phase.
First we consider the case in which the second Zhukovsky variable $\tx^\pm_{a 2}$ (where $a$ can either be L or R) is of massive type. Using the expansions in~\eqref{eq:app_expansion_massive_Zhuk_BMN_mirror} and the first line of~\eqref{eq:app_BMN_massless_Zh_mir} we obtain
\bal
\la{eq:thetaafs_mir_sec_der2}
\tilde{\theta}_{\L a}^\afs(\tx^{\pm 0}_{\L 1}, \tx^\pm_{a 2})&=h \left(\tPhi_{\L a}^\afs(\tx^{+0}_{\L 1}, \tx^+_{a 2})+\tPhi_{\L a}^\afs(\tx^{-0}_{\L 1}, \tx^-_{a 2})-\tPhi_{\L a}^\afs(\tx^{+0}_{\L 1}, \tx^-_{a 2})-\tPhi_{\L a}^\afs(\tx^{-0}_{\L 1}, \tx^+_{a 2}) \right)\\
&=\sqrt{1-q^2} \, \frac{\xi \tilde{p}_{ 1}  \tx_{a2} \tilde{\omega}_{a2} }{T} {\stackrel{\prime\prime}{\tPhi}}{}^{\afs}_{\L a}(\tx_{\L 1},\tx_{a2}) +\mathcal{O}(T^{-2})=-\frac{\tp_{ 1} \tilde{\omega}_{a2}}{2 T}+\mathcal{O}(T^{-2})\,,
\eal
where in the last equality we used the results in~\eqref{eq:app_AFS_BMN_mixed_mir}.
The limit of the even part of the mixed-mass dressing factor is then given by
\bal
\log (\Sigma_{\L a}^{01 \, \besratio} )^{-2}(u_1, u_2) \to \frac{i}{T} \tp_{\L1} \tilde{\omega}_{a2} \,.
\eal
Similarly, for the scattering of massless modes with $\tp_1>0>\tp_2$ we obtain
\bal
\label{eq:app_AFS_massless_massless_mirror_region}
&\tilde{\theta}_{\L \L}^\afs(\tx^{\pm 0}_{\L 1}, \tx^{\pm 0}_{\L 2})&= \frac{\sqrt{1-q^2}}{T} \tp_{\L 1} \tp_{\L 2} {\stackrel{\prime\prime}{\tPhi}}{}^{\afs}_{\L \L}(\xi,-1/\xi) +\mathcal{O}(T^{-2})= \frac{\tp_{\L1} \tp_{\L2}}{2T}+\mathcal{O}(T^{-2}) \,.
\eal
The BMN limit of the even part of the dressing factor for massless particles of opposite chirality is then given by
\bal
\label{eq:near_BMN_even_mirmassless_mirmassless}
\log (\Sigma_{\L \L}^{00 \, \besratio} )^{-2}(u_1, u_2) \to -\frac{i}{T} \tp_{1} \tp_{2} \,.
\eal

This concludes the computation of the near-BMN limit of BES both in the mixed-mass and massless-massless sectors. We still need to evaluate the same limit on the remaining pieces of the S-matrix. Below we show how this can be done in the two different setups of mixed-mass and massless scattering.

\paragraph{Massless-massive scattering.}

We compute the near BMN limit of the S-matrix element $S_{T^{1 \dot{1}} Y}$ in the transmission channel
\bal
\label{eq:app_S_T_Y_for_near_BMN}
\mathbf{S}\,\big| T^{1\dot{1}}_{1}Y_{2} \big\rangle=&  S_{T^{1 \dot{1}} Y}\,\big| T^{1\dot{1}}_{1}Y_{2} \big\rangle + \dots\,.
\eal
The ellipses correspond to the reflection element, which we do not consider since it is connected to the transmission element by supersymmetry.
The bosonic massless state
\bal
|T^{1 \dot{1}} \rangle= | \varphi_\L^{F} \rangle \otimes |\phi_\L^{F} \rangle
\eal
is obtained by lowering the first state of $|\chi^{\dot{1}} \rangle= | \phi_\L^{B} \rangle \otimes |\phi_\L^{F} \rangle$.
We can then obtain $S_{T^{1 \dot{1}} Y}$ by introducing the factor
\bal
\label{eq:app_DLL_factor_mir}
D_{\L \L}(u_1, u_2)=\sqrt{\frac{\tx^{+}_{\L2}}{\tx^{-}_{\L2}}} \, \frac{\tx^{-0}_{\L1} - \tx^-_{\L2}}{\tx^{-0}_{\L1} - \tx^+_{\L2}}
\eal
(as defined in appendix B of~\cite{Frolov:2023lwd}) to the normalisation in the first row of~\eqref{eq:massless_massive_normmir_bstates}:
\bal
\label{eq:app_S_TY_mirmir}
S_{T^{1 \dot{1}} Y} (u_1, u_2)&=D_{\L \L}(\tx^{\pm 0}_{\L1}, \tx^{\pm}_{\L2}) S_{\chi^{\dot{1}} Y} (u_1, u_2)\\
&={\sqrt{\ta_{\L2}^{-}} \ov \sqrt{\ta_{\L2}^{+}}}  \, \frac{\tx^{+0}_{\L1}}{\tx^{-0}_{\L1}} \, \sqrt{\frac{\tx^-_{\L2}}{\tx^+_{\L2}}}  \, \frac{(\tx^{-0}_{\L1} - \tx^{+}_{\L2}) (\tx^{-0}_{\L1} - \tx^{-}_{\L2})}{(\tx^{+0}_{\L1} - \tx^{-}_{\L2} )^2}  \left(\Sigma^{01}_{\L\L}(u_1,u_2)\right)^{-2} \,.
\eal
In the BMN limit, the different terms of the S-matrix element above become
\bal
&\log (\Sigma_{\L\L}^{01 \, \besratio} )^{-2}(u_1, u_2) \to \frac{i}{T} \tp_{1} \tilde{\omega}_{\L2} \,,\\
&\log (\Sigma_{\L\L}^{01 \, \barnes} )^{-2}(u_1, u_2)=\log \frac{R^2(\tg^{-0-}_{\L \L}) R^2(\tg^{+0+}_{\L \L})}{R^2(\tg^{-0+}_{\L \L}) R^2(\tg^{+0-}_{\L \L}) } \to \frac{i \tilde{\omega}^2_{\L2}}{2 T} \,,\\
&\log \left( {\sqrt{\ta_{\L2}^{-}} \ov \sqrt{\ta_{\L2}^{+}}}  \, \frac{\tx^{+0}_{\L1}}{\tx^{-0}_{\L1}} \, \sqrt{\frac{\tx^-_{\L2}}{\tx^+_{\L2}}}  \, \frac{(\tx^{-0}_{\L1} - \tx^{+}_{\L2}) (\tx^{-0}_{\L1} - \tx^{-}_{\L2})}{(\tx^{+0}_{\L1} - \tx^{-}_{\L2} )^2} \right) \to 
-\frac{i}{T} \tp_{1} \tp_{2} -\frac{i}{T} \tp_{1} \tilde{\omega}_{\L2} - \frac{i}{2T} \tilde{\omega}^2_{\L2} \,.
\eal
Combining the expressions above we get
\bal
\log S^{01}_{T^{1 \dot{1}} Y}(u_1,u_2) \to -\frac{i}{T} \tp_{1} \tp_{2}\,,
\eal
in agreement with the results of~\cite{Baglioni:2023zsf} in the $a=0$ gauge (see equation (3.14) of that paper).

In principle, one could repeat a similar computation for $S_{Y T^{1 \dot{1}}}(p_1, p_2)$. 
However, this is not necessary since the mirror S-matrix is parity invariant and therefore it holds
\bal
S_{Y \chi^{\dot{1}}}(\tp_1, \tp_2)=S_{\chi^{\dot{1}} Y}(-\tp_2, -\tp_1) \,.
\eal
Then the correctness of $S_{T^{1 \dot{1}} Y}$ implies that also $S_{Y T^{1 \dot{1}}}$ has the correct expansion. Similarly the S-matrix elements involving massive particles of right type must be correct since they are connected to the element above by charge conjugation (or left-right symmetry).

\paragraph{Massless-massless scattering.}

Let us now compute the massless-massless S-matrix element
\bal
\label{eq:STT_mir_mir_for_comparison_near_BMN}
S^{00}_{T^{1 \dot{1}} T^{1 \dot{1}}} (u_1, u_2)&=F_{\L \L}(u_1, u_2) S_{\chi^{\dot{1}} \chi^{\dot{1}}}(u_1, u_2)\\
&={\sqrt{\ta_{\L1}^{+0}} \ov \sqrt{\ta_{\L1}^{-0}}} \,  {\sqrt{\ta_{\L2}^{-0}} \ov \sqrt{\ta_{\L2}^{+0}}}  \ \sqrt{\frac{\tx^{-0}_{\L2}}{\tx^{+0}_{\L2}}} \, \sqrt{\frac{\tx^{+0}_{\L1}}{\tx^{-0}_{\L1}}} \, \frac{\tx^{-0}_{\L1} - \tx^{+}_{\L2}}{\tx^{+0}_{\L1} - \tx^{-}_{\L2}}  \left(\Sigma^{00}_{\L\L}(u_1,u_2)\right)^{-2}\,.
\eal
The limit on the even part of the dressing factor can be read from~\eqref{eq:near_BMN_even_mirmassless_mirmassless}, while the odd part and rational terms return
\bal
\label{eq:app_barnes_massless_mirror_BMN}
&\log  (\Sigma_{\L\L}^{00 \, \barnes} )^{-2}(u_1, u_2) =\log \frac{R^2(\tg^{-0- 0}_{\L \L}) R^2(\tg^{+0+0}_{\L \L})}{R^2(\tg^{-0+0}_{\L \L}) R^2(\tg^{+0-0}_{\L \L}) }  \to -\frac{q}{3 T} (\tp_1 - \tp_2) + i \frac{\pi}{2} +\mathcal{O}(1/T^2)\,,\\
&\log \left(-{\sqrt{\ta_{\L1}^{+0}} \ov \sqrt{\ta_{\L1}^{-0}}} \,  {\sqrt{\ta_{\L2}^{-0}} \ov \sqrt{\ta_{\L2}^{+0}}} \, \sqrt{\frac{\tx^{-0}_{\L2}}{\tx^{+0}_{\L2}}} \, \sqrt{\frac{\tx^{+0}_{\L1}}{\tx^{-0}_{\L1}}} \, \frac{\tx^{-0}_{\L1} - \tx^{+}_{\L2}}{\tx^{+0}_{\L1} - \tx^{-}_{\L2}} \right) \to \frac{q}{3 T} (\tp_1 - \tp_2) + \mathcal{O}(1/T^2) \,.
\eal
Combining all terms above, we end up with
\bal
\label{eq:app_final_STT_massless_mirror_BMN}
 S_{T^{\dot{1} 1} T^{\dot{1} 1}}(\tp_1, \tp_2) \to e^{-i \frac{\pi}{2}} \left(1  -\frac{i}{T} \tp_{1} \tp_{2} \right) + \mathcal{O}(1/T^2) \,, \quad \tp_1>0> \tp_2\,.
\eal
We stress that this corresponds to the BMN limit of the ZF S-matrix. Such an S-matrix should be connected to the physical S-matrix through equation \eqref{eq:physicalS}, involving nontrivial exchange relations in the free theory limit (as discussed in~\cite{Frolov:2025ozz} and references therein). Here we postulate that all $s_{ij}=1/4$ for the massless-massless scattering; therefore, the physical S-matrix for the scattering of massless modes is given by
\begin{equation}
\label{eq:physicalSTT_BMN}
    \mathbb{S}_{T^{\dot{1} 1} T^{\dot{1} 1}}(\tp_1, \tp_2)= e^{i \frac{\pi}{2}} \,S_{T^{\dot{1} 1} T^{\dot{1} 1}}(\tp_1, \tp_2)=1  -\frac{i}{T} \tp_{1} \tp_{2}  + \mathcal{O}(1/T^2) \,,
\end{equation}
in agreement with the results in~\cite{Baglioni:2023zsf} for the mirror-mirror massless kinematics.

\subsection{Near BMN limit in the string region for mixed-mass}
\label{app:BMN_string_mixed}

Let us consider the scattering of a massless particle coming from the left and a massive particle coming from the right, in the string kinematical region.
After continuing the expression in~\eqref{eq:app_S_TY_mirmir} to the string region, the odd part of the dressing factor transforms as follows
\bal
&\frac{R^2(\tg^{-0-}_{\L\L}) R^2(\tg^{+0+}_{\L\L})}{R^2(\tg^{-0+}_{\L\L}) R^2(\tg^{+0-}_{\L\L}) }
\quad  \xrightarrow{\tx_{\L1}^{\pm0}\to x_{\L1}^{\pm0},\  \tx_{\L2}^\pm\to x_{\L2}^\pm}
\quad 
\frac{R^2(\g^{-0-}_{\L\L}) R^2(\g^{+0+}_{\L\L}-2\pi i)}{R^2(\g^{-0+}_{\L\L}) R^2(\g^{+0-}_{\L\L}-2\pi i) } \,,
\eal
and we obtain
\bal
\label{eq:app_S_TY_strstr}
S_{T^{1 \dot{1}} Y} (u_1, u_2)
&={\sqrt{\a_{\L2}^{-}} \ov \sqrt{\a_{\L2}^{+}}}  \, \frac{x^{+0}_{\L1}}{x^{-0}_{\L1}} \, \sqrt{\frac{x^-_{\L2}}{x^+_{\L2}}}  \, \frac{(x^{-0}_{\L1} - x^{+}_{\L2}) (x^{-0}_{\L1} - x^{-}_{\L2})}{(x^{+0}_{\L1} - x^{-}_{\L2} )^2} \\
&\times \frac{R^2(\g^{-0-}_{\L\L}) R^2(\g^{+0+}_{\L\L}-2\pi i)}{R^2(\g^{-0+}_{\L\L}) R^2(\g^{+0-}_{\L\L}-2\pi i) } \, \left(\Sigma^{01,\besratio}_{\L\L}(u_1,u_2)\right)^{-2} \,.
\eal
In this appendix, we compute the near-BMN limit of the expression above in the kinematical branch $p_1>0$. We find agreement at the tree level with the perturbative computation of~\cite{Hoare:2013pma}.
We perform the limit on the case in which the massive particle is of left type since the case in which it is of right type is connected by charge conjugation.

\paragraph{Expansion of string kinematical variables.}

We start considering the expansion of the massive Zhukovsky variables, which is given by 
\bal
&
x_{\L2}^{\pm}&=x_{\L2} \left(1 \pm \frac{i p_2}{2 T} \right)+ \mathcal{O}(1/T^2)\,,
\eal
where
\bal
x_{\L2}=\frac{1+q p_2 +\omega_{\L2}}{p_2 \sqrt{1-q^2}}\,, \quad \omega_{\L2}=\sqrt{1+p_2^2+2 q p_2} \,.
\eal
The massive $\g$ variables have instead the following expansion
\bal
\g^{\pm}_{\L2}&= \log \left( \frac{x_{\L2} - \xi }{\xi x_{\L2} -1} \right) \pm \frac{i}{2 T} p^2_2+ \mathcal{O}(\frac{1}{T^2})\,.
\eal

We also list the expansion of the different massless variables on the positive momentum branch:
\bal
\label{eq:app_BMN_massless_massless_string_x1lim}
&x_{\L1}^{\pm0}=\xi \left(1\pm \frac{i p_1}{2T} - (3-q) \frac{p^2_1}{24 T^2}\right)+ \mathcal{O}(1/T^3)\,,\\
&\g^{\pm 0}_{\L1}= \log \left( \frac{\xi \,  p_1}{2 T (1+ \xi^2)} \right) \pm \frac{i \pi}{2} \mp q \frac{i p_1}{3 T} + \mathcal{O}(1/T^2) \,.
\eal

\paragraph{Deforming the contours of BES.}

To perform the limit in a smooth way, we continue the massless particle (with Zhukovsky variables $\tx^{\pm 0}_{\L1}$) to the string region leaving the massive particle (with Zhukovsky variables $\tx^{\pm}_{\L2}$) in the mirror region. As discussed in~\cite{Frolov:2025uwz} we can continue the second variable to the string region after the limit.
The continuation of BES is 
\bal
\label{eq:app_first_massless_continued_to_string}
\tilde{\theta}_{\L \L}^{\bes}(x^{\pm 0}_{\L 1}, \tx^\pm_{\L2})&=\tilde{\varPhi}_{\L \L}(x^{\pm 0}_{\L 1},\tx^{\pm }_{\L2})-\tPsi_{\L}(u_1, \tx^+_{\L2})+\tPsi_{\L}(u_1, \tx^-_{\L2})\,.
\eal
When we continue the second variable to the string region instead of picking up additional residues we deform the contour. This is equivalent to compute the BMN limit with the second variable in the mirror region and make the continuation after the limit. The interested reader can find more details in appendix J of~\cite{Frolov:2025uwz}.
The expression above is still not suitable to perform a large $h$ expansion; indeed in the BMN limit the massless variables $x_{\L1}^{+0}$ and $x_{\L1}^{-0}$, which are just above and below the real line, pinch the integration contour in the $x$ plane from opposite sides. In particular from the first line in~\eqref{eq:app_BMN_massless_massless_string_x1lim} we see that $x^{+0}_{\L1}$ and $x^{-0}_{\L1}$ are both close to $\xbr$ and have a small positive and negative imaginary part, respectively. 
To have a better expansion, we deform the contours of the functions $\tPhi_{\L\L}^{--}$ and $\tPhi_{\L\L}^{++}$ in such a way to surround the point $x^{-0}_{\L1}$ from below. In this way, after the deformations, both $x^{+0}_{\L1}$ and $x^{-0}_{\L1}$ are above the integration contour in the $x$ plane.

After deforming the contours the $\tPhi$ functions are modified as follows
\bal
&\tPhi^{--}_{\L \L}(x^{- 0}_{\L 1},\tx_{\L2})=\tPhi^{--, \rm def}_{\L \L}(x^{- 0}_{\L 1},\tx_{\L2}) - \tPsi_\L^{-}(u_1, \tx_{\L2})\,,\\
&\tPhi^{++}_{\L \L}(x^{- 0}_{\L 1},\tx_{\L2})=\tPhi^{++, \rm def}_{\L \L}(x^{- 0}_{\L 1},\tx_{\L2}) - \tPsi_\L^{+}(u_1, \tx_{\L2})\,,
\eal
where by `def' we mean that the contour is deformed below the real line in such a way to go around $x^{-0}_{\L1}$ from below.
Plugging these relations into~\eqref{eq:app_first_massless_continued_to_string} all $\tPsi$ functions simplify and we get
\bal
\label{eq:app_definition_of_bes_in_terms_of_def_Phi}
\tilde{\theta}_{\L \L}^{\bes}(x^{\pm 0}_{\L 1}, \tx^\pm_{\L2})&=\tilde{\varPhi}^{\rm def}_{\L \L}(x^{\pm 0}_{\L 1},\tx^{\pm }_{\L2})\,.
\eal

\paragraph{Deforming the contour of HL.}

Before deforming the HL contour, we start recalling that under continuation of the first variable to the string region, the HL $\tPhi$ functions are continued as follows
\bal
&\tPhi^\hl_{\L\L}(\tx^{+0}_{\L1}, \tx^\pm_{\L2}) \to  \tPhi^\hl_{ab}(x^{+0}_{\L1}, \tx^\pm_{\L2}) - \frac{1}{2i} \log (-\tx^\pm_{\L2})+\frac{1}{2i} \log \frac{x^{+0}_{\L1} - \tx^\pm_{\L2}}{x^{-0}_{\L1} - \tx^\pm_{\L2}}+ \frac{\pi}{2}\,,
\eal
and we obtain
\bal
\label{eq:app_cont_hl_first_variable_massless_to_string}
\tilde{\theta}_{\L \L}^{\hl}(x^{\pm 0}_{\L 1}, \tx^\pm_{\L2})&=\tilde{\varPhi}^\hl_{\L \L}(x^{\pm 0}_{\L 1},\tx^{\pm}_{\L2})+\frac{1}{2 i} \log \frac{\tx^-_{\L2}}{\tx^+_{\L2}} \, \frac{(x^{+0}_{\L1}- \tx^+_{\L2}) (x^{-0}_{\L1}- \tx^-_{\L2}) }{(x^{-0}_{\L1}- \tx^+_{\L2}) (x^{+0}_{\L1}- \tx^-_{\L2})} \,.
\eal
Now we deform the contour of the function $\tPhi^\hl_{\L \L}$ in the same manner as we did for BES and get
\bal
&\tPhi^{- -, \hl}_{\L \L}(x^{-0}_{\L1}, x_2)=\tPhi^{- -, \hl, {\rm def}}_{\L \L}(x^{-0}_{\L1}, x_2) - \tPsi^{- , \hl}_{\L}(u_1, x_2)\,,\\
&\tPhi^{+ +, \hl}_{\L \L}(x^{-0}_{\L1}, x_2)=\tPhi^{+ +, \hl, {\rm def}}_{\L \L}(x^{-0}_{\L1}, x_2) - \tPsi^{+, \hl}_{\L}(u_1, x_2)\,.
\eal
In the expressions above $x_2$ can either be $\tx^+_2$ or $\tx^-_2$ and
\bal
\label{eq:app_Psi_HL}
\tPsi^{-, \hl}_{\L}(u_1,x_2)&=- \lint_{\rm \widetilde{cuts}} \frac{d v}{2 \pi i} \, \frac{\tx'_{\L}(v)}{\tx_{\L}(v) - x_2} \, K^\hl_\eps (u_1-v)\,,\\
\tPsi^{+, \hl}_{\L}(u_1,x_2)&=+ \lint_{\rm \widetilde{cuts}} \frac{d v}{2 \pi i} \, \frac{\left(\frac{1}{\tx_{\R}(v)}\right)'}{\frac{1}{\tx_{\R}(v)} - x_2} \, K^\hl_\eps (u_1-v)\,.
\eal
The point $u_1=u_{\L}(x^{\pm0}_{\L1})$ is on the main string theory cut $(-\infty, \nu)$ and in the BMN limit approached $\nu$ from the left. This point is away from the integration contour, which runs around the mirror theory cuts. We can then send $\eps \to 0$ in the $\tPsi$ functions and obtain
\bal
K^\hl_\eps (u_1-v)=\frac{\pi}{2} \text{sign} \left( \Re (v - u_1) \right) \,.
\eal
Replacing the kernel into~\eqref{eq:app_Psi_HL} we obtain that
\bal
\label{eq:app_tpsi_hl_massless_massive_bmn_string}
&\tPsi^{-, \hl}_{\L}(u_1,x_2)=-\frac{1}{2 i} \log R +\frac{1}{2i} \log(-x_2) - \frac{\pi}{4}\,,\\
&\tPsi^{+, \hl}_{\L}(u_1,x_2)=-\frac{1}{2 i} \log R +\frac{1}{2i} \log(-x_2) - \frac{\pi}{4}\,,
\eal
where $R \gg1$ is a regulator defining the boundary of the integration contours in the $v$ plane (this means we are integrating between $\nu$ and $\nu +R$ around the main mirror cut and similarly for the $\ka$ cut).
After replacing the expression above into the full phase all the regulator-dependent pieces cancel and we end up with
\bal
\label{eq:app_definition_of_hl_in_terms_of_def_Phi}
\tilde{\theta}_{\L \L}^{\hl}(x^{\pm 0}_{\L 1}, \tx^\pm_{\L2})=\tilde{\varPhi}^{\hl, \rm def}_{\L \L}(x^{\pm 0}_{\L 1},\tx^\pm_{\L2})+\frac{1}{2 i} \log \frac{(x^{+0}_{\L1}- \tx^+_{\L2}) (x^{-0}_{\L1}- \tx^-_{\L2}) }{(x^{-0}_{\L1}- \tx^+_{\L2}) (x^{+0}_{\L1}- \tx^-_{\L2})}\,.
\eal

Combining~\eqref{eq:app_definition_of_bes_in_terms_of_def_Phi} and~\eqref{eq:app_definition_of_hl_in_terms_of_def_Phi} we obtain that
\bal
\label{eq:app_difference_string_even_for_BMN1}
\log (\Sigma^{0 1 \besratio}_{\L\L})^{-2}= -2 i \tilde{\varPhi}^{\rm def}_{\L\L}(x^{\pm 0}_{\L1}, x^{\pm}_{\L2}) +2 i \tilde{\varPhi}^{\hl, \rm def}_{\L\L}(x^{\pm 0}_{\L1}, x^{\pm}_{\L2})+\log \frac{(x^{+0}_{\L1}- x^+_{\L2}) (x^{-0}_{\L1}- x^-_{\L2}) }{(x^{-0}_{\L1}- x^+_{\L2}) (x^{+0}_{\L1}- x^-_{\L2})}\,.
\eal
The point $x^+_{\L2}$ is continued to the string region without picking any further residue since we take its continuation into account by deforming the contour as explained in appendix J.2 of~\cite{Frolov:2025uwz}.
Now both in BES and HL the points $x^{+0}_{\L1}$ and $x^{-0}_{\L1}$ are on the same side of the contour (the same applies to $x^{+}_{\L2}$ and $x^{-}_{\L2}$). Therefore we can compute the near BMN limit of the even dressing factor using the mixed derivatives of the phase. This was not possible before since the points would have pinched the integration contour from opposite sides. 

We also stress that the term
\begin{equation}
\log \frac{(x^{+0}_{\L1}- x^+_{\L2}) (x^{-0}_{\L1}- x^-_{\L2}) }{(x^{-0}_{\L1}- x^+_{\L2}) (x^{+0}_{\L1}- x^-_{\L2})}\,,
\end{equation}
coming from the continuation of HL starts contributing at the order $T^{-2}$; therefore from now on we will neglect sich a term, since we compare the expansion only to tree level results. One could of course perform the expansion also to one loop order. However, already for the pure Ramond-Ramond case, a disagreement was observed at this order~\cite{Frolov:2021fmj}. In that paper, the authors justified such a disagreement due to problems with UV and IR singularities in perturbative computations.

\paragraph{The mixed-mass S-matrix in the near-BMN limit.}

Following~\cite{Frolov:2025uwz} it is possible to show that for generic points $x_1$ and $x_2$ in the complex plane the following relation holds
\bal
\label{eq:afs_mixed_derivative_generic}
{\stackrel{\prime\prime}{\tPhi}}{}_{\L\L}^{\afs}(x_1,x_2)=&-{u_\L'(x_1)u_\L'(x_2)\ov 2 u_{12}} \left(\theta\left(\Im(x_1)\right) \theta\left(\Im(x_2)\right) +\theta\left(-\Im(x_1)\right) \theta\left(-\Im(x_2)\right) \right)\\
&+{1\ov 4 (x_1 -x_2)} \left( u_\L'(x_1)+  u_\L'(x_2) \right) + \frac{x_1-x_2}{4 x_1^2 x_2^2} \,.
\eal
Since when moving $x_2$ to the string region we are deforming the conotour so that $x_2$ is always below the contour then we need to evaluate the expression above for $\Im(x_2)<0$. 
Moreover the contour is deformed so that it goes below the point $x_1$; then the AFS order of BES is computed as if $\Im(x_1)>0$. Taking this into account the expression in~\eqref{eq:afs_mixed_derivative_generic} simplifies to
\bal
{\stackrel{\prime\prime}{\tPhi}}{}_{\L\L}^{\afs, {\rm def}}(x_1,x_2)={1\ov 4 (x_1 -x_2)} \left( u_\L'(x_1)+  u_\L'(x_2) \right) + \frac{x_1-x_2}{4 x_1^2 x_2^2} \,.
\eal
Finally using that at the point around which we are expanding it holds that $x_1=\xi$ (and $u'_\L(\xi)=0$) then we get
\bal
{\stackrel{\prime\prime}{\tPhi}}{}_{\L\L}^{\afs, {\rm def}}(\xi,x_2)={u_\L'(x_2) \ov 4 (\xi -x_2)} + \frac{\xi-x_2}{4 \xi^2 x_2^2} \,.
\eal

From~\eqref{eq:app_definition_of_bes_in_terms_of_def_Phi} we see that the BES phase can be written in terms of $\tPhi$ functions with deformed contours. Performing a large $h$ expansion we have
\bal
\tilde{\theta}^{\bes}_{\L\L}(x^{\pm 0}_{\L1}, x^{\pm}_{\L2})=h\tilde{\varPhi}^{\afs, \rm def}_{\L\L}(x^{\pm 0}_{\L1}, x^{\pm}_{\L2}) + \tilde{\varPhi}^{\hl, \rm def}_{\L\L}(x^{\pm 0}_{\L1}, x^{\pm}_{\L2})+ \dots \,.
\eal
Since the points are on the same side of the contour, the function $\tilde{\varPhi}^{\afs, \rm def}_{\L\L}(x^{\pm 0}_{\L1}, x^{\pm}_{\L2})$ is analytic around $\xbr$ (which is the leading order of the near-BMN expansion of $x^{\pm 0}_{\L1}$) and we can compute the expression above using the mixed derivative of the phase. We get
\bal
\label{eq:app_tilde_BES_near_BMN1}
\tilde{\theta}^{\bes}_{\L\L}(x^{\pm 0}_{\L1}, x^{\pm}_{\L2}) &\to -\frac{ p_1 p_2 \sqrt{1-q^2} x_{\L2} \xi }{T} \ {\stackrel{\prime\prime}{\tPhi}}{}_{\L\L}^{\afs}(\xi,x_{\L2})+ \tilde{\varPhi}^{\hl, \rm def}_{\L\L}(x^{\pm 0}_{\L1}, x^{\pm}_{\L2}) + \mathcal{O}(1/T^3)\\
&= \frac{p_1 p_2}{2T}+ \tilde{\varPhi}^{\hl, \rm def}_{\L\L}(x^{\pm 0}_{\L1}, x^{\pm}_{\L2}) + \mathcal{O}(1/T^3)\,.
\eal
As expected the result is the same as the one in~\eqref{eq:thetaafs_mir_sec_der2} after sending $p_1 \to i \tilde{\omega}_{\L1}=i \tilde{p}_1$ and $p_2 \to i \tilde{\omega}_{\L2}$. This implies that at the tree level, continuing and then performing the limit is the same as performing the BMN limit first and then continue to string region.

Taking finally the difference between BES and HL we obtain the following BMN limit for the even part of the dressing factor
\bal
\label{eq:app_near_BMN_even_part}
\log (\Sigma_{\L \L}^{10 \, \besratio} )^{-2}(u_1, u_2) \to -i \frac{p_1 p_2}{T} + \mathcal{O}(1/T^2) \,.
\eal

The remaining odd part and rational factors appearing in the S-matrix element have the following BMN expansion
\bal
&\log \frac{R^2(\g^{--}_{\L \L}) R^2(\g^{++}_{\L \L}- 2 \pi i)}{R^2(\g^{-+}_{\L \L}) R^2(\g^{+-}_{\L \L} - 2 \pi i) } \to -\frac{i p_2^2}{2T} \,,\\
& \log {\sqrt{\a_{\L2}^{-}} \ov \sqrt{\a_{\L2}^{+}}}  \, \frac{x^{+0}_{\L1}}{x^{-0}_{\L1}} \, \sqrt{\frac{x^-_{\L2}}{x^+_{\L2}}}  \, \frac{(x^{-0}_{\L1} - x^{+}_{\L2}) (x^{-0}_{\L1} - x^{-}_{\L2})}{(x^{+0}_{\L1} - x^{-}_{\L2} )^2} \to i \frac{p_1 p_2}{T} + i \frac{p_1 \omega_{\L2}}{T}+\frac{i p_2^2}{2T}\,.
\eal
Combining the values above into~\eqref{eq:app_S_TY_strstr} we obtain

\bal
\log S_{T^{1 \dot{1}} Y}(u_1, u_2)=& + i \frac{p_1 \omega_{\L2}}{T}\,.
\eal
This result agrees with perturbative computations at the tree level, as can be seen (for example) from equation (2.51) of~\cite{Baglioni:2023zsf}.

\subsection{Near BMN limit in the string region for massless-massless}
\label{app:BMN_string_massless}

We want to consider now the scattering of a chiral massless particle coming from the left with an antichiral massless particle coming from the right, both with string kinematics.
In order to analyse the limit, we recall that the region of positive momentum for the first particle is reached following the path sketched in figure~\ref{fig:positive-p-path}, while the negative momentum region for the second particle is reached through the path in figure~\ref{fig:negative-p-path}. We invite the reader to look at the discussion in section~\ref{sec:massless_in_string} for more details on the two continuations. 

\paragraph{Expansion of string kinematical variables.}

After the continuation, we have
\bal
&\tx^{-0}_{\L1} \to =x^{-0}_{\L}(u_1)\,, \qquad u_1<\nu\,,\\
&\tx^{+0}_{\L1} \to =x^{+0}_{\L}(u_1)\,, \qquad u_1<\nu\,,
\eal
and $u_1$ is on the main string theory cut. In the BMN limit $u_1$ appraches $\nu$ from the left. 
For the second particle we have instead
\bal
&\tx^{-0}_{\L2}(u_2 + i \ka ) \to \frac{1}{x^{+ k}_{\R2}(u_2)}\,, \qquad  u_2> -\nu\,,\\
&\tx^{+0}_{\L2}(u_2 + i \ka )=\frac{1}{x^{-k}_{\R2}(u_2)}\,, \qquad \ u_2> -\nu\,.
\eal
In this case, in the BMN limit $u_2$ approaches the point $-\nu$ from the right and $x^{\pm k}_{\R2}$ are evaluated near the branchpoints $-\nu \pm \frac{i}{h}k$.
Under continuation to the string region, the $\g$ variables are continued as follows
\bal
\label{eq:app_massless_string_BMN_g_cont}
\tg^{- 0}_{\L1} \to \g^{- 0}_{\L1} + i\frac{\pi}{2}\,, \qquad \tg^{+ 0}_{\L1} \to \g^{+ 0}_{\L1} - i\frac{3\pi}{2}\,,
\eal
\bal
\tg^{- 0}_{\L2} \to \g^{+ k}_{\R2} - i\frac{\pi}{2}\,, \qquad \tg^{+ 0}_{\L2} \to \g^{- k}_{\R2} - i\frac{\pi}{2}\,.
\eal

The near-BMN expansion of the first particle kinematics was given in~\eqref{eq:app_BMN_massless_massless_string_x1lim} and is identical also in this situation. For the second particle, we have
\bal
p^{(0)}_{\L 2}=  p^{(k)}_{\R 2} - 2 \pi  \in (-2 \pi, 0)
\eal
and we expand
\bal
p^{(k)}_{\R 2} = 2 \pi + \frac{p_2}{T}\,, \quad \text{with} \ \ p_2<0\,.
\eal
From this consideration we obtain
\bal
\label{eq:app_BMN_massless_massless_string_x2lim}
&x_{\R2}^{\pm k}= \xi \left(-1 \mp \frac{i p_2}{2 T} + (3-q) \frac{p^2_2}{24 T^2} \right)+\mathcal{O}(1/T^3)\,, \\
&\frac{1}{x^{\mp k}_{\R2}}= \frac{1}{\xi} \left( -1 \mp \frac{i p_2}{2 T} + (3+q) \frac{p^2_2}{24 T^2} \right)+\mathcal{O}(1/T^3) \,,
\eal
and
\bal
\g^{\pm k}_{\R2}= \log T \pm i \frac{\pi}{2} - \log \left(- \frac{p_2}{4} \sqrt{1-q^2} \right) \pm \frac{i q}{3T} p_2 +\mathcal{O}(1/T^2) \,.
\eal

\paragraph{Deforming the contours of BES and HL.}
As for the massless-massive scattering to facilitate the limit we deform the integration contours in the BES and HL phases so that both points $1/x^{-k}_{\R2}$ and $1/x^{+k}_{\R2}$ are on the same side of the contour (and the same applies to $x^{+0}_{\L1}$ and $x^{-0}_{\L1}$). In this case, we deform the integration contours slightly below on the lower half $x$ plane in such a way as to surround both the points $x^{-0}_{\L1}$ and $1/x^{+k}_{\R2}$ from below.
The deformation around the point $1/x^{+k}_{\R2}$ is a bit tricky for the function $\tPhi^{++}_{\L\L}$ since $\Re(1/x^{+k}_{\R2})< 0$ and the contour deformation must be performed entering the logarithmic cut of $u_{\L}(w_2)$ (see definition in~\eqref{eq:app_tPhidef}).

We start with the following  BES phase evaluated in the string region
\bal
\label{eq:app_second_massless_continued_to_string2}
\tilde{\theta}_{\L \L}^{\bes}(x^{\pm 0}_{\L 1}, \frac{1}{x^{\mp k}_{\R2}})&=\tilde{\varPhi}_{\L \L}(x^{+ 0}_{\L 1},\frac{1}{x^{\mp k}_{\R2}})+\frac{1}{2}\tPsi^+_\L(u_2-\frac{i}{h} k, x^{+0}_{\L1}) -\frac{1}{2}\tPsi^+_\L(u_2-\frac{i}{h} k, x^{-0}_{\L1})\\
&-\tPsi_{\L}(u_1, \frac{1}{x^{-k}_{\R2}})+\tPsi_{\L}(u_1, \frac{1}{x^{+k}_{\R2}})+\frac{1}{2}\tPsi^-_\L(u_2+\frac{i}{h} k, x^{+0}_{\L1}) -\frac{1}{2}\tPsi^-_\L(u_2+\frac{i}{h} k, x^{-0}_{\L1})\\
&+\frac{1}{2} K^\bes \left(u_1-u_2-\frac{i}{h} k \right) +\frac{1}{2} K^\bes \left(u_1-u_2+\frac{i}{h} k \right) \,.
\eal
After the double contour deformation, it turns out that
\bal
\label{eq:app_theta_bes_massless_massless_bmn_contdef2_step3}
&\tilde{\theta}_{\L \L}^{\bes}(x^{\pm 0}_{\L 1}, \frac{1}{x^{\mp k}_{\R2}})=\tilde{\varPhi}^{{\rm def}_{12}}_{\L \L}(x^{\pm 0}_{\L 1},\frac{1}{x^{\mp k}_{\R2}})\,.
\eal
Both integration contours are now deformed so that all Zhukovsky variables are above the integration contours in the $x$ plane.
By def$_{12}$ we mean that the contours are deformed both around the first and second particles in the way just described. 
A similar deformation on the HL phase leads to
\bal
\label{eq:app_definition_of_hl_in_terms_of_def_Phi_massless}
\tilde{\theta}^\hl_{\L\L}(x^{\pm 0}_{\L1}, \frac{1}{x^{\mp k}_{\R2}})&= \tilde{\varPhi}^{\hl, {\rm def}_{12}}_{\L\L}(x^{\pm 0}_{\L1}, \frac{1}{x^{\mp k}_{\R2}}) \,.
\eal
Combining~\eqref{eq:app_theta_bes_massless_massless_bmn_contdef2_step3} and~\eqref{eq:app_definition_of_hl_in_terms_of_def_Phi_massless} we obtain
\bal
\label{eq:app_difference_string_even_for_BMN2}
\log (\Sigma^{0 0 \besratio}_{\L\L})^{-2}= &-2 i \tilde{\varPhi}^{{\rm def}_{12}}_{\L\L}(x^{\pm 0}_{\L1}, \frac{1}{x^{\mp k}_{\R2}}) +2 i \tilde{\varPhi}^{\hl, {\rm def}_{12}}_{\L\L}(x^{\pm 0}_{\L1}, \frac{1}{x^{\mp k}_{\R2}})\,.
\eal
Now both in BES and HL the points $x^{+0}_{\L1}$ and $x^{-0}_{\L1}$ are on the same side of the contour (the same applies to $1/x^{\mp k}_{\R2}$). Therefore, we can compute the near BMN limit of the even dressing factor using the mixed derivatives of the phase.

\paragraph{The massless S-matrix in the near-BMN limit.}

Formula~\eqref{eq:afs_mixed_derivative_generic} was evaluated in~\cite{Frolov:2025uwz} for undeformed contours; however, as already mentioned, the only relevant thing is on what side of the contour the points $x_1$ and $x_2$ are located. In this case, both the first and second integration contours have been deformed so that the points $x_1$ and $x_2$ are always above the contours. Then we can apply the formula in~\eqref{eq:afs_mixed_derivative_generic} assuming $\Im(x_1)>0$ and $\Im(x_2)>0$. The actual point locations can indeed be reached by evaluating AFS for $\Im(x_1)>0$ and $\Im(x_2)>0$, and then continuing the solution to the location we want.

Since at the leading order in the large tension limit we have (see~\eqref{eq:app_BMN_massless_massless_string_x1lim} and~\eqref{eq:app_BMN_massless_massless_string_x2lim})
$$
u'_{\L}(x_1)=u'_{\L}(\xi)=0
$$
and
$$
u'_{\L}(x_2)=u'_{\L}(-\frac{1}{\xi})=u'_{\R}(-\xi)=0\,,
$$
then we get the following result for the second-order derivative of AFS
\bal
\label{eq:afs_mixed_derivative_generic3}
{\stackrel{\prime\prime}{\tPhi}}{}_{\L\L}^{\afs}(\xi,-\frac{1}{\xi})=\frac{\xi+\frac{1}{\xi}}{4}=\frac{1}{2 \sqrt{1-q^2}} \,.
\eal
Performing the large $h$ expansion on formula~\eqref{eq:app_theta_bes_massless_massless_bmn_contdef2_step3}, we obtain
\bal
\tilde{\theta}^{\bes}_{\L\L}(x^{\pm 0}_{\L1}, \frac{1}{x^{\mp k}_{\R2}})=h\tilde{\varPhi}^{\afs, {\rm def}_{12}}_{\L\L}(x^{\pm 0}_{\L1}, \frac{1}{x^{\mp k}_{\R2}}) + \tilde{\varPhi}^{\hl, {\rm def}_{12}}_{\L\L}(x^{\pm 0}_{\L1}, \frac{1}{x^{\mp k}_{\R2}})+ \dots \,.
\eal
The points are on the same side of the contour and do not trap the contour in the large $h$ limit. Then we can use the mixed derivative of the phase and obtain
\bal
\label{eq:app_tilde_BES_near_BMN2}
\tilde{\theta}^{\bes}_{\L\L}(x^{\pm 0}_{\L1}, \frac{1}{x^{\mp k}_{\R2}}) &= \frac{4}{T^2} \left( \frac{i p_1 \xi}{2} \right) \left( -\frac{i p_2}{2 \xi} \right) \, h {\stackrel{\prime\prime}{\tPhi}}{}_{\L\L}^{\afs}(\xi,- 1/\xi)+ \tilde{\varPhi}^{\hl, {\rm def}_{12}}_{\L\L}(x^{\pm 0}_{\L1}, \frac{1}{x^{\mp k}_{\R2}}) + \mathcal{O}(1/T^3)\\
&= \frac{p_1 p_2}{2T}+ \tilde{\varPhi}^{\hl, {\rm def}_{12}}_{\L\L}(x^{\pm 0}_{\L1}, \frac{1}{x^{\mp k}_{\R2}}) + \mathcal{O}(1/T^3)\,.
\eal
The result is the same as the one obtained in~\eqref{eq:app_AFS_massless_massless_mirror_region} for the mirror theory, after continuing 
$$
p_1 \to i \tilde{\omega}_{\L1}=i \tilde{p}_1\,, \quad p_2 \to i \tilde{\omega}_{\L2}=-i \tilde{p}_2 \,.
$$ 
We conclude that performing the BMN limit in the mirror region and continuing the result to the string region is equivalent to first continuing to the string region and performing the BMN limit there.

Let us finally compute the S-matrix elements for the scattering of massless bosons $T^{\dot{1} 1}$ with opposite velocities in the string region. 
Continuing~\eqref{eq:STT_mir_mir_for_comparison_near_BMN} to the string region (where the $\g$ variables are continued as in~\eqref{eq:app_massless_string_BMN_g_cont}) we end up with
\bal
S^{00}_{T^{1 \dot{1}} T^{1 \dot{1}}} (u_1, u_2)=& {\sqrt{\a_{\L1}^{+0}} \ov \sqrt{\a_{\L1}^{-0}}} \,  {\sqrt{\a_{\R2}^{+k}} \ov \sqrt{\a_{\R2}^{-k}}}  \ \sqrt{\frac{x^{+k}_{\R2}}{x^{-k}_{\R2}}} \, \sqrt{\frac{x^{+0}_{\L1}}{x^{-0}_{\L1}}} \, \frac{1-\tx^{-0}_{\L1} \tx^{-k}_{\R2}}{1-\tx^{+0}_{\L1} \tx^{+k}_{\R2}}  \\
&\times \frac{R^2(\tg^{-0 +k}_{\L \R} + i \pi) R^2(\tg^{+0 -k}_{\L \R} - i \pi)}{R^2(\tg^{-0 -k}_{\L \R} + i \pi) R^2(\tg^{+0 +k}_{\L \R} -i \pi) } \,(\Sigma_{\L\L}^{00 \, \besratio} )^{-2} \,.
\eal
The even part of the dressing factor is obtained by substituting~\eqref{eq:app_tilde_BES_near_BMN2} in~\eqref{eq:app_difference_string_even_for_BMN2}:
\bal
\label{eq:app_even_dressing_massless_string_BMN}
\log (\Sigma^{0 0 \besratio}_{\L\L})^{-2}= -i \frac{p_1 p_2}{T}+ \mathcal{O}(1/T^3) \,.
\eal
The limit of the odd part of the dressing factor is
\bal
\label{eq:app_barnes_massless_string_BMN}
\log  \frac{R^2(\tg^{-0 +k}_{\L \R} + i \pi) R^2(\tg^{+0 -k}_{\L \R} - i \pi)}{R^2(\tg^{-0 -k}_{\L \R} + i \pi) R^2(\tg^{+0 +k}_{\L \R} -i \pi) } \to \frac{i q}{3 T} (p_1 + p_2) + i \frac{\pi}{2} + \mathcal{O}(1/T^2)\,.
\eal
The remaining rational terms instead return
\bal
\label{eq:app_rat1_massless_string_BMN}
\log \left( {\sqrt{\a_{\L1}^{+0}} \ov \sqrt{\a_{\L1}^{-0}}} \,  {\sqrt{\a_{\R2}^{+k}} \ov \sqrt{\a_{\R2}^{-k}}} \, \sqrt{\frac{x^{+k}_{\R2}}{x^{-k}_{\R2}}} \, \sqrt{\frac{x^{+0}_{\L1}}{x^{-0}_{\L1}}} \, \frac{1-\tx^{-0}_{\L1} \tx^{-k}_{\R2}}{1-\tx^{+0}_{\L1} \tx^{+k}_{\R2}} \right) \to -\frac{i q}{3 T} (p_1 + p_2) - i \pi +\mathcal{O}(1/T^2) \,.
\eal
Combining~\eqref{eq:app_even_dressing_massless_string_BMN}, \eqref{eq:app_barnes_massless_string_BMN}, and~\eqref{eq:app_rat1_massless_string_BMN} all the $q$ dependence disappears and we end up with
\bal
S^{00}_{T^{\dot{1} 1} T^{\dot{1} 1}}(p_1, p_2) \to e^{-i \frac{\pi}{2}} \left(1-\frac{i}{T} p_{1} p_{2} \right) + \mathcal{O}(1/T^2) \,, \quad p_1>0> p_2\,,
\eal
This is the same expression as in~\eqref{eq:app_final_STT_massless_mirror_BMN} if we continue $p_1 \to i\tilde{\omega}_1= i \tp_1$ and $p_2 \to i\tilde{\omega}_2= -i \tp_2$. The same consideration applies term by term to all the different pieces composing the S-matrix element.
As for the mirror theory (see the discussion below equation~\eqref{eq:app_final_STT_massless_mirror_BMN}) the factor $-i$ comes from having non-trivial exchange relations between creation operators in the free-theory limit. This factor connects the ZF S-matrix to the physical one, which in the near BMN limit is therefore given by
\bal
&\mathbb{S}_{T^{\dot{1} 1} T^{\dot{1} 1}}(p_1, p_2) \to 1 -\frac{i}{T} p_1 p_2 \,, \quad p_1>0> p_2\,.
\eal 
This result agrees with perturbative computations at the tree level in the gauge $a=0$ (see, e.g., equation (2.52) in~\cite{Baglioni:2023zsf}).

\section{Relativistic limit}
\label{app:rel_limit}

In this appendix, we show that in the relativistic limit of~\cite{Frolov:2023lwd}, our proposal for the massless-massless S-matrix of equal chirality reduces to~\eqref{eq:app_dressing_rel_paper}.

\subsection{The limit of the kinematics and crossing}

If we consider $h \ll1$ and expand the momenta and energies of chiral massless particles around the minimum of the dispersion relation with fluctuations of order $h$ (as done in~\cite{Frolov:2023lwd}), we obtain the following relations
\bal
p_i=\frac{2 \pi}{k} h e^{\theta_i}\,, \qquad E_i=h e^{\theta_i}\,, \qquad h\ll1\,, \ k \in \mathbb{N}\,.
\eal
The Zhukovsky and $u$ variables become
\bal
\label{eq:app_rel_lim_Zhk_var}
&x^{\pm 0}_{\L i}= \xbr - \frac{h \pi}{3 k} e^{2 \theta_i} \pm i e^{\theta_i}  +\mathcal{O}(h^2)\,,\\
&u_i \equiv u_\L(x^{\pm 0}_{\L i})=\nu - \frac{h \pi}{2 k} e^{2 \theta_i} + \mathcal{O}(h^2) \,,
\eal
where
\bal
\label{eq:app_rel_lim_xi}
&\xi = \frac{k}{\pi h} + \frac{\pi h}{k}+ \mathcal{O}(h^2) \,,\\
&\nu=u_\L(\xbr)=- \frac{k}{\pi h} \log \frac{k}{\pi h e} + \frac{h \pi}{k} + \mathcal{O}(h^2) \,.
\eal
The $\g$ variables become
\bal
\label{eq:app_rel_g_variables}
\g^{\pm 0}_{\L i}= - 2 \log \frac{k}{\pi h}+\theta_i \pm \frac{i \pi}{2} + \mathcal{O}(h) \,.
\eal
Under crossing in the full model the $\g_\L$ variables transform as follows
\bal
\g^{+0}_{\L i} &\to \g_{\R i}^{+0} + i\pi=\g_{\L i}^{-0} +2 i\pi\,,\\
\g^{-0}_{\L i} &\to \g_{\R i}^{-0} + i\pi=\g_{\L i}^{+0}\,.
\eal
Since the relativistic limit of the $\g$ variables is given by~\eqref{eq:app_rel_g_variables} then
the transformation above corresponds to send $\theta_i \to \theta_i + i\pi$.
This is exactly the type of crossing equation used in~\cite{Frolov:2023lwd} and we should expect that the relativistic limit of the massless-massless S-matrices of this paper should match the S-matrices found in~\cite{Frolov:2023lwd}.
In the same way from~\eqref{eq:app_rel_lim_Zhk_var} we see that the transformation $\theta_i \to \theta_i + i \pi$ transform $x^{\pm 0}_{\L i}\to x^{\mp0}_{\L i}$, as expected from crossing in the full theory. 
Once we are sure that the crossing symmetry used in the limit is compatible with the limit of the crossing paths used in the full model, we can match the solution for the dressing factor.

\subsection{The relativistic limit of BES}

The continuation of BES to the positive-positive momentum branch of the string theory is
\bal
\label{eq:app_thetaBES_split_rel_limit}
\tilde{\theta}_{\L \L}^{\bes}(x^{\pm 0}_{\L 1}, x^{\pm 0}_{\L 2})&=\tilde{\varPhi}_{\L \L}(x^{\pm 0}_{\L 1},x^{\pm 0}_{\L 2})-\tPsi_{\L}(u_1, x^{+0}_{\L2})+\tPsi_{\L}(u_1, x^{-0}_{\L2})\\
&+\tPsi_\L(u_2, x^{+0}_{\L1}) -\tPsi_\L(u_2, x^{-0}_{\L1})+ K^\bes \left(u_1-u_2\right) \,.
\eal
From the second line in~\eqref{eq:app_rel_lim_Zhk_var} in the limit we have
\bal
K^\bes \left(u_1-u_2\right) \to 0 \,.
\eal
Moreover, from~\eqref{eq:app_rel_lim_Zhk_var} we notice that 
\bal
\label{eq:app_rel_limit_generation_delta}
\frac{1}{\tx - x^{- 0}_{\L1}}- \frac{1}{\tx - x^{+ 0}_{\L1}}&= -\frac{2 \pi i}{\xbr} \delta(1- \frac{\tx}{\xbr}- \frac{h \pi}{3 k \xbr} e^{2 \theta_1} )\\
&=-2 \pi i \delta \left(\tx- (\xbr - \frac{\pi h}{3 k} e^{2 \theta_1}) \right)\,.
\eal
Then we obtain
\bal
\tPhi^{\a \a }_{\L \L}(x^{+ 0}_{\L 1},x_2) - \tPhi^{\a \a }_{\L \L}(x^{- 0}_{\L 1}, x_2) &=  \lint_{\pa \cR_\a} \frac{{\rm d} w_2}{2 \pi i} \, \frac{1}{w_2 - x_2} i \log \frac{\Gamma \left(1+\frac{i h}{2} (u_\L(\xbr) - u_\L(w_2) )  \right)}{\Gamma \left(1-\frac{i h}{2} (u_\L(\xbr) - u_\L(w_2) )  \right)}\\
&=-\tPsi^\a_\L(\nu, x_2)\,,
\eal
(we recall from the definition in~\eqref{eq:app_tPhidef} that $\a$ can either be $+$ or $-$ according with the integration contours used) from which we get
\bal
\tilde{\varPhi}_{\L \L}(x^{\pm 0}_{\L 1},x^{\pm 0}_{\L 2})=-\tPsi_\L(\nu, x^{+0}_{\L2}) + \tPsi_\L(\nu, x^{-0}_{\L2}) \,.
\eal
Using again the relation in~\eqref{eq:app_rel_limit_generation_delta} the expression above becomes
\bal
\tilde{\varPhi}_{\L \L}(x^{\pm 0}_{\L 1},x^{\pm 0}_{\L 2})&=-\tPsi_\L(\nu, x^{+0}_{\L2}) + \tPsi_\L(\nu, x^{-0}_{\L2})= -i \log \frac{\Gamma \left(1+\frac{i h}{2} (\nu - u_\L(\xbr) )  \right)}{\Gamma \left(1-\frac{i h}{2} (\nu - u_\L(\xbr) )  \right)}=0\,.
\eal
The other differences of $\tPsi$ functions in~\eqref{eq:app_thetaBES_split_rel_limit}
also give a vanishing result in the relativistic limit and in the end we obtain
\bal
\label{eq:app_rel_limit_BES}
\tilde{\theta}_{\L \L}^{\bes}(x^{\pm 0}_{\L 1}, x^{\pm 0}_{\L 2})=0 \implies \left( \Sigma^{00 \bes}_{\L\L}(u_1 ,u_2) \right)^2 \to 1\,.
\eal
This agrees with the fact that BES satisfies homogeneous crossing equations, and in the relativistic limit, we should expect at most a simple CDD factor.

\subsection{The relativistic limit of HL}

The simplest way to evaluate the relativistic limit of the HL phase is to start with the expression \eqref{eq:HLvsBarnesstr}, corresponding to the HL phase in the string region written in terms of Barnes Gamma functions. Replacing into such a function the limit of the $\g$ variables~\eqref{eq:app_rel_g_variables} we obtain
\bal
\label{eq:app_rel_lim_HLfinalLimit}
\Sigma^{00, \hl}_{\L\L}(u_1,u_2)^{2} & \to - {R^4(\theta_{12}-i \pi) R^4(\theta_{12}+i \pi) \ov R^4(\theta_{12})}\,.
\eal
Since the result~\eqref{eq:HLvsBarnesstr} was only checked numerically to agree with the integral representation of HL, it is instructive to see how the relativistic limit of HL can also be generated by its integral representation. This provides a further confirmation that for massless particles the expression in~\eqref{eq:HLvsBarnesstr} is indeed equivalent to the integral representation of HL. In the remaining of this subsection we will check this fact explicitly.

After continuation to the positive momentum region of the string theory the HL phase is given by
\bal
2\tilde{\theta}^\hl_{\L\L}(x^{\pm 0}_{\L1}, x^{\pm 0}_{\L2})&= 2\tilde{\varPhi}^\hl_{\L\L}(x^{\pm 0}_{\L1}, x^{\pm 0}_{\L2})+ \frac{1}{i} \log \frac{x^{+ 0}_{\L1}}{x^{- 0}_{\L1}} \frac{x^{- 0}_{\L2}}{x^{+ 0}_{\L2}} + \pi\,.
\eal
Since in the relativistic limit it holds
\bal
\frac{x^{+ 0}_{\L1}}{x^{- 0}_{\L1}} \frac{x^{- 0}_{\L2}}{x^{+ 0}_{\L2}} \to 1
\eal
then we obtain
\bal
\label{eq:app_rel_limit_HL1}
2\tilde{\theta}^\hl_{\L\L}(x^{\pm 0}_{\L1}, x^{\pm 0}_{\L2}) \to 2\tilde{\varPhi}^\hl_{\L\L}(x^{\pm 0}_{\L1}, x^{\pm 0}_{\L2})+ \pi\,.
\eal
Let us now perform the limit of the $\tilde{\varPhi}^\hl_{\L\L}$ function. The computation of this contribution cannot be performed on the same route of appendix K in~\cite{Frolov:2025uwz} since for massless particles the points $x_1$ and $x_2$ become singular in the limit.
Since in the relativistic limit 
\bal
\label{eq:app_rel_limit_Rexpm0}
\Re(x^{\pm 0}_{\L1})=\Re(x^{\pm 0}_{\L2}) \simeq \frac{k}{\pi h} \gg 1
\eal
then we can approximate the HL $\tPhi$ functions as follows
\bal
&\tPhi_{\L \L}^{--, \hl}(x^{n_1 0}_{\L1}, x^{n_2 0}_{\L2}) = \tPhi_{\L \L}^{++, \hl}(x^{n_1 0}_{\L1}, x^{n_2 0}_{\L2}) =\tPhi_{\L \L}^{\hl}(x^{n_1 0}_{\L1}, x^{n_2 0}_{\L2})=\\
&-\lint^{+\infty}_{0} \frac{{\rm d} w_1}{2 \pi i} \, \lint^{+\infty}_{0} \frac{{\rm d} w_2}{2 \pi i} \, \frac{1}{w_1 - x^{n_1 0}_{\L1}} \, \frac{1}{w_2 - x^{n_2 0}_{\L2}}\\
& \frac{\log \left(\eps -i (u_\L (w_1)-u_\L (w_2)) \right) -\log \left(\eps +i (u_\L (w_1)-u_\L (w_2)) \right) }{2i}\,,
\eal
with $n_1 = \pm$ and $n_2= \pm$. Indeed the integrand is suppressed on the negative real axis because of~\eqref{eq:app_rel_limit_Rexpm0}. Then we can apply the following change of integration variables
\bal
w_1=\frac{k}{\pi h} + \hat{w}_1\,, \qquad w_2=\frac{k}{\pi h} + \hat{w}_2\,,
\eal
from which in the limit $h\ll 1$ we obtain
\bal
u_\L(w_i)=-\frac{k}{\pi h} \log \frac{k }{\pi h e}+ \frac{\pi h}{k}+\frac{\pi h}{2 k} \hat{w}^2_i\ + \mathcal{O}(h^2), \quad i=1,2\,.
\eal
Then in the limit $h\ll 1 $ we can write
\bal
\tPhi_{\L \L}^{\hl}(x^{n_1 0}_{\L1}, x^{n_2 0}_{\L2})=&-\lint^{+\infty}_{-\infty} \frac{{\rm d} \hat{w}_1}{2 \pi i} \, \lint^{+\infty}_{-\infty} \frac{{\rm d} \hat{w}_2}{2 \pi i} \, \frac{1}{\hat{w}_1 - n_1 i e^{\theta_1}} \, \frac{1}{\hat{w}_2 - n_2 i e^{\theta_2}}\\
&  \qquad \times\frac{\log \left(\eps -i \frac{\pi h}{2 k} (\hat{w}^2_{1}- \hat{w}^2_2) \right) -\log \left(\eps +i \frac{\pi h}{2 k} (\hat{w}^2_{1}- \hat{w}^2_2) \right) }{2i}\,.
\eal
Sending the regulator $\eps \to 0$ we get
\bal
&\tPhi_{\L \L}^{\hl}(x^{n_1 0}_{\L1}, x^{n_2 0}_{\L2})=\frac{\pi}{2} \lint^{+\infty}_{-\infty} \frac{{\rm d} \hat{w}_1}{2 \pi i} \, \lint^{+\infty}_{-\infty} \frac{{\rm d} \hat{w}_2}{2 \pi i} \, \frac{1}{\hat{w}_1 - n_1 i e^{\theta_1}} \, \frac{1}{\hat{w}_2 - n_2 i e^{\theta_2}}    \text{sgn} (\hat{w}^2_{1} -\hat{w}^2_{2}) \\
&=\frac{\pi}{2} \lint^{+\infty}_{-\infty} \frac{{\rm d} \hat{w}_2}{2 \pi i}  \, \frac{1}{\hat{w}_2 - n_2 i e^{\theta_2}} \, \left( \lint_{|\hat{w}_1|>|\hat{w}_2|} \frac{{\rm d} \hat{w}_1}{2 \pi i} \, \frac{1}{\hat{w}_1 - n_1 i e^{\theta_1}} - \lint_{|\hat{w}_1|<|\hat{w}_2|}\frac{{\rm d} \hat{w}_1}{2 \pi i} \, \frac{1}{\hat{w}_1 - n_1 i e^{\theta_1}}    \right) \,.
\eal
It is simpler to perform the integration after deriving wrt $\theta_1$. By doing so we get
\bal
&\frac{d}{d \theta_1}\tPhi_{\L \L}^{\hl}(x^{n_1 0}_{\L1}, x^{n_2 0}_{\L2})=\frac{i \pi n_1}{2 (2 \pi i)^2} e^{\theta_1}\lint^{+\infty}_{-\infty} {\rm d} \hat{w}_2\, \frac{1}{\hat{w}_2 - n_2 i e^{\theta_2}}\\
&\qquad\times  \, \left( \lint_{|\hat{w}_1|>|\hat{w}_2|} {\rm d} \hat{w}_1 \, \frac{1}{(\hat{w}_1 - n_1 i e^{\theta_1})^2} - \lint_{|\hat{w}_1|<|\hat{w}_2|} {\rm d} \hat{w}_1 \, \frac{1}{(\hat{w}_1 - n_1 i e^{\theta_1})^2}    \right)\,.
\eal
Performing the integral wrt both $w_1$ and $w_2$ we obtain
\bal
\frac{d}{d \theta_1}\tPhi_{\L \L}^{\hl}(x^{n_1 0}_{\L1}, x^{n_2 0}_{\L2})=& +\frac{n_1}{8 \pi i} \frac{1}{\sinh \theta_{12}} \left[ \pi \left(e^{\theta_{12}} -1 \right) + 2 i n_2 \theta_{12} \right] \\
&-\frac{n_1}{8 \pi i} \frac{1}{\sinh \theta_{12}} \left[ \pi \left(e^{\theta_{12}} -1 \right) - 2 i n_2 \theta_{12} \right] =\frac{n_1 n_2}{2 \pi} \frac{\theta_{12}}{\sinh{\theta_{12}}}\,.
\eal
From this relation we can read
\bal
\label{eq:app_rel_limit_tPhiHL}
\frac{d}{d \theta_1} \tilde{\varPhi}_{\L \L}^{\hl}(x^{\pm 0}_{\L1}, x^{\pm 0}_{\L2})=\frac{2}{\pi} \frac{\theta_{12}}{\sinh{\theta_{12}}}=\frac{1}{i} \frac{d}{d\theta_{12}} \log \frac{R^2(\theta_{12}-i \pi) \, R^2(\theta_{12}+i \pi) }{R^4(\theta_{12})}\,.
\eal
From~\eqref{eq:app_rel_limit_tPhiHL} and~\eqref{eq:app_rel_limit_HL1} we see that the relativistic limit of HL is~\eqref{eq:app_rel_lim_HLfinalLimit}.

\subsection{The limit of the full S-matrix element}

The massless S-matrix for particles with positive momenta in the string region is given by (see equation~\eqref{eq:S_massless_massless_pos_pos_final})
\begin{equation}
\label{eq:app_rel_limit_starting_S_matrix}
    \begin{aligned}
    S_{\chi^{\dot{1}} \chi^{\dot{1}}} (u_1, u_2)=+
    {\sqrt{\a_{\L1}^{-0}} \ov \sqrt{\a_{\L1}^{+0}}} \,  {\sqrt{\a_{\L2}^{+0}} \ov \sqrt{\a_{\L2}^{-0}}} \,    \left(\Sigma^{\barnes}_{\L \L} (\g^{\pm0}_{\L1}, \g^{\pm0}_{ \L 2}) \right)^{-2} \left(\Sigma^{00, \besratio}_{\L\L}(u_1,u_2)\right)^{-2}  \, .
    \end{aligned}
\end{equation}
In the limit it is easy to show that
\bal
\label{eq:app_rel_limit_norm}
 {\sqrt{\a_{\L1}^{-0}} \ov \sqrt{\a_{\L1}^{+0}}} \,  {\sqrt{\a_{\L2}^{+0}} \ov \sqrt{\a_{\L2}^{-0}}}  \to 1\,.
\eal
and
\bal
\label{eq:app_rel_limit_odd}
\left(\Sigma^{\barnes}_{\L \L} (\g^{\pm0}_{\L1}, \g^{\pm0}_{ \L 2}) \right)^{-2} =\frac{R^2(\g^{-0-}_{\L \L}) R^2(\g^{+0+}_{\L \L})}{R^2(\g^{-0+}_{\L \L}) R^2(\g^{+0-}_{\L \L}) } \to \frac{R^4(\theta_{12})}{R^2(\theta_{12}- i \pi) R^2(\theta_{12}+ i \pi) } \,.
\eal
Substituting into~\eqref{eq:app_rel_limit_starting_S_matrix} the expressions above together with the limits of BES and HL (see~\eqref{eq:app_rel_limit_BES} and~\eqref{eq:app_rel_lim_HLfinalLimit}) we obtain the result in~\eqref{eq:app_dressing_rel_paper}.

\section{New vs old auxiliary TBA kernels}
\label{app:aux_kernels}

In this appendix we discuss the relation between 
the auxiliary S matrices and kernels we are using here and the ones used in the RR case in 
\cite{Arutyunov:2009ur,Frolov:2021bwp}. 

A natural generalisation of the S matrices in 
\cite{Arutyunov:2009ur,Frolov:2021bwp} is given by
\bal
S_{-}^{Q y}(u, v)&=\frac{\tx_\L\left(u-i \frac{Q}{h}\right)-\tx_\L(v)}{\tx_\L\left(u+i \frac{Q}{h}\right)-\tx_\L(v)} \,
\frac{\sqrt{\tx_\L\left(u+i \frac{Q}{h}\right)}}{\sqrt{\tx_\L\left(u-i \frac{Q}{h}\right)}}\,,\qquad S_{-}^{yQ}(u, v)={1\ov S_{-}^{Q y}(v, u)}\,,
\eal
\bal
S_{+}^{Q y}(u, v)&=\frac{\tx_\L\left(u-i \frac{Q}{h}\right)-\frac{1}{\tx_\R(v)}}{\tx_\L\left(u+i \frac{Q}{h}\right)-\frac{1}{\tx_\R(v)}}\, \frac{\sqrt{\tx_\L\left(u+i \frac{Q}{h}\right)}}{\sqrt{\tx_\L\left(u-i \frac{Q}{h}\right)}}\,,\qquad  S_{+}^{yQ }(u, v)= { S_{+}^{Q y}(v, u)}\,,
\eal
\bal
S_{\pm}^{0 y}(u, v)&=\lim_{Q\to 0}S_{\pm}^{Q y}(u, v)\,,
\eal
\bal
\overline{S}_{-}^{Q y}(u, v)&=\frac{\tx_\R\left(u-i \frac{Q}{h}\right)-\tx_\R(v)}{\tx_\R\left(u+i \frac{Q}{h}\right)-\tx_\R(v)} \,
\frac{\sqrt{\tx_\R\left(u+i \frac{Q}{h}\right)}}{\sqrt{\tx_\R\left(u-i \frac{Q}{h}\right)}}\,,\qquad \overline{S}_{-}^{yQ}(u, v)={1\ov \overline{S}_{-}^{Q y}(v, u)}\,,
\eal
\bal
\overline{S}_{+}^{Q y}(u, v)&=\frac{\tx_\R\left(u-i \frac{Q}{h}\right)-\frac{1}{\tx_\L(v)}}{\tx_\R\left(u+i \frac{Q}{h}\right)-\frac{1}{\tx_\L(v)}}\, \frac{\sqrt{\tx_\R\left(u+i \frac{Q}{h}\right)}}{\sqrt{\tx_\R\left(u-i \frac{Q}{h}\right)}}\,,\qquad  \overline{S}_{+}^{yQ }(u, v)= { \overline{S}_{+}^{Q y}(v, u)}
\,.
\eal
In the RR case 
$\overline{S}_{\pm}^{Q y}={S}_{\pm}^{Q y}$ and $S_+^{y0}=S_-^{y0}$ but in the mixed-flux case we have to distinguish between them.
These S matrices and associated kernels are related to the new ones 
\eqref{eq:new_aux_kernels} as follows
\bal
S_{-}^{Q y}(u, v)&={1\ov S_{\L\L}^{Q y}(u, v)}\,,\qquad &&K_{-}^{Q y}(u, v)=-K_{\L\L}^{Q y}(u, v)\,,
\\
S_{-}^{yQ}(u, v)&={S_{\L\L}^{yQ}(u, v)}\,,\qquad &&K_{-}^{yQ}(u, v)=+K_{\L\L}^{yQ}(u, v)\,,
\\
S_{+}^{Q y}(u, v)&={1\ov S_{\L\R}^{Q y}(u, v)}\,,\qquad &&K_{+}^{Q y}(u, v)=-K_{\L\R}^{Q y}(u, v)\,,
\\
S_{+}^{yQ}(u, v)&={S_{\R\L}^{yQ}(u, v)}\,,\qquad &&K_{+}^{yQ}(u, v)=+K_{\R\L}^{yQ}(u, v)\,,
\\
\overline{S}_{-}^{Q y}(u, v)&={S_{\R\R}^{Q y}(u, v)}\,,\qquad &&\overline{K}_{-}^{Q y}(u, v)=+K_{\R\R}^{Q y}(u, v)\,,
\\
\overline{S}_{-}^{yQ}(u, v)&={S_{\R\R}^{yQ}(u, v)}\,,\qquad &&\overline{K}_{-}^{yQ}(u, v)=+K_{\R\R}^{yQ}(u, v)\,,
\\
\overline{S}_{+}^{Q y}(u, v)&={S_{\R\L}^{Q y}(u, v)}\,,\qquad &&\overline{K}_{+}^{Q y}(u, v)=+K_{\R\L}^{Q y}(u, v)\,,
\\
\overline{S}_{+}^{yQ}(u, v)&={S_{\L\R}^{yQ}(u, v)}\,,\qquad &&\overline{K}_{+}^{yQ}(u, v)=+K_{\L\R}^{yQ}(u, v)\,.
\eal
In terms of these kernels the TBA equations take the form
\begin{equation}\label{TBA_L2}
	\begin{aligned}
		\log {Y}_Q & =-L \widetilde{\mathcal{E}}_Q+\log \left(1+{Y}_{Q^{\prime}}\right) \star K_{ Y Y}^{Q^{\prime} Q}  +\log \left(1+ \overline{Y}_{Q^{\prime}}\right) \star {K}_{\bar Z Y}^{Q^{\prime} Q} + \log \left(1+Y_0\right) \check{\star} {K}^{0 Q}_{\chi  Y}\\
		& +\sum_{\alpha=1,2} \log \left(1-\frac{e^{i \mu_\alpha}}{Y_{-}^{(\alpha)}}\right) \hat{\star} {K}_{-}^{y Q}+\sum_{\alpha=1,2} \log \left(1-\frac{e^{i \mu_\alpha}}{Y_{+}^{(\alpha)}}\right) \hat{\star} {K}_{+}^{y Q} \,,
	\end{aligned}
\end{equation}
\begin{equation}\label{TBA_R2}
	\begin{aligned}
		\log  \overline{Y}_Q & =-L \widetilde{\mathcal{E}}_Q + \log \left(1+ \overline{Y}_{Q^{\prime}}\right) \star K_{\bar Z\bar Z}^{Q^{\prime} Q}  +\log \left(1+{Y}_{Q^{\prime}}\right) \star {K}_{ Y  \bar Z}^{Q^{\prime} Q}  +\log \left(1+Y_0\right) \check{\star} K^{0 Q}_{\chi  \bar{Z}} \\
		& +\sum_{\alpha=1,2} \log \left(1-\frac{e^{i \mu_\alpha}}{Y_{-}^{(\alpha)}}\right) \hat{\star} \overline{K}_{+}^{y Q}+\sum_{\alpha=1,2} \log \left(1-\frac{e^{i \mu_\alpha}}{Y_{+}^{(\alpha)}}\right) \hat{\star} \overline{K}_{-}^{y Q}\,,
	\end{aligned}
\end{equation}
\begin{equation}\label{TBA_02}
	\begin{aligned}
		\log Y_0^{(\dot{\alpha})} & =-L \widetilde{\mathcal{E}}_0 + \log \left(1+Y_0\right) \check{\star} K^{00}_{\chi\chi}  +\log \left(1+Y_Q\right) \star K^{Q 0}_{Y\chi}+\log \left(1+\overline{Y}_Q\right) \star {K}^{Q 0}_{\bar Z\chi} \\
		& +\sum_{\alpha=1,2} \log \left(1-\frac{e^{i \mu_\alpha}}{Y_{-}^{(\alpha)}}\right) \hat{\star} K_{-}^{y 0}+\sum_{\alpha=1,2} \log \left(1-\frac{e^{i \mu_\alpha}}{Y_{+}^{(\alpha)}}\right) \hat{\star} K_{+}^{y 0}\,,
	\end{aligned}
\end{equation}
\begin{equation}\label{TBA_ym2}
\begin{aligned}
	-\log Y_{-}^{(\alpha)}&=-\log \left(1+Y_Q\right) \star K_{-}^{Q y}+\log \left(1+\overline{Y}_Q\right) \star \overline{K}_{+}^{Q y}- \log \left( 1+Y_0\right) \check{\star} K_{-}^{0 y} \,,
\end{aligned}
\end{equation}
\begin{equation}\label{TBA_yp2}
\begin{aligned}
	-\log Y_{+}^{(\alpha)} &= - \log \left(1+Y_Q\right) \star K_{+}^{Q y} + \log \left(1+\overline{Y}_Q\right) \star \overline{K}_{-}^{Q y}- \log \left( 1+Y_0\right) \check{\star} K_{+}^{0 y} \,.
\end{aligned}
\end{equation}
To compare the TBA equations with those in \cite{Frolov:2021bwp} one should do the following exchanges
\bal
Y_Q\ \leftrightarrow\ \overline{Y}_Q\,,\qquad Y_+^{(\alpha)}\ \leftrightarrow\ Y_-^{(\alpha)}\,,
\eal
and use the relations between the S matrices and kernels in the RR case mentioned above.

\bibliographystyle{JHEP}
\bibliography{refs}
\end{document}